\documentclass[a4paper,11pt]{article}
\pdfoutput=1
\usepackage[T1]{fontenc}
\usepackage{xcolor}
\usepackage{jheppub}

\usepackage[normalem]{ulem}

\usepackage{bm}
\usepackage{mathtools}
\usepackage{epstopdf}
\usepackage{setspace}
\usepackage{dsfont}
\usepackage{latexsym}
\usepackage{psfrag}
\usepackage{booktabs}
\usepackage{color}
\usepackage{subfig}
\usepackage{tensor}
\usepackage{lscape}
\usepackage{tikz}
\usepackage[inline]{enumitem}

\usepackage{cleveref}

\usetikzlibrary{shapes}
\usetikzlibrary{patterns}
\usetikzlibrary{matrix,arrows.meta} 				% For commutative diagram
\usetikzlibrary{positioning}				% For "above of=" commands
\usetikzlibrary{calc,through}				% For coordinates
\usetikzlibrary{decorations.pathreplacing}  % For curly braces
\usetikzlibrary{decorations.pathmorphing}	% For Feynman Diagrams

\pgfmathsetmacro{\h}{1.5}
\pgfmathsetmacro{\hp}{2}
\pgfmathsetmacro{\w}{1.5}
\pgfmathsetmacro{\blb}{0.25}
\pgfmathsetmacro{\ar}{0.2}

\usetikzlibrary{decorations.markings}
\tikzset{
    fermion/.style={draw=black, postaction={decorate},
        decoration={markings,mark=at position .55 with {\arrow[draw=black]{>}}}},
    fermionbar/.style={draw=black, postaction={decorate},
        decoration={markings,mark=at position .55 with {\arrow[draw=black]{<}}}},
    graviton/.style={decorate, draw=black,
        decoration={coil,aspect=0,amplitude=1.5pt, segment length=5pt}}
}

\tikzstyle{block} = [draw, rectangle, 
    minimum height=3em, minimum width=6earticlem]

\definecolor{darkblue}{rgb}{0.2, 0, 0.8}

\newcommand{\ang}[1]{\langle #1\rangle}

\renewcommand{\r}{\rho}
\newcommand{\s}{\sigma}

\def\be{\begin{equation}}
\def\ee{\end{equation}}
\def\bea{\begin{eqnarray}}
\def\eea{\end{eqnarray}}
\def\ba{\begin{array}}
\def\ea{\end{array}}
\def\bd{\begin{displaymath}}
\def\ed{\end{displaymath}}

\def\Tr{{\rm Tr}}

\def\a{\alpha}
\def\b{\beta}
\def\c{\gamma}

\def\l{\lambda}

\def\r{\rho}                                     %     \varrho
\def\s{\sigma}                                   %     \varsigma

\def\x{\xi}

\def\J{\Psi}

\def\>{\rangle} %right angle
\def\<{\langle} %left angle

\def\J{{\cal J}}

\def\sf{\mathsf{f}}
\def\lm{\ell}
\def\lmb{\pmb{\ell}}
\def\qbd{\pmb{q}}
\def\dd{\textrm{d}}

\def\pb{\bar{p}}
\def\ub{\bar{u}}
\def\mb{\bar{m}}

\def\ie{i0}
\def\hdelta{\hat{\delta}}
\def\eIR{\epsilon_{\text{IR}}}
\def\opO{\hat{\mathcal{O}}}
\def\opS{\hat{S}}
\def\opT{\hat{T}}
\def\opN{\hat{N}}
\def\Mfull{\mathcal{M}}
\def\Mred{M}
\def\Mtot{\mathsf{M}}
\newcommand{\real}[1]{\mathcal{R}_{#1}}
\newcommand{\imaginary}[1]{\mathcal{I}_{#1}}

\DeclareMathOperator{\arccosh}{arccosh}
\DeclareMathOperator{\arcsinh}{arcsinh}

\crefname{figure}{figure}{figures}
\Crefname{figure}{Figure}{Figures}

\title{The Sub-Leading Scattering Waveform from Amplitudes}

\author[a]{Aidan Herderschee,}
\author[b,c]{Radu Roiban}
\author[b,c]{and  Fei Teng}

\affiliation[a]{Leinweber Center for Theoretical Physics, Randall Laboratory of Physics\\ The University of Michigan, Ann Arbor, MI 48109-1040, USA}
\affiliation[b]{Department of Physics, Pennsylvania State University, University Park, PA 16802, USA}
\affiliation[c]{Kavli Institute for Theoretical Physics,
University of California, Santa Barbara, CA 93106, USA}

\emailAdd{aidanh@umich.edu}
\emailAdd{radu@phys.psu.edu}
\emailAdd{fei.teng@psu.edu}

\abstract{
  We compute the next-to-leading order term in the scattering waveform of uncharged black holes in classical general relativity and of half-BPS black holes in $\mathcal{N}=8$ supergravity.
  We propose criteria, generalizing explicit calculations at next-to-leading order, for determining the terms in amplitudes that contribute to local observables.  
  For general relativity, we construct the relevant classical integrand through generalized unitarity in two distinct ways, (1) in a heavy-particle effective theory and (2) in general relativity minimally-coupled to scalar fields. With a suitable prescription for the matter propagator in the former, we find agreement between the two methods, thus demonstrating the absence of interference of quantum and classically-singular contributions.
  The classical $\mathcal{N}=8$ integrand for massive scalar fields is constructed through dimensional reduction of the known five-point one-loop integrand. 
  Our calculation exhibits novel features compared to conservative calculations and inclusive observables, such as the appearance of master integrals with intersecting matter lines and the appearance of a  classical infrared divergence whose absence from classical observables requires a suitable definition of the retarded time. 
  
}

\preprint{LCTP-23-04}

\begin{document} 

\addtocontents{toc}{\protect\setcounter{tocdepth}{2}}

\maketitle

%%%%%%%%%%%%

%%%%%%%%%%%%%%%%%%%%%%%%%%%%%%
%%%%%%%%%%%%%%%%%%%%%%%%%%%%%%
%%%%%%%%%%%%%%%%%%%%%%%%%%%%%%%%%%%%%%%%%%%%%%%%%%%%%%%%%%%%%%%%%%%%%%%%%%%%%%%%%%%%%%%%%%%%%%%%%%%%%%%%%%%%%%%%%%%%%%%%%%%%%%%%%%%%%%%%%%%%%%%%%%%%%%%%%%%%%%%%%%%%%%%%%%%%%%%%%%%%%%%%%%%%%%%%%%%%%%%%%%%%%%%%%%%%%%%%%%%%%%%%%%%%%%%%%%%%%%%%%%%%%%%%%%%%%%%%%%%%%
\section{Introduction}
\label{sec:intro}

Much like accelerated charges in Maxwell's theory emit electromagnetic radiation, accelerated masses in gravitational theories emit gravitational radiation. Increasingly more precise waveform models, yielding the asymptotic space-time metric, play a critical role in the detection and analysis of gravitational wave signals from compact binaries~\cite{LIGOScientific:2021djp}.
Current waveform computations for binaries in bound orbits use effective-one-body methods~\cite{Buonanno:1998gg, Damour:2000we,Buonanno:2000ef,Damour:2001tu,Damour:2014sva} and numerical-relativity approaches~\cite{Pretorius:2005gq,Pretorius:2007nq},  in addition to direct solutions of Einstein's equations sourced by the motion of the bodies in the post-Newtonian approximation~\cite{Blanchet:2013haa,Schafer:2018kuf}, and computations in the effective-field theory approach pioneered by Goldberger and Rothstein Ref.~\cite{Goldberger:2004jt, Goldberger:2006bd, Goldberger:2009qd, Porto:2012as, Blanchet:1994ez}, see \cite{Porto:2016pyg, Levi:2018nxp} for reviews.

The post-Minkowskian (PM) approximation, keeping exact velocity dependence for each power of Newton's constant, is natural for bound binary systems on eccentric orbits and for unbound binaries~\cite{Khalil:2022ylj}.
Inclusive dissipative observables through ${\cal O}(G^3)$ and waveforms to leading order in Newton's constant have been discussed with traditional methods in e.g.~\cite{Kovacs:1stInSeries, Crowley:1977us, Kovacs:1977uw, Kovacs:1978eu, Amati:1990xe, Ciafaloni:2018uwe, Damour:2020tta} and 
amplitude and worldline methods in Ref.~\cite{Kalin:2020mvi, Kalin:2020fhe, Dlapa:2021npj, Herrmann:2021tct, Riva:2021vnj, Mougiakakos:2022sic, Riva:2022fru, Manohar:2022dea, Jakobsen:2022psy,Jakobsen:2021smu,Jakobsen:2021lvp} and 
at ${\cal O}(G^4)$ in \cite{Kalin:2022hph}.
The properties of scattering waveforms -- such as the absence of periodicity, short duration, and low amplitude -- pose a challenge for current gravitational-wave detectors. They may, however, be interesting goals for future terrestrial and space-based detectors. From a theoretical standpoint, scattering waveforms are an important part of the program to leverage scattering amplitudes and quantum field theory methods for precision predictions of gravitational wave physics, complementing the effort to determine the conservative motion and inclusive dissipative observables. 
It is an important challenge for the future to identify the analytic continuation that connects them to bound-state waveforms, as it is the case for certain (parts of) inclusive observables such as the scattering angle vs. the periastron advance and the energy loss~\cite{Kalin:2019rwq, Kalin:2019inp, Bini:2020hmy, Cho:2021arx}.

The observable-based formalism of Kosower, Maybee and O'Connell (KMOC)~\cite{Kosower:2018adc} establishes a direct link between scattering amplitudes and scattering waveforms~\cite{Cristofoli:2021vyo}. Having only scattering amplitudes as input, it builds on the vast technical advances used to construct them, such as the generalized unitarity \cite{UnitarityMethod, BernMorgan, Fusing, Bern:1997sc, Britto:2004nc, Bern:2007ct} and the double copy~\cite{BCJ, BCJLoop, KLT}, as well modern techniques for the evaluation of the relevant Feynman and phase space integrals, such as differential equations \cite{Kotikov:1990kg,Bern:1993kr,Remiddi:1997ny,Gehrmann:1999as} and integration-by-parts identities \cite{Chetyrkin:1981qh,Smirnov:2008iw,Smirnov:2019qkx}. 
This link has been used to evaluate the classical energy loss \cite{Herrmann:2021tct} and angular momentum loss \cite{Manohar:2022dea} at ${\cal O}(G^3)$.
Modeling compact spinless bodies as scalar fields, waveforms are determined by certain parts of the five-point amplitude~\cite{Cristofoli:2021vyo}. 

A lesson from the calculations of conservative effects is that amplitudes exhibit vast simplifications when expanded in the classical limit. An important feature of this expansion is that the leading term is not always the classical one. The first $L$ terms at $L$ loops are classically singular, or super-classical, and can be interpreted as iterations of lower-loop terms. 
Since the KMOC formalism contains terms bilinear in amplitudes, one may naively expect to find classical contributions that arise from the interference between superclassical terms in one factor and classically subleading (quantum) contributions from the other. 
Intuitively however such terms should not contribute at least to the classical amplitude and we demonstrate explicitly that at one-loop order this is indeed the case. This property however does not extend to the cut contribution to the gravitational waveform, whose importance was emphasized in \cite{Caron-Huot:2023vxl}.
Assuming the exponential representation of the scattering matrix proposed in Refs.~\cite{Bern:2021dqo, Damgaard:2021ipf} we derive a necessary condition for a given term to be included in the classical amplitude. 

The heavy-particle effective theory (HEFT) of Refs.~\cite{Brandhuber:2021kpo, Brandhuber:2021eyq} provides a strategy to isolate the classical amplitude, discarding the super-classical and quantum operators from the outset. It has been demonstrated in Ref.~\cite{Brandhuber:2021eyq} that this approach indeed gives the correct four-point classical amplitudes through two-loop order. 
The use of this framework for the construction of the one-loop five-point amplitude with outgoing radiation requires that the $\ie$ prescription of the uncut linearized matter propagators be clearly stated.
For four-point amplitudes specifying a presciption is not required through two loops because the uncut matter propagators are always off-shell due to kinematic constraints.
This is no longer the case in the presence of an outgoing graviton; sewing a 
one-loop five-point amplitude into a four-point higher-loop one suggests that a similar feature may appear at three loops for this multiplicity.
In this paper, we find, by comparing the HEFT and direct calculation of the classical one-loop five-point amplitude in GR coupled to scalars, that the correct prescription at this order for these uncut matter propagators is that they are principal-valued. 
With this prescription, the HEFT construction reproduces the classical expansion of the full theory amplitude once the super-classical terms are subtracted from the latter.  

We also compute the classical amplitude and scattering waveform for half-BPS black holes in $\mathcal{N}=8$ supergravity \cite{Arkani-Hamed:2008owk,Caron-Huot:2018ape}.
We use the existing $d$ dimensional one-loop five-point amplitude of maximally supersymmetric Yang-Mills theory~\cite{Mafra:2014gja,He:2017spx}, the double copy \cite{BCJLoop} and dimensional reduction to construct the relevant parts of the four-massive-scalar-one-graviton amplitude in four dimensions. Their remarkable simplicity leads to a (relatively) 
compact integral representation for the waveform.

We encounter several interesting features, not present in the amplitudes-based approach to inclusive observables such as the impulse and the energy and angular momentum loss. 
One of the requirements of the classical limit is that the matter particles are always separated. In the four-point classical amplitudes relevant for inclussive observables, this requirement leads to the absence of diagrams with intersecting matter lines.
As we will see in \cref{sec:TheAmplitude}, the five-point amplitude relevant for the scattering waveform does receive contributions from certain graphs with this topology. We will understand how their presence is consistent with the separation of the matter particles.
Furthermore, we find that the presence of the outgoing graviton introduces a certain asymmetry between matter lines in contributing diagrams and enhances the importance of the $\ie$ prescription.
Last, but not least, the classical five-point amplitudes in both GR and $\mathcal{N}=8$ supergravity are infrared divergent. The structure of infrared divergences in quantum gravitational amplitudes was understood long ago in Ref.~\cite{Weinberg:1965nx}. We remarkably find that the IR divergence of the classical amplitude is a pure phase that can be absorbed into the definition of the retarded time. This indicates that the asymptotic waveform contains no information about the lapsed time between the scattering event and observation. 

We directly integrate the resulting master integrals and find the complete classical amplitude. It was pointed out in \cite{Caron-Huot:2023vxl} that an additional (cut) contribution is needed to obtain the asymptotic waveform; we comment on these terms in Sec.~\ref{cutcontrib}.
The construction of the asymptotic spectral waveform also requires a further Fourier transform to impact parameter space while that of the time-domain waveform requires a Fourier transform of the outgoing graviton frequency.

The late-time properties of the waveform can be inferred directly from its integral representation. We find that, while the classical amplitude contributes to the ${\cal O}(G^2)$ correction to the gravitational wave memory in GR, such a correction is absent in ${\cal N}=8$ supergravity. As we will see, these contributions to the memory are proportional to the scattering angle at $\mathcal{O}(G^2)$. 
Thus, we trace the absence of the classical amplitude contribution to the memory in $\mathcal{N}=8$ supergravity to the absence of an ${\cal O}(G^2)$ correction to the scattering angle or, equivalently, to the absence of one-loop triangle integrals in this theory. 
The cut contribution however contributes to the gravitational wave memory in both GR and ${\cal N}=8$ supergravity.
For both ${\cal N}=8$ supergravity and GR, we analytically evaluate the frequency integral and all but one of the integral transforms to impact-parameter space. The integral was evaluated numerically. We leave a complete analytic evaluation of the spectral and time-domain waveform to future work.

Our paper is organized as follows. 
In \cref{sec:HEFT} we review the classical limit of amplitudes and the HEFT of Refs.~\cite{Brandhuber:2021kpo, Brandhuber:2021eyq}.
In \cref{sec:KMOC} we review observable-based formalism for waveforms, demonstrate  the cancellation of two-matter-particle-reducible (2MPR) contributions, and spell out the relation between the two-matter-particle-irreducible part of the amplitude and the waveform. We postpone to \cref{cutcontrib} a brief discussion of certain two-matter-particle-reducible contributions originating in the cut part of the observable-based formalism and pointed out in \cite{Caron-Huot:2023vxl}.
In \cref{sec:TheAmplitude}, we obtain the HEFT prediction for the classical 
one-loop five-point amplitude and compare it with the result of direct calculation in GR coupled to massive scalar fields.  
In \cref{sec:N8res}, we construct the classical one-loop five-point amplitude in $\mathcal{N}=8$ supergravity using the double copy. 
In \cref{integration} we discuss aspects of the integration of the relevant one-loop master integrals while leaving further details to \cref{sec:integrals}.  
In \cref{sec:Waveforms} we numerically compute the classical amplitude contribution to the waveform for ${\cal N}=8$ supergravity and discuss our results.
In \cref{cutcontrib} we summarize the cut contribution to the gravitational waveform in GR and ${\cal N}=8$ supegravity.
In \cref{sec:Conclusions}, we discuss our conclusions.
\Cref{sec:selfconheft} contains an argument that if a term in the classical HEFT integrand factorizes on a two-matter-particle into the product of lower point HEFT integrands, then it requires quantum information about the HEFT tree amplitudes used in the unitarity construction

\paragraph{Note added: } While this paper was in preparation, we became aware
of Refs.~\cite{Brandhuber:2023hhy} and \cite{Elkhidir:2023dco}, which partly overlap with aspects of 
our analysis. We thank the authors for communicating and for sharing copies 
of their drafts prior to publication.

\section{The classical limit and HEFT amplitudes}
\label{sec:HEFT}

Integrands of scattering amplitudes simplify considerably in the classical limit~\cite{Cheung:2018wkq, Bern:2019nnu, Bern:2019crd}. It is therefore advantageous to 
take this limit as early as possible and weed out terms that do not contribute to classical observables from the outset.

One approach to this limit uses the correspondence principle, according to which classical physics emerges in the limit of large charges. Thus, considering the scattering of two massive spinless bodies, the classical regime emerges for masses much larger than the Planck mass and for orbital angular momenta much larger than unity (in natural units). This limit also corresponds to the inter-particle separation being much larger than the de Broglie wavelength. As the separation of particles is Fourier-conjugate to the change in the momentum of each particle, it follows that the momentum transfer $q_i$ is much smaller than the momenta of the two particles. If there is any massless radiation in the initial or final state, its momentum should be much smaller than the massive particles.
This is implemented by the rescaling
\begin{align}\label{rescaling1}
(q,k,\lm)\rightarrow (\lambda q,\lambda k,\lambda\lm)\,,
\end{align}
where $k$ and $\lm$ are, respectively, the momenta of external and internal gravitons, and expanding at small $\lambda$. This process is later called the \emph{soft expansion}.
Accounting for the classical nature of Newton's potential, classical $L$-loop four-scalar amplitudes depend on Newton's constant and $\lambda$
as ${\cal M}^{\text{cl.}}_{4+n_g,L\text{-loop}}\sim G^{L+1+\frac{1}{2}n_g} \lambda^{-2+L}$, where $n_g$ is the number of external gravitons.

Another perspective on the classical limit was taken in Ref.~\cite{Kosower:2018adc} and involves a suitable restoration of the dependence on Planck's constant and an expansion at small $\hbar$. From this perspective, external momenta and masses pick up a factor of $\hbar^{-1}$, which is equivalent to messenger momenta picking up a factor of $\hbar$ through a change of variables. Thus, Planck's constant effectively plays the role of the momentum transfer and its restoration in four dimensions, 
\begin{equation}
\label{rescaling}
(q,k, \lm) \rightarrow (\hbar q, \hbar k,\hbar\lm) \ , \qquad \kappa \rightarrow \kappa/\hbar^{1/2}
\end{equation}
realizes the classical limit as a limit of small messenger momenta. 
With this scaling, the classical four-point amplitudes scale as ${\cal M}_{4,L\text{-loop}}^{\text{cl.}}\sim \hbar^{-3}$, which can also be recovered from the correspondence principle perspective by identifying $\lambda$ and $\hbar$ and further rescaling $G\rightarrow G/\lambda$, so that the contributions to the classical amplitude have the same scaling at all loop orders. 

The generalization to amplitudes with four scalars and any number of gravitons is straightforward based on the observation that the emission of arbitrarily-many low-energy messengers is a classical process, so it should not involve additional quantum suppression.
Thus, allowing for an arbitrary (even) number of scalars $n_\phi$ and messengers $n_g$, the amplitude scales as
\begin{equation}
\label{classical_scaling}
{\cal M}^{\text{cl.}}_{n_\phi+n_g,L\text{-loop}}\sim G^{L+\left(\frac{1}{2}n_\phi-1\right)+\frac{1}{2}n_g} \lambda^{-2\left(\frac{1}{2}n_\phi-1\right)+L} \sim \hbar^{-3\left(\frac{1}{2}n_\phi-1\right)-\frac{1}{2}n_g} \ ,
\end{equation}
in four dimensions. Indeed, eq. (\ref{classical_scaling}) can be verified for the tree-level five-point amplitude of Ref.~\cite{Luna:2017dtq} and we will use it to extract the classical limit of the one-loop four-scalar-one-graviton amplitude.  As for the four-point amplitude, the method of regions provides a systematic way of isolating the relevant contributions to the amplitude \cite{Beneke:1997zp}. Interestingly, unlike the one-loop four-scalar amplitude, not all internal messengers need to be in the potential region. However, this fact is unsurprising as, through the unitarity method, the one-loop five-point amplitude is part of the three-loop four-point amplitude, which receives contributions from messenger momenta in the radiation region.

The identification of the classical limit as a large-mass expansion accompanied by small messenger momenta establishes a connection with heavy-quark effective theory~\cite{Georgi:1990um, Luke:1992cs, Neubert:1993mb, Manohar:2000dt}, first utilized for classical gravitational scattering in Ref.~\cite{Damgaard:2019lfh} and further developed in Ref.~\cite{Aoude:2020onz, Haddad:2020tvs}. One approach to constructing the heavy-particle effective theory starts with the action of a scalar field coupled to gravity, 
\begin{equation}
\begin{split}
	 \mathcal{L}_{\text{sc-grav}}&=\frac{1}{2}(g^{\mu\nu}\partial_{\mu}\phi\partial_{\nu}\phi-m^{2}\phi^{2}) \ .
	\label{scgravl}
	\end{split}
\end{equation}
Building on the assumption that the messenger momenta are soft, one considers a process in which scalar fields exchange some gravitons. 
Decomposing the soft part of the scalar momenta 
and redefining the fields so that they only depend on the soft momenta,
\begin{equation}\label{decompos}
p^{\mu}=m u^{\mu}+ p^{\mu}_{\text{soft}}\,,
\qquad
\phi=e^{-i {m}u\cdot x}\chi+e^{i {m}u\cdot x}\chi^{*} \ ,
\end{equation}
leads to the new action
\begin{equation}\label{massprod}
\mathcal{L}_{\text{sc-grav}}\longrightarrow \chi^{\star} \left( 2i m (g^{\mu\nu}u_{\mu}\partial_{\nu})+(g^{\mu\nu}\partial_{\mu}\partial_{\nu}) \right) \chi \ ,
\end{equation}
where we have neglected the terms with a highly oscillatory phase $e^{\pm 2imu\cdot x}$. 
Denoting the momentum of $\chi$ as $q$, the propagator is\footnote{This is equivalent to taking a Taylor series expansion of the quadratic propagator at the level of the amplitude.}
\begin{equation}
\label{2pf_soft_scalar}
\langle\chi^*(-q)\chi(q)\rangle = 
\frac{i}{-2m q\cdot u - q^2} \simeq \frac{-i}{2m q\cdot u} 
\left(1-\frac{q^2}{2m q\cdot u}+\dots\right) \ .
\end{equation}
As one might expect from the propagator of a massive field, the leading term 
is ${\cal O}(\hbar^{-1})$, the next to leading term is ${\cal O}(\hbar^0)$, etc.
Upon constructing scattering amplitudes from the Lagrangian in eq.~\eqref{massprod} extended with the Einstein-Hilbert action, all vertices on a matter line must be symmetrized as a consequence of Bose symmetry. Therefore, all leading-order propagators are replaced by Dirac delta functions through the identity
\begin{align}
    \frac{i}{2 m q\cdot u + \ie }+\frac{i}{-2 m   q\cdot u + \ie}= 2\pi \delta(2 m   q\cdot u)\equiv \hdelta(2 m  q\cdot u)\,,
\end{align}
and its multi-propagator generalization~\cite{Akhoury:2013yua}. This phenomenon can be visualized as 
\begin{equation}
\begin{tikzpicture}[baseline={([yshift=-.5ex]current bounding box.center)}]
    \pgfmathsetmacro{\dis}{1.75};
    \pgfmathsetmacro{\ang}{70};
    \pgfmathsetmacro{\len}{1.5};
    \pgfmathsetmacro{\db}{4.5};
    \draw [very thick] (-\dis,0) -- (\dis,0);
    \draw [graviton] (-\dis/2,0) -- ++(-\ang:\len);
    \draw [graviton] (-\dis/2,0) -- ++(-180+\ang:\len);
    \draw [graviton] (\dis/2,0) -- ++(-\ang:\len);
    \draw [graviton] (\dis/2,0) -- ++(-180+\ang:\len);
    \node at (-\dis/2,0) (bub1) [draw, thick, fill=black!20, circle, inner sep=4] {$A$};
    \node at (\dis/2,0) (bub2) [draw, thick, fill=black!20, circle, inner sep=4] {$B$};
    \node at (\dis/2,-2*\len/3) {\small$\ldots$};
    \node at (-\dis/2,-2*\len/3) {\small$\ldots$};
    \node at (-\dis/2,-\len) {$\lm_A$};
    \node at (\dis/2,-\len) {$\lm_B$};
    \begin{scope}[xshift=\db cm]
    \draw [very thick] (-\dis,0) -- (\dis,0);
    \draw [graviton] (-\dis/2,0) -- ++(-\ang:\len);
    \draw [graviton] (-\dis/2,0) -- ++(-180+\ang:\len);
    \draw [graviton] (\dis/2,0) -- ++(-\ang:\len);
    \draw [graviton] (\dis/2,0) -- ++(-180+\ang:\len);
    \node at (-\dis/2,0) (bub1) [draw, thick, fill=black!20, circle, inner sep=4] {$B$};
    \node at (\dis/2,0) (bub2) [draw, thick, fill=black!20, circle, inner sep=4] {$A$};
    \node at (\dis/2,-2*\len/3) {\small$\ldots$};
    \node at (-\dis/2,-2*\len/3) {\small$\ldots$};
    \node at (-\dis/2,-\len) {$\lm_B$};
    \node at (\dis/2,-\len) {$\lm_A$};
    \end{scope}
    \begin{scope}[xshift=2*\db cm]
    \draw [very thick] (-\dis,0) -- (\dis,0);
    \draw [graviton] (-\dis/2,0) -- ++(-\ang:\len);
    \draw [graviton] (-\dis/2,0) -- ++(-180+\ang:\len);
    \draw [graviton] (\dis/2,0) -- ++(-\ang:\len);
    \draw [graviton] (\dis/2,0) -- ++(-180+\ang:\len);
    \node at (-\dis/2,0) (bub1) [draw, thick, fill=black!20, circle, inner sep=4] {$A$};
    \node at (\dis/2,0) (bub2) [draw, thick, fill=black!20, circle, inner sep=4] {$B$};
    \node at (\dis/2,-2*\len/3) {\small$\ldots$};
    \node at (-\dis/2,-2*\len/3) {\small$\ldots$};
    \node at (-\dis/2,-\len) {$\lm_A$};
    \node at (\dis/2,-\len) {$\lm_B$};
    \end{scope}
    \node at (\db/2,-\len/2) {$+$};
    \node at (3*\db/2,-\len/2) {$=$};
    \draw [draw=red] (2*\db,1/3) -- (2*\db,-1/3);
\end{tikzpicture} \,,
\end{equation}
where the red vertical line represents the cut massive propagator, and the two blobs connected by the cut commute
\begin{equation}\label{irrelvanorder}
\begin{tikzpicture}[baseline={([yshift=-.5ex]current bounding box.center)}]
    \pgfmathsetmacro{\dis}{1.75};
    \pgfmathsetmacro{\ang}{70};
    \pgfmathsetmacro{\len}{1.5};
    \pgfmathsetmacro{\db}{4.5};
    \draw [very thick] (-\dis,0) -- (\dis,0);
    \draw [graviton] (-\dis/2,0) -- ++(-\ang:\len);
    \draw [graviton] (-\dis/2,0) -- ++(-180+\ang:\len);
    \draw [graviton] (\dis/2,0) -- ++(-\ang:\len);
    \draw [graviton] (\dis/2,0) -- ++(-180+\ang:\len);
    \node at (-\dis/2,0) (bub1) [draw, thick, fill=black!20, circle, inner sep=4] {$A$};
    \node at (\dis/2,0) (bub2) [draw, thick, fill=black!20, circle, inner sep=4] {$B$};
    \node at (\dis/2,-2*\len/3) {\small$\ldots$};
    \node at (-\dis/2,-2*\len/3) {\small$\ldots$};
    \node at (-\dis/2,-\len) {$\lm_A$};
    \node at (\dis/2,-\len) {$\lm_B$};
    \draw [draw=red] (0,1/3) -- (0,-1/3);
    \node at (\db/2,-\len/2) {$=$};
    \begin{scope}[xshift=\db cm]
    \draw [very thick] (-\dis,0) -- (\dis,0);
    \draw [graviton] (-\dis/2,0) -- ++(-\ang:\len);
    \draw [graviton] (-\dis/2,0) -- ++(-180+\ang:\len);
    \draw [graviton] (\dis/2,0) -- ++(-\ang:\len);
    \draw [graviton] (\dis/2,0) -- ++(-180+\ang:\len);
    \node at (-\dis/2,0) (bub1) [draw, thick, fill=black!20, circle, inner sep=4] {$B$};
    \node at (\dis/2,0) (bub2) [draw, thick, fill=black!20, circle, inner sep=4] {$A$};
    \node at (\dis/2,-2*\len/3) {\small$\ldots$};
    \node at (-\dis/2,-2*\len/3) {\small$\ldots$};
    \node at (-\dis/2,-\len) {$\lm_B$};
    \node at (\dis/2,-\len) {$\lm_A$};
    \draw [draw=red] (0,1/3) -- (0,-1/3);
    \end{scope}
\end{tikzpicture} \,.
\end{equation}
To leading order in the classical expansion, the exposed propagators of heavy scalar states can be treated as on-shell in the gravitational heavy-particle effective theory. At loop level, we also encounter propagators that form a principal value combination,
\begin{align}
    \frac{i}{2m q\cdot u +\ie}-\frac{i}{-2m q\cdot u+\ie}=\text{PV}\frac{i}{m q\cdot u}\,.
\end{align}
As we will see later, they often show up in classical amplitudes.

We take a classical expansion of the full tree amplitudes to compute the HEFT tree amplitudes. The HEFT amplitudes are naturally organized as a sum of terms manifesting the factorization properties of Feynman diagrams with the on-shell matter propagator being linearized. Doubled (and perhaps higher powers of) linearized propagators, arising from the expansion in eq.~\eqref{2pf_soft_scalar}, are also present. This is analogous to the structure obtained by directly expanding in the soft region. A more systematic way to construct the classical parts of tree-level HEFT two-scalar-graviton amplitudes without reference to a Lagrangian (though probably equivalent to one) that also exhibits double-copy properties was proposed in Ref.~\cite{Brandhuber:2021kpo}. For example, the three and four-point gravitational Compton amplitudes, expanded up to the classical order, are given by
\begin{align}\label{eq:comp_3}
    \vcenter{\hbox{\begin{tikzpicture}[scale=0.8,every node/.style={font=\small}]
        \draw [very thick] (0,0) node[left=0]{$1$} -- (2,0) node[right=0]{$3$};
        \draw [graviton] (1,0) -- (1,1) node[above=0]
        {$k_2$};
        \filldraw [fill=gray!30!white,thick] (1,0) circle (0.25);
    \end{tikzpicture}
    }}&=-\kappa m_1^2(u_1\cdot\varepsilon_2)^2\,,\\
    \label{eq:comp_4}
    \vcenter{\hbox{\begin{tikzpicture}[scale=0.8,every node/.style={font=\small}]
        \draw [very thick] (0,0) node[left=0]{$1$} -- (2,0) node[right=0]{$4$};
        \draw [graviton] (1,0) -- ++(135:1) node[above=0]
        {$k_2$};
        \draw [graviton] (1,0) -- ++(45:1) node[above=0]
        {$k_3$};
        \filldraw [fill=gray!30!white,thick] (1,0) circle (0.25);
    \end{tikzpicture}
    }}&=i\kappa^2m_1^3\hdelta(2u_1\cdot k_2)(u_1\cdot\varepsilon_2)^2(u_1\cdot\varepsilon_3)^2-\frac{\kappa^2m_1^2}{2k_2\cdot k_3}\Big[\frac{u_1\cdot f_2\cdot f_3\cdot u_1}{u_1\cdot k_3}\Big]^2\,,
\end{align}
where $f_i^{\mu\nu}=k_i^{\mu}\varepsilon_i^{\nu}-k_i^{\nu}\varepsilon_i^{\mu}$ is the momentum space linearized field strength. The three-point amplitude scales classically under \cref{rescaling} as $\mathcal{O}(\hbar^{-1/2})$. The first term in the four-point amplitude exhibits super-classical $\mathcal{O}(\hbar^{-2})$ scaling, which contains the characteristic delta function that localizes it on a special momentum configuration, while the second term scales classically as $\mathcal{O}(\hbar^{-1})$. The classical part of the amplitude is referred to as the ``HEFT amplitude''. 

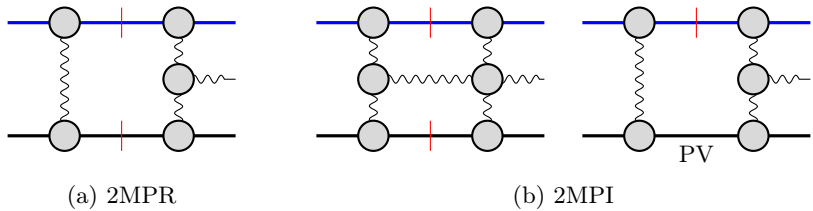
\begin{figure}
    \centering
    \subfloat[2MPR]{
    \begin{tikzpicture}
        \draw [very thick, blue] (-\w,\h/2) -- (\w,\h/2) ;
        \draw [very thick] (-\w,-\h/2) -- (\w,-\h/2) ;
        \draw [graviton] (-\w/2,-\h/2) -- (-\w/2,\h/2) (\w/2,-\h/2) -- (\w/2,\h/2) (\w/2,0) -- (\w,0);
        \filldraw [fill=gray!30!white,thick] (-\w/2,-\h/2) circle (0.2cm) (\w/2,-\h/2) circle (0.2cm) (-\w/2,\h/2) circle (0.2cm) (\w/2,\h/2) circle (0.2cm) (\w/2,0) circle (0.2cm);
        \draw [draw=red] (0,\h/2+0.2) -- (0,\h/2-0.2) (0,-\h/2+0.2) -- (0,-\h/2-0.2);
        \node at (0,-\h/2) [below=0pt]{\vphantom{\footnotesize PV}};
    \end{tikzpicture}
    }
    \qquad
    \subfloat[2MPI]{
    \begin{tikzpicture}
        \draw [very thick, blue] (-\w,\h/2) -- (\w,\h/2) ;
        \draw [very thick] (-\w,-\h/2) -- (\w,-\h/2) ;
        \draw [graviton] (-\w/2,-\h/2) -- (-\w/2,\h/2) (\w/2,-\h/2) -- (\w/2,\h/2) (-\w/2,0) -- (\w,0);
        \filldraw [fill=gray!30!white,thick] (-\w/2,-\h/2) circle (0.2cm) (\w/2,-\h/2) circle (0.2cm) (-\w/2,\h/2) circle (0.2cm) (\w/2,\h/2) circle (0.2cm) (\w/2,0) circle (0.2cm) (-\w/2,0) circle (0.2cm);
        \draw [draw=red] (0,\h/2+0.2) -- (0,\h/2-0.2) (0,-\h/2+0.2) -- (0,-\h/2-0.2);
        \begin{scope}[xshift=3.5cm]
        \draw [very thick, blue] (-\w,\h/2) -- (\w,\h/2) ;
        \draw [very thick] (-\w,-\h/2) -- (\w,-\h/2) ;
        \draw [graviton] (-\w/2,-\h/2) -- (-\w/2,\h/2) (\w/2,-\h/2) -- (\w/2,\h/2) (\w/2,0) -- (\w,0);
        \filldraw [fill=gray!30!white,thick] (-\w/2,-\h/2) circle (0.2cm) (\w/2,-\h/2) circle (0.2cm) (-\w/2,\h/2) circle (0.2cm) (\w/2,\h/2) circle (0.2cm) (\w/2,0) circle (0.2cm) ;
        \draw [draw=red] (0,\h/2+0.2) -- (0,\h/2-0.2) ;
        \node at (0,-\h/2) [below=0pt]{\footnotesize PV};
        \end{scope}
    \end{tikzpicture}
    }
    \caption{Examples of 2MPR and 2MPI diagrams.}
    \label{fig:2mpr}
\end{figure}

At loop level, we will focus on amplitudes with four scalars (two distinct massive scalar lines). The classical expansion naturally decomposes amplitudes into two matter-particle reducible (2MPR) and two matter-particle irreducible (2MPI) contributions. We define the 2MPR contribution as follows:
\begin{itemize}
    \item The diagram becomes disconnected by cutting two matter propagators.
    \item The two cut matter propagators are both represented by $\delta(p_i\cdot\lm)$, and the residue on the cut follows from the factorization 
    of the amplitude's integrand.
    %unitarity principle.
\end{itemize}
Being complementary to 2MPR, the 2MPI contributions include the following two classes of diagrams:
\begin{enumerate}
    \item The diagram remains connected after cutting any two matter propagators.
    \item If the diagram becomes disconnected after cutting two matter propagators, then at least one of the matter lines exhibits a principal-value propagator. Consequently, this cut has zero residue.
\end{enumerate}
Therefore, by construction, the 2MPR diagrams are given by the product of their 2MPI components on the support of the explicit delta functions that enforce the two-matter cuts. Some simple examples of 2MPR and 2MPI diagrams are given in \cref{fig:2mpr}.

It was further demonstrated in Ref.~\cite{Brandhuber:2021eyq} that, together with a suitable application of generalized unitarity, the classical HEFT tree amplitudes, such as \cref{eq:comp_3} and the second term of \cref{eq:comp_4}, yield the classical part of the two-loop four-scalar amplitude, which can be identified with the radial action. As expected, the classical contributions are all 2MPI. More generally, the prescription of Ref.~\cite{Brandhuber:2021eyq} manifests the 2MPR versus 2MPI classification by construction, such that the 2MPI contributions have the correct classical scaling while the 2MPR parts are super-classical. The only way to make the 2MPR diagrams have a classical contribution is to include local quantum contributions in the HEFT tree amplitudes. Such terms are, however, discarded from the outset; since they can appear in a complete calculation, one may wonder whether such $\hbar/\hbar$ contributions can appear in classical scattering observables.
By comparing the HEFT calculation of the five-point amplitude with the analogous result GR coupled with scalar fields we will see that such contributions are in fact absent. We postpone for \cref{sec:selfconheft} a more general discussion of these points.

Finally, an essential aspect of using the HEFT tree amplitudes in a unitarity-based 
construction of loop-level amplitudes is fixing the $\ie$ prescription for the uncut linearized matter propagators.
We find that for the one-loop five-point amplitude considered in this work, we can treat all the uncut linearized matter propagators as principal-valued. We justify this approach by showing that, with this prescription, the HEFT result agrees with the classical part of the full quantum amplitude. It is beyond the scope of this work to identify a general $\ie$ prescription for the uncut matter propagators when applying the HEFT construction at loop levels.

%We find that for the one-loop five-point problem considered in this work, we can treat all the uncut linearized matter propagators as principal-valued.

\section{Waveforms in the observable-based formalism}
\label{sec:KMOC}

We begin this section by briefly reviewing the observable-based approach of Kosower, Maybee, and O'Connell \cite{Kosower:2018adc} (KMOC) to inclusive and local scattering observables. We then discuss certain cancellations present in this formalism, which can be made manifest through the use of the HEFT organization of amplitudes.

The KMOC formalism constructs quantum scattering observables, described by some operator $\opO$, by comparing the expectation value of $ \opO$ in the final and initial states of the scattering process:
\begin{equation}
\langle \opO\rangle=\langle \psi_{\textrm{out}}|\opO|\psi_{\textrm{out}}\rangle-\langle \psi_{\textrm{in}}|\opO|\psi_{\textrm{in}}\rangle  
\label{KMOCgeneral}
\end{equation}
where $\psi_{\textrm{out}}$ and $\psi_{\text{in}}$ are the outgoing and incoming state respectively. Further using that the incoming and outgoing states are connected by the time-evolution operator whose matrix elements form the scattering matrix, $\opS$, 
\begin{equation}
|\psi_{\textrm{out}}\rangle= \opS|\psi_{\textrm{in}} \rangle \quad \text{ where } \quad \opS=\mathbb{I}+i\opT  \ ,
\end{equation}
and $\hat{T}$ is the transition matrix, \cref{KMOCgeneral} becomes~\cite{Kosower:2018adc}
\begin{equation}
\label{KMOCformula1}
\langle \opO\rangle=
i\langle \psi_{\textrm{in}}  |{\hat {\mathcal O}} \opT - \opT^\dagger {\hat {\mathcal O}} |\psi_{\textrm{in}} \rangle+
  \langle \psi_{\textrm{in}} |\opT^{\dagger}\opO\opT|\psi_{\textrm{in}} \rangle \ .
\end{equation}
Thus, using eq.~\eqref{KMOCformula1}, scattering observables 
%corresponding to operators ${\hat {\mathcal O}}$ 
are expressed in terms of (i.e. phase space integrals of) scattering amplitudes (i.e. matrix elements of $\hat{T}$) and their cuts, dressed with~${\hat {\mathcal O}}$. 
The operators ${\hat {\mathcal O}}$ may correspond to global (inclusive) observables such as the impulse of the matter particles or the radiated momentum~\cite{Kosower:2018adc}, to local (exclusive) observables of the waveform of the outgoing radiation~\cite{Cristofoli:2021vyo}, or to a combination of both. The corresponding classical observables can be obtained in the appropriate classical limit.

Impact parameter space provides one convenient means to taking this limit. The classical limit corresponds to an impact parameter significantly larger than the de~Broglie wavelengths of the scattered particles and the horizon radii of black holes of masses equal to their masses; these, in turn, imply that the orbital angular momentum is large, making contact with Bohr's correspondence principle. 
To make use of this, and assuming that the initial state contains no incoming radiation, the initial two-particle state is built in terms of wave packets over the tensor product of a single-particle phase-space of measure $\dd\Phi[p]$:
\begin{align}\label{impactparameterspace}
|\psi_{\textrm{in}}\rangle&=\int  \dd\Phi[p_1]\dd\Phi[p_2] \phi(p_1)\phi(p_2) e^{ip_1\cdot b_1+ip_2\cdot b_2}|p_1p_2\rangle\,,\\
%\label{phasespace}
\dd\Phi[p]&\equiv\frac{\dd^4p}{(2\pi)^4} 2\pi \Theta(p^0)\delta(p^2-m^2) \,.\nonumber
\end{align}
The combination $b_1 - b_2$ is the impact parameter (i.e. the separation between the incoming particles),  while $b_1+b_2$ is the conjugate of the momentum of the center of mass.
The wave packets $\phi_{i}$ are sufficiently localized not to interfere with the classical limit conditions~\cite{Kosower:2018adc}, i.e. their widths $ \ell_{\phi(p_i)}$ obey $\ell_\text{horizon}\ll \ell_{\phi(p_i)} \ll \sqrt{-(b_1-b_2)^2}$, namely, it is much greater than the horizon size $\ell_\text{horizon}$ of the black holes but much smaller than their separation.

\subsection{A brief review of the waveform in the observable-based formalism}\label{KMOCform}

The waveform in the infinite future, obtained by measuring the 
linearized Riemann curvature tensor~\cite{Cristofoli:2021vyo} for $g_{\mu\nu} = \eta_{\mu\nu}+\kappa h_{\mu\nu}$, is the focus of this paper. The relevant operator, written in terms of graviton creation and annihilation operators, is 
\begin{align}\label{Ropertfor}
\mathbb{R}_{\mu\nu\rho\sigma}(x) = \frac{\kappa}{2} \,
\sum_{h = \pm} \int \dd \Phi(k) \, & \Big[\,  k_{[\mu}^{\vphantom{*}} \varepsilon_{\nu],h}^{*}(k)  \, k_{[\rho}^{\vphantom{*}}\varepsilon_{\sigma],h}^{*}(k) \, e^{-i k \cdot x} \, \hat{a}_{hh}(k)
\\
&
+ k_{[\mu}\varepsilon_{\nu],h}(k)  \, k_{[\rho}\varepsilon_{\sigma],h}(k) \, e^{+i k \cdot x} \, \hat{a}^\dagger_{hh}(k) \Big] \, ,
\nonumber
\end{align}
where the antisymmetrization has strength 2 (i.e. it does not include a division by the number of terms), and $\dd\Phi(k)$ is given by \cref{impactparameterspace} at zero mass, $\varepsilon_{h}(k)$ are polarization vectors normalized as  
\begin{align}\label{eq:pols}
    \varepsilon_{h}(k)\cdot\varepsilon_{h}^{*}(k)=-1\,,\qquad \varepsilon_{+}^{*}(k)=\varepsilon_{-}(k)\,.
\end{align}
In the second relation, we choose $h=\pm$ to represent the $\pm 1$ helicity state. The operator $\hat{a}_{hh}(k)$ annihilates a graviton with polarization $\varepsilon^{\mu\nu}_{h}=\varepsilon^{\mu}_{h}\varepsilon^{\nu}_{h}$. We note that, even though this operator superficially depends on an arbitrary space-time point, $x$, the formulation of the measurement process through scattering theory in \cref{KMOCgeneral} implicitly assumes that this point is both in the far future and at spatial infinity.

Assuming temporarily that the observable is measured at finite distance and there is no gravitational radiation in the infinite past, the waveform is given by~\cite{Cristofoli:2021vyo, Kosower:2022yvp}
\begin{align}
\langle \mathbb{R}_{\mu\nu\rho\sigma}(x) \rangle &= i \int \dd \Phi(k)
\left[e^{-ik\cdot x} {\tilde J}_{\mu\nu\rho\sigma}(k)
- e^{ik\cdot x}{\tilde J}^\dagger_{\mu\nu\rho\sigma}(k)\right] \ ,
\label{Jdef}
\\
{\tilde J}_{\mu\nu\rho\sigma}(k)&= \, \frac{\kappa}{2} \sum_{h = \pm} \, k_{[\mu}^{\vphantom{*}} \varepsilon_{\nu],h}^{*}(k)  \, k_{[\rho}^{\vphantom{*}} \varepsilon_{\sigma],h}^{*}(k) (-i) \langle\psi_\text{in} |\opS^\dagger \hat a_{hh}(k)\opS|\psi_\text{in}\rangle \ ,
\\
{\tilde J}^\dagger_{\mu\nu\rho\sigma}(k)&=\frac{\kappa}{2} \sum_{h = \pm} 
k_{[\mu} \varepsilon_{\nu],h}(k)  \, k_{[\rho} \varepsilon_{\sigma],h}(k) (+i) \langle\psi_\text{in} |\opS^\dagger \hat a^\dagger_{hh}(k)\opS|\psi_\text{in}\rangle \ . 
\end{align}
%%%%%%%%%%%%%%%%%%%%%%%
%
At large distances, $|x^0| \rightarrow \infty$ and $|\bm x|\rightarrow\infty$, the integral over the angular directions of the on-shell graviton momentum,
\begin{align}
\label{onshellgraviton}
k=\omega(1, \bm{n}_{\bm{k}})\,,\quad \text{with } \omega>0\text{ and } \bm{n}^2_{\bm{k}}=1 \ ,
\end{align}
can be evaluated through various methods~\cite{Cristofoli:2021vyo, Kosower:2022yvp} and each of the exponentials in \cref{Jdef} leads to a linear combination of advanced and retarded propagators while $\bm{n}_{\bm{k}}$ is localized to the spatial unit vector $\bm n \equiv \bm x/|\bm x|$ at the observation point $x$,
\begin{align}\label{eq:ret_adv}
\langle \mathbb{R}_{\mu\nu\rho\sigma}(x) \rangle &= 
\frac{1}{4\pi |\bm x|} \int_0^\infty \frac{d\omega}{2\pi}
\Big[e^{-i\omega (x^0 - |\bm x|)}{\tilde J}_{\mu\nu\rho\sigma}(\omega, \omega \bm n)+ e^{+i\omega (x^0 - |\bm x|)}{\tilde J}^\dagger_{\mu\nu\rho\sigma}(\omega, \omega \bm n) \nonumber\\[2pt]
& \qquad
-e^{-i\omega (x^0 + |\bm x|)}{\tilde J}_{\mu\nu\rho\sigma}(\omega, -\omega \bm n)- e^{+i\omega (x^0 + |\bm x|)}{\tilde J}^\dagger_{\mu\nu\rho\sigma}(\omega, -\omega \bm n) \Big] \ .
\end{align}
In the infinite future, only the terms depending on the retarded time $t=x^0-|\bm x|$ are relevant. We thus drop the second line of \cref{eq:ret_adv}. The waveform $W_{\mu\nu\rho\sigma}(t,\bm{n})$ for the curvature tensor and the spectral waveform, $f_{\mu\nu\rho\sigma}(\omega,\bm{n})$, is given by 
\begin{subequations}
\label{eq:WFs}
\begin{align}
 \label{spectralWF}
\langle \mathbb{R}_{\mu\nu\rho\sigma}(x) \rangle \Big|_{|{\bm x}|\rightarrow \infty} &=
\frac{1}{|\bm x|} W_{\mu\nu\rho\sigma}(t, \bm n)  \equiv 
\frac{1}{ |\bm x|} 
\int_{-\infty}^{+\infty} \frac{d\omega}{2\pi} f_{\mu\nu\rho\sigma}(\omega, \bm n)e^{-i\omega t}\,,
%\\
%W(t, \bm n, x)_{\mu\nu\rho\sigma}&= \frac{1}{4\pi} 
%\int d\omega {\tilde J}(\omega, \omega \bm n)_{\mu\nu\rho\sigma}e^{-i\omega (x^0-|{\bm x}|)}
\\[2pt]
\label{just_spectralWF}
 f_{\mu\nu\rho\sigma}(\omega, \bm n)&=
 \frac{1}{4\pi} \left[
 \Theta(\omega){\tilde J}_{\mu\nu\rho\sigma}(\omega, \omega \bm n)
 +
 \Theta(-\omega){\tilde J}^\dagger_{\mu\nu\rho\sigma}(|\omega|, |\omega| \bm n)\right]\,.
\end{align}
\end{subequations}

A convenient presentation of the curvature tensor (and consequently of the gravitational waveform) is in terms of Newman-Penrose scalars~\cite{Newman:1961qr}. They are constructed as projections of the curvature tensor on a complex basis of null vectors. Following Ref.~\cite{Cristofoli:2021vyo} we choose these vectors to be
\begin{align}
\label{vector_basis}
   L^\mu = \frac{k^{\mu}}{\omega} = (1,\bm{n}_{\bm{k}})
   \ ,\quad
   N^\mu = \zeta^\mu
   \ ,\quad
   M^\mu = \varepsilon_{+}^{\mu} 
   \ ,\quad
   M^*{}^\mu = \varepsilon_{-}^{\mu} \ ,
\end{align} 
where $M\cdot M^* = -1$ following \cref{eq:pols}, and $\zeta$ is a gauge choice such that $\zeta\cdot \varepsilon_{\pm}=0$ and $L\cdot N = L\cdot \zeta=1$. The independence of $L$ on the frequency of the outgoing graviton makes eq.~\eqref{vector_basis} a suitable basis both for the waveform and the spectral waveform. 
The Newman-Penrose scalars are defined by the independent contractions of the Weyl tensor with the vectors in eq.~\eqref{vector_basis}. The one with the slowest decay at large distances, typically denoted by $\Psi_4$, describes the transverse radiation propagating along $L$~\cite{Newman:1961qr},
\begin{align}
\label{NPscalar_t_space}
\Psi_4(x) &= -N^\mu M^*{}^\nu N^\rho M^*{}^\sigma \langle \mathbb{R}_{\mu\nu\rho\sigma}(x) \rangle =
\frac{1}{|\bm x|} \Psi_4^{\infty}+\dots 
\end{align}
where the ellipsis stands for terms suppressed at large distance. Using the transversality and null property of $M^{\mu}$ and \cref{eq:ret_adv}, we can write the spectral representation of $\Psi_4^{\infty}$ as
\begin{align}
{\widetilde \Psi}_4^{\infty}(\omega, {\bm n})&=
 -\frac{\kappa}{4\pi} \Big[\Theta(\omega) \omega^2 (-i)
\langle \psi_\text{in}|\opS^\dagger {\hat a}_{--}(\omega,\omega \bm n)\opS|\psi_\text{in}\rangle
\cr
&\qquad\qquad +
\Theta(-\omega) \omega^2 (+i)
\langle \psi_\text{in}|\opS^\dagger {\hat a}^\dagger_{++}(|\omega|,|\omega|\bm n)\opS|\psi_\text{in}\rangle
\Big]\,.
%{}_\text{in}\langle p'_1p'_2 |\opS^\dagger \hat a_{--}(k)S|p_1
%p_2\rangle{}_\text{in}
\label{NPscalar}
\end{align}
For an asymptotically flat spacetime, outgoing radiation at large distances is described by linearized general relativity in transverse traceless gauge. 
Using that $k\cdot N = \omega$, the Newman-Penrose scalar $\Psi_4$ takes the form 
\begin{align}
\label{psi4NPscalar_linearized}
\Psi_4 = \frac{1}{2}\kappa\, \varepsilon_{-}^{\mu}\varepsilon_{-}^{\nu}\ddot h_{\mu\nu}
\equiv 
\frac{1}{2}\kappa(\ddot h_{+} + i \ddot h_\times) = \frac{\kappa}{8\pi |\bm x|}
(\ddot {h}^{\infty}_{+} + i \ddot h^{\infty}_\times )\ ,
\end{align}
For $k=(\omega,0,0,\omega)$, the negative-helicity polarization vector is $\varepsilon_{-}=-\frac{1}{\sqrt{2}}(0,1,i,0)$ and the $+$ and $\times$ graviton polarizations are 
\begin{align}
    h_{+} = \frac{1}{2} (h_{11} - h_{22})\,, \qquad h_\times = h_{12}=h_{21} \ .
\end{align}
In general, the $\times$ and $+$ polarization are defined with respect to the vector $L^\mu$ pointing along the graviton momentum. They are related to the real and imaginary parts of the outgoing negative helicity polarization tensor, see \cref{sec:Waveforms}. 
The metric perturbation $h_{\mu\nu}$ is in transverse traceless gauge and is normalized in such a way that, at spatial infinity, it falls off as
\begin{align}\label{eq:metric_pert}
    g_{\mu\nu}\Big|_{|\bm x|\rightarrow\infty}=\eta_{\mu\nu}+\frac{\kappa}{4\pi|\bm x|}h_{\mu\nu}^{\infty}\,.
\end{align}
Therefore, we may directly identify the Fourier transform of the metric at infinity in terms of the frequency-space Newman-Penrose scalar:
\begin{align}
\label{metric_infinity_omegaspace}
\tilde {h}^{\infty}_{+} + i  \tilde {h}^{\infty}_\times & = 
%-\frac{i}{8\pi} 
\Big[\, \Theta(\omega) (-i)
\langle \psi_\text{in}|\opS^\dagger {\hat a}_{--}(\omega,\omega \bm n)\opS|\psi_\text{in}\rangle
\cr
&\qquad\quad 
+\Theta(-\omega) (+i)
\langle \psi_\text{in}|\opS^\dagger {\hat a}^\dagger_{++}(|\omega|,|\omega|\bm n)\opS|\psi_\text{in}\rangle
\Big] \ .
\end{align}
Thus, in frequency space, the standard gravitational-wave polarizations, $h_{+,\times}$, are given directly in terms of scattering amplitudes. 

Let us now discuss the matrix elements
$\langle \psi_\text{in}|\opS^\dagger {\hat a}_{--}(k)\opS|\psi_\text{in}\rangle$ and $\langle \psi_\text{in}|\opS^\dagger {\hat a}^\dagger_{++}(k)\opS|\psi_\text{in}\rangle$, which define ${\tilde J}$ 
and ${\tilde J}^\dagger$ in Eq.~\eqref{just_spectralWF}. We will focus on the 
former and obtain the latter by complex conjugation. We first go to momentum space and consider the matrix element with an initial state with momenta $p_1$ and $p_2$, 
\begin{align}\label{iniMatrixElement}
    \langle\psi_\text{in} |\opS^\dagger \hat a_{hh}(k)\opS|\psi_\text{in}\rangle=\int &\dd\Phi[p_1]\dd\Phi[p_2]\dd\Phi[p'_1]\dd\Phi[p'_2]\phi(p_1)\phi(p_2)\phi^{*}(p'_1)\phi^{*}(p'_2)\nonumber\\
    &\times \langle p'_1p'_2 |\opS^\dagger \hat a_{hh}(k)\opS|p_1 p_2\rangle e^{i(p_1-p'_1)\cdot b_1}e^{i(p_2-p'_2)\cdot p_2}
\end{align}
We then expand $\opS$ in terms of the transition matrix $\opT$, which leads to
\begin{align}\label{SaS}
    \langle p'_1p'_2 |\opS^\dagger \hat a_{hh}( k)\opS|p_1 p_2\rangle&=\langle p'_1p'_2 k^h| i\opT|p_1p_2\rangle+\langle p'_1p'_2|\opT^{\dagger} \hat{a}_{hh}(k)\opT|p_1p_2\rangle\,.
\end{align}
The first term is simply the 2-to-3 scattering amplitude
\begin{align}\label{matrixElement}
    \langle p'_1p'_2k^h|i\opT|p_1p_2\rangle &= i\Mfull(p_1p_2\rightarrow p'_1p'_2 k^h)\nonumber\\
    & = i \Mred(p_1p_2\rightarrow p'_1p'_2k^h)\underbrace{(2\pi)^d\delta^{(d)}(p_1+p_2-p'_1-p'_2-k)}_{\equiv\hdelta^{(d)}(p_1+p_2-p'_1-p'_2-k)}\,,
\end{align}
where $\Mfull$ contains implicitly the momentum-conserving delta function, and $\Mred$ is the reduced amplitude with this delta function stripped off. The second term gives the $s$-channel unitarity cuts of this virtual amplitude,
\begin{align}
    \langle p'_1p'_2|\opT^{\dagger} \hat{a}_{hh}(k)\opT|p_1p_2\rangle&=\sum_{X}\int\dd\Phi[r_1]\dd\Phi[r_2]\langle p_1'p_2'|T^{\dagger}|r_1r_2 X\rangle \langle r_1 r_2 k X| T|p_1 p_2\rangle\\
    &=\sum_{X}\int\dd\Phi[r_1]\dd\Phi[r_2]\Mfull^*(p_1' p_2'\rightarrow r_1 r_2, X) \Mfull(p_1 p_2\rightarrow r_1 r_2 k^h, X)\,,\nonumber
\end{align}
where we have judiciously inserted a complete set of states between $\opT^{\dagger}$ and $\hat{a}_{hh}(k)$, and $X$ stands for graviton exchanges.
This term can be graphically represented by
\begin{align}
    \langle p'_1p'_2|\opT^{\dagger} \hat{a}_{hh}(k)\opT|p_1p_2\rangle= \sum_{X}\int\dd\Phi[r_1]\dd\Phi[r_2]\vcenter{\hbox{\begin{tikzpicture}
        \pgfmathsetmacro{\w}{1.75}
        \node [left=0pt] (p1) at (-\w,0) {$1$};
        \node [right=0pt] (p4) at (\w,0) {$1'$};
        \node [left=0pt] (p2) at (-\w,\h) {$2$};
        \node [right=0pt] (p3) at (\w,\h) {$2'$};
        \node [right=0pt] (p5) at (0,{3*\h/2}) {$k$};
        \draw [very thick] (p1) -- (p4);
        \draw [very thick,blue] (p2) -- (p3);
        \draw [graviton] (-\w/2,\h/2) -- (\w/2,\h/2) (-\w/2,\h/2+0.3) -- (\w/2,\h/2+0.3) (-\w/2,\h/2-0.3) -- (\w/2,\h/2-0.3);
        \draw [graviton] (-\w/2,\h) to[bend left=45] (p5.west);
        \filldraw [fill=gray!30!white,thick] (-\w/2,\h/2) ellipse (0.25cm and 1cm);
        \filldraw [fill=gray!30!white,thick] (\w/2,\h/2) ellipse (0.25cm and 1cm);
        \draw [line width=3pt,draw=white] (0,{3*\h/2 + 0.2}) -- (0,-\h/2 - 0.2);
        \draw [dashed,draw=red] (0,{3*\h/2 + 0.2}) -- (0,-\h/2 - 0.2);
        \node [fill=white,inner sep=0.5mm] at (0,\h/2) {$X$};
        \node [fill=white,inner sep=0.5mm] at (0,\h) {$r_2$};
        \node [fill=white,inner sep=0.5mm] at (0,0) {$r_1$};
        \node [below=0.25cm] at (-\w/2,0) {$\mathcal{M}$};
        \node [below=0.25cm] at (\w/2,0) {$\mathcal{M}^{*}$};
    \end{tikzpicture}}}\,.
\end{align}
We note that $\sum_{X}$ implicitly contains both the integration over the graviton phase spaces and the sum over polarizations, together with symmetry factors for identical particles.

The phase factor in \cref{iniMatrixElement} can be reorganized using the momentum-conserving delta functions. We introduce $q_i=p_i-p'_i$, which are related to the momentum of the outgoing graviton as $q_1+q_2=k$. The phase factor thus becomes $e^{iq_1\cdot b_1+iq_2\cdot b_2}=e^{iq_1\cdot (b_1-b_2)+ik\cdot b_2}$. The second term can be absorbed into the $e^{ikx}$ factor in \cref{Jdef} by redefining the position $x$. Since the $b_i$ are finite, this redefinition is irrelevant at large distances. We can thus choose the impact parameter to be $b = b_1-b_2$, which is the Fourier-conjugate of $q_1$.\footnote{Other choices are possible, but differ only by a phase whose argument is linear in $ k$.} Consequently, it is equivalent to performing the Fourier transform in \cref{iniMatrixElement} as
% \begin{align}
% \langle\psi_\text{in} |\opS^\dagger \hat a_{hh}(\bar k)S|\psi_\text{in}\rangle = \int [\phi d\Phi]
% e^{(p_1 - p_1')\cdot b }\;
% {}_\text{in}\langle p'_1p'_2 |\opS^\dagger \hat a_{hh}(k)S|p_1 p_2\rangle{}_\text{in} \ .
% \end{align}
\begin{align}\label{toInfinityX}
    \langle\psi_\text{in} |\opS^\dagger \hat a_{hh}(k)\opS|\psi_\text{in}\rangle=\int &\dd\Phi[p_1]\dd\Phi[p_2]\dd\Phi[p'_1]\dd\Phi[p'_2]\phi(p_1)\phi(p_2)\phi^{*}(p'_1)\phi^{*}(p'_2)\nonumber\\
    &\times \langle p'_1p'_2 |\opS^\dagger \hat a_{hh}(k)\opS|p_1 p_2\rangle e^{i(p_1-p'_1)\cdot b}\,.
\end{align}
Finally, the soft expansion in the classical limit, $q_i\ll p_i$, introduces substantial simplifications. After introducing $\pb_i\equiv p_i+(1/2)q_i$, we can rewrite the measure $\dd\Phi(p'_i)$ as
\begin{align}
    \dd\Phi(p'_i)
    =\frac{\dd^4 p'_i}{(2\pi)^4}(2\pi)\delta(p'_i{}^2-m_i^2)
    =\frac{\dd^4 q_i}{(2\pi)^4}(2\pi)\delta(2\pb_i\cdot q_i)\equiv\frac{\dd^4 q_i}{(2\pi)^4}\hdelta(2\pb_i\cdot q_i)\,,
\end{align}
where $\hdelta(x)\equiv 2\pi\delta(x)$, and we have used the fact that external physical particles always have positive energy. We further identify the initial and final momentum space wave packets in this limit, $\phi(p_i)\rightarrow\phi(\pb_i)$ and $\phi^{*}(p'_i)\rightarrow\phi^{*}(\pb_i)$.
Accounting for these features of the classical limit, \cref{toInfinityX} now takes its final form,
\begin{align}
\label{modMatrixElement}
& (-i)\langle\psi_\text{in} |\opS^\dagger \hat a_{hh}(k)\opS|\psi_\text{in}\rangle
= \int 
\dd\Phi[p_1] \dd\Phi[p_2]   |\phi({\bar p}_1)|^2|\phi({\bar p}_2)|^2 {\cal B}_{h}\,,
\\
&{\cal B}_{h}  \equiv \int \frac{\dd^dq_1}{(2\pi)^d} \frac{\dd^dq_2}{(2\pi)^d} 
{\hat \delta}(2{\bar p}_1\cdot q_1)
{\hat \delta}(2{\bar p}_2 \cdot q_2)
e^{ i q_1\cdot b}(-i)
\langle p_1 - q_1, p_2 - q_2 |\opS^\dagger \hat a_{hh}(k)\opS|p_1 p_2\rangle \ ,
\nonumber
\end{align}
where the matrix element $\langle p_1 - q_1, p_2 - q_2 |\opS^\dagger \hat a_{hh}(k)\opS|p_1 p_2\rangle$ should also be evaluated in the classical limit. The matrix element $\langle\psi_\text{in} |\opS^\dagger \hat a^\dagger_{hh}(k)\opS|\psi_\text{in}\rangle$ follows by complex conjugation.
The reason we included an explicit factor of $(-i)$ is to make $\mathcal{B}_h$ be given by amplitudes directly, which are defined as matrix elements of $\opT$, see \cref{SaS,matrixElement}.
We will now discuss some important properties of these matrix elements.

\subsection{On the structure of local (inclusive) observables}\label{kmoconeloop}

The matrix element $\langle p'_1 p'_2 |\opS^\dagger \hat a_{hh}(k)\opS|p_1 p_2\rangle$ determining the spectral  waveform is written out explicitly 
in terms of scattering amplitudes in \cref{SaS}.  
As discussed in Refs.~\cite{Herrmann:2021lqe, Herrmann:2021tct} the two different $\ie$ prescriptions that appear in the last term in \cref{SaS}
pose no difficulty to the evaluation of inclusive observables. For example, in the related KMOC calculation of inclusive observables~\cite{Herrmann:2021lqe, Herrmann:2021tct}, the terms bilinear in amplitudes 
were evaluated through reverse unitarity~\cite{Anastasiou:2002yz, Anastasiou:2002qz, Anastasiou:2003yy}. In the full quantum theory, we may simply construct the five-point virtual amplitude $\Mfull(p_1 p_2\rightarrow p'_1 p'_2 k^h)$ and then take an s-channel cut to find the second term in \cref{SaS}.\footnote{We note that here the term bilinear in amplitudes are literally the $s$-channel cut of the term linear in amplitudes without any additional dressing factors. This is a consequence of the structure of the observable, which only ``measures'' properties of outgoing particles and thus does not actively participate in the phase space integration. As another side note, we allow disconnected matrix elements to apppear across the $s$-channel cuts.}
However, such direct calculations can obscure additional simplifications as propagators with opposite $\ie$ conventions can exhibit nontrivial cancellations. In particular, we argue that the cancellation of superclassical terms is a consequence of such cancellations to all loop orders.

At one-loop, one can directly show, with a little effort, from \cref{SaS} that the {superclassical} contributions cancel in the classical waveform. The second term in \cref{SaS}, $\langle p'_1p'_2|\opT^{\dagger} \hat{a}_h(k)\opT|p_1p_2\rangle$, evaluates to the 2MPR contributions plus one-loop 3-particle cuts,
\begin{align}
    \vcenter{\hbox{\begin{tikzpicture}[scale=0.9]
        \draw [very thick, blue] (-\w/2,0) -- (-\w,\h/2) (-\w/2,0) to[bend left=60] (\w/2,0) (\w/2,0) -- (\w,\h/2) ;
        \draw [very thick] (-\w/2,0) -- (-\w,-\h/2) (-\w/2,0) to[bend right=60] (\w/2,0) (\w/2,0) -- (\w,-\h/2) ;
        \draw [graviton] (-\w/2,0) to[bend left=45] (0,\h/2);
        \filldraw [fill=gray!30!white,thick] (-\w/2,0) circle (0.25cm) (\w/2,0) circle (0.25cm);
        \draw [line width=3pt,draw=white] (0,\h/2+0.1) -- (0,-\h/2-0.1);
        \draw [dashed,draw=red] (0,\h/2+0.1) -- (0,-\h/2-0.1);
    \end{tikzpicture}}}+\left[
    \vcenter{\hbox{\begin{tikzpicture}[scale=0.9]
        \draw [very thick, blue] (0,\h/4) -- (-\w,\h/4);
        \draw [very thick, blue] (0,\h/4) to[bend left=25] (\w/2,0) (\w/2,0) -- (\w,\h/2) ;
        \draw [very thick] (0,-\h/4) -- (-\w,-\h/4);
        \draw [very thick] (0,-\h/4) to[bend right=25] (\w/2,0) (\w/2,0) -- (\w,-\h/2) ;
        \draw [graviton] (-\w/2,\h/4) to[bend left=45] (0,\h/2);
        \draw [graviton] (-\w/2,\h/4) to[bend right=20] (\w/2,0);
        \filldraw [fill=gray!30!white,thick] (-\w/2,\h/4) circle (0.25cm) (\w/2,0) circle (0.25cm);
        \draw [line width=3pt,draw=white] (0,\h/2+0.1) -- (0,-\h/2-0.1);
        \draw [dashed,draw=red] (0,\h/2+0.1) -- (0,-\h/2-0.1);
    \end{tikzpicture}}} + \vcenter{\hbox{
    \begin{tikzpicture}[scale=0.9]
        \draw [very thick, blue] (-\w/2,0) -- (-\w,\h/2) (-\w/2,0) to[bend left=25] (0,\h/4) -- (\w,\h/4) ;
        \draw [very thick] (-\w/2,0) -- (-\w,-\h/2) (-\w/2,0) to[bend right=25] (0,-\h/4) -- (\w,-\h/4) ;
        \draw [graviton] (-\w/2,0) to[bend left=45] (0,\h/2) ;
        \draw [graviton] (-\w/2,0) to[bend right=20] (\w/2,\h/4) ;
        \filldraw [fill=gray!30!white,thick] (-\w/2,0) circle (0.25cm) (\w/2,\h/4) circle (0.25cm);
        \draw [line width=3pt,draw=white] (0,\h/2+0.1) -- (0,-\h/2-0.1);
        \draw [dashed,draw=red] (0,\h/2+0.1) -- (0,-\h/2-0.1);
    \end{tikzpicture}}}+\text{up-down flip}\right]\,,
\end{align}
while the first diagram, being 2MPR and superclassical, manifestly cancels the corresponding contribution in $\langle p'_1p'_2 k^h| i\opT|p_1p_2\rangle$, the first term of \cref{SaS}. The diagrams in the bracket only have zero energy support, and they further subtract out certain $s$-channel cuts. 
We now demonstrate this feature at all loop. Namely, the classical waveform does not contain super-classical contributions.

To see this, it is convenient to use the exponential representation of the classical $S$-matrix found in Refs.~\cite{Bern:2021dqo,Damgaard:2021ipf}, which we now briefly review. Ref.~\cite{Bern:2021dqo} argued that, in the classical limit, the conservative 2-to-2 $S$-matrix has an 
exponential representation,
\begin{align}
S = e^{iI_r} 
\label{RadAction}
\end{align}
where the exponential is defined via its series expansion with the product being the integral over the two-matter-particle phase 
space. It was further argued in Ref.~\cite{Damgaard:2021ipf} that this exponential structure continues to hold as the exponent is promoted to an operator that has $2+n\rightarrow 2+m$ matrix elements, where $n$ and $m$ are initial and final state graviton emission. That is,  
\begin{align}
{\hat S} = e^{i \opN}  \equiv \sum_{n=0}^\infty  \frac{(i\opN)^n}{n!} \ , 
\label{expWrad}
\end{align}
where $\opN$ is a Hermitian operator and the product of operators is defined by inserting the identity operator in the complete Hilbert space of states, 
\begin{align}
\label{identityResolution}
    \mathbb{I}&=\underbrace{|r_1r_2\rangle\langle r_1r_2|}_{\hat{P}_{2,0}} + \underbrace{|r_1r_2k_1\rangle\langle r_1r_2k_1|}_{\hat{P}_{2,1}}+\underbrace{|r_1r_2k_1k_2\rangle\langle r_1r_2k_1k_2|}_{\hat{P}_{2,2}} + \ldots=\sum_{m=0}^{\infty}\hat{P}_{2,m}
\end{align}
where ${\hat P}_{2,m}$ is the identity operators in the $(2+m)$-particle Hilbert space. We can perturbatively expand $\hat{N}$ in $\kappa$,
\begin{align}
    \opN=\kappa^2 \opN_0^{} + \kappa^3 \opN_{0}^{\text{rad}} + \kappa^4 \opN_1^{} + \kappa^5 \opN_{1}^{\text{rad}} + \kappa^6 \opN_2^{} + \ldots 
\end{align}
where the subscript denotes the loop order, and the superscript signals the presence of graviton emission. We note that graviton emission also contributes to even orders of $\kappa$, which we suppress here for simplicity. We can perform a similar expansion for $\opT$ and solve $N_i$, the matrix element of $\opN_i$, order by order. For example, when restricted to two-particle initial and final states, we have~\cite{Damgaard:2021ipf},
\begin{align}\label{eq:Nmatrix}
    N_0 = \vcenter{\hbox{\begin{tikzpicture}
        \draw [very thick] (-150:1) -- (0,0) -- (-30:1);
        \draw [very thick,blue] (150:1) -- (0,0) -- (30:1);
        \filldraw [fill=gray!30!white,thick] (0,0) circle (0.25);
    \end{tikzpicture}}}\,,\qquad
    N_1 = \vcenter{\hbox{\begin{tikzpicture}
        \draw [very thick] (-150:1) -- (0,0) -- (-30:1);
        \draw [very thick,blue] (150:1) -- (0,0) -- (30:1);
        \filldraw [fill=gray!30!white,thick] (0,0) circle (0.5);
        \filldraw [fill=white,thick] (0,0) circle (0.25);
    \end{tikzpicture}}}-\frac{i}{2}\,\vcenter{\hbox{\begin{tikzpicture}
        \draw [very thick] (-150:1) -- (0,0);
        \draw [very thick,blue] (150:1) -- (0,0);      
        \draw [very thick] (1.25,0) -- ++(-30:1);
        \draw [very thick] (0,0) to[bend right=60] (1.25,0);
        \draw [very thick,blue] (0,0) to[bend left=60] (1.25,0);
        \draw [very thick,blue] (1.25,0) -- ++(30:1);
        \filldraw [fill=gray!30!white,thick] (0,0) circle (0.25);
        \filldraw [fill=gray!30!white,thick] (1.25,0) circle (0.25);
        \draw [red,dashed] (0.625,-0.5) -- (0.625,0.5);
    \end{tikzpicture}}}\,,\qquad\text{etc,}
\end{align}
where the blobs represent virtual amplitudes, which are matrix elements of $\opT$, and the cut propagator is integrated with measure $\dd\Phi[p]$.

We now substitute \cref{expWrad} into \cref{KMOCgeneral}, finding a sum of nested commutators,
\begin{align}\label{commutatorexpans}
\langle \opO\rangle&=\langle\psi_{\text{in}}|e^{-i\opN}[\opO,e^{i\opN}]|\psi_{\text{in}}\rangle=\langle \psi_{\text{in}}|\sum_{k=1}^{\infty}\frac{i^{k}}{k!}\underbrace{[[[[\opO,\opN],\opN],\ldots],\opN]}_{k \textrm{ times}}|\psi_{\text{in}}\rangle\,.
\end{align}
The projectors $P_{2,m}$ introduced in eq.~\eqref{identityResolution} contain the complete on-shell condition for the two matter fields. 
We consider first the leading order in the soft expansion of the cut matter propagator,
\begin{equation}\label{propexp}
\delta\big((p+\ell)^2-m^2\big) = \delta(2p\cdot\ell+\ell^{2})\sim \delta(2p\cdot\ell)+\mathcal{O}(\ell^{0}) \ .
\end{equation}
Here $\ell$ is a momentum of the same $\hbar$ order as the momentum transfer and it is usually a loop momentum. 
We denote by $P_{2,m}^{(0)}$ the projectors in \cref{identityResolution} but with only this linear constraint. All the super-classical terms come from such linearized projectors $P_{2,m}^{(0)}$ inserted into \cref{commutatorexpans}.
Further restricting ourselves to the waveform operator, or more generally those that only measure properties of particular external states, we find additional identities among the matrix elements in the classical limit when $\hat{P}_{2,0}^{(0)}$ is inserted,
\begin{align}\label{simplifyide}
    \langle\psi_{\text{in}}|[[[\opO,\opN],\ldots],\opN]\hat{P}_{2,0}^{(0)}\opN|\psi_{\text{in}}\rangle=\langle\psi_{\text{in}}|\opN\hat{P}_{2,0}^{(0)}[[[\mathcal{O},\opN],\ldots],\opN]|\psi_{\text{in}}\rangle\,.
\end{align}
This relation follows from reformulating \cref{irrelvanorder} in operator language, which states that the difference between the two matrix elements in \cref{simplifyide} is subleading in the momentum transfer $q$. It is crucial that we are considering an operator $\opO$ that does not participate in the integrals implied by the phase-space projector and from the linearity of the on-shell condition. Therefore, all insertions of $P^{(0)}_{2m}$, which corresponds to the super-classical terms, cancel in \cref{commutatorexpans}.
As a direct consequence of this result, only 2MPI diagrams and possibly 2MPR diagrams with derivatives of the linearized mass-shell conditions (subleading terms in \cref{propexp}) contribute to the classical waveform in the KMOC formalism\footnote{The importance of the latter contributions was incorrectly overlooked in earlier versions of this paper and was pointed out in \cite{Caron-Huot:2023vxl}.}. This generalizes the intuitive picture that the iteration part of the scattering amplitudes (often superclassical) should not contribute. We will evaluate the 2MPI part of the classical five-point amplitude at one loop order in \cref{sec:TheAmplitude,sec:N8res,integration}. At this order there are nontrivial contributions from diagrams with derivatives of the linearized mass-shell conditions, as was pointed out in \cite{Caron-Huot:2023vxl}. We will evaluate them in \cref{cutcontrib}.

\subsection{Infrared divergences of amplitudes in the classical limit}
\label{IRdivANDwaveform}

Five-point (and in general $n$-point) gravitational amplitudes are typically IR divergent. In cross-section calculations, some divergences exponentiate to a harmless total phase while others must be removed by summing over final state radiation \cite{Kinoshita:1962ur,Lee:1964is} or dressing the external states \cite{Dirac:1955uv,Kulish:1970ut,Grammer:1973db}. 
Some of these divergences are in the super-classical contributions, and thus cancel out in observables such as the waveform. It is important to understand the IR divergences of the surviving diagrams in the classical amplitude, e.g. \cref{spectralWF,modMatrixElement}, as these are relevant for classical observables. We discuss this here in the context of the classical amplitude and find that they factorize as an overall phase that can be safely absorbed into a linear re-definition of $t$ in \cref{eq:WFs}.
Similar treatment was first discussed in Refs.~\cite{Goldberger:2009qd, Porto:2012as} in the context of PN expansion. 
The classical 2MPR contributions to the waveform, originating in the bilinear-in-$\opT$ terms in \cref{matrixElement}, also exhibit IR divergences~\cite{Caron-Huot:2023vxl}; we will discuss them at one-loop order in Sec.~\ref{cutcontrib}.

Ref.~\cite{Weinberg:1965nx} famously showed that the virtual IR divergences of gravitational amplitudes come from loop-momentum integration regions in which a graviton $\lm$ connecting the external particles with momenta $p_a$ and $p_b$ becomes soft.
They factorize and exponentiate as~\cite{Weinberg:1965nx}
\begin{align}
\label{IRfactorization}
\frac{\Mfull(\alpha\rightarrow\beta)}{\Mfull^{0}(\alpha\rightarrow\beta)} &= 
\exp\Bigg[
4\pi G \sum_{a, b} \left((p_a\cdot p_b)^2-\frac{1}{2}m_a^2 m_b^2\right)
\eta_a\eta_b J_{ab}
\Bigg]\,,
\\
J_{ab} &=-i\mu^{2\epsilon}\int^{\Lambda}\frac{\dd^d\lm}{(2\pi)^d}\frac{1}{(\lm^2+\ie)(p_b\cdot\lm+i\eta_b0)(p_a\cdot\lm-i\eta_a0)} \ ,
\label{Jmn}
\end{align}
where $\mathcal{M}(\alpha\rightarrow\beta)$ is the all-order $\alpha\rightarrow\beta$ amplitude, and $\mathcal{M}^{0}(\alpha\rightarrow\beta)$ is its counterpart without the virtual soft gravitons. In the exponent, the factor $(p_a\cdot p_b)^2-\frac{1}{2}m_a^2m_b^2$ comes from the contraction of two stress-energy tensors with the numerator of the graviton propagator. The summation $\sum_{a,b}$ runs over all the unordered pairs $(a,b)$ of external particles and $\eta_{a}=\pm 1$ depending on whether $p_a$ is outgoing or incoming. We follow Ref.~\cite{Ware:2013zja} and dimensionally regularize $J_{ab}$ by using $d=4-2\epsilon$, $\mu$ is the scale of dimensional regularization, and $\Lambda$ is the cutoff $|\ell^2|< \Lambda^2$ that defines the virtual soft momenta.\footnote{These momenta should note be confused with the momenta in the soft region as defined in eq.~\eqref{rescaling1}. }

The IR-divergent integral $J_{ab}$ has both a real and an imaginary part~\cite{Weinberg:1965nx}.
One option to eliminate them and define IR-finite S-matrix elements is by choosing suitable asymptotic states \cite{Kulish:1970ut, Chung:1965zza, Ware:2013zja}. While we will not pursue this approach here, it would be interesting to understand if it can be realized by judiciously choosing the wave packets $\phi(p_i)$ in \cref{impactparameterspace}. 
Instead, we will show here that, in the classical limit, the 2MPI diagrams do not contribute to the real IR divergence. The remaining IR-divergent phase can be absorbed into the definition of the time variable of the waveform.

Following Ref.~\cite{Weinberg:1965nx}, let us consider the scattering of two massive particles (labeled as $1$ and $2$) with graviton emission in the final state. 
In the classical limit (that is, expanding in the soft region), all the matter propagators are linearized. The $\hbar$ counting further implies that, for a connected amplitude, there can be at most one graviton (labeled as $k$) in the final state which can be relevant to classical obserables~\cite{Cristofoli:2021jas, Britto:2021pud}.\footnote{We note that disconnected amplitudes at higher points may still contribute to KMOC-type observables on the support of zero graviton energy. However, as we will see later, such configurations are not relevant for the waveform.} In the following discussion, the ``virtual soft graviton'' is even softer than the soft region, namely, $\Lambda\ll|\bm q|$ and $|\bm k|$. Under this setup, the IR divergence receives contributions from the following three configurations shown in \cref{fig:IR}:

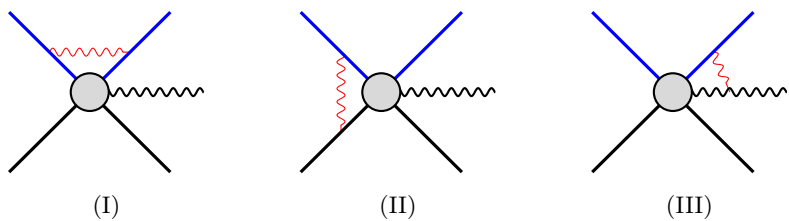
\begin{figure}
    \pgfmathsetmacro{\r}{1.5}
    \pgfmathsetmacro{\c}{0.25}
    \pgfmathsetmacro{\shift}{0.2}
    \pgfmathsetmacro{\y}{1}
    \pgfmathsetmacro{\l}{0.5}
    \centering
    \renewcommand{\thesubfigure}{\Roman{subfigure}}
    \subfloat[]{\label{fig:IR1}
    \begin{tikzpicture}
    \coordinate (o) at (0,0);
    \node (p3) at (45:\r) {};
    \node (p4) at (-45:\r) {};
    \node (p2) at (135:\r) {};
    \node (p1) at (-135:\r) {};
    \node (p5) at (0:\r) {};
    \draw [graviton,red] (135:\r/2) -- (45:\r/2);
    \draw [very thick] (p1.center) -- (o) -- (p4.center);
    \draw [very thick,blue] (p2.center) -- (o) -- (p3.center);
    \draw [graviton,thick] (o) -- (p5.center);
    \filldraw [fill=gray!30!white,thick] (o) circle (\c);
    \end{tikzpicture}
    }\qquad
    \subfloat[]{\label{fig:IR2}
    \begin{tikzpicture}
    \coordinate (o) at (0,0);
    \node (p3) at (45:\r) {};
    \node (p4) at (-45:\r) {};
    \node (p2) at (135:\r) {};
    \node (p1) at (-135:\r) {};
    \node (p5) at (0:\r) {};
    \draw [graviton,red] (135:\r/2) -- (-135:\r/2);
    \draw [very thick] (p1.center) -- (o) -- (p4.center);
    \draw [very thick,blue] (p2.center) -- (o) -- (p3.center);
    \draw [graviton,thick] (o) -- (p5.center);
    \filldraw [fill=gray!30!white,thick] (o) circle (\c);
    \end{tikzpicture}
    }\qquad
    \subfloat[]{\label{fig:IR3}
    \begin{tikzpicture}
    \coordinate (o) at (0,0);
    \node (p3) at (45:\r) {};
    \node (p4) at (-45:\r) {};
    \node (p2) at (135:\r) {};
    \node (p1) at (-135:\r) {};
    \node (p5) at (0:\r) {};
    \draw [graviton,red] (45:\r/2) -- (\r/2,0);
    \draw [very thick] (p1.center) -- (o) -- (p4.center);
    \draw [very thick,blue] (p2.center) -- (o) -- (p3.center);
    \draw [graviton,thick] (o) -- (p5.center);
    \filldraw [fill=gray!30!white,thick] (o) circle (\c);
    \end{tikzpicture}
    }
    \caption{Typical contribution to the IR divergence. The soft graviton is shown in red. }
    \label{fig:IR}
\end{figure}

\paragraph{\protect\subref{fig:IR1} The virtual soft graviton starting and ending on the same particle:}
For such a configuration, the soft expansion implies that the corresponding loop integral is either scaleless or its propagators are linearly-dependent. The former has the topology of an external bubble, which  integrates to zero in dimensional regularization.\footnote{A regulator, corrresponding e.g. to an external particle being slightly off shell, may be required to prevent matter propagators from being on shell and to allow the determination of contributions from graphs with the topology of bubbles on external lines. In the complete amplitude the potentially-singular propagator cancels out and the regulator can be removed explicitly. While the bubble integral is not scaleless in the presence of the regulator, it is so -- and thus vanishes in dimensional regularization -- after the regulator is removed, so we may take it to vanish from the outset. } The latter requires partial fractioning~\cite{Herrmann:2021tct, Bern:2021yeh} and the resulting integrals are either scaleless or finite. Thus, this graviton configuration does not lead to an IR divergence in the classical limit.

\paragraph{\protect\subref{fig:IR2} The virtual soft graviton connecting different massive particles:}
In the classical amplitudes, the incoming and outgoing momenta of the same massive particle are actually equal. This is because, in the full amplitude, their difference is quantum, and the soft expansion homogenizes the $\hbar$ counting in each diagram. As a result, the exponent in \cref{IRfactorization} becomes
\begin{align}
    &\sum_{a, b} \left((p_a\cdot p_b)^2-\frac{1}{2}m_a^2 m_b^2\right)
\eta_a\eta_b J_{ab}\Bigg|_{a\text{ and }b\text{ are different massive particles}}\nonumber\\
    &=-i\big(2(p_1\cdot p_2)^2-m_1^2m_2^2\big)\mu^{2\epsilon}\sum_{\eta_{a,b}=\pm 1}\int^{\Lambda}\frac{\dd^d\lm}{(2\pi)^d}\frac{\eta_a\eta_b}{(\lm^2+\ie)(p_b\cdot\lm+i\eta_b0)(p_a\cdot\lm-i\eta_a0)}\nonumber\\
    &=i\big(2(p_1\cdot p_2)^2-m_1^2m_2^2\big)\mu^{2\epsilon}\int^{\Lambda}\frac{\dd^d\lm}{(2\pi)^d}\frac{(4\pi)^2\delta(p_1\cdot\lm)\delta(p_2\cdot\lm)}{\lm^2+\ie}\,.
\end{align}
This term contributes a real IR divergence, but it belongs to the 2MPR part of the amplitude. Therefore, this configuration does not contribute to the IR divergence of the 2MPI amplitude. This also shows that in the $2\rightarrow 2$ scatterings, the IR divergence is real and captured by 2MPR diagrams, while the 2MPI contributions are finite.

\paragraph{\protect\subref{fig:IR3} The virtual soft graviton connecting a massive and a massless particle:}
For this configuration, the exponent of \cref{IRfactorization} is given by
\begin{align}\label{eq:Jint}
    &\sum_{a, b} \left((p_a\cdot p_b)^2-\frac{1}{2}m_a^2 m_b^2\right)
\eta_a\eta_b J_{ab}\Bigg|_{(a,b)\in(\text{one massless, one massive})}\\
    &=-i\mu^{2\epsilon} \sum_{b=1,2}\int^{\Lambda}\frac{\dd^d\lm}{(2\pi)^d}\left[\frac{2(p_b\cdot k)^2}{(\lm^2+\ie)(p_b\cdot\lm+\ie)(k\cdot\lm-\ie)}-\frac{2(p_b\cdot k)^2}{(\lm^2+\ie)(p_b\cdot\lm-\ie)(k\cdot\lm-\ie)}\right].\nonumber
\end{align}
The evaluation proceeds by first integrating over $\lm^0$, following~\cite{Weinberg:1965nx, Ware:2013zja}. For the first integral in \cref{eq:Jint}, we close the $\lm^{0}$ contour from above, picking up the pole $\lm^0=-|\pmb{\lm}|+\ie$ and $\lm^0=\frac{1}{\omega}\pmb{k}\cdot\pmb{\lm}+\ie$. For the second integral, we close the contour from below, picking up the pole $\lm^0=|\pmb{\lm}|-\ie$. The contribution from $-|\pmb{\lm}|+\ie$ and $|\pmb{\lm}|-\ie$ cancel each other, leaving only an imaginary IR divergence from the pole $\lm^0=\frac{1}{\omega}\pmb{k}\cdot\pmb{\lm}+\ie$,
\begin{align}
    &-i\mu^{2\epsilon} \sum_{b=1,2}\int^{\Lambda}\frac{\dd^d\lm}{(2\pi)^d}\left[\frac{2(p_b\cdot k)^2}{(\lm^2+\ie)(p_b\cdot\lm+\ie)(k\cdot\lm-\ie)}-\frac{2(p_b\cdot k)^2}{(\lm^2+\ie)(p_b\cdot\lm-\ie)(k\cdot\lm-\ie)}\right]\nonumber\\
    %&=-\frac{i}{4\pi\epsilon}(p_1\cdot k+p_2\cdot k)+\text{IR finite terms}\,,
    &=-\frac{i}{4\pi}(p_1\cdot k+p_2\cdot k)\left[\frac{1}{\epsilon}-\log\frac{\Lambda^2}{\mu^2}+\mathcal{O}(\epsilon)\right]\,,
\end{align}
where we have used the fact that $p_b\cdot k>0$ for physical processes. If there are more external gravitons, then the internal soft graviton connecting two external gravitons may also contribute to the IR divergence. However, as we have discussed before, amplitudes that are relevant to classical physics can only contain at most one external graviton~\cite{Cristofoli:2021jas, Britto:2021pud}.

Therefore, we have shown that for classical 2MPI amplitudes, the IR divergence is purely imaginary. For the four-scalar-one-graviton amplitude, it is a pure phase
\begin{align}
\label{IRdivGeneral}
\Mfull^{\text{2MPI}, \text{cl.}}_5(p_1 p_2\rightarrow p_1' p_2' k) &= 
\exp\left[ - 
i G (p_1\cdot k + p_2\cdot k)\left(\frac{1}{\epsilon}-\log\frac{\Lambda^2}{\mu^2}\right)\right]\nonumber\\
&\quad\times\Mfull^{0, \text{2MPI}, \text{cl.}}_5(p_1 p_2\rightarrow p_1' p_2' k) \ .
\end{align}
To leading order in Newton's constant $\Mfull^{0, \text{2MPI}, \text{cl.}}(p_1p_2\rightarrow p_1'p_2' k)$ is the tree-level five-point amplitude. We will indeed verify the one-loop part of this relation in \cref{integratedamplitudes}.

The five-point amplitude corresponds to the $\langle p'_1p'_2k^h|i\opT|p_1p_2\rangle$ part of the waveform. 
We can now understand the fate of the IR-divergent phase in the 
amplitude contribution to the waveform.
Indeed, after the cancellation of the super-classical part of the five-point amplitude by the bilinear-in-$\opT$ contribution $\langle p'_1p'_2|\opT^{\dagger} \hat{a}_{hh}(k)\opT|p_1p_2\rangle$, the remaining 2MPI part, and consequently the amplitude contribution to the spectral waveform~\eqref{just_spectralWF}, have the same IR-divergent phase. 
Using the solution for the on-shell condition in \cref{onshellgraviton} for the outgoing graviton, we can write its argument as
\begin{align}
-G (p_1\cdot k + p_2\cdot k)\left(\frac{1}{\epsilon}-\log\frac{\Lambda^2}{\mu^2}\right) = - G
\big(E_1+E_2-(\bm p_1+\bm p_2)\cdot \bm{n}_{\bm{k}}\big)\omega\left(\frac{1}{\epsilon}-\log\frac{\Lambda^2}{\mu^2}\right)\, ,
\end{align}
which may be removed by defining the observation time as
\begin{equation}
\tau = t + G \big(E_1+E_2-(\bm p_1+\bm p_2)\cdot \bm n\big)
%\left(1+\frac{\sigma(\sigma^2-Y)}{(\sigma^2-1)^{3/2}}\right)
\left(\frac{1}{\epsilon}-\log\frac{\Lambda^2}{\mu^2}\right)\,,
\label{tredef}
\end{equation}
where $t$ is the retarded time first defined in \cref{eq:ret_adv}. We thus conclude that the IR-divergent phase of 
the amplitude contribution 
can be removed by choosing a suitable origin of the observation time or, alternatively, focusing on observation-time intervals.
Similar IR divergences appear in the far-zone EFT; in the PN expansion they were discussed in Refs.~\cite{Goldberger:2009qd, Porto:2012as} where they were also absorbed in the definition of the retarded time. 

We note that the bilinear-in-$\opT$ matrix element $\langle p'_1p'_2|\opT^{\dagger} \hat{a}_h(k)\opT|p_1p_2\rangle$ also contributes a similar IR divergence in the classical limit, which can be removed by a similar time shift~\cite{Caron-Huot:2023vxl}. Schematically, the full IR divergence of the matrix element $\langle p'_1p'_2|\opS^{\dagger}\hat{a}_{hh}(k)\opS|p_1p_2\rangle$ is
\begin{align}\label{eq:waveformIR}
    \langle p'_1p'_2|\opS^{\dagger}\hat{a}_{hh}(k)\opS|p_1p_2\rangle &= \exp\left[ - 
    i G (p_1\cdot k + p_2\cdot k)\left(1+\phi(p_1,p_2)\right)\left(\frac{1}{\epsilon}-\log\frac{\Lambda^2}{\mu^2}\right)\right]\nonumber\\
    &\quad\times\langle p'_1p'_2|\opS^{\dagger}\hat{a}_{hh}(k)\opS|p_1p_2\rangle^{0}\,,
\end{align}
where the superscript $0$ stands for the IR finiteness. The additional phase $\phi(p_1,p_2)$ comes from the $\opT$-bilinear part, and we will study it in more detail at one loop in \cref{cutcontrib}.

\subsection{Waveforms from KMOC: summary and further comments \label{KMOCloopSummary}}

We will discuss the calculation of waveforms 
at leading order and next-to-leading order by applying this formalism to classical ${\cal N}=8$ supergravity and GR in \cref{sec:Waveforms,cutcontrib}.
To facilitate this application, we collect here the relevant formulae and further organize them, using the properties of amplitudes in the classical limit to streamline their connection to waveforms in the time domain.

The spectral waveform (or the frequency-space curvature tensor), the frequency-space Newman-Penrose scalar, and the frequency-space metric in a transverse-traceless gauge are given by \cref{just_spectralWF,NPscalar,metric_infinity_omegaspace}. After the IR divergence is absorbed in the definition of time, they become
\begin{align}
\label{fX}
f_{\mu\nu\rho\sigma}(\omega, \bm n)&=
 \frac{ \kappa}{8\pi}  \sum_{h = \pm} \Big[
 \;  \Theta(\omega) \, k_{[\mu}^{\vphantom{*}} \varepsilon_{\nu],h}(k)  \, k_{[\rho}^{\vphantom{*}} \varepsilon_{\sigma],h}^{*}(k) \, 
 (-i)\langle\psi_\text{in} |\opS^\dagger \hat a_{hh}(k)\opS|\psi_\text{in}\rangle^{0} \big|_{k=(\omega, \omega\bm n)} 
 \nonumber\\
& \quad
+\Theta(-\omega) k_{[\mu} \varepsilon_{\nu],h}(k) k_{[\rho}\varepsilon_{\sigma],h}(k)
(+i)\langle\psi_\text{in} |\opS^\dagger \hat a^\dagger_{hh}(k)\opS|\psi_\text{in}\rangle^{0}\big|_{ k=(|\omega|, |\omega| \bm n)}  \Big]\,,
 \\
 \label{PsiX}
 {\widetilde \Psi}_4^\infty (\omega, {\bm n})&=
 -\frac{\kappa}{8\pi} \, \omega^2 (\tilde {h}^{\infty}_{+} + i  \tilde {h}^{\infty}_\times)\,,
\\
\label{hX}
\tilde h^{\infty}_{+} + i  \tilde {h}^{\infty}_\times & =   \Big[\Theta(\omega) \;(-i)
\langle \psi_\text{in}|\opS^\dagger {\hat a}_{--}(k)\opS|\psi_\text{in}\rangle^{0}\big|_{k=(\omega, \omega\bm n)} 
\\
& 
\quad+  \Theta(-\omega) \;(+i)
\langle \psi_\text{in}|\opS^\dagger {\hat a}^\dagger_{++}(k)\opS|\psi_\text{in}\rangle^{0}\big|_{k=(|\omega|, |\omega|\bm n)}
\Big] + {\rm M} \delta(\omega)+ {\rm S} \delta'(\omega)\ ,
\nonumber
\end{align}
where the superscript $0$ indicates that the IR divergences have been absorbed in the 
definition of the observation time $\tau$, see \cref{tredef}.
The term  proportional to $\delta'(\omega)$ is a gauge degree of freedom, and we will ignore it in the following.
The remaining $\delta(\omega)$ term in \cref{hX} can be present due to specific contributions that only have support on zero graviton energy and will, at most, lead to a time-independent background that the initial condition will fix. Additionally, such terms are irrelevant specifically for the evaluation of the asymptotic Newman-Penrose scalar $\widetilde{\Psi}_4^{\infty}$ because of the additional factor of $\omega^2$ in \cref{PsiX}.
The time-domain observables follow from the Fourier-transform in \cref{spectralWF},
\begin{align}
\label{omegaFT}
{\cal F}(\tau) = \int_{-\infty}^{+\infty} \frac{d\omega}{2\pi} \widetilde {\cal F}(\omega) e^{-i\omega \tau} \ ,\quad {\cal F}\in\{f_{\mu\nu\rho\sigma}, {\widetilde \Psi}_4^\infty,\tilde h_{+} + i  \tilde h_\times\}\,.
\end{align}
The matrix element $(-i)
\langle \psi_\text{in}|\opS^\dagger {\hat a}_{hh}(k)\opS|\psi_\text{in}\rangle$
and its conjugate $(+i)
\langle \psi_\text{in}|\opS^\dagger {\hat a}^\dagger_{hh}(k)\opS|\psi_\text{in}\rangle$
are then computed using \cref{modMatrixElement},
\begin{align}
\label{modMatrixElementX}
& (-i)\langle\psi_\text{in} |\opS^\dagger \hat a_{hh}(k)\opS|\psi_\text{in}\rangle^{0}
= \int 
\dd\Phi[p_1] \dd\Phi[p_2]   |\phi({\bar p}_1)|^2|\phi({\bar p}_2)|^2 {\cal B}_h
\\
&{\cal B}_h  = \int \frac{\dd^dq_1}{(2\pi)^d} \frac{\dd^dq_2}{(2\pi)^d} 
{\hat \delta}(2{\bar p}_1\cdot q_1)
{\hat \delta}(2{\bar p}_2 \cdot q_2)
e^{ i q_1\cdot b}\;(-i)
\langle p_1 - q_1, p_2 - q_2 |\opS^\dagger \hat a_{hh}(k)\opS|p_1 p_2\rangle^{0}
\nonumber
\\[3pt]
& (+i)\langle\psi_\text{in} |\opS^\dagger \hat a^\dagger_{hh}( k)\opS|\psi_\text{in}\rangle^{0}
= \left( (-i)\langle\psi_\text{in} |\opS^\dagger \hat a_{hh}( k)\opS|\psi_\text{in}\rangle^{0} \right)^* \ ,
\label{MEconjugate}
\end{align}
where, as for frequency-domain observables, the superscript $0$ indicates that the IR divergent phase has been removed from the matrix element and absorbed into the definition of the observation time.
The gravitational memory of the observable $\mathcal{F}$ is defined as the difference between its initial and final values,
\begin{align}\label{eq:memory}
    \Delta\mathcal{F}\equiv\mathcal{F}(+\infty)-\mathcal{F}(-\infty)=\int_{-\infty}^{+\infty}\dd\omega(-i\omega)\widetilde{\mathcal{F}}(\omega)\delta(\omega)\,,
\end{align}
which can be derived by integrating the derivative of $\mathcal{F}$ between $\tau=\pm\infty$. Therefore, the memory is determined solely by the residue of $\widetilde{\mathcal{F}}$ at zero frequency.

To evaluate frequency-domain observables, it is necessary to evaluate the $q_1$ and $q_2$ integrals in \cref{modMatrixElementX}; only one of them is nontrivial because of the momentum-conserving constraint $q_1+q_2=k$ implicit in $\langle p_1-q_1, p_2-q_2 |\opS^\dagger \hat a_{hh}(k)\opS|p_1 p_2\rangle$. The two explicit delta functions, as well as the phase factor, suggest that it is convenient to decompose the integration variable into components along ${\bar u}_1$, ${\bar u}_2$, $b$, and a fourth vector orthogonal on these, as described in Ref.~\cite{Cristofoli:2021vyo}.

To evaluate classical time-domain observables, it is convenient first to evaluate the $\omega$ integral because it localizes parts of the remaining integrals. To expose this structure, we use properties of classical amplitudes -- and thus of the matrix elements $\langle\psi_\text{in} |\opS^\dagger \hat a_{hh}(k)\opS|\psi_\text{in}\rangle$ -- under the soft-region rescaling in eq.~\eqref{rescaling1}, 
\begin{align}
\label{pulloutOmega}
&\langle p_1-q_1, p_2-q_2 |\opS^\dagger \hat a_{hh}(k)\opS|p_1 p_2\rangle^{0}
\big|_{L\text{ loops}}\\
& \qquad = 
\lambda^{2-L+d} \langle p_1-\lambda q_1, p_2-\lambda q_2 |\opS^\dagger \hat a_{hh}(\lambda k)\opS|p_1 p_2\rangle^{0}\big|_{L\text{ loops}} \,,
\nonumber
\end{align}
where $\lambda^{2-L}$ comes from the scaling of the $L$-loop classical amplitude, and $\lambda^{d}$ comes from the momentum conserving delta function $\delta^{(d)}(q_1+q_2-k)$ implicit in the matrix element.
Choosing $\lambda = \omega^{-1}$ and changing integration variables 
$q_i = \omega \tilde q_i$ we may therefore isolate the $\omega$ dependence 
of ${\cal B}$ to the Fourier phase and overall factors:
\begin{align}
{\cal B}_h(k) &= \sum_{L\ge 0} {\cal B}_{h, L}(k)=
\sum_{L\ge 0}\omega^{L+d-4} \widetilde {\cal B}_{h, L}(\tilde k) 
\\
\widetilde {\cal B}_{h, L} &=
\int \frac{\dd^d \tilde q_1}{(2\pi)^d} \frac{\dd^d \tilde q_2}{(2\pi)^d} 
{\hat \delta}(2{\bar p}_1\cdot \tilde q_1)
{\hat \delta}(2{\bar p}_2 \cdot \tilde q_2)
e^{ i \omega \tilde q_1\cdot b}(\real{L} + i\, \imaginary{L})\nonumber\\
\real{L} + i\,\imaginary{L}&=
(-i)
\langle p_1-\tilde q_1, p_2-\tilde q_2 |\opS^\dagger \hat a_{hh}(\tilde k)\opS|p_1 p_2\rangle^{0} \big|_{L\text{ loops}} \ ,
\label{ReImSplit}
\end{align}
where $\tilde k = k/\omega=(1,\bm n)$. The tildes on $q_i$ can now be dropped as they are dummy integration variables. $\real{L}$ and $\imaginary{L}$ are the real and imaginary parts of the matrox element $(-i)
\langle p_1-\tilde q_1, p_2-\tilde q_2 |\opS^\dagger \hat a_{hh}(\tilde k)\opS|p_1 p_2\rangle^{0} \big|_{L\text{ loops}}$ {\em with the polarization tensors stripped off.} 
We note here that $\real{L}$ receives contributions only from the virtual five-point amplitude, while $\imaginary{L}$ receives contributions from both the virtual five-point amplitude and the bilinear-in-$\opT$ (cut) terms of this matrix element.
In the following, we will simply refer to $\real{L}$ and $\imaginary{L}$ as the real and imaginary parts of the $L$-loop matrix elements. 
In the conjugate matrix element, the relation between $q_i$ and ${\tilde q}_i$ is $q_i = (-\omega) {\tilde q}_i$ because $\Theta(-\omega)$ localizes the integrals to the domain $(-\omega)>0$. 

We can now explore the structure of \cref{omegaFT} given integrands of the form in \cref{fX,PsiX,hX}. Since all IR divergences have been removed, we may set $d=4$. The relevant integral to compute the time-domain observables at $L$-loop order is
\begin{align}
\label{typical}
\mathcal{J}_{n,L} &\equiv \int_{-\infty}^{+\infty} \frac{d\omega}{2\pi} e^{-i\omega(\tau- q_1\cdot b)} \omega^{2n}
\left[\Theta(\omega) \omega^{L} (\real{L} + i\, \imaginary{L}) + \Theta(-\omega) (-\omega)^L(\real{L} - i\, \imaginary{L})\right]
\\
&=\frac{i^{L+2n+1}}{2\pi}\Gamma(L+2n+1) \; \real{L}  
\left[\frac{1}{(\tau-{q}_1\cdot b + i 0)^{L+2n+1}} - \frac{(-1)^{L+2n}}{(\tau-{q}_1\cdot b - i 0)^{L+2n+1}}\right]
\nonumber\\
&
\quad -\frac{i^{L+2n+2}}{2\pi}\Gamma(L+2n+1) \; \imaginary{L}  \left[\frac{1}{(\tau-{q}_1\cdot b + i 0)^{L+2n+1}} + \frac{(-1)^{L+2n}}{(\tau-{q}_1\cdot b - i 0)^{L+2n+1}}\right] \,,\nonumber
\end{align}
where the additional factor of $\omega^{2n}$ accounts for and generalizes such factors in \cref{fX,PsiX}. 
For now we keep the exponent $n$ to be a real number. As we will see in 
sections \ref{integratedamplitudes} and \ref{sec:Waveforms}, the amplitude contains a logarithmic dependence on 
$\omega$; we may find the relevant Fourier transform by simply differentiating with respect to $n$. This logarithmic dependence  
on the outgoing-graviton frequency yields the so-called gravitational-wave tail first studied in~\cite{Blanchet:1992br,Blanchet:1993ec} and represents the effect of the scattering of the leading-order gravitational wave off the gravitational field of the source.
For the integer part of the exponent, depending on the parity of the loop order $L$ the contribution to the waveform from the real or imaginary part of the matrix element \eqref{ReImSplit}
localize because:
\begin{align}
\frac{1}{(x + i 0)^{s}} 
- \frac{1}{(x - i 0)^{s}}&=-2\pi i (-1)^{s-1} \delta^{(s-1)}(x)\,, 
\nonumber\\
\frac{1}{(x + i 0)^{s}} 
+ \frac{1}{(x - i 0)^{s}}&=2(-1)^{s-1} {\rm PV}^{(s-1)}\left[\frac{1}{x}\right] \ .
\end{align}
%%%%%%%%%%%%%%%%%%%%%%%%%
For example, the localization implied by the delta function occurs at tree level for the real part of the matrix element~\eqref{ReImSplit} ($L=0$ and $n=1$), where $\real{L=0}$ is just the tree-level classical five-point amplitude.
It also occurs at one loop for the imaginary part of the matrix element \eqref{ReImSplit} ($L=1$ and $n=1$), where $\imaginary{L=1}$ is sum of the imaginary part of the classical one-loop five-point amplitude with subtracted IR divergences and a bilinear-in-$\opT$ (cut) contribution. 
In terms of $\mathcal{J}_{n,L}$, the time-domain Newman-Penrose scalar and waveform have very compact expressions,
\begin{align}\label{eq:t-obs}
    \Psi_4^{\infty}&=-\frac{\kappa}{8\pi}\sum_{L\geq 0}\int\frac{\dd^d q_1}{(2\pi)^d}\frac{\dd^d q_2}{(2\pi)^d}\hdelta(2\pb_1\cdot q_1)\hdelta(2\pb_2\cdot q_2)\mathcal{J}_{2,L}\,,\nonumber\\
    h_{+}^{\infty}+i h_{\times}^{\infty}&=\sum_{L\geq 0}\int\frac{\dd^d q_1}{(2\pi)^d}\frac{\dd^d q_2}{(2\pi)^d}\hdelta(2\pb_1\cdot q_1)\hdelta(2\pb_2\cdot q_2)\mathcal{J}_{0,L}\,,
\end{align}
where we have assumed that the wavepackets for the massive states are highly localized.

Let us now spell out the ingredients necessary for the evaluation of the waveform $h_{+}^{\infty}+i h_{\times}^{\infty}$ at leading order, ${\cal O}(\kappa^3)$, and next-to-leading order, ${\cal O}(\kappa^5)$. At ${\cal O}(\kappa^3)$, the matrix element determining ${\cal B}_h$ in \cref{modMatrixElementX} is simply the tree-level $2\rightarrow 3$ amplitude evaluated at $\tilde{k}=(1,\bm n)$,
\begin{align}\label{eq:LOme}
(-i)
\langle p_1- q_1, p_2- q_2 |\opS^\dagger \hat a_{hh}(\tilde k)\opS|p_1 p_2\rangle^{0} \Big|_{\text{tree}}&=
\Mfull^\text{\text{cl.}}_{5, \text{tree}}(p_1p_2\rightarrow p_1-q_1,p_2-q_2,\tilde{k}^h) \ ,
\end{align}
where we do not decorate the right-hand side with the ``2MPI'' designation because it is irrelevant at tree level. Thus we have $\mathcal{R}_{L=0}=\Mfull^{\text{cl.}}_{5,\text{tree}}$ and $\mathcal{I}_{L=0}=0$ at the leading order.

At $\mathcal{O}(\kappa^5)$, we separate the matrix element determing $\mathcal{B}_h$ in \cref{modMatrixElementX} into three parts:
\begin{align}\label{one_and_two}
    (-i)\langle p_1{-} q_1, p_2{-} q_2 |\opS^\dagger \hat a_{hh}(\tilde k)\opS|p_1 p_2\rangle^{0}\Big|_{1\text{ loop}} =  \mathcal{W}_{\text{amp}}^{\text{1 loop}} + \mathcal{W}_{\text{cut}}^{\text{1 loop}} + \mathcal{W}_{\text{disc}}^{\text{1 loop}}\,.
\end{align}
For the first part, we have
\begin{align}
    \mathcal{W}_{\text{amp}}^{\text{1 loop}} = \Mfull^{0,\text{2MPI}, \text{cl.}}_{5, \text{1 loop}} (p_1p_2\rightarrow p_1-q_1,p_2-q_2,k^h)\,,
\end{align}
which is just the 2MPI part of the one-loop five-point amplitude in which the IR-divergent contribution due to soft virtual gravitons has been removed per \cref{tredef}. The second part, $\mathcal{W}_{\text{cut}}^{\text{1 loop}}$, denotes the classical and \emph{connected} part of the bilinear-in-$\opT$ matrix element $(-i)\langle p_1{-}q_1,p_2{-}q_2|\opT^{\dagger}\hat{a}_{hh}(k)\opT|p_1p_2 \rangle$ with the IR divergence removed per \cref{tredef}. We note that this contribution is purely imaginary, and postpone a more detailed discussion to \cref{cutcontrib}. The third part is the \emph{disconnected} contribution of the bilinear-in-$\opT$ matrix element. It is given by 
\begin{align}
    \mathcal{W}_{\text{disc}}^{\text{1 loop}} &= -i  \int  \dd\Phi[r_1]\dd\Phi[r_2]\dd\Phi[\lm]\nonumber\\
    &\qquad\quad\times\sum_{a=\pm}\Mfull^{\text{cl.}\,*}_{5, \text{tree}}(p_1 {-} q_1, p_2 {-} q_2\rightarrow r_1 r_2 \lm^{a})\Mfull^{\text{disc.}}_{6, \text{tree}}(p_1 p_2\rightarrow r_1 r_2 k^h \lm^{a}) 
\end{align}
%
%%%%%%%%%%%%%%
%
%
%
%%%%%%%%%%%%%%
%
where $\lm$ here is a single graviton whose polarization $a$ is summed over. We only keep terms at $\mathcal{O}(\kappa^5)$ order. It consists of a five-point classical amplitude $\Mfull^{\text{cl.} \, *}_{5, \text{tree}}$, which contributes at $\mathcal{O}(\kappa^3)$, and the disconnected pieces of the six-point amplitude $\Mfull^{\text{disc.}}_{6, \text{tree}}$, which contributes at $\mathcal{O}(\kappa^2)$. Diagrammatically, these cut terms are
\begin{align}
\vcenter{\hbox{
    \begin{tikzpicture}
        \draw [very thick, blue] (-\w,\h/4) -- (0,\h/4) to[bend left=20] (\w/2,0) -- (\w,\h/2);
        \draw [very thick] (-\w,-\h/4) -- (0,-\h/4) to[bend right=20] (\w/2,0) -- (\w,-\h/2) ;
        \draw [graviton] (-\w/2,\h/4) to[bend left=45] (0,\h/2) ;
        \draw [graviton] (-\w/2,-\h/4) to[bend left=20] (\w/2,0) ;
        \filldraw [fill=gray!30!white,thick] (-\w/2,\h/4) circle (0.25cm) (-\w/2,-\h/4) circle (0.25cm) (\w/2,0) circle (0.25cm);
        \draw [line width=3pt,draw=white] (0,\h/2+0.1) -- (0,-\h/2-0.1);
        \draw [dashed,draw=red] (0,\h/2+0.1) -- (0,-\h/2-0.1);
    \end{tikzpicture}}}    
+
\vcenter{\hbox{
    \begin{tikzpicture}
        \draw [very thick, blue] (-\w,\h/4) -- (0,\h/4) to[bend left=20] (\w/2,0) -- (\w,\h/2);
        \draw [very thick] (-\w,-\h/4) -- (0,-\h/4) to[bend right=20] (\w/2,0) -- (\w,-\h/2) ;
        \draw [graviton] (-\w/2,\h/4) to[bend left=45] (0,\h/2) ;
        \draw [graviton] (-\w/2,\h/4) to[bend right=20] (\w/2,0) ;
        \filldraw [fill=gray!30!white,thick] (-\w/2,\h/4) circle (0.25cm) (\w/2,0) circle (0.25cm);
        \draw [line width=3pt,draw=white] (0,\h/2+0.1) -- (0,-\h/2-0.1);
        \draw [dashed,draw=red] (0,\h/2+0.1) -- (0,-\h/2-0.1);
    \end{tikzpicture}}}    
+\text{up-down flip}\,.    
\end{align}
It is not difficult to see that such cuts are kinematically forbidden unless the two outgoing gravitons have zero energy. They would at most contribute to the $\delta(\omega)$ terms in the metric~\eqref{hX}, which corresponds to a time-independent background that can be subtracted. The factors of graviton energy $\omega$ in \cref{PsiX} imply that such configurations do not contribute to the Newman-Penrose scalar and the spectral waveform.
Thus, for our calculation we may neglect $\mathcal{W}_{\text{disc}}^{\text{1 loop}}$ and write 
\begin{align}
& (-i)
\langle p_1{-} q_1, p_2{-} q_2 |\opS^\dagger \hat a_{--}( k)\opS|p_1 p_2\rangle ^{0}\Big|_{1\text{ loop}}\equiv \Mfull^{0,\text{2MPI},\text{\text{cl.}}}_{5, \text{1 loop}}(k^{--}) + \mathcal{W}_{\text{cut}}^{\text{1 loop}}(k^{--}) \ ,\\
& (+i)
\langle p_1{-} q_1, p_2{-} q_2 |\opS^\dagger \hat a_{++}^\dagger( k)\opS|p_1 p_2\rangle^{0} \Big|_{1\text{ loop}}\equiv \Big[ \Mfull^{0,\text{2MPI}, \text{cl.}}_{5, \text{1 loop}}(k^{++}) + \mathcal{W}_{\text{cut}}^{\text{1 loop}}(k^{++}) \Big]^* \, .\nonumber
\end{align}
It was pointed out in \cite{Caron-Huot:2023vxl} that $\mathcal{W}_{\text{cut}}^{\text{1 loop}}$ contributes nontrivially to the scattering waveform. As mentioned, we will summarize its evaluation in \cref{cutcontrib}.
Upon the rescaling \eqref{pulloutOmega}, it identifies $\real{L=1}$ and $\imaginary{L=1}$ as the real and imaginary parts of $\mathcal{W}_{\text{amp}}^{\text{1 loop}}+\mathcal{W}_{\text{cut}}^{\text{1 loop}}$ evaluated at $\tilde{k}=(1,\bm n)$,
\begin{align}\label{eq:ReIm1l}
    \Mfull^{0,\text{2MPI},\text{\text{cl.}}}_{5, \text{1 loop}}(k^{--}) + \mathcal{W}_{\text{cut}}^{\text{1 loop}}(k^{--}) &= \real{L=1} + i\,\imaginary{L=1}\,,\nonumber\\
    \Big[ \Mfull^{0,\text{2MPI}, \text{cl.}}_{5, \text{1 loop}}(k^{++}) + \mathcal{W}_{\text{cut}}^{\text{1 loop}}(k^{++}) \Big]^* &= \real{L=1} - i\,\imaginary{L=1}\,.
\end{align}
Note that we have used the fact that $\varepsilon_{+}^{*}=\varepsilon_{-}$. The amplitude contributes to both the real and imaginary parts, while the cut only contributes to the imaginary part.

In the next two sections, 
%Secs.~\ref{sec:TheAmplitude} and \ref{integration} to 
we will evaluate the integrand and then the integrals of this one-loop classical amplitude. We collect them and discuss their properties in \cref{integratedamplitudes}. In \cref{sec:Waveforms}, we proceed to discuss their contribution to waveform observables for ${\cal N}=8$ supergravity and GR. We will then discuss the cut contribution in \cref{cutcontrib}.

\section{The five-point classical integrand in minimally scalar-coupled GR}
\label{sec:TheAmplitude}

The integrand for the complete one-loop four-scalar-one-graviton amplitude in general 
relativity coupled to self-interacting scalars and in ${\cal N}=0$ supergravity was constructed in Ref.~\cite{Carrasco:2021bmu} through double-copy methods.
In this section, we construct the corresponding one-loop HEFT amplitude and compare it with the 2MPI part of the classical limit of the full theory amplitude, which we find convenient to re-derive directly from generalized unitarity considerations.
We isolate from the classical field-theory amplitude the 2MPR diagrams, in which both matter lines are cut. As discussed in the previous sections, these contributions are super-classical and cancel in the waveform. 
The remaining diagrams, which have exactly one matter line cut, reproduce the HEFT amplitude, thus demonstrating the absence of $\hbar/\hbar$ contributions to the classical five-point amplitude.\footnote{See Ref.~\cite{Brandhuber:2021eyq} for the analogous result for the four-point amplitude and related observables through two loops.}

\subsection{Preliminaries}

We begin by setting up the notation and variables for five-point kinematics.
While we are ultimately interested in physical kinematics, with two incoming and three outgoing particles, it is convenient  to take all momenta to be outgoing, 
\begin{equation}
\label{kinematics}
\vcenter{\hbox{\begin{tikzpicture}
    \pgfmathsetmacro{\r}{1.5}
    \pgfmathsetmacro{\c}{0.6}
    \pgfmathsetmacro{\shift}{0.2}
    \pgfmathsetmacro{\y}{1}
    \pgfmathsetmacro{\l}{0.5}
    \coordinate (o) at (0,0);
    \node (p3) at (45:\r) [label={[label distance=-8pt]45:$p_3$}] {};
    \node (p4) at (-45:\r) [label={[label distance=-8pt]-45:$p_4$}] {};
    \node (p2) at (135:\r) [label={[label distance=-8pt]135:$p_2$}] {};
    \node (p1) at (-135:\r) [label={[label distance=-8pt]-135:$p_1$}] {};
    \node (p5) at (0:\r) [label={[label distance=-5pt]0:$k$}] {};
    \draw [very thick] (p1.center) -- (o) -- (p4.center);
    \draw [very thick,blue] (p2.center) -- (o) -- (p3.center);
    \draw [graviton,thick] (o) -- (p5.center);
    \filldraw [fill=gray!30!white,thick] (o) circle (\c);
    \foreach \x in {45,-45,135,-135} {
        \draw [-Stealth] (o) ++ (\x:\shift) ++ (-\x:\y) -- ++ (-\x:\l);
    }
    \draw [-Stealth] (o) ++ (0,-\shift) ++ (\y,0) -- ++(\l,0);
\end{tikzpicture}}}
\hspace{2cm}
\renewcommand{\arraystretch}{1.2}
\begin{array}{l}
p_1^2 = p_4^2 = m_1^2 \\
p_2^2 = p_3^2 = m_2^2 \\
k^2 = 0
\end{array} \ .
\end{equation}
To cleanly separate different orders in the soft expansion, it is convenient to introduce $\pb_{1,2}$ variables that are respectively orthogonal on $q_{1,2}$, the momentum transfer from particle $1$ and $2$~\cite{Luna:2017dtq},
\begin{align}\label{barredmomenta}
\left.\begin{array}{l}
p_{1}^{\mu}=\bar{p}_{1}^{\mu}-q_{1}^{\mu}/2  \qquad p_{4}^{\mu}=-\bar{p}_{1}^{\mu}-q_{1}^{\mu}/2  \\
p_{2}^{\mu}=\bar{p}_{2}^{\mu}-q_{2}^{\mu}/2  \qquad p_{3}^{\mu}=-\bar{p}_{2}^{\mu}-q_{2}^{\mu}/2 
\end{array}\quad\right\}\;\Longrightarrow\;\pb_1\cdot q_1=\pb_2\cdot q_2=0\,.
\end{align}
The external graviton momentum $k$ is related to the momentum transfers as $k=q_1+q_2$. The on-shell conditions expressed in terms of the shifted matter momenta are
\begin{align}
    \pb_1^2 = \mb_1^2 = m_1^2 - \frac{q_1^2}{4}\,,\qquad \pb_2^2 = \mb_2^2 = m_2^2 - \frac{q_2^2}{4}\,.
\end{align}
We define barred four-velocity $\ub_1=\pb_1/\mb_1$ and $\ub_2=\pb_2/\mb_2$ such that $\ub_1^2=\ub_2^2=1$. It is also convenient to define $y=\ub_1\cdot\ub_2$. We will also use normal four-velocity $u_1=p_1/m_1$ and $u_2=p_2/m_2$, and define $\sigma=u_1\cdot u_2$. The difference between $y$ and $\sigma$ is of the order $\mathcal{O}(q^2)$.

The classical four-scalar-one-graviton tree-level amplitude in GR was given in Ref.~\cite{Luna:2017dtq}. In the notation above and for real kinematics, %(i.e., with $p_{1, 2}$ incoming), 
it can be written as
\begin{align}
\label{fullGS1grav}
\Mred_{5, \text{tree}}^{\text{cl.} \text{GR}}
%(\ub_1,\ub_2,q_1,q_2,k^{\varepsilon}) 
&= -\frac{\kappa^3}{4}  
{\bar m}_1^2 {\bar m}_2^2 \; \varepsilon^*(k)_{\mu\nu} \left[
\frac{4P_{12}^\mu P_{12}^\nu}{q_1^2 q_2^2} + \frac{2y}{q_1^2 q_2^2} \left( Q_{12}^\mu P_{12}^\nu + Q_{12}^\nu P_{12}^\mu\right) \right.  \\
&\qquad\qquad
\left. + \left(y^2 - \frac{1}{d-2} \right) \left(\frac{Q_{12}^\mu Q_{12}^\nu}{q_1^2 q_2^2} - \frac{P_{12}^\mu P_{12}^\nu}{(k \cdot {\bar u}_1)^2 (k\cdot {\bar u}_2)^2} \right) 
\right] \ ,
\nonumber\\
P_{12}^\mu &\equiv k \cdot {\bar u}_1 \; {\bar u}_2^\mu - k \cdot {\bar u}_2 \; {\bar u}_1^\mu\,,\quad Q_{12}^\mu \equiv (q_1-q_2)^\mu- \frac{q_1^2}{k \cdot {\bar u}_1} {\bar u}_1^\mu+ \frac{q_2^2}{k \cdot {\bar u}_2} {\bar u}_2^\mu,
\end{align}
where  $\varepsilon(k)_{\mu \nu}=\varepsilon(k)_{\mu}\varepsilon(k)_{\nu} $ is the graviton polarization tensor, and the conjugation indicates that it is an outgoing graviton. The coupling $\kappa$ is related to Newton's constant through $\kappa^2=32\pi G$. One can easily verify that both $\varepsilon^*\cdot P_{12}$ and $\varepsilon\cdot Q_{12}$ are gauge invariant. We will use this expression in \cref{integratedamplitudes} to compare the IR divergences of the classical amplitude with the prediction of Weinberg's analysis \eqref{IRdivGeneral} and in \cref{sec:Waveforms} to construct the leading-order waveform. Here is a list of tree amplitudes that will be used in unitarity cut constructions,
\begin{align}
\vcenter{\hbox{
\begin{tikzpicture}[scale=1]
    \draw [very thick] (0,0) node[left=0pt]{$1$} -- (2,0) node[right=0pt]{$4$};
    \draw [graviton] (1,0) -- ++ (135:1) node[left=0pt]{$k_1$}; 
    \draw [graviton] (1,0) -- ++ (45:1) node[right=0pt]{$k_2$};
    \filldraw [fill=gray!30!white,thick] (1,0) circle (0.25cm);
    \node[above=0pt] at (1,1) {$\vphantom{k_2}$};
    \node[scale=0.85] at (4.1,0.35) {$\begin{array}{l}
        \Mred^{\text{Comp.}}_{4,\text{tree}}(p_1,p_4,k_1,k_2) \\[4pt] 
        \Mred^{\text{cl.Comp.}}_{4,\text{tree}}(\ub_1,k_1,k_2)
    \end{array}$};
    \begin{scope}[yshift=-2cm]
    \draw [very thick] (1,0) -- ++(-45:1) node[right=0pt]{$4$};
    \draw [very thick] (1,0) -- ++(-135:1) node[left=0pt]{$1$};
    \draw [very thick,blue] (1,0) -- ++(45:1) node[right=0pt,black]{$3$};
    \draw [very thick,blue] (1,0) -- ++(135:1) node[left=0pt,black]{$2$};
    \draw [graviton] (1,0) -- (2,0) node[right=0pt]{$k$};
    \filldraw [fill=gray!30!white,thick] (1,0) circle (0.25cm);
    \node[scale=0.85] at (4.2,0) {$\begin{array}{l}
        \Mred^{\text{GR}}_{5,\text{tree}}(p_1,p_4,p_2,p_3,k) \\[4pt] 
        \Mred^{\text{cl.GR}}_{5,\text{tree}}(\ub_1,\ub_2,q_1,q_2,k)
    \end{array}$};
    \end{scope}
\end{tikzpicture}
}}\!\!\!\!\!\!   
\vcenter{\hbox{
\begin{tikzpicture}[scale=1]
    \draw [very thick] (0,0) node[left=0pt]{$1$} -- (2,0) node[right=0pt]{$4$};
    \draw [graviton] (1,0) -- ++ (135:1) node[left=0pt]{$k_1$};
    \draw [graviton] (1,0) -- (1,1) node[above=0pt]{$k_2$};
    \draw [graviton] (1,0) -- ++ (45:1) node[right=0pt]{$k_3$};
    \filldraw [fill=gray!30!white,thick] (1,0) circle (0.25cm);
    \node[scale=0.85] at (4.4,0.35) {$\begin{array}{l}
        \Mred^{\text{Comp.}}_{5,\text{tree}}(p_1,p_4,k_1,k_2,k_3) \\[4pt] 
        \Mred^{\text{cl.Comp.}}_{5,\text{tree}}(\ub_1,k_1,k_2,k_3)
    \end{array}$};
    \begin{scope}[yshift=-2cm]
    \draw [very thick] (1,0) -- ++(-45:1) node[right=0pt]{$4$};
    \draw [very thick] (1,0) -- ++(-135:1) node[left=0pt]{$1$};
    \draw [very thick,blue] (1,0) -- ++(45:1) node[right=0pt,black]{$3$};
    \draw [very thick,blue] (1,0) -- ++(135:1) node[left=0pt,black]{$2$};
    \draw [graviton] (1,0) -- ++(15:1) node[right=0pt]{$k_1$};
    \draw [graviton] (1,0) -- ++(-15:1) node[right=0pt]{$k_2$};
    \filldraw [fill=gray!30!white,thick] (1,0) circle (0.25cm);
    \node[scale=0.85] at (4.7,0) {$\begin{array}{l}
        \Mred^{\text{GR}}_{6,\text{tree}}(p_1,p_4,p_2,p_3,k_1,k_2) \\[4pt] 
        \Mred^{\text{cl.GR}}_{6,\text{tree}}(\ub_1,\ub_2,q_1,q_2,k_1,k_2)
    \end{array}$};
    \end{scope}
\end{tikzpicture}
}}   
\label{eq:tree_amps}
\end{align}
including the three-point amplitudes that are uniform in $\hbar$,
\begin{align}
    & \Mred_{3,\text{tree}}^{\text{Comp.}}(p_1,p_4,k)=\Mred_{3,\text{tree}}^{\text{Comp.}}(\ub_1,k)=-\kappa^2\mb_1^2(\ub_1\cdot\varepsilon)^2\,,\nonumber\\
    & \Mred_{3,\text{tree}}^{\text{graviton}}(k_1,k_2,k_3)=-\kappa^2(\varepsilon_1\cdot\varepsilon_2\,\varepsilon_3\cdot k_1+\varepsilon_2\cdot\varepsilon_3\,\varepsilon_1\cdot k_2+\varepsilon_3\cdot\varepsilon_1\,\varepsilon_2\cdot k_3)^2\,.
\end{align}
In \cref{eq:tree_amps}, the full quantum amplitudes are given in the first entries, which we compute through the standard Feynman diagram approach. The amplitudes with a superscript ``cl'' are classical amplitudes, which are obtained from the full amplitudes through a soft expansion and keeping the terms with the classical scaling~\eqref{classical_scaling}. These classical tree amplitudes will be used later to construct HEFT cuts.

The sum over the physical graviton states is a common ingredient in both HEFT and full amplitude calculations, as it enters in the evaluation of generalized unitarity cuts. Generally, it is
\begin{equation}
\sum_{h}\varepsilon_h(k)^{\mu\nu}\varepsilon_h^*(k)^{\alpha\beta}
= \frac{1}{2}\left(\mathcal{P}^{\mu\alpha}\mathcal{P}^{\nu\beta}
+ \mathcal{P}^{\mu\beta}\mathcal{P}^{\nu\alpha}
- \frac{2}{d-2}\mathcal{P}^{\mu\nu}\mathcal{P}^{\alpha\beta}\right) \,,
\label{CompletenessRelationGravity}
\end{equation}
where 
\begin{equation}
\mathcal{P}^{\mu\nu}(k)=\eta^{\mu\nu}-\frac{r^\mu k^\nu+r^\nu k^\mu}{r\cdot k} \, 
\label{PhysicalStateProjector}
\end{equation}
is the physical-state projector for a vector field, and $r^\mu$ is an arbitrary null reference vector that should drop out of physical expressions.
This sum simplifies considerably if the amplitudes being sewn obey generalized Ward identities, i.e., they obey the Ward identity for external leg $i$ without using the transversality of the polarization vectors for any of the other external gravitons~\cite{Kosmopoulos:2020pcd,KoemansCollado:2019ggb,Bern:2019crd}.
For such amplitudes, all terms proportional to the momentum of the sewn legs drop out so we can effectively use the much simplified (and manifestly covariant) graviton state sum
\begin{equation}
\label{simple_sewing}
\sum_h \varepsilon_h(k){}^{\mu\nu}\varepsilon_h^*(k)^{\alpha\beta}=
\frac{1}{2}\left ( \eta^{\mu\alpha}\eta^{\nu\beta}+\eta^{\mu\beta}\eta^{\nu\alpha}
-\frac{2}{d-2}\eta^{\mu\nu}\eta^{\alpha\beta} \right )   \ .  
\end{equation}
Ref.~\cite{Kosmopoulos:2020pcd} showed that, through simple manipulations, it is always possible to put scattering amplitudes into a form that obeys the generalized Ward identities. In fact, by  being manifestly written in terms of linearized field strengths, HEFT amplitudes already obey such generalized Ward identities without any additional manipulations. We will use such amplitudes in our loop calculations.

\begin{figure}
    \centering
    \begin{tikzpicture}
        \node [left=0pt] (p1) at (-\w,0) {$1$};
        \node [right=0pt] (p4) at (1.4*\w,0) {$4$};
        \node [left=0pt] (p2) at (-\w,1.3*\h) {$2$};
        \node [right=0pt] (p3) at (1.4*\w,1.3*\h) {$3$};
        \node [right=0pt] (p5) at (1.4*\w,1.3*\h/2) {$5$};
        \node [right=0pt] (p6) at (0.8*\w,1.3*\h/4) {\small $(\lm-q_{1})^{2}$};
        \node [right=0pt] (p6) at (0.8*\w,1.3*3*\h/4) {\small $(\lm+q_{2})^{2}$};
        \node [left=0pt] (p6) at (-0.3*\w,1.2*\h/2) {\small $\lm^{2}$};
        \node [above=0pt] (p6) at (0.25*\w,1.3*\h) {\small $\hdelta(2\ub_{2}\cdot\lm)$};
        \node [below=0pt] (p6) at (0.25*\w,0) {\small $(2\ub_{1}\cdot\lm)$};
        \draw [very thick] (p1) -- (p4);
        \draw [very thick] (p2) -- (p3);
        \draw [very thick] (-0.3*\w,1.3*\h) -- (-0.3*\w,0);
        \draw [very thick] (0.8*\w,1.3*\h) -- (0.8*\w,0);
        \draw [very thick] (0.8*\w,1.3*\h/2) -- (p5);
        %\draw [dashed,draw=red] (-\w/6,\h/2+0.3) -- (-\w/6,\h/2-0.3);
        %\draw [dashed,draw=red] (\w/2-0.25*0.5*\w,0.75*\h+0.25*0.6*\h) -- (\w/2+0.25*0.5*\w,0.75*\h-0.25*0.6*\h);
    \end{tikzpicture}
    \caption{Topology of integrals given in \cref{integrals_definition}.}
    \label{topolgy}
\end{figure}
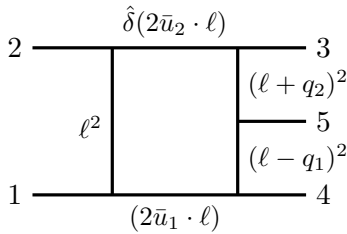

The expressions for the 2MPI amplitudes, both in HEFT and the full theory calculation,
are naturally expressed in terms of scalar integrals of pentagon topology and with two linear propagators. One of our results, which is natural in the HEFT approach, is that one matter line is always cut;
thus, all integrals will be of the special cases of
\begin{align}
\label{integrals_definition}
    I^{\pm}_{a_1,a_2,a_3,a_4}=\int\frac{\dd^d\lm}{(2\pi)^d}\frac{\hdelta(2\ub_2\cdot\lm)}{[\lm^{2}]^{a_1}[(\lm+q_2)^{2}]^{a_2}[(\lm-q_1)^{2}]^{a_3}(2\ub_1\cdot\lm\pm\ie)^{a_4}} \ ,
\end{align}
where the delta function realized as
\begin{align}
\label{propagator_cut_identity}
    \frac{i}{x+\ie}+\frac{i}{-x+\ie}=2\pi\delta(x)\equiv\hdelta(x) \ .
\end{align}
We will omit the $\pm$ superscript if the linearized matter propagator is absent, i.e., when $a_4=0$. 
This is the generalization to the five-point case of the analogous feature present in four-point amplitudes~\cite{Cheung:2018wkq, Bern:2019nnu, Bern:2019crd}. The diagrammatic representation of the master integrals are shown in \cref{topolgy}.

Unlike the four-point case, the five-point classical amplitudes depend on master integrals with contact matter vertices, namely, $a_1=0$ in \cref{integrals_definition}. They come from the IBP reduction of integrands with higher topologies, and their coefficients contain non-local dependence on $q_i^2$ through the Gram determinants generated by IBP. As a result, these contributions are relevant to the classical long-range interaction in the position space after a Fourier transform.\footnote{In contrast, at four points, the coefficients of such contact master integrals can only have a polynomial dependence on $q^2$, while the integrals are independent of $q$. Therefore, they can only result in delta-function interactions in position space.}

\subsection{The five-point 2MPI HEFT integrand}\label{generalizedunitarity}

\begin{figure}
    \centering
    \captionsetup[subfigure]{labelformat=simple,labelsep=colon,font=normalsize}
    \renewcommand{\thesubfigure}{HEFT Cut 1}
    \subfloat[$\mathcal{C}^{(1,\text{H})}_{m_1^2m_2^3}$]{\label{fig:cut1h}
    \begin{tikzpicture}
        \node [left=0pt] (p1) at (-\w,0) {$1$};
        \node [right=0pt] (p4) at (\w,0) {$4$};
        \node [left=0pt] (p2) at (-\w,\h) {$2$};
        \node [right=0pt] (p3) at (\w,\h) {$3$};
        \node [right=0pt] (p5) at (\w,{\h/2}) {$5$};
        \draw [very thick] (p1) -- (p4);
        \draw [very thick,blue] (p2) -- (p3);
        \draw [graviton] (-\w/2,\h/2) -- (\w/2,\h) node[pos=0.5,below=0pt]{$\lm_2$};
        \draw [graviton] (\w/2,\h) -- (p5.west);
        \filldraw [fill=gray!30!white,thick] (-\w/2,\h/2) ellipse (0.25cm and 1cm) (\w/2,\h) circle (\blb);
        %\node [above=4pt] at (0,\h) {$\lm_3$};
        \draw [draw=red] (0,\h+0.2) -- (0,\h-0.2);
        %\draw [dashed,draw=red] (0,\h-0.1) -- (0.3*\w/2,\h-0.1-0.3*\h);
    \end{tikzpicture}
    }\qquad\renewcommand{\thesubfigure}{HEFT Cut 2}
    \subfloat[$\mathcal{C}^{(2,\text{H})}_{m_1^2m_2^3}$]{\label{fig:cut2h}
    \begin{tikzpicture}
    \node [left=0pt] (p1) at (-\w,0) {$1$};
        \node [right=0pt] (p4) at (\w,0) {$4$};
        \node [left=0pt] (p2) at (-\w,\h) {$2$};
        \node [right=0pt] (p3) at (\w,\h) {$3$};
        \node [right=0pt] (p5) at (\w,{\h/2}) {$5$};
        \draw [very thick] (p1) -- (p4);
        \draw [very thick,blue] (p2) -- (p3);
        \draw [graviton] (-\w/2,\h/2) -- (\w/2,\h) node[pos=0.5,below=0pt]{$\lm_3$};
        \draw [graviton] (-\w/2,\h/2) to[bend right=25] (p5.west);
        \filldraw [fill=gray!30!white,thick] (-\w/2,\h/2) ellipse (0.25cm and 1cm) (\w/2,\h) circle (\blb);
        %\node [above=4pt] at (0,\h) {$\lm_3$};
        \draw [draw=red] (0,\h+0.2) -- (0,\h-0.2);
        %\draw [dashed,draw=red] (0,\h-0.1) -- (0.3*\w/2,\h-0.1-0.3*\h);
    \end{tikzpicture}
    }
    \caption{The spanning cuts of the five-point one-loop amplitude. All exposed lines are cut. The loop momentum $\lm_2=\lm-q_1$ and $\lm_3=\lm+q_2$ follow the clockwise direction. }
    \label{cuts}
\end{figure}
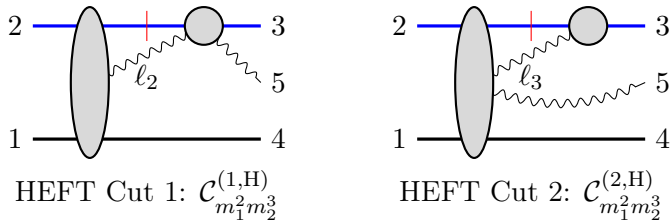

We use generalized unitarity in $d$ dimensions~\cite{UnitarityMethod, Fusing, BernMorgan} to construct the 2MPI HEFT amplitude integrand from the spanning set of generalized cuts in \cref{cuts}. Unlike generalized unitarity in the full quantum theory, the cut matter line, denoted by a red vertical line, is permanently cut and dressed by a delta function $\hdelta(2\pb_i\cdot\lm)$. In addition, the input tree amplitude for each blob contains only terms with classical scaling in the full tree amplitude. Such terms are free of the delta-function contributions that are super-classical, as reviewed in \cref{sec:HEFT}. We can safely ignore the super-classical delta-function dependent terms because they ultimately cancel out in the KMOC formalism. 
We will discuss in Sec.~\ref{cutcontrib} the classical terms that survive this cancellation from the bilinear-in-$T$ contributions. 
Finally, here we focus on integrands where matter line 2 is cut as the classical amplitude's integrands where matter line 1 is cut can be derived from a relabelling. We note that cuts with no cut matter lines, for example, 
\begin{equation}
    \vcenter{\hbox{\begin{tikzpicture}
        \node [left=0pt] (p1) at (-\w,0) {$1$};
        \node [right=0pt] (p4) at (\w,0) {$4$};
        \node [left=0pt] (p2) at (-\w,\h) {$2$};
        \node [right=0pt] (p3) at (\w,\h) {$3$};
        \node [right=0pt] (p5) at (\w,{\h/2}) {$5$};
        \draw [very thick] (p1) -- (p4);
        \draw [very thick,blue] (p2) -- (p3);
        \draw [graviton] (0,\h) to[out=-30,in=30] (0,0);
        \draw [graviton] (0,\h) to[out=-150,in=150] (0,0);
        \draw [graviton] (0,\h) -- (p5.west);
        \filldraw [fill=gray!30!white,thick] (0,\h) circle (\blb) (0,0) circle (\blb);
        %\draw [dashed,draw=red] (-\w/2,\h/2) -- (\w/2,\h/2);
    \end{tikzpicture}}}
\end{equation}
exhibit quantum $\hbar$ scaling.

In the HEFT spanning cuts in \cref{cuts} all exposed lines are cut. The red vertical line indicates that the matter propagator remains cut in the final expression.\footnote{This is equivalent to removing all contributions in which both matter lines are collapsed; direct counting indicates that such graviton loops do not contribute to the classical limit, in analogy with e.g.~\cite{Cheung:2018wkq, Bern:2019crd}.} These HEFT cuts are given by:
\begin{align}\label{eq:heftcuts}
\mathcal{C}^{(1,\text{H})}_{m_{1}^{2}m_{2}^{3}}&=\sum_{h_2} 
\Mred_{5,\text{tree}}^{\text{cl.GR}}(\ub_1,\ub_2,q_1,q_2,-\lm_2^{h_2})
\Mred_{4,\text{tree}}^{\text{cl.Comp.}}(\ub_2,\lm_2^{-h_2},k^{\varepsilon})  \, , \\
\mathcal{C}^{(2,\text{H})}_{m_{1}^{2}m_{2}^{3}}&=\sum_{h_3} 
\Mred_{6,\text{tree}}^{\text{cl.GR}}(\ub_1,\ub_2,q_1,q_2,-\lm_3^{h_3},k^{\varepsilon})
\Mred_{3,\text{tree}}^{\text{Comp}}(\ub_2,\lm_3^{-h_3})- \text{divergent bubbles.}\nonumber
\end{align}
As first mentioned in \cref{sec:HEFT}, we make all the uncut linear propagators symmetric in $\ie$ and thus principal-valued in the HEFT cuts. 
The \subref*{fig:cut2h} is technically divergent due to external bubble contributions such as 
\begin{equation}
    \vcenter{\hbox{\begin{tikzpicture}
        \node [left=0pt] (p1) at (-\w,0) {$1$};
        \node [right=0pt] (p4) at (1.5*\w,0) {$4$};
        \node [left=0pt] (p2) at (-\w,\h) {$2$};
        \node [right=0pt] (p3) at (1.5*\w,\h) {$3$};
        \node [right=0pt] (p5) at (1.5*\w,{\h/2}) {$5$};
        \draw [very thick] (p1) -- (p4);
        \draw [very thick,blue] (p2) -- (p3);
        \draw [graviton] (\w/5,\h) to[out=-90,in=-90] (\w,\h);
        \draw [graviton] (-\w/2,\h/2) to[bend right=15] (p5.west);
        \filldraw [fill=gray!30!white,thick] (-\w/2,\h/2) ellipse (0.25cm and 1cm) (\w/5,\h) circle (\blb) (\w,\h) circle (\blb);
        \node [below=11pt] at (0.6*\w,\h) {$\lm_3$};
        %\node [above=4pt] at (0.6*\w,\h) {$\lm_3$};
        %\node [above=0pt] at (2.4*\w,0.2*\h) {$\in\mathcal{C}^{(2)}_{m_{1}^{2}m_{2}^{3}}$ .};
        \draw [draw=red] (0.6*\w,\h+0.2) -- (0.6*\w,\h-0.2);
        %\draw [dashed,draw=red] (0.6*\w,\h+0.15-0.35) -- (0.6*\w,\h-0.15-0.35);
    \end{tikzpicture}}}\in\mathcal{C}^{(2,\text{H})}_{m_{1}^{2}m_{2}^{3}}\,.
    \label{divbub}
\end{equation}
Thus we need to construct the cut with an additional regulator, such as taking external scalars off-shell or not imposing momentum conservation. The regularization will break the generalized Ward identity. As a result, we need to use the full graviton state projector~\eqref{CompletenessRelationGravity} to compute \subref*{fig:cut2h}. Since such external bubbles are scaleless after the regulator is removed, we can simply subtract them from \subref*{fig:cut2h} to reach a finite result.

\begin{figure}
    \centering
    \begin{tikzpicture}
        \node [left=0pt] (p1) at (-\w,0) {$1$};
        \node [right=0pt] (p4) at (1.4*\w,0) {$4$};
        \node [left=0pt] (p2) at (-\w,\h) {$2$};
        \node [right=0pt] (p3) at (1.4*\w,\h) {$3$};
        \node [right=0pt] (p5) at (1.4*\w,{\h/2}) {$5$};
        \draw [very thick] (p1) -- (p4);
        \draw [very thick,blue] (p2) -- (p3);
        \draw [graviton] (-\w/2,\h/2) -- (0.2*\w,\h/2) node[pos=0.5,below=0pt,scale=0.9]{$\lm_2$};
        \draw [graviton] (0.2*\w,\h/2) -- (0.8*\w,\h) node[pos=0.35,right=3pt,scale=0.9]{$\lm_3$};
        \draw [graviton] (0.2*\w,\h/2) -- (p5.west);
        \filldraw [fill=gray!30!white,thick] (-\w/2,\h/2) ellipse (0.25cm and 1cm) (0.8*\w,\h) circle (\blb) (0.2*\w,\h/2) circle (\blb);
        \draw [draw=red] (0.15*\w,\h+0.2) -- (0.15*\w,\h-0.2);
        %\draw [dashed,draw=red] (-\w/6,\h/2+0.3) -- (-\w/6,\h/2-0.3);
        %\draw [dashed,draw=red] (\w/2-0.25*0.5*\w,0.75*\h+0.25*0.6*\h) -- (\w/2+0.25*0.5*\w,0.75*\h-0.25*0.6*\h);
    \end{tikzpicture}
    \caption{The overlap between \protect\subref*{fig:cut1h} and \protect\subref*{fig:cut2h}, denoted as $\mathcal{C}^{(12,\text{HEFT})}_{m_1^2m_2^3}$.}
    \label{overlapcuts}
\end{figure}
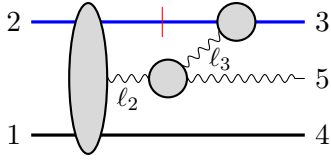

We now merge the cuts in \cref{eq:heftcuts} into the HEFT integrand. To this end, it is convenient to follow Ref.~\cite{Bern:2004cz} and simply add them together and subtract the overlap shown in \cref{overlapcuts} and given by
\begin{equation}\label{eq:overlap_heft}
\begin{split}
\mathcal{C}^{(12,\text{H})}_{m_1^2m_2^3}=\sum_{h_2,h_3}&\Mred_{5,\text{tree}}^{\text{cl.GR}}(\ub_1,\ub_2,q_1,q_2,-\lm_2^{h_2})\\
&\times\Mred_{3,\text{tree}}^{\text{graviton}}(\lm_2^{-h_2},k^{\varepsilon},-\lm_3^{h_3})\Mred_{3,\text{tree}}^{\text{Comp}}(\ub_2,\lm_3^{-h_3})\,.
\end{split}
\end{equation}
%Putting together \cref{eq:heftcuts,eq:overlap_heft}, we 
%get the one-loop HEFT 2MPI integrand 
Thus, the resulting one-loop HEFT 2MPI integrand is
\begin{align}
\label{final1}
\Mred_{5,m_1^2m_2^3}^\text{H-2MPI}&=-\int \frac{\dd^{d}\lm}{(2\pi)^{d}}\hdelta(2\mb_2\lm\cdot\ub_{2})\Bigg[
\frac{\mathcal{C}^{(1,\text{H})}_{m_{1}^{2}m_{2}^{3}}}{(\lm-q_1)^2}
+\frac{\mathcal{C}^{(2,\text{H})}_{m_{1}^{2}m_{2}^{3}}}{(\lm+q_2)^2}
+\frac{\,\mathcal{C}^{(12,\text{H})}_{m_{1}^{2}m_{2}^{3}}}{(\lm-q_1)^2(\lm+q_2)^{2}} \Bigg] \, , \nonumber \\
\Mred_{5, \text{1 loop}}^\text{H-2MPI}&=
\Mred_{5,m_1^2m_2^3}^\text{H-2MPI}+
(\ub_1\leftrightarrow \ub_2, \mb_1\leftrightarrow \mb_2, q_1\leftrightarrow q_2) \,, 
\end{align}
which we will later IBP-reduce to master integrals. 
The relative signs between the cut contributions, $\mathcal{C}^{(1,\text{H})}$ and $\mathcal{C}^{(2,\text{H})}$, and the overlap contribution,  $\mathcal{C}^{(12,\text{H})}$, are a consequence of the factors of $i$ in the definition of matrix elements and of propagators. 

\subsection{The five-point classical integrand from the quantum integrand}
\label{fullGR5points}

In this section, we construct the classical limit of the four-scalar-one-graviton amplitude in GR coupled to two scalar fields. The result will verify the completeness of the HEFT amplitude and expose the fate of terms of $\hbar/\hbar$ type that naturally appear in the classical expansion of cuts of the full theory.

\begin{figure}
    \centering
    \captionsetup[subfigure]{labelformat=simple,labelsep=colon,font=normalsize}
    \renewcommand{\thesubfigure}{GR Cut 1}
    \subfloat[$\mathcal{C}^{(1)}_{m_1^2m_2^3}$]{\label{fig:cut1GR}
    \begin{tikzpicture}
        \node [left=0pt] (p1) at (-\w,0) {$1$};
        \node [right=0pt] (p4) at (\w,0) {$4$};
        \node [left=0pt] (p2) at (-\w,\h) {$2$};
        \node [right=0pt] (p3) at (\w,\h) {$3$};
        \node [right=0pt] (p5) at (\w,{\h/2}) {$5$};
        \draw [very thick] (p1) -- (p4);
        \draw [very thick,blue] (p2) -- (p3);
        \draw [graviton] (0,0) -- (-\w/2,\h) node[pos=0.5,left=0pt]{$\lm_1$};
        \draw [graviton] (0,0) -- (\w/2,\h) node[pos=0.5,right=0pt]{$\lm_2$};
        \draw [graviton] (\w/2,\h) -- (p5.west);
        \filldraw [fill=gray!30!white,thick] (0,0) circle (\blb) (\w/2,\h) circle (\blb) (-\w/2,\h) circle (\blb);
        \node [above=0pt] at (0,\h) {$\lm_4$};
    \end{tikzpicture}
    }\quad
    \renewcommand{\thesubfigure}{GR Cut 2}
    \subfloat[$\mathcal{C}^{(2)}_{m_1^2m_2^3}$]{\label{fig:cut2GR}
    \begin{tikzpicture}
        \node [left=0pt] (p1) at (-\w,0) {$1$};
        \node [right=0pt] (p4) at (\w,0) {$4$};
        \node [left=0pt] (p2) at (-\w,\h) {$2$};
        \node [right=0pt] (p3) at (\w,\h) {$3$};
        \node [right=0pt] (p5) at (\w,{\h/2}) {$5$};
        \draw [very thick] (p1) -- (p4);
        \draw [very thick,blue] (p2) -- (p3);
        \draw [graviton] (0,0) -- (-\w/2,\h) node[pos=0.5,left=0pt]{$\lm_1$};
        \draw [graviton] (\w/2,\h) -- (0,0) node[pos=0.5,right=0pt]{$\lm_3$};
        \draw [graviton] (0,0) -- (p5.west);
        \filldraw [fill=gray!30!white,thick] (0,0) circle (\blb) (\w/2,\h) circle (\blb) (-\w/2,\h) circle (\blb);
        \node [above=0pt] at (0,\h) {$\lm_4$};
    \end{tikzpicture}
    }\quad
    \renewcommand{\thesubfigure}{GR Cut 3}
    \subfloat[$\mathcal{C}^{(3)}_{m_1^2m_2^3}$]{\label{fig:cut3GR}
    \begin{tikzpicture}
        \node [left=0pt] (p1) at (-\w,0) {$1$};
        \node [right=0pt] (p4) at (\w,0) {$4$};
        \node [left=0pt] (p2) at (-\w,\h) {$2$};
        \node [right=0pt] (p3) at (\w,\h) {$3$};
        \node [right=0pt] (p5) at (\w,{\h/2}) {$5$};
        \draw [very thick] (p1) -- (p4);
        \draw [very thick,blue] (p2) -- (p3);
        \draw [graviton] (-\w/2,\h/2) -- (\w/2,\h) node[pos=0.5,below=0pt]{$\lm_2$};
        \draw [graviton] (\w/2,\h) -- (p5.west);
        \filldraw [fill=gray!30!white,thick] (-\w/2,\h/2) ellipse (0.25cm and 1cm) (\w/2,\h) circle (\blb);
        \node [above=0pt] at (0,\h) {$\lm_4$};
    \end{tikzpicture}
    }
    \caption{The spanning cuts of the five-point one-loop amplitude in GR coupled to scalars. All exposed lines are cut. The loop momentum $\lm_i$ follows the clockwise direction. Contributions captured solely by \protect\subref*{fig:cut3GR} involve intersecting matter lines which do not contribute in the classical limit but whose possible appearance is related to the absence of four-scalar contact terms in the classical action. To construct the integrand we consider all relabelings of external legs.}
    \label{cutsGR}
\end{figure}
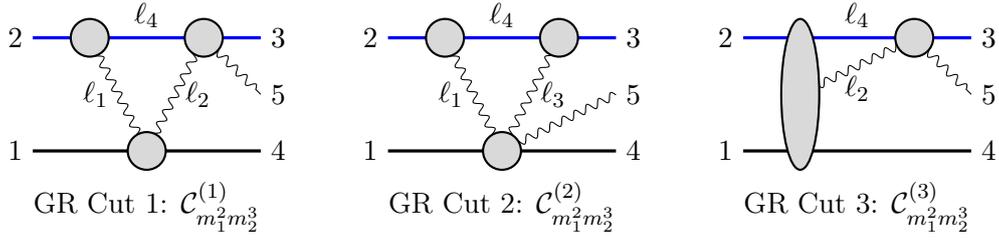
%%%%%%%%%%%%%%%%%%%%%%%%%%%%%%%

The spanning set of generalized unitarity cuts determines the (classical part) of the five-point amplitude in GR and is given in \cref{cutsGR}. The terms in the integrand determined solely by \subref*{fig:cut3GR} contain (1) 1PR mushroom graphs and (2) graphs with intersecting matter lines with numerators that are polynomial in external and loop momenta. The former is required by gauge invariance; the latter, while not having a Feynman vertex counterpart, may appear depending on choices made in the construction of the integrand. 
The integral corresponding to the latter does not depend on the momentum transfer $(q_{1}-q_2)$ and as such these contact terms contribute only $\delta(b)$ terms to the waveform and we could ignore it. The same integral also appears in the IBP reduction of terms determined by the first three cuts; their coefficients turn out to have a rather nontrivial dependence on the momentum transfer $(q_{1}-q_2)$ and thus contribute nontrivially to the waveform. We will find that for the integrand we construct, the contact terms with the topology of \subref*{fig:cut3GR} vanish identically. 

In our construction, we will ignore 2MPI contributions to the classical amplitude which are not captured by these cuts, because the corresponding (master) integrals are scaleless and thus vanish in dimensional regularization. We use $d$-dimensional generalized unitarity~\cite{UnitarityMethod, Fusing, BernMorgan} to construct the five-point integrand. In terms of tree amplitudes, the cuts in \cref{cutsGR} are 
\begin{align}
\mathcal{C}^{(1)}_{m_{1}^{2}m_{2}^{3}}&=\sum_{h_1, h_2}
\Mred^{\text{Comp.}}_{3,\text{tree}}(p_2, \lm_4,-\lm_{1}^{h_1})
\Mred^{\text{Comp.}}_{4,\text{tree}}(-\lm_4, p_3,\lm^{h_2}_{2}, k^{\varepsilon})
\Mred^{\text{Comp.}}_{4,\text{tree}}(p_1,p_4,\lm^{-h_1}_{1},-\lm^{-h_2}_{2})\,,
\nonumber\\
\label{newcuts5p}
\mathcal{C}^{(2)}_{m_{1}^{2}m_{2}^{3}}&=\sum_{h_1, h_3} 
\Mred^{\text{Comp.}}_{3,\text{tree}}(p_2,\lm_4, -\lm^{h_1}_{1})
\Mred^{\text{Comp.}}_{3,\text{tree}}(-\lm_4, p_3,\lm^{h_3}_{3})
\Mred^{\text{Comp.}}_{5,\text{tree}}(p_1,p_4,\lm^{-h_1}_{1},-\lm^{-h_3}_{3}, k^{\varepsilon})\,,
\nonumber\\
\mathcal{C}^{(3)}_{m_{1}^{2}m_{2}^{3}}&=\sum_{h_2} 
\Mred^{\text{GR}}_{5,\text{tree}}(p_1, p_4, p_2, \lm_4,-\lm^{h_2}_{2})
\Mred^{\text{Comp.}}_{4,\text{tree}}(-\lm_4,p_3, \lm^{h_2}_{2},k^{\varepsilon})  \ ,
\end{align}
where the $\lm_i$ are defined according to \cref{cutsGR}.
We use complete tree-level amplitudes of GR minimally-coupled to scalar fields that obey the generalized Ward identities~\cite{Kosmopoulos:2020pcd,KoemansCollado:2019ggb,Bern:2019crd}, so the sum over the internal graviton physical states, labeled here by $h_{1,2,3}$, is given by \cref{simple_sewing}. 
The resulting cuts reproduce those used in the construction of the quantum five-point integrand in Ref.~\cite{Carrasco:2021bmu} up to the contributions of four-scalar contact terms which do not contribute in the classical limit but are natural in the double copy construction used there.\footnote{Since we are interested in the classical limit, we do not include all cuts required to construct the complete quantum five-point amplitude which was considered in~Ref.~\cite{Carrasco:2021bmu}.}

\begin{figure}
    \centering
    \subfloat[]{\label{contacttopologies-a}\begin{tikzpicture}
    \draw [line width=8pt,in=180,out=-90,draw=white] (-\w/2,\h) to (\w,\h/2);
    \draw [graviton] (-\w/2,0) -- (-\w/2,\h);
    \draw [graviton] (\w/2,0) -- (\w/2,\h);
    %\draw [line width=8pt,in=180,out=-90,draw=white] (0,\h) to (\w,\h/2);
    \draw [graviton] (\w/2,\h/2) -- (\w,\h/2);
    \draw [thick] (-\w,\h) -- (\w,\h) (-\w,0) -- (\w,0);
    \end{tikzpicture}}\;
    \subfloat[]{\label{contacttopologies-b}\begin{tikzpicture}
    \draw [line width=8pt,in=180,out=-90,draw=white] (-\w/2,\h) to (\w,\h/2);
    \draw [graviton] (-\w/2,0) -- (-\w/2,\h);
    \draw [graviton] (\w/2,0) -- (\w/2,\h);
    \draw [line width=8pt,in=180,out=-90,draw=white] (0,\h) to (\w,\h/2);
    \draw [graviton,in=180,out=-90] (0,\h) to (\w,\h/2);
    \draw [thick] (-\w,\h) -- (\w,\h) (-\w,0) -- (\w,0);
    \end{tikzpicture}}\;
    \subfloat[]{\label{contacttopologies-c}\begin{tikzpicture}
    \draw [line width=8pt,in=180,out=-90,draw=white] (-\w/2,\h) to (\w,\h/2);
    \draw [graviton] (-\w/2,0) -- (-\w/2,\h);
    \draw [graviton] (\w/2,0) -- (\w/2,\h);
    \draw [graviton] (\w/2,\h) to (\w,\h/2);
    \draw [thick] (-\w,\h) -- (\w,\h) (-\w,0) -- (\w,0);
    \end{tikzpicture}}\;
    \subfloat[]{\label{contacttopologies-d}\begin{tikzpicture}
    \draw [line width=8pt,in=180,out=-90,draw=white] (-\w/2,\h) to (\w,\h/2);
    \draw [graviton] (0,0) -- (-\w/2,\h);
    \draw [graviton] (0,0) -- (\w/2,\h);
    \draw [line width=8pt,in=180,out=-90,draw=white] (0,\h) to (\w,\h/2);
    \draw [graviton,in=180,out=-90] (0,\h) to (\w,\h/2);
    \draw [thick] (-\w,\h) -- (\w,\h) (-\w,0) -- (\w,0);
    \end{tikzpicture}} \\
    \subfloat[]{\label{contacttopologies-e}\begin{tikzpicture}
    \draw [line width=8pt,in=180,out=-90,draw=white] (-\w/2,\h) to (\w,\h/2);
    \draw [graviton] (0,0) -- (-\w/2,\h);
    \draw [graviton] (0,0) -- (\w/2,\h);
    %\draw [line width=8pt,in=180,out=-90,draw=white] (0,\h) to (\w,\h/2);
    \draw [graviton] (\w/4,\h/2) -- (\w,\h/2);
    \draw [thick] (-\w,\h) -- (\w,\h) (-\w,0) -- (\w,0);
    \end{tikzpicture}}\;
    \subfloat[]{\label{contacttopologies-f}\begin{tikzpicture}
    \draw [line width=8pt,in=180,out=-90,draw=white] (-\w/2,\h) to (\w,\h/2);
    \draw [graviton] (0,0) -- (-\w/2,\h);
    \draw [graviton] (0,0) -- (\w/2,\h);
    %\draw [line width=8pt,in=180,out=-90,draw=white] (0,\h) to (\w,\h/2);
    \draw [graviton] (\w/2,\h) -- (\w,\h/2);
    \draw [thick] (-\w,\h) -- (\w,\h) (-\w,0) -- (\w,0);
    \end{tikzpicture}}\;
    \subfloat[]{\label{contacttopologies-g}\begin{tikzpicture}
    \draw [line width=8pt,in=180,out=-90,draw=white] (-\w/2,\h) to (\w,\h/2);
    \draw [graviton] (0,0) -- (-\w/2,\h);
    \draw [graviton] (0,0) -- (\w/2,\h);
    %\draw [line width=8pt,in=180,out=-90,draw=white] (0,\h) to (\w,\h/2);
    \draw [graviton] (0,0) -- (\w,\h/2);
    \draw [thick] (-\w,\h) -- (\w,\h) (-\w,0) -- (\w,0);
    \end{tikzpicture}}\;
    \subfloat[]{\label{contacttopologies-h}\begin{tikzpicture}
    \draw [line width=8pt,in=180,out=-90,draw=white] (-\w/2,\h) to (\w,\h/2);
    \draw [graviton] (-\w/2,0) -- (-\w/2,\h);
    \draw [graviton,in=-90,out=-90] (-\w/2,\h) to (\w/2,\h);
    \draw [graviton] (\w/2,\h) -- (\w,\h/2);
    \draw [thick] (-\w,\h) -- (\w,\h) (-\w,0) -- (\w,0);
    \end{tikzpicture}}
    \caption{Topologies of the contact terms that contribute to 
    the classical limit of the five-point amplitude before reduction to master integrals. The complete basis includes all inequivalent permutations of these diagrams.}
    \label{contacttopologies}
\end{figure}
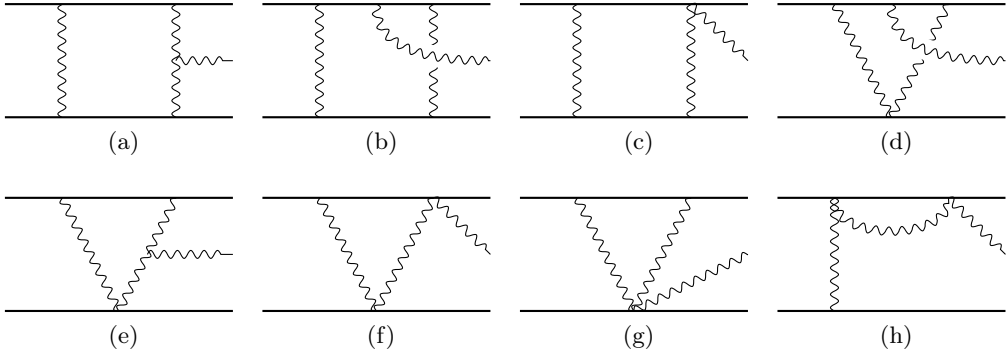
%%%%%%%%%%%%%%%%%%%%%%%%%%%%%%%                        

Merging these cuts using the method of maximal cuts~\cite{Bern:2007ct} while maintaining quadratic propagators for the matter fields leads to the relevant part of the one-loop integrand, which includes only graphs with three, four, and five propagators with at least one matter line in the loop. The relevant topologies are shown in \cref{contacttopologies}.
Diagrams with fewer propagators either do not have internal matter lines or intersecting matter lines, neither of which contribute to the classical amplitude.

We then expand it in the soft region limit~\cite{Parra-Martinez:2020dzs}, $q_{1,2}, k, \ell\ll p_{1, 2}$ , e.g. 
\be
\frac{1}{(p+\ell)^2-m^2+i0} = \frac{1}{2p\cdot \ell + \ell^2+i0}=
\frac{1}{2p\cdot \ell+i0}\sum_{n=0}^{\infty} \left(-\frac{\ell^2}{2p\cdot \ell+i0}\right)^n \ .
\ee
In practice, we also convert to $\pb_i$ variables defined in \cref{barredmomenta} at this step, which will introduce additional $q_i$ dependence in the above expansion.

The leading soft-region scaling of the five-point one-loop amplitude is super-classical, as expected from the existence of graphs with 
two-particle matter cuts. 
Direct inspection of contributing diagrams suggests that one of them, \cref{contacttopologies-b}, scales as $q^{-3}$ while the diagrams in  
\cref{contacttopologies-a,contacttopologies-c,contacttopologies-d,contacttopologies-e} scale as $q^{-2}$, and the remaining two triangle graphs exhibit classical scaling, $q^{-1}$,  at leading order.
After soft expanding to ${\cal O}(q^{-1})$, in which the diagrams with topology \cref{contacttopologies-b} must be expanded to second order, all matter propagators have linear dependence in loop momenta~\cite{Parra-Martinez:2020dzs} and may be raised to a power higher than one. 
Upon soft expanding the integrand, the two matter propagators in the top matter line of \cref{contacttopologies-b,contacttopologies-d} become linearly dependent because the momentum of the outgoing graviton is of the 
same order, ${\cal O}(\hbar)$, as the loop momentum. Linear dependence of propagators prevents a direct IBP reduction. These integrands are first partial-fractioned, and the resulting terms are assigned to the box diagrams in which the graviton is attached to the left and right vertex on that matter line,
\begin{align}
\vcenter{\hbox{\begin{tikzpicture}
    \draw [line width=8pt,in=180,out=-90,draw=white] (-\w/2,\h) to (\w,\h/2);
    \draw [graviton] (-\w/2,0) -- (-\w/2,\h);
    \draw [graviton] (\w/2,0) -- (\w/2,\h);
    \draw [line width=8pt,in=180,out=-90,draw=white] (0,\h) to (\w,\h/2);
    \draw [graviton,in=180,out=-90] (0,\h) to (\w,\h/2);
    \draw (-\w,\h) -- (\w,\h) (-\w,0) -- (\w,0);
\end{tikzpicture}}}
\xrightarrow{\text{partial fraction}}
\vcenter{\hbox{\begin{tikzpicture}
    \draw [line width=8pt,in=180,out=-90,draw=white] (-\w/2,\h) to (\w,\h/2);
    \draw [graviton] (-\w/2,0) -- (-\w/2,\h);
    \draw [graviton] (\w/2,0) -- (\w/2,\h);
    \draw [graviton] (\w/2,\h) to (\w,\h/2);
    \draw (-\w,\h) -- (\w,\h) (-\w,0) -- (\w,0);
\end{tikzpicture}}}
\;
+
\;
\vcenter{\hbox{\begin{tikzpicture}
    \draw [graviton] (\w/2,0) -- (\w/2,\h);
    \draw [line width=8pt,in=180,out=-90,draw=white] (-\w/2,\h) to (\w,\h/2);
    \draw [graviton] (-\w/2,0) -- (-\w/2,\h);
    \draw [graviton,in=180,out=-60] (-\w/2,\h) to (\w,\h/2);
    \draw (-\w,\h) -- (\w,\h) (-\w,0) -- (\w,0);
\end{tikzpicture}}}\,.
\end{align}
A similar feature occurs in the calculation of radiative observables 
at two and higher loops~\cite{Herrmann:2021tct} and in the calculation of the tail effect at three loops~\cite{Bern:2021yeh}.

To expose cut matter propagators and make contact with the HEFT integrand we separate each diagram into its symmetric and antisymmetric parts with respect to permutations of vertices on the two matter lines.  Using the identity in eq.~\eqref{propagator_cut_identity}, the symmetric part effectively cuts a matter propagator~\cite{Akhoury:2013yua}. 
In the analogous four-point one-loop calculation, it suffices to carry out this procedure for only one matter line in each diagram; if the diagram has a second matter line, it gets cut by changing integration variables and using the same identity.
Here, because ${\bar p}_1\cdot q_2 \ne 0 \ne {\bar p}_2\cdot q_1 $, we need to actively decompose each diagram into symmetric and antisymmetric parts with respect to \emph{both} matter lines.\footnote{We remove the $\ie$ prescription in the linearized matter propagators that are independent of the loop momentum. They will at most contribute terms proportional to $\delta(\ub_i\cdot k)\sim\delta(\omega)$, which are irrelevant to the waveform computation as discussed in \cref{KMOCloopSummary}.}

The antisymmetric part can then be reinterpreted as the principal value of the matter propagator. Interestingly, the terms in which both matter propagators are replaced by 
their principal values cancel out for the classical amplitude and, as one might expect, 
the resulting classical integrand has at least one cut matter line. This cut, which we will shortly identify with the cut present in all HEFT diagrams, implies that at least one of the gravitons present in each diagram is in the potential region.\footnote{For example, in an integral with the topology of \cref{contacttopologies-f}, the cut matter line implies that the left graviton is in the potential region while the right graviton could be in the radiation region. As we will see in the next section, the integral $I_{1,0,1,0}$ described here indeed has both a real and an imaginary part.}
We note that the principal-valued propagators appear naturally from a first principle unitarity calculation without any assumptions. 

In the quantum theory one may construct the bilinear-in-$T$ part of the waveform by simply taking the $s_{12}$-channel cut of the five-point amplitude. The appearance of the principal-valued propagators, which have no pole or imaginary part, make this step difficult. With hindsight which we will justify in Sec.~\ref{cutcontrib} and noticing that classical diagrams with one principal-valued propagator may be written as sums of two diagrams differing by either $p_2\leftrightarrow p_3$ or $p_1\leftrightarrow p_4$, we may still extract (at this loop order) the bilinear-in-$T$ part of the waveform from the classical amplitude.

\subsection{Integrand reduction}

We have computed the integrand using generalized unitarity at the level of HEFT and the full quantum amplitude. We now reduce the integrand to a basis of master integrals with kinematic coefficients. We first expand the polarization tensor, $\varepsilon^{\mu\nu} = \varepsilon^{\mu}\varepsilon^{\nu}$ and $\varepsilon^2 = 0$, in a basis of external momenta:
\begin{align}
\label{EPSdecomposition}
\varepsilon^{\mu} & = \alpha_1 {\bar p}^{\mu}_1 +\alpha_2 {\bar p}^{\mu}_2 +\alpha_3 q^{\mu}_1+\alpha_4 q^{\mu}_2 \ .
\end{align}
This decomposition, equivalent to Passarino-Veltman reduction~\cite{Passarino:1978jh}, introduces spurious poles in the form of Gram determinants, $G[\pb_1,\pb_2, q_1, q_2]$, which should cancel in the final integrated expression. The resulting separated and soft-expanded integrands are reduced to master integrals using FIRE \cite{Smirnov:2008iw, Smirnov:2019qkx}.
We keep pentagon, box, triangle, and bubble integrals that do not vanish in the classical limit. While bubble integrals are independent of the momentum transfer, their coefficients, which are generated by the IBP reduction, can exhibit such 
a dependence and thus can contribute nontrivially in the classical limit.

Integrands with different numbers of cut propagators form distinct sectors under IBP reduction. 
Diagrams with two cut matter lines are the 2MPR contributions are not included (though of course computable) in the HEFT amplitude. Factorization of the one-loop amplitude of the full theory identifies these terms as the product of the classical limit of four-point and five-point tree amplitudes.
Integrals with a single cut matter line, including mushroom graphs, are the same as in the reduction of the HEFT 2PMI amplitude~\eqref{MIpart1} and we have verified that the coefficients are also the same.
This demonstrates the complete cancellation of the $\hbar/\hbar$ terms in the full-theory calculation; such terms appear at intermediate stages, with the numerator coming from quantum terms in one tree-level factor in a cut and the denominator from superclassical terms in another.

Ultimately, the classical 2MPI part of the amplitude becomes a linear combination of the master integrals
\begin{gather}
     I_{1,1,0,0}\,,\quad I_{0,1,0,1}^{+}+I_{0,1,0,1}^{-}\,,\quad I_{1,1,0,1}^{+}+I_{1,1,0,1}^{-}\,,\nonumber\\ 
     I_{0,0,1,0}\,,\quad I_{1,0,1,0}\,,\quad I_{0,0,1,1}^{+}+I_{0,0,1,1}^{-}\,,\quad I_{1,1,1,0}\,,\nonumber\\
     I_{1,0,1,1}^{+}+I_{1,0,1,1}^{-}\,,\quad I_{0,1,1,1}^{+}+I_{0,1,1,1}^{-}\,,\quad  I_{1,1,1,1}^{+}+I_{1,1,1,1}^{-}\,,\label{eq:master}
\end{gather}
and their image under exchanging the matter lines,
\begin{align}
\Mred_{5,m_1^2m_2^3}^\text{2MPI,cl.GR}&=
c_{1}I_{0,0,1,0}+c_{2}I_{1,1,0,0}+c_{3}I_{1,0,1,0} +c_{4} (I_{0,1,0,1}^{+}+I_{0,1,0,1}^{-}) \nonumber\\
&+c_{5}(I^{+}_{0,0,1,1}+I^{-}_{0,0,1,1})+c_{6}I_{1, 1, 1, 0} +c_{7}(I_{1, 1, 0, 1}^{+}+I_{1, 1, 0, 1}^{-})\nonumber\\
& + c_{8}(I^{+}_{1,0,1,1}+I^{-}_{1,0,1,1})+c_{9}(I_{0,1,1,1}^{+}+I_{0,1,1,1}^{-})+c_{10}(I_{1, 1, 1, 1}^{+} + I_{1, 1, 1, 1}^{-} )\,,
\\
\Mred_{5, \text{1 loop}}^\text{2MPI,cl.GR}&=\Mred_{5,m_1^2m_2^3}^\text{2MPI,cl.GR}+(\ub_1\leftrightarrow \ub_2, \mb_1\leftrightarrow \mb_2, q_1\leftrightarrow q_2) \ ,
\label{MIpart1}
\end{align}
where the integrals are defined in \cref{integrals_definition}. The master integral coefficients $c_{i}$ are lengthy and are included in the ancillary Mathematic-readable file. The symmetric combinations $I_{1, 1, 0, 1}^{+}+I_{1, 1, 0, 1}^{-}$ and $I_{1, 1, 1, 1}^{+} + I_{1, 1, 1, 1}^{-} $ correspond to replacing the linear matter propagators of these integrals with their principal value. We have checked with the authors of Ref.~\cite{Brandhuber:2023hhy} and we find full agreement on the master integral coefficients for GR. The seven master integrals in the second and third line of \cref{eq:master} have contributions from radiation region gravitons. They are thus complex and contain the radiation reaction as discussed in Ref.~\cite{Elkhidir:2023dco}.

\section{The classical five-point integrand in \texorpdfstring{${\cal N}=8$}{N=8} supergravity}
\label{sec:N8res}

The ${\cal N}=8$ supergravity provides an important laboratory to explore properties of gravitational theories in a setting where amplitudes have somewhat simpler expressions. In this section, we construct the classical 
five-point amplitude in this theory, with the massive scalars being the lightest Kaluza-Klein modes for scalar modes of gravitons in the compact dimensions while all other particles are zero-compact-momentum modes, following~\cite{Parra-Martinez:2020dzs}.

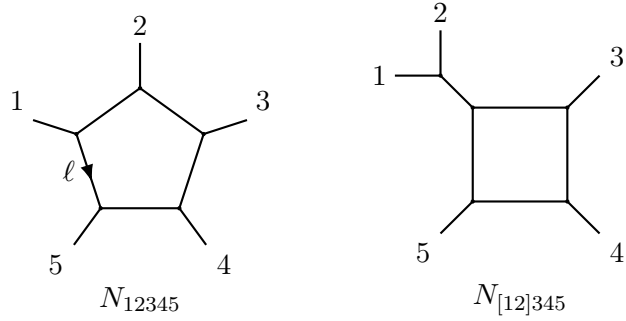
\begin{figure}
    \centering
    \begin{tikzpicture}[decoration={markings,mark=at position 0.65 with {\arrow{Latex}}}]
        \pgfmathsetmacro{\r}{1.1}
        \pgfmathsetmacro{\l}{0.75}
        \pgfmathsetmacro{\y}{-2.4}
        \pgfmathsetmacro{\s}{0.8}
        \begin{scope}[xshift=0cm,scale=\s]
            \foreach \x in {1,2,...,5} {
                \pgfmathsetmacro{\a}{162-72*(\x-1)}
                \pgfmathsetmacro{\b}{126-72*(\x-1)}
                \coordinate (\x) at ( \a : \r);
                \fill (\x) circle (1pt);
                \draw[thick] (\x) -- ++ ( \a : \l) node[label={[label distance=-5pt]\a:\x}] {};
                %\node at ( \b: {\r + 0.15} ) {$\ell_{\x}$};
            }
            \draw[thick] (2) -- (1);
            \draw[thick] (3) -- (2);
            \draw[thick] (4) -- (3);
            \draw[thick] (5) -- (4);
            \draw[thick,postaction={decorate}] (1) -- (5) node[pos=0.5,left=1pt]{$\lm$};
            \node at (0,\y) {$N_{12345}$};
        \end{scope}
        \begin{scope}[xshift=5cm,scale=\s]
            \foreach \x in {1,2,...,4} {
                \pgfmathsetmacro{\a}{135-90*(\x-1)}
                \coordinate (\x) at ( \a : \r);
                \fill (\x) circle (1pt);
            }
            \draw[thick] (2) -- (1);
            \draw[thick] (2) -- ++( 45 : \l) node[label={[label distance=-6pt]45:3}] {};
            \draw[thick] (3) -- (2);
            \draw[thick] (3) -- ++( -45 : \l) node[label={[label distance=-6pt]-45:4}] {};
            \draw[thick] (4) -- (3);
            \draw[thick] (4) -- ++( -135 : \l) node[label={[label distance=-6pt]-135:5}] {};
            \draw[thick] (1) -- (4);
            \draw[thick] (1) -- ++( 135 : \l) -- ++( 180 : \l) node[left=-1pt]{$1$};
            \draw[thick] (1) ++( 135 : \l ) -- ++( 90 : \l) node[above=-1pt]{$2$};
            \fill (1) ++( 135 : \l ) circle (1pt);
            \node at (0,\y) {$N_{[12]345}$};
        \end{scope}
    \end{tikzpicture}
    \caption{Nonvanishing numerators in five-point maximally-supersymmetric Yang-Mills and supergravity amplitudes.}
    \label{fig:Neq8}
\end{figure}

We construct massless maximally-supersymmetric supergravity amplitude in generic dimensions via the double copy. The classical tree-level five-point amplitude was constructed in Ref.~\cite{Luna:2017dtq}. With the notation introduced in \cref{fullGS1grav}, it is given by,
\begin{align}
\label{fullNeq81grav}
\Mred_{5, \text{tree}}^{\text{cl.} {\cal N}=8} = -\frac{\kappa^3}{4} 
{\bar m}_1^2 {\bar m}_2^2 \; \varepsilon^*(k)_{\mu\nu}  &\left[ \frac{4P_{12}^\mu P_{12}^\nu}{q_1^2 q_2^2} + \frac{2y}{q_1^2 q_2^2} \left( Q_{12}^\mu P_{12}^\nu + Q_{12}^\nu P_{12}^\mu\right) \right.  \nonumber\\
&
\left. + y^2 \left(\frac{Q_{12}^\mu Q_{12}^\nu}{q_1^2 q_2^2} - \frac{P_{12}^\mu P_{12}^\nu}{(k \cdot {\bar u}_1)^2 (k\cdot {\bar u}_2)^2} \right) 
\right] \ .
\end{align}
It differs from that of GR result in eq.~\eqref{fullGS1grav} only by the inclusion of the dilaton exchange here.

At one loop, we start with the one-loop five-gluon BCJ numerators of maximally-supersymmetric Yang-Mills theory~\cite{Mafra:2014gja,He:2017spx}, which consist of only pentagon and box topology, shown in \cref{fig:Neq8},
\begin{align}\label{eq:Nsym}
    N_{12345}^{\text{SYM}}&=\frac{1}{2}\lm_{\mu}t_8^{\mu}(1,2,3,4,5)-\frac{1}{4}\Big[t_8(\sf_{12},\sf_3,\sf_4,\sf_5)+t_8(\sf_{13},\sf_2,\sf_4,\sf_5)+t_8(\sf_{14},\sf_2,\sf_3,\sf_5)\nonumber\\
    &\quad+t_8(\sf_{15},\sf_2,\sf_3,\sf_4)+t_8(\sf_{23},\sf_1,\sf_4,\sf_5)+t_8(\sf_{24},\sf_1,\sf_3,\sf_5)+t_8(\sf_{25},\sf_1,\sf_3,\sf_4)\nonumber\\
    &\quad+t_8(\sf_{34},\sf_1,\sf_2,\sf_5)+t_8(\sf_{35},\sf_1,\sf_2,\sf_4)+t_8(\sf_{45},\sf_1,\sf_2,\sf_3)\Big] \ ,\\
    N_{[12]345}^{\text{SYM}}&=-\frac{1}{2}t_8(\sf_{12},\sf_3,\sf_4,\sf_5)\,,\nonumber
\end{align}
where the scalar and vector $t_8$ tensors are defined as
\begin{align}
t_8(\sf_1,\sf_2,\sf_3,\sf_4)&=\Tr[\sf_1 \sf_2 \sf_3 \sf_4]-\frac{1}{4}\Tr[\sf_1 \sf_2]\Tr[\sf_3 \sf_4]+\text{cyclic}(2,3,4)
\\
t^\mu_8(1,2,3,4,5)&= \mathsf{e}_1^\mu t_8(\sf_2,\sf_3,\sf_4,\sf_5) + \mathsf{e}_2^\mu t_8(\sf_1,\sf_3,\sf_4,\sf_5) + \mathsf{e}_3^{\mu}t_8(\sf_1,\sf_2,\sf_4,\sf_5)\nonumber\\
&\quad +\mathsf{e}_4^{\mu}t_8(\sf_1,\sf_2,\sf_3,\sf_5)+\mathsf{e}_5^{\mu}t_8(\sf_1,\sf_2,\sf_3,\sf_4)\ .
\end{align}
The traces entering the definition of $t_8$ are over the Lorentz indices and the linearized one- and two-particle field strengths, $\sf_i^{\mu\nu}$ and $\sf_{ij}^{\mu\nu}$, are
\begin{align}
\sf_i^{\mu\nu} &= \mathsf{k}_i^\mu \mathsf{e}_i^\nu - \mathsf{k}_i^\nu \mathsf{e}_i^\mu \,, \nonumber
\\
\sf_{ij}^{\mu\nu}&= 
(\mathsf{e}_i\cdot \mathsf{k}_j) \sf_j^{\mu\nu} - (\mathsf{e}_j\cdot \mathsf{k}_i) \sf_i^{\mu\nu} 
+ \sf_i^\mu{}_\lambda \sf_j^{\lambda\nu} - \sf_j^\mu{}_\lambda \sf_i^{\lambda\nu} \,,
\end{align}
where $\mathsf{k}_i$ and $\mathsf{e}_i$ are respectively the massless momentum and polarization in higher dimensions. From \cref{eq:Nsym}, we can get the numerators of maximal supergravity through the double copy $N^{\text{SUGRA}}=(N^{\text{SYM}})^2$.

We introduce four compact dimensions and use a dimensional reduction in which the compact momenta responsible for the scalar masses are in orthogonal directions,\footnote{We may relax this assumption and essentially rotate one of the two compact momenta by an arbitrary angle, as in Ref.~\cite{Parra-Martinez:2020dzs}. }
\begin{align}\label{eq:dim_red}
    & \mathsf{k}_1=(p_1,m_1,0,0,0)\,, & & \mathsf{k}_2=(p_2,0,m_2,0,0)\,, \nonumber\\
    & \mathsf{k}_4=(p_4,-m_1,0,0,0)\,,& & \mathsf{k}_3=(p_3,0,-m_2,0,0)\,,\\
    & \mathsf{k}_5=(k,0,0,0,0)\,. & & \nonumber
\end{align}
Since the masses arise from higher-dimensional momenta, they obey conservation relations, i.e., they change signs with the orientation of the corresponding momentum. The on-shell condition $\mathsf{k}_{1,2,3,4}^2=0$ for higher dimensional massless momenta thus lead to the massive on-shell condition $p_1^2=p_4^2=m_1^2$ and $p_2^2=p_3^2=m_2^2$. The kinematic configuration in \cref{eq:dim_red} gives the following reduction rules for Mandelstam variables,
\begin{align}
    \begin{array}{llll} \mathsf{k}_1\cdot \mathsf{k}_4=p_1\cdot p_4+m_1^2 & \quad \mathsf{k}_1\cdot \mathsf{k}_2=p_1\cdot p_2 & \quad \mathsf{k}_1\cdot \mathsf{k}_3=p_1\cdot p_3 & \quad \mathsf{k}_5\cdot \mathsf{k}_i=k\cdot p_i\\
    \mathsf{k}_2\cdot \mathsf{k}_3=p_2\cdot p_3+m_2^2 & \quad \mathsf{k}_2\cdot \mathsf{k}_4=p_2\cdot p_4 & \quad \mathsf{k}_3\cdot \mathsf{k}_4=p_3\cdot p_4 &
    \end{array}\,,
\end{align}
while $\lm\cdot \mathsf{k}_i=\lm\cdot p_i$ for $i=1,2,3,4$, and $\lm\cdot \mathsf{k}_5=\lm\cdot k$. The massive scalars are realized as the scalar graviton modes in the compact dimensions,
\begin{align}
    \mathsf{e}_1=\mathsf{e}_4=(0,0,0,1,0)\,,\qquad\mathsf{e}_2=\mathsf{e}_3=(0,0,0,0,1)\,,\qquad\mathsf{e}_5=(\varepsilon,0,0,0,0)\,,
\end{align}
such that dot products involving polarization vectors are reduced by
\begin{align}
    &\mathsf{e}_1\cdot\mathsf{e}_4=\mathsf{e}_2\cdot\mathsf{e}_3=-1\text{ and }\mathsf{e}_i\cdot\mathsf{e}_j=0\text{ otherwise,}\nonumber\\
    &\mathsf{e}_i\cdot \mathsf{k}_j=\left\{\begin{array}{ll}
    0 & \text{ for }i=1,2,3,4\text{ and arbitrary }j \\ 
    \varepsilon\cdot p_j & \text{ for }i=5\text{ and }j=1,2,3,4
    \end{array}\right.\,.
\end{align}
The diagrams that survive as these relations are plugged into the one-loop integrand are the ones that are consistent with higher-dimensional momentum conservation at each vertex. Namely, we keep the diagrams in which the two matter lines connecting $\{p_1,p_4\}$ and $\{p_2,p_3\}$ do not cross each other. They are all captured by the spanning set of cuts in \cref{cutsGR}.

As in the GR calculation outlined in \cref{fullGR5points}, we expand the resulting integrand in the soft limit. By using eq.~\eqref{propagator_cut_identity}, we can expose all the 2MPR diagrams. The remaining 2MPI diagrams have one cut matter propagator.
Upon reduction to master integrals, the uncut linear matter propagators turn into principal-value propagators. The 2MPI part of the classical amplitude takes the same form as \cref{MIpart1}, with vanishing values for the coefficients $c^{{\cal N}=8}_2$ and $c^{{\cal N}=8}_3$, i.e.
\begin{align}
\Mred_{5,m_1^2m_2^3}^{\text{2PMI,cl.}\mathcal{N}=8}&=
c^{{\cal N}=8}_{1}I_{0,0,1,0}+c^{{\cal N}=8}_{4} (I_{0,1,0,1}^{+}+I_{0,1,0,1}^{-}) 
+c^{{\cal N}=8}_{5}(I^{+}_{0,0,1,1}+I^{-}_{0,0,1,1})
\nonumber\\
&\quad +c^{{\cal N}=8}_{6}I_{1, 1, 1, 0} +c^{{\cal N}=8}_{7}(I_{1, 1, 0, 1}^{+}+I_{1, 1, 0, 1}^{-})
+ c^{{\cal N}=8}_{8}(I^{+}_{1,0,1,1}+I^{-}_{1,0,1,1})
\nonumber\\
&\quad +c^{{\cal N}=8}_{9}(I_{0,1,1,1}^{+}+I_{0,1,1,1}^{-})+c^{{\cal N}=8}_{10}(I_{1, 1, 1, 1}^{+} + I_{1, 1, 1, 1}^{-} )\,,
\\
\Mred_{5, \text{1 loop}}^{\text{2MPI,cl.}{\cal N}=8}&=\mathcal{M}_{5,m_1^2m_2^3}^{\text{2PMI,cl.}\mathcal{N}=8}+ (\ub_1\leftrightarrow \ub_2, \mb_1\leftrightarrow \mb_2, q_1\leftrightarrow q_2) \ .
\label{MIpart1Neq8}
\end{align}
While the coefficients are somewhat unwieldy, it is not difficult to verify that each of them is separately gauge-invariant, as they should be.
Interestingly, in the classical limit, the ${\cal N}=8$ amplitude includes triangle and bubble integrals, unlike the quadratic-propagator counterpart~\cite{Bjerrum-Bohr:2006xbk}. Similar to the GR five-point amplitude, they contribute nontrivially to the waveform because their coefficients exhibit nontrivial dependence on the momentum transfer $(q_1-q_2)$.

\section{Integrating in the rest frame}
\label{integration}

In this section, we evaluate, for physical kinematic configurations, the bubble, triangle, and pentagon master integrals \cref{eq:master} that appear in GR and ${\cal N}=8$ supergravity one-loop five-point classical amplitudes. We also list the expressions for the box integrals and relegate the details of their evaluation to \cref{sec:integrals}. We note that the results given in this section are in full agreement with Ref.~\cite{Brandhuber:2023hhy}.

For all master integrals, it is convenient to work in the rest frame of particle~2, in which $\ub_2=(1,0,0,0)$, and integrate out $\hdelta(2\ub_2\cdot\lm)$ in the numerators. This projects out the temporal component of the loop momentum such that $\lm=(0,\lmb)$. We are thus left with a Euclidean loop integral in $3-2\epsilon$ dimensions, expressed in terms of non-covariant quantities. We can uplift the result back into a generic frame by using
\begin{align}\label{eq:rest_to_generic}
    & \ub_1^0=y\,,
    & & \pmb{\ub}_1^2=y^2-1\,,
    & & E_{q_1}=\ub_2\cdot q_1\,,
    & & \qbd_1^2=(\ub_2\cdot q_1)^2-q_1^2\,,\nonumber\\
    & \qbd_2^2=-q_2^2\,,
    & & \pmb{\ub}_1\cdot\qbd_1=y(\ub_2\cdot q_1)\,,
    & &\pmb{\ub}_1\cdot\qbd_2=-\ub_1\cdot q_2\,,
    & & \qbd_1\cdot\qbd_2=\frac{q_1^2+q_2^2}{2}\,.
\end{align}
where the left hand side comes from the components of $\ub_1$, $q_1$ and $q_2$ written in the $\ub_2$ rest frame, $\ub_1=(\ub_1^0,\pmb{\ub}_1)$, $q_1=(E_{q_1},\qbd_1)$ and $q_2=(0,\qbd_2)$. In the physical region, we have $\ub_1\cdot q_2>0$ and $\ub_2\cdot q_1>0$ because the outgoing graviton has positive energy.\footnote{The reason for $\ub_2\cdot q_1>0$ is that in the $\ub_2$ rest frame, $\ub_2\cdot q_1=\ub_2\cdot k_5=(k_5)^0$. Since the graviton is outgoing, its energy $(k_5)^0$ is always positive. We also have $\ub_1\cdot q_2>0$ for the same reason.} We also have $q_1^2<0$ and $q_2^2<0$, because the momentum transfer of particle $1$ and $2$ is spatial-like.

With these preparations, let us first discuss the bubble integral $I_{0,0,1,0}$ as the simplest example,
\begin{align}
    I_{0,0,1,0}=\int\frac{\dd^{4-2\epsilon}\lm}{(2\pi)^{4-2\epsilon}}\frac{\hdelta(2\ub_2\cdot\lm)}{(\lm-q_1)^2+\ie}&=-\frac{(4\pi^2)^{\epsilon}}{16\pi^3}\int\frac{\dd^{3-2\epsilon}\lmb}{\lmb^2-(\ub_2\cdot q_1)^2-\ie}\nonumber\\
    &=-\frac{(4\pi^2)^{\epsilon}\Gamma(\epsilon-1/2)}{16\pi^{3/2+\epsilon}[-(\ub_2\cdot q_1)^2-\ie]^{\epsilon-1/2}}\,.
\end{align}
In the classical amplitude, the bubble integral usually comes with a divergent coefficient $\frac{1}{d-4}$ that encodes part of the IR divergence. We expand this combination up to the terms finite in $\epsilon$,
\begin{align}\label{eq:IRbubble}
    \frac{1}{d-4}I_{0,0,1,0}=-\frac{\ub_2\cdot q_1}{16}-\frac{i(\ub_2\cdot q_1)}{16\pi}\left[-\frac{1}{\eIR}-2+\log(4\pi(\ub_2\cdot q_1)^2)\right]+\mathcal{O}(\epsilon)\,,
\end{align}
where we have defined for convenience,
\begin{align}
    -\frac{1}{\eIR}\equiv-\frac{1}{\epsilon}+\gamma_{\text{E}}-\log(4\pi^2)\,,
\end{align}
and $\gamma_{\text{E}}$ is the Euler constant. It is crucial to track the $\ie$ prescription to determine the analytic continuation into the physical region of external kinematics.\footnote{To derive \cref{eq:IRbubble}, we have used the analytic continuation $[-(\ub_2\cdot q_1)^2-\ie]^{1/2}=-i(\ub_2\cdot q_1)$ and $\log(-(\ub_2\cdot q_1)^2-\ie)=\log(\ub_2\cdot q_1)^2-i\pi$ with $\ub_2\cdot q_1>0$.} We will mainly consider the results in $d=4$ only, so we will omit the $\mathcal{O}(\epsilon)$ terms in the following.

\subsection{Triangle integrals}

Next, we consider the triangle integral $I_{1,1,0,0}$. We first introduce the Feynman parameterization to combine the denominator while going to the $\ub_2$ rest frame. The integral is then straightforward to work out,
\begin{align}
    I_{1,1,0,0}=\int\frac{\dd^4\lm}{(2\pi)^4}\frac{\hdelta(2\ub_2\cdot\lm)}{\lm^2(\lm+q_2)^2}=\frac{1}{16\pi^3}\int_{0}^{1}\dd x\int\frac{\dd^3\pmb{\lm}}{[\pmb{\lm}^2+x(1-x)\pmb{q}_2^2]^2}=\frac{1}{16\sqrt{-q_2^2}}\,.
\end{align}
This integral is purely real as expected because further cutting either $\lm^2$ or $(\lm+q_2)^2$ leads to vanishing results due to on-shell three-point kinematics. In contrast, the other triangle master integral $I_{1,0,1,0}$ is complex. We start with the same Feynman parameterization, and the integration proceeds as before,
\begin{align}
    I_{1,0,1,0}=\int\frac{\dd^4\lm}{(2\pi)^4}\frac{\hdelta(2\ub_2\cdot\lm)}{\lm^2[(\lm-q_1)^2+\ie]}&=\frac{1}{16\pi^3}\int_{0}^{1}\dd x\int\frac{\dd^3\pmb{\lm}}{[\pmb{\lm}^2+x(1-x)\pmb{q}_1^2-x E_{q_1}^2-\ie]^2}\nonumber\\
    &=\frac{1}{8\pi\sqrt{\pmb{q}_1^2}}\arcsin\left(\sqrt{\frac{\pmb{q}_1^2}{\pmb{q}_1^2-E_{q_1}^2}}+\ie\right)\nonumber\\
    &=\frac{1}{16\sqrt{\pmb{q}_1^2}}+\frac{i}{8\pi\sqrt{\pmb{q}_1^2}}\arccosh\sqrt{\frac{\pmb{q}_1^2}{\pmb{q}_1^2-E_{q_1}^2}}\,.
\end{align}
We note that the argument of $\arcsin$ in the second line is greater than $1$ in the physical region. The analytic continuation in $+\ie$ prescription is
\begin{align}
    \arcsin(x+\ie)=\frac{\pi}{2}+i\arccosh(x)\quad\text{for}\quad x>1\,.
\end{align}
By using \cref{eq:rest_to_generic}, we can uplift the result to a generic frame,
\begin{align}
    % I_{1,0,1,0}=\frac{1}{16\sqrt{(\ub_2\cdot q_1)^2-q_1^2}}+\frac{i}{8\pi\sqrt{(\ub_2\cdot q_1)^2-q_1^2}}\arccosh\sqrt{1+\frac{(\ub_2\cdot q_1)^2}{-q_1^2}}\,.
    I_{1,0,1,0}=\frac{1}{16\sqrt{(\ub_2\cdot q_1)^2-q_1^2}}+\frac{i}{8\pi\sqrt{(\ub_2\cdot q_1)^2-q_1^2}}\arcsinh\frac{\ub_2\cdot q_1}{\sqrt{-q_1^2}}\,,
\end{align}
where we have also used the identity $\arccosh\sqrt{x+1}=\arcsinh\sqrt{x}$.

To compute the integrals with a linear matter propagator, we use a different parameterization to combine the denominator,
\begin{align}\label{eq:x_parameter}
    \frac{1}{ab}=\int_{0}^{\infty}\frac{\dd x}{(a+xb)^2}\,.
\end{align}
Applying eq.~(\ref{eq:x_parameter}) to $I^{+}_{0,0,1,1}$, we get
\begin{align}
    I^{+}_{0,0,1,1}=\int\frac{\dd^4\lm}{(2\pi)^4}\frac{\hdelta(2\ub_2\cdot\lm)}{(\lm-q_1)^2(2\ub_1\cdot\lm+\ie)}&=\int_{0}^{\infty}\dd x\int\frac{\dd^4\lm}{(2\pi)^4}\frac{\hdelta(2\ub_2\cdot\lm)}{[(\lm-q_1)^2+2x\ub_1\cdot\lm+\ie)]^2}\nonumber\\
    &=\frac{i}{16\pi\sqrt{y^2-1}}\int_{-\alpha}^{\infty}\frac{\dd x}{(x^2+\beta)^{1/2}}\,,
\end{align}
where $\alpha=\frac{y(\ub_2\cdot q_1)}{y^2-1}$ and $\beta=-\frac{(\ub_2\cdot q_1)^2}{(y^2-1)^2}+\ie$. The integration over $x$ is divergent at $x\rightarrow\infty$, which encodes the IR divergence due to the linear matter propagator $2\ub_1\cdot\lm$. However, in the classical amplitude, the linear matter propagator only appears in the principal-valued combination $I^{+}_{0,0,1,1}+I^{-}_{0,0,1,1}$. After including $I^{-}_{0,0,1,1}$,
\begin{align}
    I^{-}_{0,0,1,1}=\int\frac{\dd^4\lm}{(2\pi)^4}\frac{\hdelta(2\ub_2\cdot\lm)}{(\lm-q_1)^2(2\ub_1\cdot\lm-\ie)}&=-\int\frac{\dd^4\lm}{(2\pi)^4}\frac{\hdelta(2\ub_2\cdot\lm)}{(\lm+q_1)^2(2\ub_1\cdot\lm+\ie)}\nonumber\\
    &=-I^{+}_{0,0,1,1}\Big|_{q_1\rightarrow-q_1}\,
\end{align}
we find that the range of $x$ gets truncated and we get a finite result,
\begin{align}
    I^{+}_{0,0,1,1}+I^{-}_{0,0,1,1}=\frac{i}{8\pi\sqrt{y^2-1}}\int_{0}^{\alpha}\frac{\dd x}{(x^2+\beta)^{1/2}}=\frac{i}{8\pi\sqrt{y^2-1}}\arcsinh\frac{\alpha}{\sqrt{\beta}}\,.
\end{align}
The $+\ie$ prescription in $\beta$ leads to the following analytic continuation of the $\arcsinh$ function in the physical region,
\begin{align}
    \arcsinh\frac{\alpha}{\sqrt{\beta}}=-\frac{i\pi}{2}+\arccosh(y)\,.
\end{align}
Therefore, the final result of this triangle integral is
\begin{align}\label{eq:I0011pv}
    I^{+}_{0,0,1,1}+I^{-}_{0,0,1,1}=\frac{1}{16\sqrt{y^2-1}}+\frac{i}{8\pi\sqrt{y^2-1}}\arccosh(y)\,.
\end{align}
We can apply the same technique to compute
\begin{align}
    I^{\pm}_{0,1,0,1}=\int\frac{\dd^d\lm}{(2\pi)^d}\frac{\hdelta(2\ub_2\cdot\lm)}{(\lm+q_2)^2(2\ub_1\cdot\lm\pm\ie)}\,.
\end{align}
In particular, the principal value combination is given by
\begin{align}
    I^{+}_{0,1,0,1}+I^{-}_{0,1,0,1}=\frac{i}{8\pi\sqrt{y^2-1}}\int_{0}^{\gamma}\frac{\dd x}{(x^2-\gamma^2+\ie)^{1/2}}=\frac{1}{16\sqrt{y^2-1}}\,,
\end{align}
where $\gamma=\frac{\ub_1\cdot q_2}{y^2-1}$.

\subsection{Box and pentagon integrals}\label{sec:box}

The box master integrals are all individually IR divergent. For $I_{1,1,1,0}$, the IR divergence is due to the presence of a massless three-point vertex,
\begin{align}\label{eq:I1110re}
    & I_{1,1,1,0}
    =\frac{1}{32q_2^2(\ub_2\cdot q_1)}+\frac{i}{32 \pi q_2^2(\ub_2\cdot q_1)}\left[-\frac{1}{\eIR}+\log(4\pi(\ub_2\cdot q_1)^2)+2\log\frac{q_2^2}{q_1^2}\right]\,,
\end{align}
The other box integrals, $I^{\pm}_{1,1,0,1}$, $I^{\pm}_{1,0,1,1}$ and $I^{\pm}_{0,1,1,1}$, all have an IR divergence due to the linear matter propagator $2\ub_1\cdot\lm$. This IR divergence cancels in the classical amplitude because the linear matter propagator always appears as a principal value,
\begin{subequations}
\begin{align}
    \label{eq:I1101pv}
    & I^{+}_{1,1,0,1}+I^{-}_{1,1,0,1}=\frac{1}{16q_2^2\sqrt{y^2-1}}\,,\\
    \label{eq:I1011re}
    & I^{+}_{1,0,1,1}+I^{-}_{1,0,1,1}=\frac{1}{16 q_1^2 \sqrt{y^2-1}}+\frac{i}{8\pi q_1^2\sqrt{y^2-1}}\arccosh(y)\,.
\end{align}
\end{subequations}
On the other hand, $I^{+}_{0,1,1,1}+I^{-}_{0,1,1,1}$ remains IR divergent due to the presence of the same massless three-point vertex as in $I_{1,1,1,0}$,
\begin{align}\label{eq:I0111pv}
    I^{+}_{0,1,1,1}+I^{-}_{0,1,1,1}&=-\frac{1}{32(\ub_1\cdot q_2)(\ub_2\cdot q_1)}\\
    & \quad -\frac{i}{32\pi (\ub_1\cdot q_2)(\ub_2\cdot q_1)}\left[-\frac{1}{\eIR}+\log(4\pi(\ub_2\cdot q_1)^2)+2\log\frac{\ub_1\cdot q_2}{\ub_2\cdot q_1}\right]\,.\nonumber
\end{align}
We leave the derivations of these integrals to \cref{sec:integrals}.

Scalar pentagon integrals with quadratic propagators in $d=4-2 \epsilon$ dimensions can be written as a sum of box integrals~and a six-dimensional pentagon integral~\cite{vanNeerven:1983vr, Bern:1992em, Bern:1993kr}. 
We derive here an analogous decomposition for our pentagon integral with linear matter propagators one of which is cut, $I^{\pm}_{1,1,1,1}$. 
We first decompose the loop momentum $\lm$ into its four-dimensional and extra-dimensional component, $\lm^2=\lm_4^2+\mu^2$, where $\lm_4$ can be expressed as a linear combination of external kinematic data, $\lm_4=\alpha_1 \ub_1+\alpha_2\ub_2+\alpha_3 q_1+\alpha_4 q_2$. The coefficients $\alpha_i$ contain the $\lm$ dependence through scalar products $v\cdot \lm_4=v\cdot\lm$ with $v\in\{\ub_1,\ub_2,q_1,q_2\}$. We then plug the above relation for $\lm_4$ into the identity
\begin{align}
    0=\int\frac{\dd^{4-2\epsilon}\lm}{(2\pi)^{4-2\epsilon}}\frac{\hdelta(2\ub_2\cdot\lm)}{\lm^2(\lm+q_2)^2(\lm-q_1)^2(2\ub_1\cdot\lm)}(\lm^2-\lm_4^2-\mu^2)\,,
\end{align}
express all the $v\cdot\lm$ in terms of inverse propagators, and perform the standard tensor reduction. This will lead to a linear relation that expresses the pentagon integral in terms of the box integrals in $d=4-2\epsilon$ and another pentagon integral $I^{\pm,d=6-2\epsilon}_{1,1,1,1}$ in $d=6-2\epsilon$,
\begin{align}\label{eq:pentagon_expansion}
    I^{\pm}_{1,1,1,1}=\beta_1 I^{\pm}_{1,1,0,1}+\beta_2 I^{\pm}_{1,0,1,1}+\beta_3 I^{\pm}_{0,1,1,1}+\beta_4 I_{1,1,1,0}+\beta_5\,\epsilon\, I^{\pm,d=6-2\epsilon}_{1,1,1,1}\,,
\end{align}
where $\beta_i$ only depend on external Mandelstam variables. The  pentagon integral $I^{\pm,d=6-2\epsilon}_{1,1,1,1}$ comes from evaluating the $\mu^2$ term contribution by the dimension shift relation~\cite{BernMorgan},
\begin{align}
    \int\dd^{4-2\epsilon}\lm\frac{\mu^2\hdelta(2\ub_2\cdot\lm)}{\lm^2(\lm+q_2)^2(\lm-q_1)^2(2\ub_1\cdot\lm)}=-\frac{\epsilon}{\pi}\int\dd^{6-2\epsilon}\lm\frac{\hdelta(2\ub_2\cdot\lm)}{\lm^2(\lm+q_2)^2(\lm-q_1)^2(2\ub_1\cdot\lm)}\,.
\end{align}
In the principal value combination $I^{+}_{1,1,1,1}+I^{-}_{1,1,1,1}$, this term is finite in $d=6-2\epsilon$ such that it contributes at most to $\mathcal{O}(\epsilon)$. The coefficients $\beta_i$ in \cref{eq:pentagon_expansion} are straightforward to compute. Here we just give the final result of the pentagon integral,
\begin{align}\label{pentagonintegral}
    I^{+}_{1,1,1,1}+I^{-}_{1,1,1,1}&=\frac{1}{32 q_{2}^{2}}\left (\frac{2}{q_{1}^{2}\sqrt{y^{2}-1}}-\frac{1}{(\ub_{1}\cdot q_{2})(\ub_{2}\cdot q_{1})} \right )\\
    & \quad +\frac{i}{32\pi q_2^2 (\ub_1\cdot q_2)(\ub_2\cdot q_1)}\left[\frac{1}{\eIR}-\log(4\pi(\ub_2\cdot q_1)^2)-2\log\frac{q_2^2}{q_1^2}\right]\nonumber\\
    & \quad +\frac{i}{16\pi}\left[\frac{C_1}{(\ub_1\cdot q_2)(\ub_2\cdot q_1)} \log\frac{(\ub_1\cdot q_2)q_1^2}{(\ub_2\cdot q_1)q_2^2}+\frac{C_2 \arccosh(y)}{q_1^2\sqrt{y^2-1}}\right]\,,\nonumber
\end{align}
where $C_{1}$ and $C_2$ are given by
\begin{align}
    C_1&=\frac{y(\ub_2\cdot q_1)(\ub_1\cdot q_2)(q_1^2+q_2^2)-(\ub_1\cdot q_2)^2q_1^2-(\ub_2\cdot q_1)^2q_2^2+2(\ub_1\cdot q_2)^2(\ub_2\cdot q_1)^2}{(\ub_2\cdot q_1)^2q_2^4-2y(\ub_1\cdot q_2)(\ub_2\cdot q_1)q_1^2q_2^2+(\ub_1\cdot q_2)^2q_1^4}\,,\nonumber\\
    C_2&=\frac{2(\ub_2\cdot q_1)^2q_2^2-2y(\ub_1\cdot q_2)(\ub_2\cdot q_1)q_1^2-(y^2-1)q_1^2(q_1^2-q_2^2)}{(\ub_2\cdot q_1)^2q_2^4-2y(\ub_1\cdot q_2)(\ub_2\cdot q_1)q_1^2q_2^2+(\ub_1\cdot q_2)^2q_1^4}\, .
\end{align}
%We include the details in Appendix~\ref{sec:integrals}.

\section{The five-point amplitude and its infrared divergences}
\label{integratedamplitudes}

Having evaluated all the relevant master integrals, we can assemble the GR and ${\cal N}=8$ classical amplitudes and carry our various consistency checks. 
When writing down the classical amplitude we can also remove the distinction 
between $\{m_i, p_i, u_i, \sigma=u_1\cdot u_2\}$ and $\{\mb_i, \pb_i, \ub_i, y=\ub_1\cdot \ub_2\}$, as they differ only by positive powers $q_1$ and $q_2$, i.e. by terms with quantum scaling in the soft expansion.

By inspecting the non-rational terms in the integrals evaluated in \cref{integration} and \cref{sec:integrals} it is straightforward to see that they are all linear combinations of the functions
\begin{gather}
\frac{1}{\sqrt{-q_1^2}}\,
,\quad
\frac{1}{\sqrt{-q_2^2}}\,
,\quad
\frac{1}{\sqrt{(k\cdot {u}_2)^2-q_1^2}}\,
,\quad
\frac{1}{\sqrt{(k\cdot {u}_1)^2-q_2^2}}\,
,\quad
\log\frac{(k\cdot {u}_1)^2(k\cdot {u}_2)^2}{\mu^4}\,
,\nonumber\\
\log\frac{q_2^2}{q_1^2}\,
,\quad
\log\frac{k\cdot{u}_1}{k\cdot{u}_2}\,  
,\quad
\frac{\arcsinh\frac{k\cdot {u}_2}{\sqrt{-q_1^2}}}
{\sqrt{(k\cdot {u}_2)^2-q_1^2}}\,
,\quad
\frac{\arcsinh\frac{k\cdot {u}_1}{\sqrt{-q_2^2}}}
{\sqrt{(k\cdot {u}_1)^2-q_2^2}}\,
,\quad
\frac{\arccosh \sigma }{(\sigma^2-1)^{3/2}}\,
,
\label{functions}
\end{gather}
with rational coefficients. Here $\mu$ is the scale of dimensional regularization, and this $\mu$-dependent logarithm is intimately related to IR divergence of amplitudes. 

In both GR and ${\cal N}=8$ supergravity, the classical two-scalar-one-graviton five-point amplitude has the general form 
\begin{align} \label{integratedAmplitude}
\Mred_{5, \text{1 loop}}^{\text{2MPI}, \text{cl.}} &= 
-\frac{i\, \kappa^2}{32\pi} (k\cdot {p}_1 + k\cdot {p}_2)\left(\frac{1}{\epsilon} 
- \frac{1}{2}\log\frac{16 \pi^2 (k\cdot {u}_1)^2 (k\cdot {u}_2)^2}{\mu^4}\right) \Mred_{5, \text{tree}}^{\text{cl.}}
\nonumber\\
&\quad
+\kappa^5\left(A^R_\text{rat} 
+ \frac{A^R_1}{\sqrt{(k\cdot {u}_2)^2-q_1^2}}
+ \frac{A^R_2}{\sqrt{(k\cdot {u}_1)^2-q_2^2}}+\frac{A_3^{R}}{\sqrt{-q_1^2}}+\frac{A_4^{R}}{\sqrt{-q_2^2}}\right)\nonumber\\
& \quad
+ i \,\kappa^5 \,\Bigg( A^I_\text{rat} 
%+ i \left( \frac{\kappa}{2}\right)^5\left( A^I_\text{rat} 
+ A^{I}_1\frac{\arcsinh\frac{k\cdot u_2}{\sqrt{-q_1^2}}}{\sqrt{(k\cdot {u}_2)^2-q_1^2}}
+ A^{I}_2\frac{\arcsinh\frac{k\cdot u_1}{\sqrt{-q_2^2}}}{\sqrt{(k\cdot {u}_1)^2-q_2^2}}
%\right.\\
%& \left. \qquad
+A^I_3 \log\frac{q_2^2}{q_1^2}\nonumber\\
& \qquad\qquad\qquad + A^I_4 \log\frac{k\cdot{u}_1}{k\cdot{u}_2}  
+ A^I_5 \frac{\arccosh \sigma}{(\sigma^2-1)^{3/2}} \Bigg) \ ,
\end{align}
%%%%%%%%%%%%%%%%%%%
where the coefficient functions $A^{R,I}_{i}$ are rational combinations of momentum invariants and polarization vector $\varepsilon$. Their scaling in the soft limit is homogeneous and it is such that $\Mred_{5, \text{1 loop}}^{\text{2MPI}, \text{cl.}}$ has classical $q^{-1}$ scaling.
They are gauge invariant as a consequence of the gauge invariance of the
master integral coefficients in \cref{MIpart1,MIpart1Neq8}.
The explicit expressions of the coefficient functions $A^{R,I}_{i}$ are included in two ancillary Mathematica-readable files, {\tt GR\_Coeffs.m} and {\tt Neq8\_Coeffs.m}, for GR and ${\cal N}=8$ supergravity, respectively.

The first line of \cref{integratedAmplitude}, reproducing expectations based on Weinberg's theorem \eqref{IRdivGeneral}, requires a nontrivial interplay between the $c_i$ and $c_i^{{\cal N}=8}$ coefficients determining the GR and ${\cal N}=8$ supergravity classical integrands, \cref{MIpart1,MIpart1Neq8} respectively, is a check of our calculation.
As discussed in \cref{IRdivANDwaveform}, the definition of the observation time $\tau$ in \cref{tredef} absorbs the IR divergence such that the part of the one-loop five-point amplitude 
that contributes to the (spectral) waveform and the Newman-Penrose scalar
is
\begin{align} 
\label{integratedAmplitudeX}
\Mred_{5, \text{1 loop}}^{0, \text{2MPI}, \text{cl.}}  &= 
\kappa^5 \left(A^R_\text{rat} 
+ \frac{A^R_1}{\sqrt{(k\cdot {u}_2)^2-q_1^2}}
+ \frac{A^R_2}{\sqrt{(k\cdot {u}_1)^2-q_2^2}}+\frac{A_3^{R}}{\sqrt{-q_1^2}}+\frac{A_4^{R}}{\sqrt{-q_2^2}}\right)\nonumber\\
&
\quad + i \,\kappa^5 \, \Bigg( A^I_\text{rat} 
%+ i \left( \frac{\kappa}{2}\right)^5\left( A^I_\text{rat} 
+ A^{I}_1\frac{\arcsinh\frac{k\cdot u_2}{\sqrt{-q_1^2}}}{\sqrt{(k\cdot {u}_2)^2-q_1^2}}
+ A^{I}_2\frac{\arcsinh\frac{k\cdot u_1}{\sqrt{-q_2^2}}}{\sqrt{(k\cdot {u}_1)^2-q_2^2}}
%\right.\\
%& \left. \qquad
+A^I_3 \log\frac{q_2^2}{q_1^2} \nonumber\\
& \qquad\qquad\qquad
+ A^I_4 \log\frac{k\cdot{u}_1}{k\cdot{u}_2}  
+ A^I_5 \frac{\arccosh\sigma}{(\sigma^2-1)^{3/2}} \Bigg) \,, \nonumber\\
& \quad + \frac{i\,\kappa^2}{64\pi}(k\cdot p_1+k\cdot p_2)\left[\log\frac{16\pi^2(k\cdot u_1)^2(k\cdot u_2)^2}{\Lambda^4}\right]\Mred_{5,\text{tree}}^{\text{cl.}}\,,
\end{align}
where the dependence on the dimensional regularization scale $\mu$ has been replaced by cutoff defining the virtual IR gravitons $\Lambda$ due to \cref{tredef}.

We have also explored the fate of spurious singularities. 
The Gram determinant $G[\pb_1,\pb_2, q_1, q_2]$ arising from the decomposition \eqref{EPSdecomposition} of the graviton polarization tensor into a basis of external momenta cancels in \cref{integratedAmplitude} within each coefficient $A_i^{R,I}$ with the help of four dimensional identities involving the vanishing Gram determinant $G[\pb_1,\pb_2, q_1, q_2,\varepsilon]$. 
Other spurious singularities occur for kinematic configurations that set to zero denominator factors arising from the IBP reduction. They are solutions to the equations
\begin{align}\label{eq:Deltas}
\Delta_1 &= -2 \sigma (u_1 \cdot k) (u_2 \cdot k) q_1^2 q_2^2 + (u_2 \cdot k)^2 (q_2^2){}^2 +(u_1 \cdot k)^2  (q_1^2){}^2  = 0 \ ,
\nonumber\\
\Delta_2 &= -2 \sigma (u_1 \cdot k) (u_2 \cdot k) + 
(u_2 \cdot k)^2 +(u_1 \cdot k)^2 = 0\,,\nonumber\\
\Delta_3 &= (q_1^2 - q_2^2)^2 + 4 (u_2\cdot k)^2 q_2^2 = 0\,,\\
\Delta_4 &= (q_1^2 - q_2^2)^2 + 4 (u_1\cdot k)^2 q_1^2 = 0 \,.\nonumber
\end{align}
It is not difficult to check that when either of these relations is satisfied, the logarithmic functions in \cref{functions} are no longer linearly-independent. 
Therefore, these singularities can cancel only in the complete expression, which they indeed do. While all four $\Delta_i$'s appear in the GR amplitude, only $\Delta_1$ and $\Delta_2$ appear in the $\mathcal{N}=8$ amplitude.

In the real part of the amplitude, the rational term $A^R_\text{rat}$ has a very simple structure. For the GR amplitude it reads
\begin{align}
    A^{R, \text{GR}}_{\text{rat}}=\frac{1}{32}\left(1-\frac{\sigma(\sigma^2-3/2)}{(\sigma^2-1)^{3/2}}\right)
    (k\cdot p_1+k\cdot p_2)\Mred_{5,\text{tree}}^{\text{cl.GR}}\Big|_{\kappa=1} \,,
    \label{GRrat}
\end{align}
where we set $\kappa=1$ in the tree amplitude because the overall $\kappa$ dependence has been pulled out. 
%On the other hand, 
The square-root functions in \cref{functions} originate only from the triangle integrals $I_{1,1,0,0}$, $I_{1,0,1,0}$ and their up-down flip; 
%while 
their coefficients are more complicated in GR. 
%However, 
Such master integrals are absent for $\mathcal{N}=8$ supergravity, see \cref{MIpart1Neq8}, and thus $A_{1,2,3,4}^{R,\mathcal{N}=8}=0$. This makes the real part of the five-point $\mathcal{N}=8$ classical amplitude much simpler than \cref{integratedAmplitudeX} implies. The complete real part is given only by $A_{\text{rat}}^{R, {\cal N}=8}$ and is similar to its GR counterpart, 
\begin{align}
\label{realpartNeq8}
\varepsilon^{\mu}\varepsilon^{\nu}\,{\rm Re}\left[\Mred_{5, \text{1 loop}}^{0, \text{2MPI}, \text{cl.} {\cal N}=8}\right]_{\mu\nu} 
&= A^{R, {\cal N}=8}_\text{rat}\nonumber\\
&= \frac{1}{32}\left(1-\frac{\sigma(\sigma^2-2)}{(\sigma^2-1)^{3/2}}\right) (k\cdot p_1+ k\cdot p_2)
\Mred_{5, \text{tree}}^{\text{cl.} {\cal N}=8}\Big|_{\kappa=1} \ .
\end{align}
We note that this is the complete amplitude contribution to $\real{L=1}$ that is used to compute the waveform for $\mathcal{N}=8$ in \cref{typical}.
Curiously, the $\sigma$-dependent factor in \cref{realpartNeq8} vanishes in the ultrarelativistic limit, $\sigma\rightarrow\infty$, implying, according to \cref{typical}, that in this limit only the imaginary part of the amplitude contributes 
to the waveform at ${\cal O}(G^2)$ in ${\cal N}=8$ supergravity.
Even though the real part of the ${\cal N}=8$ classical one-loop amplitude is proportional to the tree-level amplitude, its contribution to the real part of the spectral waveform is different from the tree-level amplitude because the $\omega$-dependent distributional factor of $\real{L=1}$ is $\Theta(\omega)-\Theta(-\omega)$, see \cref{typical}.

Finally, we have verified that our one-loop amplitudes have the expected behavior in the soft limit $k\rightarrow 0$. At the leading order the imaginary part vanishes, and the real part factorizes into the four-point amplitude and the Weinberg soft factor,
\begin{align}\label{eq:softlimit}
    & \lim_{k\rightarrow 0}\Mred_{5, \text{1 loop}}^{0, \text{2MPI,cl.GR}} = \frac{\kappa}{2}\big[\varepsilon^{\mu}\varepsilon^{\nu}\mathcal{S}(k, q)_{\mu\nu}\big]\Mred_{4,\text{1 loop}}^{\text{2MPI,cl.GR}}\,,\nonumber\\
    & \lim_{k\rightarrow 0}\Mred_{5, \text{1 loop}}^{0, \text{2MPI,cl.} {\cal N}=8} = 0\,.
\end{align}
The leading soft limit for the five-point 2MPI amplitude of $\mathcal{N}=8$ vanishes because the corresponding four-point classical amplitude vanishes.
Meanwhile,  the $2\rightarrow 2$ amplitude in GR 
%receives contributions from triangles, and it 
is given by
\begin{align}\label{eq:M4GR1L}
    \Mred_{4,\text{1 loop}}^{\text{2MPI,cl.GR}}=\frac{3\kappa^4m_1^2m_2^2(m_1+m_2)(5\sigma^2-1)}{512\sqrt{-q^2}}\,,
\end{align}
where $q$ is the momentum transfer for this four-point %$2\rightarrow 2$ 
scattering. At the leading order, we can take $q=q_1=-q_2$, such that the soft factor $\mathcal{S}$ is given by
\begin{align}
\label{soft_factor_ini}
    \mathcal{S}(k, q)^{\mu\nu}&=\sum_{i=1}^{4}\frac{p_i^{\mu}p_i^{\nu}}{p_i\cdot k}=-\sum_{i=1}^{2}\left[\frac{p_i^{\mu}q_i^{\nu}+p_i^{\nu}q_i^{\mu}}{p_i\cdot k}-\frac{p_i^{\mu}p_i^{\nu}q_i\cdot k}{(p_i\cdot k)^2}\right]+\mathcal{O}(\hbar)\,,\\
 \varepsilon^{\mu}\varepsilon^{\nu}\mathcal{S}(k, q)_{\mu\nu}   &= -\frac{2\varepsilon\cdot p_1 \varepsilon\cdot q}{p_1\cdot k}+\frac{(\varepsilon\cdot p_1)^2 k\cdot q}{(p_1\cdot k)^2}+\frac{2\varepsilon\cdot p_2\varepsilon\cdot q}{p_2\cdot k}-\frac{(\varepsilon\cdot p_2)^2k\cdot q}{(p_2\cdot k)^2}+\mathcal{O}(k^0)\,.\nonumber
\end{align}
The gravitational-wave memory receives contributions only from the leading soft limit of the spectral waveform. Thus, the classicial $\mathcal{N}=8$ amplitude gives a vanishing contribyution at NLO.
We will explicitly compute the GR amplitude contribution to gravitational-wave memory in the next section and in \cref{cutcontrib} that of the bilinear-in-$T$ terms for both GR and ${\cal N}=8$ supergravity.\

\section{Amplitude contribution to the LO and NLO waveforms}
\label{sec:Waveforms}

Having constructed the relevant part of the one-loop five-point amplitude, we now use the formulae summarized in \cref{KMOCloopSummary} to construct its contribution to the leading order and next-to-leading order waveform observables focusing on ${\cal N}=8$ supergravity. 
The time-domain leading-order asymptotic metric was first discussed in Ref.~\cite{Kovacs:1978eu} 
and more recently from the worldline QFT perspective in Ref.~\cite{Jakobsen:2021smu}.
%
%The general relation between time-domain asymptotic metric and the tree 
%amplitude \eqref{fX} is the same as in Ref.~\cite{Jakobsen:2021smu} so 
%\draftnote{presumably we agree with their results. }

The last ingredient that we need are polarization vectors $\varepsilon_\pm(k)$ for a massless particle with momentum $k$. We may use spinor-helicity notations for it; for numerical evaluation, however, it is more convenient to start with the special outgoing momentum $k_0$ 
\begin{align}
k_0 = (1, 0, 0, 1)\qquad \varepsilon_{\pm}(k_0) = -\frac{1}{\sqrt{2}} (0, 1, \mp i, 0)
\end{align}
and obtain the general angle-dependent polarization vector through a rotation:
\begin{align}
\varepsilon_{\pm}(k) = R(\theta, \phi)\varepsilon_{\pm}(k_0) 
\qquad
R(\theta, \phi) = 
{\tiny{
\left(
\begin{array}{cccc}
    1 & 0 & 0 & 0 \cr
    0 & \cos\phi & -\sin\phi & 0 \cr
    0 & \sin\phi & \cos \phi & 0 \cr
    0 & 0 & 0 & 1
\end{array}\right)
\cdot
\left(\begin{array}{cccc}
    1 & 0 & 0 & 0 \cr
    0 & \cos\theta & 0 & \sin\theta \cr
    0 & 0 & 1 & 0 \cr
    0 & -\sin\theta & 0 & \cos \theta 
\end{array}\right)
}}
\ .
\end{align}
The third angle of a general rotation simply multiplies $\varepsilon$ by a phase, so we will ignore it. This rotation also maps $k_0$ to $\tilde{k} = (1, \bm n _{\bm k})$ with 
$\bm{n}_{\bm k}$ a general unit vector.
We stress that, as discussed before, the complex nature of these polarization tensors does not change the definition of $\real{L}$ and $\imaginary{L}$ in \cref{ReImSplit}, because they are stripped off before taking the real and imaginary part of the amplitude and of the bilinear-in-$T$ terms. Indeed, we may decompose the outgoing polarization tensor 
$\varepsilon_{--} = \varepsilon_{-} \otimes \varepsilon_{-}$ as 
\begin{align}
\varepsilon_{--} = \varepsilon_{+} + i \varepsilon_{\times}  \ ,
\end{align}
and use separately the real polarizations $\varepsilon_{+}$ and $\varepsilon_{\times}$ to define the real and imaginary parts of the matrix element ${\cal B}_-$ and subsequently the waveforms.
The asymptotic metric $h_{\mu\nu}^{\infty}$ is defined in \cref{eq:metric_pert}. We further parametrize it as
\begin{align}\label{eq:norm_h}
%g_{\mu\nu} &= \eta_{\mu\nu} + \frac{\kappa^2 \Mtot}{8\pi |\bm x|} {\hat h}_{\mu\nu} 
h_{\mu\nu}^{\infty}=(\kappa\Mtot)\hat{h}_{\mu\nu}\,,
\qquad
{\hat h}_{\mu\nu}= \frac{\kappa^2 \Mtot}{\sqrt{-b^2}} {\hat h}^{(1)}_{\mu\nu} + 
\left(\frac{\kappa^2 \Mtot}{\sqrt{-b^2}}\right)^2 {\hat h}^{(2)}_{\mu\nu}  +\dots
\end{align}
where $\Mtot=m_1+m_2$ is the total mass. 
With this notation, the asymptotic metric is
\begin{align}
g_{\mu\nu} = \eta_{\mu\nu}+\frac{8G\Mtot}{|\bm x|} \, {\hat h}_{\mu\nu} \ .
\label{fullmetric}
\end{align}
Here $\hat{h}_{\mu\nu}^{(1)}$ and $\hat{h}_{\mu\nu}^{(2)}$ are respectively the reduced waveform at 1PM and 2PM order. 
Similar to the previous section, we will not distinguish barred and unbarred variables since their difference is quantum.

We will perform explicit calculations in the center-of-mass (COM) frame in which the black holes move along the $z$ axis and $b^{\mu}$ is along the $x$ axis,\footnote{Technically, $u_1$ and $u_2$ are outgoing four-velocities in our setup. However, in the classical limit, we can treat them as incoming since the difference is quantum.}
\begin{align}
    u_1^{\mu} = (\gamma ,0,0,-\gamma v)\,,\qquad u_2^{\mu} = (\gamma,0,0,\gamma v)\,,\qquad b^{\mu} = (0,|\bm b|,0,0)\,,
\end{align}
where $\gamma=\frac{1}{\sqrt{1-v^2}}$. We will plot amplitude contribution to the waveforms at various locations $(\theta,\phi)$ at spatial infinity for different velocities $v$.

\subsection{Gravitational memory}

By plugging \cref{hX,omegaFT,modMatrixElementX} into \cref{eq:memory}, we can write the gravitational-wave memory as
\begin{align}\label{Deltah_start}
    \Delta(h_{+}^{\infty}+ih^{\infty}_{\times})&=\lim_{\omega\rightarrow 0^+}(-i\omega)\Theta(\omega)(-i)\langle\psi_{\text{in}}|\opS^{\dagger}\hat{a}_{--}(k)\opS|\psi_{\text{in}}\rangle^0\big|_{k=(\omega,\omega\bm n)}\nonumber\\
    &\quad +\lim_{\omega\rightarrow 0^-}(-i\omega)\Theta(-\omega)(+i)\langle\psi_{\text{in}}|\opS^{\dagger}\hat{a}^{\dagger}_{++}(k)\opS|\psi_{\text{in}}\rangle^0\big|_{k=(|\omega|,|\omega|\bm n)}\,.
\end{align}
Following the decomposition \eqref{one_and_two} of the matrix element, we also decompose 
\begin{align}
\label{one_and_two_metric}
\Delta(h_{+}^{\infty}+ih^{\infty}_{\times}) = 
\Delta(h_{+}^{\infty}+ih^{\infty}_{\times}) \Big|_{\text{amp}}
+ 
\Delta(h_{+}^{\infty}+ih^{\infty}_{\times}) \Big|_{\text{cut}} \ .
\end{align}
In the soft limit, we have seen explicitly in \cref{eq:softlimit} through one loop order, that the amplitude contribution to the matrix element \eqref{one_and_two} becomes 
\begin{align}\label{eq:me_soft}
    %&\lim_{\omega\rightarrow 0^{+}}(-i\omega)\Theta(\omega)(-i)\langle\psi_{\text{in}}|\opS^{\dagger}\hat{a}_{--}(k)\opS|\psi_{\text{in}}\rangle^0\Big|^{(1)}_{k=(\omega,\omega\bm n)}\nonumber\\
    &\lim_{\omega\rightarrow 0^{+}}(-i\omega)\Theta(\omega)\mathcal{W}_{\text{amp}}^{\text{1 loop}}\Big|_{k=(\omega,\omega\pmb{n})}\nonumber\\
    &\qquad = - \frac{i\kappa}{4}\int\frac{\dd^d q}{(2\pi)^d}\hdelta(2p_1\cdot q)\hdelta(2p_2\cdot q)e^{iq\cdot b}\,\varepsilon_{-}^{\mu}\varepsilon_{-}^{\nu}\mathcal{S}(\tilde{k},q)_{\mu\nu}\Mred_{4}^{\text{2MPI.cl.}}\,,
\end{align}
where $\tilde{k}=(1,\pmb{n})$ and we have used $\Theta(0)=1/2$. The classical four-point one-loop amplitude for GR is given in \cref{eq:M4GR1L} while it vanishes for $\mathcal{N}=8$.
The soft-graviton theorem implies this result holds at all orders. A similar calculation can be applied to the amplitude contribution to the conjugate matrix element.\footnote{We note that in this calculation, cancelling $\omega$ in $k$ leads to an extra minus sign because $k=(-\omega,-\omega\bm{n})$ and $\omega<0$. This sign will be absorbed by redefining $q\rightarrow-q$ to align the exponential factor $e^{iq\cdot b}$ because the soft factor $\mathcal{S}_{\mu\nu}$ is odd in $q$. The final result differs from \cref{eq:me_soft} only by a complex conjugation on $\Mred_4^{\text{2MPI.cl.}}$.} Therefore, the amplitude contribution to \cref{Deltah_start} becomes
\begin{align}
&\Delta (h_+^{\infty} +i h_\times^{\infty}) \Big|_{\text{amp}} = - \frac{i\kappa}{2} 
\int \frac{\dd^dq}{(2\pi)^d}
{\hat \delta}(2p_1\cdot q)
{\hat \delta}(2p_2 \cdot q)
e^{ i q\cdot b}\,\varepsilon_{-}^\mu \varepsilon_{-}^\nu {\cal S}({\tilde k}, q)_{\mu\nu}\,
{\rm Re}\big[M_4^{\text{2MPI.cl}}\big]\,.
\label{Deltah}
\end{align}
The typical integral in \cref{Deltah} is
\begin{align}
    I^{\mu}=\int\frac{\dd^d q}{(2\pi)^d}\hdelta(2p_1\cdot q)\hdelta(2p_2\cdot q)e^{iq\cdot b}q^{\mu}f(q^2)\,.
\end{align}
We can fix $d=4$ and decompose $q$ in terms of Ref.~\cite{Cristofoli:2021vyo} 
\begin{align}\label{eq:qbasis}
    q^{\mu}=z_1 u_1^{\mu} + z_2 u_2^{\mu} + z_b b^{\mu} + z_v v^{\mu}\,,\quad v^{\mu}\equiv 4\epsilon^{\mu\nu\rho\sigma}u_{1\nu}u_{2\rho}b_{\sigma}\,.
\end{align}
We then integrate over the coefficients $\{z_1,z_2,z_b,z_v\}$. It is easy to see that the two delta functions fix $z_1=z_2=0$. The integration over $z_v$ vanishes because the integrand is odd in $z_v$. As a result, $I^{\mu}$ must be proportional to $b^{\mu}$.  
$\varepsilon_{-}^\mu \varepsilon_{-}^\nu {\cal S}({\tilde k}, q)_{\mu\nu}$ can therefore be pulled out of the integral by replacing $q^\mu \rightarrow -i \partial/\partial b_\mu$. 
Recalling that, up to two loops, the constrained Fourier-transform of the real part of the 2MPI four-point amplitude is expected to be the radial action \cite{Bern:2021yeh,Brandhuber:2021eyq}, we find  
\begin{align}
\Delta (h_+^{\infty} +i h_\times^{\infty}) \Big|_{\text{amp}} & = - \frac{\kappa}{2} 
\left[\varepsilon_{-}^\mu \varepsilon_{-}^\nu {\cal S}({\tilde k}, \partial/\partial b)_{\mu\nu}\right] I_r(b)\,,\nonumber \\
I_r(b) &= \int \frac{\dd^dq}{(2\pi)^d}
{\hat \delta}(2p_1\cdot q)
{\hat \delta}(2p_2 \cdot q)
e^{ i q \cdot b}\; 
{\rm Re}\left[M_4^{\text{2MPI.cl}}\right]\,.
\label{Deltah1}
\end{align}
The action of $\partial/\partial b_{\mu}$ on the radial action is proportional to the scattering angle $\chi$,
\begin{align}\label{eq:action_angle}
\frac{\partial I_r}{\partial b_\mu} = 
-\frac{m_1m_2 \sqrt{\sigma^2-1}}{\Mtot\sqrt{1+2\nu (\sigma-1)}} \frac{b^\mu}{\sqrt{-b^2}}\frac{\partial I_r}{\partial J}=\frac{m_1m_2 \sqrt{\sigma^2-1}}{\Mtot\sqrt{1+2\nu (\sigma-1)}} \tilde{b}^{\mu}\chi\,,
\end{align}
where $\nu=m_1 m_2/\Mtot^2$, $\tilde{b}^{\mu}=b^{\mu}/\sqrt{-b^2}$ and $\chi=-\partial I_r/\partial J$. The angular momentum 
$J$ is given by $J=p_{\infty}\sqrt{-b^2}$, and $p_{\infty}$ is the norm of the initial COM momentum, which is given by the prefactor in \cref{eq:action_angle}.
Therefore, the amplitude contribution to the gravitational-wave memory is also proportional to $\chi$,
\begin{align}
\Delta (h_+^{\infty} +i h^{\infty}_\times) \Big|_{\text{amp}} & = - \frac{\kappa m_1 m_2 \sqrt{\sigma^2-1}}{2\Mtot\sqrt{1+2\nu (\sigma-1)}}
\left(\varepsilon_{-}^\mu \varepsilon_{-}^\nu {\cal S}({\tilde k}, \tilde{b})_{\mu\nu} \right) \chi\,, 
\\
\varepsilon^{\mu}\varepsilon^{\nu}\mathcal{S}(\tilde k, \tilde b)_{\mu\nu}   &=-\frac{2\varepsilon\cdot p_1 \varepsilon\cdot \tilde b}{p_1\cdot \tilde k}+\frac{(\varepsilon\cdot p_1)^2 \tilde k\cdot \tilde b}{(p_1\cdot \tilde k)^2}+\frac{2\varepsilon\cdot p_2\varepsilon\cdot \tilde b}{p_2\cdot \tilde k}-\frac{(\varepsilon\cdot p_2)^2\tilde k\cdot \tilde b}{(p_2\cdot \tilde k)^2}\,.\nonumber
\end{align}
Importantly, the angular dependence of the memory is completely encoded in the soft factor. We now plug in the 1PM and 2PM scattering angles for GR~\cite{Damour:2016gwp,Damour:2017zjx} and $\mathcal{N}=8$ supergravity~\cite{Parra-Martinez:2020dzs}, 
\begin{align}
    &\chi^{(1)}=\frac{\kappa^2\Mtot}{\sqrt{-b^2}}\frac{\sqrt{1+2\nu(\sigma-1)}(2\sigma^2-X)}{16\pi(\sigma^2-1)}\,,\nonumber\\
    &\chi^{(2)}=\left(\frac{\kappa^2\Mtot}{\sqrt{-b^2}}\right)^2\frac{3\sqrt{1+2\nu(\sigma-1)}(5\sigma^2-1)X}{4096\pi(\sigma^2-1)}\,,
\end{align}
where $X=1$ for GR and $X=0$ for $\mathcal{N}=8$ supergravity. 
The final result for amplitude contribution to the gravitational-wave memory is
\begin{align}
\label{LOmemoryA}
    \Delta(\hat{h}_{+}^{(1)}+i\hat{h}_{\times}^{(1)})\Big|_{\text{amp}} &= -\frac{m_1 m_2 (2\sigma^2-X)}{32\pi\Mtot^2\sqrt{\sigma^2-1}}\left(\varepsilon_{-}^\mu \varepsilon_{-}^\nu {\cal S}({\tilde k}, \tilde{b})_{\mu\nu} \right)\,,\\
    \Delta(\hat{h}_{+}^{(2)}+i\hat{h}_{\times}^{(2)})\Big|_{\text{amp}} &= -\frac{3m_1m_2(5\sigma^2-1)X}{8192\pi\Mtot^2\sqrt{\sigma^2-1}}\left(\varepsilon_{-}^\mu \varepsilon_{-}^\nu {\cal S}({\tilde k}, \tilde{b})_{\mu\nu} \right)\,,
\label{NLOmemoryA}    
\end{align}
where $\hat{h}$ is defined in \cref{eq:norm_h}. In particular, there is no amplitude contribution to the gravitational-wave memory at 2PM in ${\cal N}=8$ supergravity. 

\subsection{Leading order (LO)}
\label{LO}

As discussed in \cref{eq:LOme}, the LO waveform is determined by the tree-level five-point amplitude. From \cref{eq:t-obs,eq:LOme}, we get 
\begin{align}
(h_+^{\infty}+ih^{\infty}_\times)(\tau, \bm n)\Big|_{\text{LO}} &= 
\int_{-\infty}^{+\infty} \frac{d\omega}{2\pi} e^{-i\omega\tau}
\int \frac{\dd^dq_1}{(2\pi)^d} \frac{\dd^dq_2}{(2\pi)^d} 
{\hdelta}(2p_1\cdot q_1)
{\hdelta}(2p_2 \cdot q_2) e^{ i q_1\cdot b}
\nonumber\\
&\quad
\times{\hdelta}^{(d)}(q_1+q_2-k)M^\text{\text{cl.}}_{5, \text{tree}}(p_1p_2\rightarrow p_1-q_1,p_2-q_2,k^{--})\big|_{k=\omega(1, \bm n)}
\nonumber
\\
&=\int \frac{\dd^dq_1}{(2\pi)^d} \frac{\dd^dq_2}{(2\pi)^d} 
{\hdelta}(2p_1\cdot q_1)
{\hdelta}(2p_2 \cdot q_2) e^{ i q_1\cdot b} \delta(\tau - q_1\cdot b)
\nonumber\\
&\quad
\times{\hdelta}^{(d)}(q_1+q_2-\tilde{k})M^\text{\text{cl.}}_{5, \text{tree}}(p_1p_2\rightarrow p_1-q_1,p_2-q_2,\tilde{k}^{--})\,,
\end{align}
where $\tilde{k}=(1,\bm n)$. After plugging in the explicit form of the tree amplitudes, \cref{fullGS1grav} for GR and \cref{fullNeq81grav} for $\mathcal{N}=8$ supergravity, we find that the contribution to the reduced LO waveform, defined in \cref{eq:norm_h}, has the following unified form,
\begin{align}\label{eq:LOwf}
(\hat h{}^{(1)}_++i\hat h{}^{(1)}_\times)(\tau, \bm n) =
-
\frac{{m}_1 {m}_2}{16\Mtot^2} & \big(
  B_1 \J^{0,0}_{1, 0}+B_2\J^{1,0}_{1, 0}+B_3 \J^{0,1}_{1, 0}
+ B_4 \J^{0,0}_{0, 1}+B_5\J^{1,0}_{0, 1}
\nonumber\\
&+B_6 \J^{0,1}_{0, 1}
+ B_7 \J^{0,0}_{1, 1}+B_8\J^{1,0}_{1, 1}+B_9 \J^{2,0}_{1, 1} \big)\,,
\end{align}
where the integrals $\mathcal{J}_{\beta_1,\beta_2}^{\alpha_1,\alpha_2}$ are defined as
\begin{align}\label{eq:Jintegral}
\mathcal{J}_{\beta_1, \beta_2}^{\alpha_1, \alpha_2}
&= \int \frac{\dd^d q_1}{(2\pi)^d}
\hat \delta(u_1\cdot q_1)
\hat\delta(u_2 \cdot  (q_1-\tilde k))
 \delta(\tau/\sqrt{-b^2} - q_1\cdot \tilde b)
\frac{(\varepsilon_-\cdot q_1)^{\alpha_1}({\tilde k}\cdot q_1)^{\alpha_2}}
{(q_1^2)^{\beta_1}((q_1-\tilde k)^2)^{\beta_2}} \,,
\end{align}
and the coefficients $B_{1,\dots,9}$ given by
\begin{align}
B_1 &= \frac{4 \sigma (\varepsilon_- \cdot {u}_2)}
{(\tilde k \cdot {u}_2)} ({u}_1\cdot \tilde{f}_{-}\cdot {u}_2)
%((\varepsilon \cdot {u}_2) (k \cdot {u}_1) - 
%   (\varepsilon \cdot {u}_1) (k \cdot {u}_2)))
&B_2 &= 2(2\sigma^2 - X) \frac{(\varepsilon_- \cdot {u}_2)}
{(\tilde k \cdot {u}_2)} 
&B_3 &=-(2\sigma^2 - X) \frac{ (\varepsilon_- \cdot {u}_2)^2}
{(\tilde k \cdot {u}_2)^2}
\nonumber\\
B_4 &= \frac{4 \sigma (\varepsilon_- \cdot {u}_1)}
{(\tilde k \cdot {u}_1)}({u}_2\cdot \tilde{f}_{-}\cdot {u}_1)
%((\varepsilon \cdot {u}_1) (k \cdot {u}_2) - 
%   (\varepsilon \cdot {u}_2) (k \cdot {u}_1)))  
&B_5 &= -2(2\sigma^2 - X) \frac{(\varepsilon_- \cdot {u}_1)}
{(\tilde k \cdot {u}_1)} 
&B_6 &=(2\sigma^2 - X) \frac{ (\varepsilon_- \cdot {u}_1)^2}
{(\tilde k \cdot {u}_1)^2}
\nonumber\\
B_7 &= 4 ({u}_2\cdot \tilde{f}_{-}\cdot {u}_1)^2
%((\varepsilon \cdot {u}_2) (k \cdot {u}_1) - 
%   (\varepsilon \cdot {u}_1) (k \cdot {u}_2)))^2
&B_8 &= 8\sigma ({u}_1\cdot \tilde{f}_{-}\cdot  {u}_2)
%((\varepsilon \cdot {u}_2) (k \cdot {u}_1) - 
%   (\varepsilon \cdot {u}_1) (k \cdot {u}_2)))
&B_9 &= 2(2\sigma^2-X) \ .
\end{align}
In these expressions, $\tilde{f}_-^{\mu\nu} = {\tilde k}^\mu\varepsilon_{-}^{\nu} - {\tilde k}^\nu \varepsilon_{-}^{\mu}$, and $X=1, 0$ for GR and ${\cal N}=8$ supergravity, respectively.

\begin{figure}[tb]
\centering
\includegraphics[width=\textwidth]{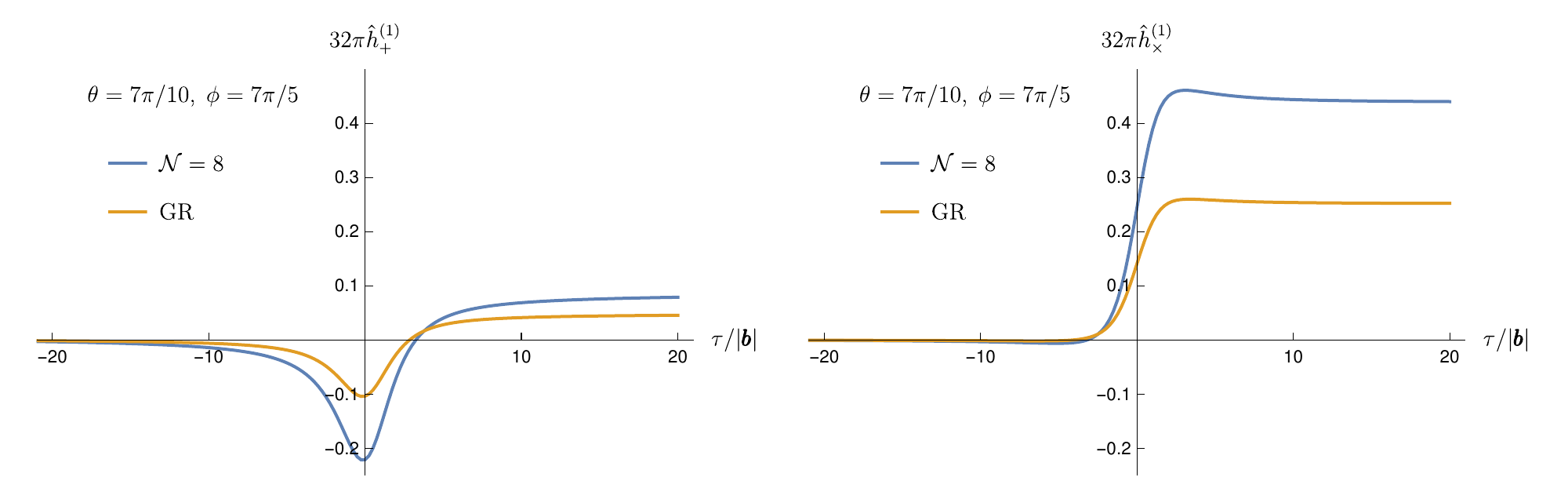}
\caption{The evolution of the LO waveform at a particular location at the spatial infinity. We work in the COM frame with $v=1/5$ and $m_1=m_2$.} 
\label{LO_plot}
\end{figure}

With foresight on the integrals required by the evaluation of time-domain observables, it is convenient to follow the strategy of Ref.~\cite{Cristofoli:2021vyo} and compute the $\mathcal{J}_{\beta_1,\beta_2}^{\alpha_1,\alpha_2}$ integrals in $d=4$ by decomposing the integration variable $q_{1}$ along four orthogonal fixed vectors as in \cref{eq:qbasis}. 
The integrals over $z_1$, $z_2$, and $z_b$ are localized due to the three delta functions in \cref{eq:Jintegral}. This leaves the integral over $z_v$ as the only nontrivial integral.
The first six integrals in \cref{eq:LOwf} have been evaluated in Ref.~\cite{Cristofoli:2021vyo}. Using this method, the last three integrals can be brought to a one-parameter integral over a finite range. The background at $\tau\rightarrow-\infty$ is also subtracted.
To demonstrate the final result at LO, we plot the evolution of the waveform at a particular location at the spatial infinity in \cref{LO_plot}. The GR and $\mathcal{N}=8$ waveforms have the same qualitative features. The difference is approximately an overall scale. Here and after, all the plots are made with $m_1=m_2$.

%%%%%%%%%%%%%% FIGURE %%%%%%%%%
%%%%%%%%%%%%%%%%%%%%%%%%%%%%%%%

\subsection{Next-to-leading order (NLO)}
\label{amplitudeNLO}

The NLO time-domain waveform follows from the $L=1$ component of \cref{eq:t-obs}. Here we first consider the amplitude contribution. 
We further separate out the tail contribution from the IR finite one-loop amplitude contribution~\eqref{integratedAmplitudeX},
\begin{align}\label{eq:tail_separate}
    \Mred_{5,\text{1 loop}}^{0,\text{2MPI, cl}} &= \overline{\Mred}_{5,\text{1 loop}}^{0,\text{2MPI, cl}} + \Mred_{\text{tail}}\,,\nonumber\\
    \Mred_{\text{tail}} &= \frac{i\,\kappa^2}{64\pi}(k\cdot p_1+k\cdot p_2)\left[\log\frac{16\pi^2(k\cdot u_1)^2(k\cdot u_2)^2}{\Lambda^4}\right]\Mred_{5,\text{tree}}^{\text{cl.}}\, ,
    \\
(h^{\infty}_{+} + i {h}^{\infty}_\times)(t, \bm n)\Big|^{\text{amp}}_\text{NLO}&
    %=\int\frac{\dd^d q_1}{(2\pi)^d}\frac{\dd^d q_2}{(2\pi)^d}\hdelta(2\pb_1\cdot q_1)\hdelta(2\pb_2\cdot q_2)\mathcal{J}_{0,1}
    =\mathcal{H}_{R}+i\,\mathcal{H}_{I}+(h^{\infty}_{+} + i {h}^{\infty}_\times)(t, \bm n)\Big|_\text{NLO}^{\text{amp tail}}\, .
\end{align}
We first compute $\mathcal{H}_{R,I}$ which contain the contribution of $\overline{\Mred}_{5,\text{1 loop}}^{0,\text{2MPI, cl}}$. 
They are obtained by combining \cref{typical,eq:ReIm1l}: 
\begin{align}
    \mathcal{H}_{R} &= -\frac{1}{2\pi}\int \frac{\dd^d q_1}{(2\pi)^d} \hdelta(2p_1\cdot q_1) \hdelta\big(2p_2\cdot (q_1-\tilde{k})\big)\,\varepsilon^{\mu}_{-}\varepsilon^{\nu}_{-}{\rm Re}\left[\overline{\Mred}_{5,\text{1 loop}}^{0,\text{2MPI, cl}}\right]_{\mu\nu}\nonumber\\
    &\hspace{3cm}\times
    \left[\frac{1}{(\tau-{q}_1\cdot b + i 0)^{2}} + \frac{1}{(\tau-{q}_1\cdot b - i 0)^{2}}\right], \\
    \mathcal{H}_{I} &= \frac{1}{2\pi}\int \frac{\dd^d q_1}{(2\pi)^d} \hdelta(2p_1\cdot q_1) \hdelta\big(2p_2\cdot (q_1-\tilde{k})\big)\,\varepsilon^{\mu}_{-}\varepsilon^{\nu}_{-}{\rm Im}\left[\overline{\Mred}_{5,\text{1 loop}}^{0,\text{2MPI, cl}}\right]_{\mu\nu}\nonumber\\
    &\hspace{3cm}\times
    \left[\frac{1}{(\tau-{q}_1\cdot b + i 0)^{2}} - \frac{1}{(\tau-{q}_1\cdot b - i 0)^{2}}\right].
    \label{HI_no_log}
\end{align}
We note that $\mathcal{H}_{R}$ and $\mathcal{H}_{I}$ do not correspond to the real and imaginary part of the amplitude contribution to the waveform, because the polarization vectors in these expressions are still complex. The tail contribution will be computed later in this section. We emphasize that we will only consider the amplitude contribution to the waveform, namely, the Fourier transform of $\mathcal{W}_{\text{amp}}^{\text{1 loop}}$ in \cref{one_and_two}. We will study $\mathcal{W}_{\text{cut}}^{\text{1 loop}}$, which is purely imaginary, in the next section.

As at leading order, we compute the integral over $q_1$ in $d=4$ by decomposing this variable in the basis of 4d vectors in \cref{eq:qbasis}.
For both GR and $\mathcal{N}=8$ supergravity, it is easier to compute $\mathcal{H}_{I}$ due to the presence of 
\begin{align}
\label{deltaprimeHI_no_log}
    \frac{1}{(\tau-{q}_1\cdot b + i 0)^{2}} - \frac{1}{(\tau-{q}_1\cdot b - i 0)^{2}}=2\pi i \delta'(\tau-q_1\cdot b)\,.
\end{align}
Thus we can use the three delta functions in $\mathcal{H}_{I}$ to localize the integrals over the $z_1$, $z_2$ and $z_b$ variables. For fixed $\tau$ the integrand is now an algebraic function of $z_v$. The resulting integral over $z_v$ is convergent and performed numerically. The background at $\tau\rightarrow-\infty$ is also subtracted at the end. 

In contrast, there are only two delta functions in $\mathcal{H}_{R}$, which localize $z_1$ and $z_2$. For $\mathcal{N}=8$ supergravity, the integrand is a rational function in $z_b$ and $z_v$ according to \cref{realpartNeq8}. Therefore, we can integrate $z_b$ using the residue theorem, resulting in an algebraic function of $z_v$, which will be integrated numerically. For $\mathcal{N}=8$ supergravity, the subtraction of background at $\tau\rightarrow-\infty$ can be done before or after the final $z_v$ integral since it does not affect the convergence.

However, for GR, after localizing $z_1$ and $z_2$, the integrand contains square roots according to \cref{integratedAmplitudeX}. For this case, we still first integrate over $z_b$. After some rescaling, the integral has the following generic form,
\begin{align}\label{eq:zb_int}
    \int_{-\infty}^{\infty}\dd z_b \frac{f(z_b)}{(z_b^2+1)^{1/2}}=\int_{0}^{\infty}\dd w\frac{f(\frac{w^2-1}{2w})}{w}=- \sum_{w_i}\text{Res}_{w=w_i}\frac{f(\frac{w^2-1}{2w})\log w}{w}\,,
\end{align}
where $f(z_b)$ is a rational function, and we have changed the variable to $z_b=\frac{w^2-1}{2w}$. Since the original $z_b$ integral is convergent, there are no poles at $w=0$ and $w=\infty$ after the change of variable. We can evaluate this integral by summing over all the residues, which leads to the final answer in \cref{eq:zb_int}. Now if we subtract the background at $\tau=-\infty$, the final $z_v$ integral is convergent and we integrate it numerically. Unlike the previous cases, the background subtraction needs to be done before the $z_v$ integrals because the integral is divergent otherwise. Alternatively, if one keeps $d=4-2\epsilon$, the integration and background subtraction can be done in any order, the difference being at most some local $\delta(|\bm b|)$ terms in the waveform and memory. 

\begin{figure}
    \centering
    \includegraphics[width=\textwidth]{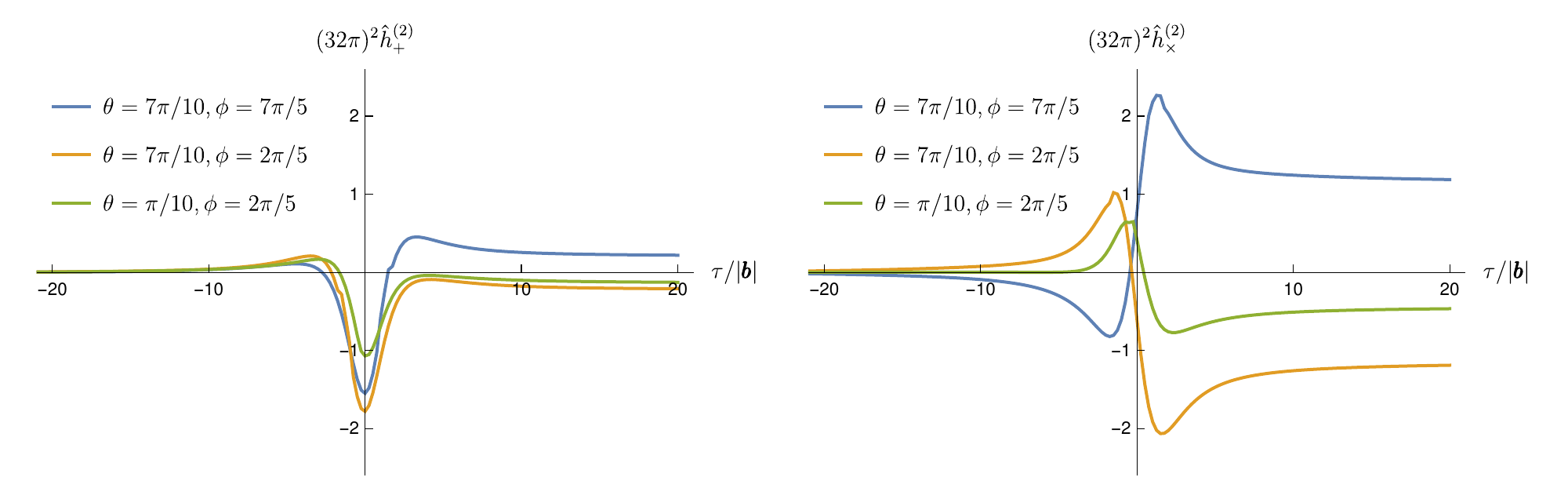}
    \caption{The NLO amplitude contribution to waveform in GR with COM velocity $v=1/5$.}
    \label{fig:NLO_GR_1}
\end{figure}

\begin{figure}
    \centering
    \includegraphics[width=\textwidth]{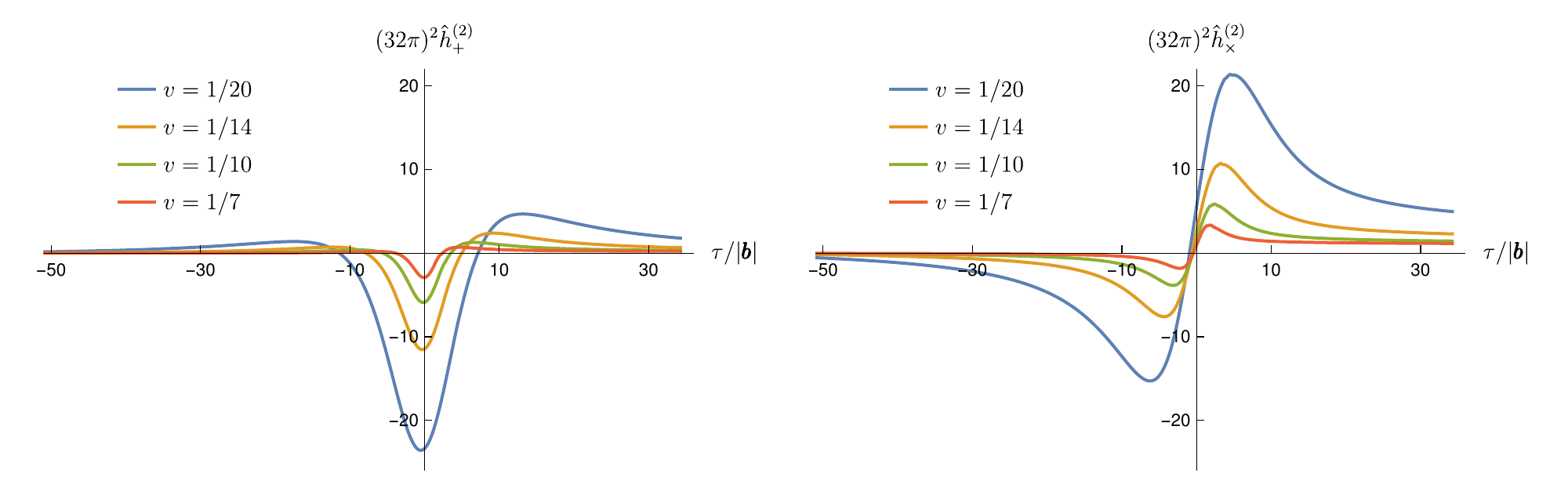}
    \caption{The dependence of the NLO amplitude contribution to waveform in GR on the COM velocities at the fixed angle $\theta=7\pi/10$ and $\phi=7\pi/5$.}
    \label{fig:NLO_GR_2}
\end{figure}

\begin{figure}
    \centering
    \includegraphics[width=\textwidth]{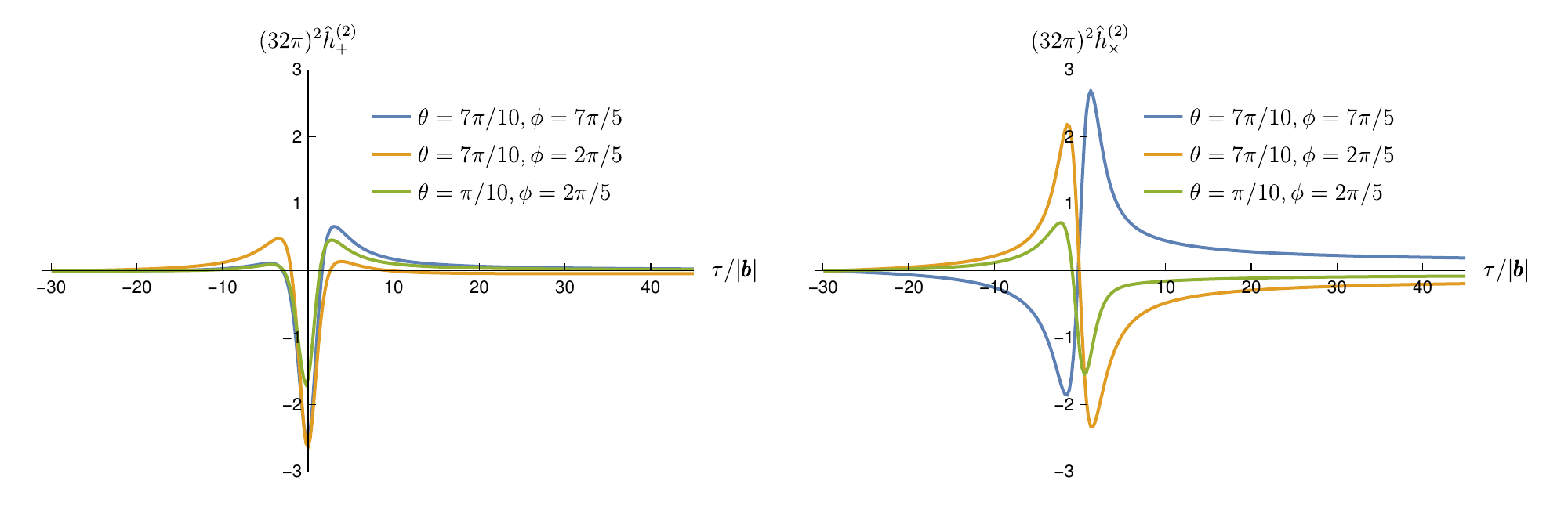}
    \caption{The NLO amplitude contribution to waveform in $\mathcal{N}=8$ supergravity with the COM velocity $v=1/5$. Both $\hat{h}_{+}^{(2)}$ and $\hat{h}_{\times}^{(2)}$ converge to zero at large $\tau$.}
    \label{fig:NLO_Neq8}
\end{figure}

\begin{figure}
    \centering
    \includegraphics[width=\textwidth]{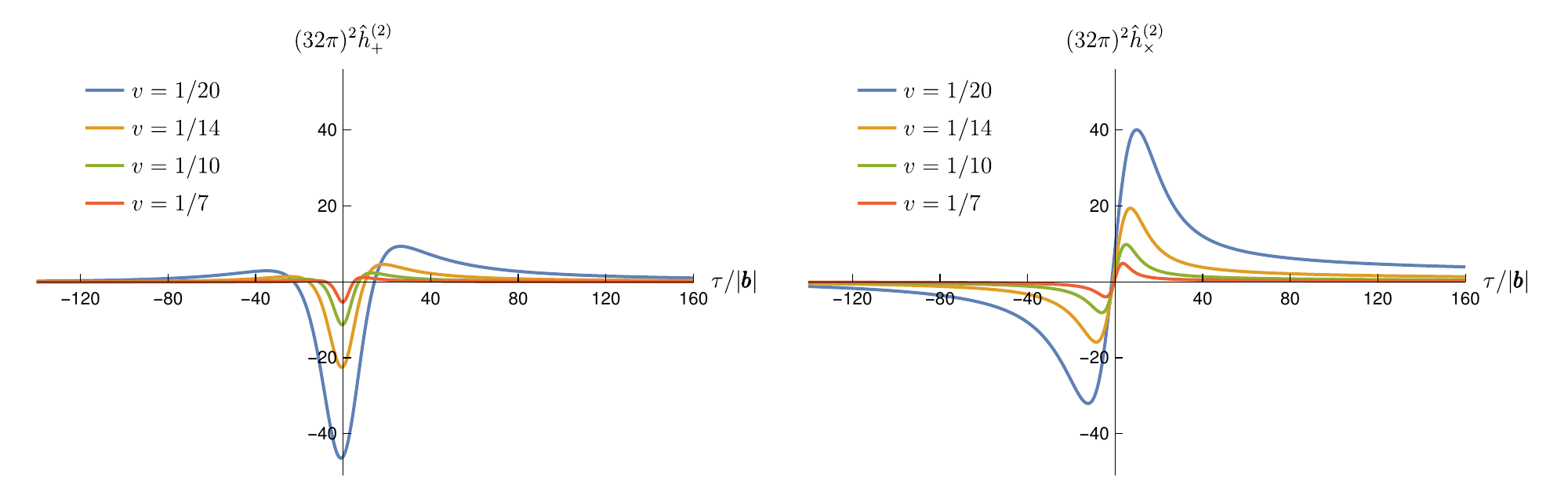}
    \caption{The NLO amplitude contribution to waveform in $\mathcal{N}=8$ supergravity on the COM velocities at the fixed angle $\theta=7\pi/10$ and $\phi=7\pi/5$. Both $\hat{h}_{+}^{(2)}$ and $\hat{h}_{\times}^{(2)}$ converge to zero at large $\tau$.}
    \label{fig:NLO_Neq8_2}
\end{figure}

We plot the amplitude contribution to the NLO waveforms in GR observed at several angles at the spatial infinity in \cref{fig:NLO_GR_1}. In this figure, we fix the COM velocity of the two black holes to be $v=1/5$. Then, in \cref{fig:NLO_GR_2}, we fix the observation angle to be $(\theta,\phi)=(\frac{7\pi}{10},\frac{7\pi}{5})$ and vary the velocities of the black holes. At low velocities, the waveforms have a much larger magnitude and are more spread out in the time domain; we may understand this intuitively by noting that at lower velocities, the two particles have a smaller minimal separation, so they experience larger accelerations and therefore radiate more. We also note that at lower velocities, we are closer to the boundary of the PM regime, which requires that $Jv^2$ be sufficiently large.

Finally, for comparison, we include in \cref{fig:NLO_Neq8,fig:NLO_Neq8_2}
the amplitude contribution to the NLO waveforms in $\mathcal{N}=8$ supergravity. As expected, this contribution to the $\mathcal{N}=8$ waveforms do not display any memory effect at NLO. At low velocities, for example, $v=1/20$, the imaginary part exhibits a small residual at $\tau\rightarrow\infty$; this is merely an artifact of our initial and final times not being infinite. We note that the profile is spread over a larger time interval and that they have a larger amplitude than the corresponding GR one. 
A possible explanation is that fields of the ${\cal N}=8$ supergravity 
yield an additional net attractive force increasing the acceleration 
experienced by the two particles relative to GR.

We remark that at certain angles, the amplitude contribution displays features which, for lack of a better name, we refer to as ``kinks.'' For example, such kinks can be seen in \cref{fig:NLO_GR_1} for $\hat{h}_{+}^{(2)}$ at $(\theta,\phi)=(\frac{7\pi}{10},\frac{7\pi}{5})$ and $(\frac{7\pi}{10},\frac{2\pi}{5})$. At these locations, several terms in the integrand become sharply peaked (but finite), and the resulting large values of the integral cancel over many orders of magnitude.
Our numerical evaluation further suggests that more prominent kinks exist at some other angles, such as $(\theta,\phi)=(\frac{\pi}{2},0)$, which corresponds to an observation direction parallel to $\bm b$ in the plane of the two incoming particles. 
It is unclear whether these features are merely numerical artifacts due to cancelation between large numbers, or they are physical. A definitive answer can be provided by an analytic evaluation of the waveform, which we leave for future work.

We now turn to the evaluation of the gravitational-wave tails. To this end 
we plug $\Mred_{\text{tail}}$ of \cref{eq:tail_separate} into $\mathcal{I}_L$ of \cref{typical}. As discussed in \cref{KMOCloopSummary}, to evaluate the Fourier transform to the time domain in the presence of the logarithmic dependence on $\omega$ we simply differentiate \cref{typical} with respect to $n$. Thus, 
\begin{align}
    (h^{\infty}_{+} + i {h}^{\infty}_\times)(t, \bm n)\Big|_\text{NLO}^{\text{amp tail}}&
    =\frac{i\,\kappa^2}{32\pi}\int \frac{\dd^d q_1}{(2\pi)^d} \hdelta(2p_1\cdot q_1) \hdelta\big(2p_2\cdot (q_1-\tilde{k})\big)(\tilde{k}\cdot p_1+\tilde{k}\cdot p_2)\Mred_{5,\text{tree}}^{\text{cl.}}\nonumber\\
    &\quad \times\int_{-\infty}^{+\infty}\frac{\dd\omega}{2\pi}e^{-i\omega(\tau-q_1\cdot b)}\omega\log\frac{4\pi\omega^2(u_1\cdot\tilde{k})(u_2\cdot\tilde{k})}{\Lambda^2}\\
    &=\frac{i\,\kappa^2}{16\pi}\int \frac{\dd^d q_1}{(2\pi)^d} \hdelta(2p_1\cdot q_1) \hdelta\big(2p_2\cdot (q_1-\tilde{k})\big)(\tilde{k}\cdot p_1+\tilde{k}\cdot p_2)\Mred_{5,\text{tree}}^{\text{cl.}}\nonumber\\
    &\quad\times\!\left[ \frac{\gamma_{\text{E}}-1+\log \frac{i(\tau-q_1\cdot b-\ie)\Lambda}{\sqrt{4\pi(u_1\cdot\tilde{k})(u_2\cdot\tilde{k})}}}{(\tau-q_1\cdot b-\ie)^2}-\frac{\gamma_{\text{E}}-1+\log \frac{-i(\tau-q_1\cdot b+\ie)\Lambda}{\sqrt{4\pi(u_1\cdot\tilde{k})(u_2\cdot\tilde{k})}}}{(\tau-q_1\cdot b+\ie)^2}\right].\nonumber
\end{align}
Since the integral is finite, we compute it directly in four dimensions. To this end we parametrize $q_1$ as in \eqref{eq:qbasis} and evaluate the $z_{1,2}$ integrals using the explicit $\delta$-functions. Unlike eqs.~\eqref{HI_no_log} and \eqref{deltaprimeHI_no_log}, the logarithmic dependence on $\omega$ prevents the appearnce a third $\delta$ function. Since the integrand does not exhibit branch cuts in $z_v$ we evaluate this integral analytically using Cauchy's residue theorem. The last integral, over $z_b$, is evaluated numerically.

\begin{figure}
    \centering
    \includegraphics[width=\textwidth]{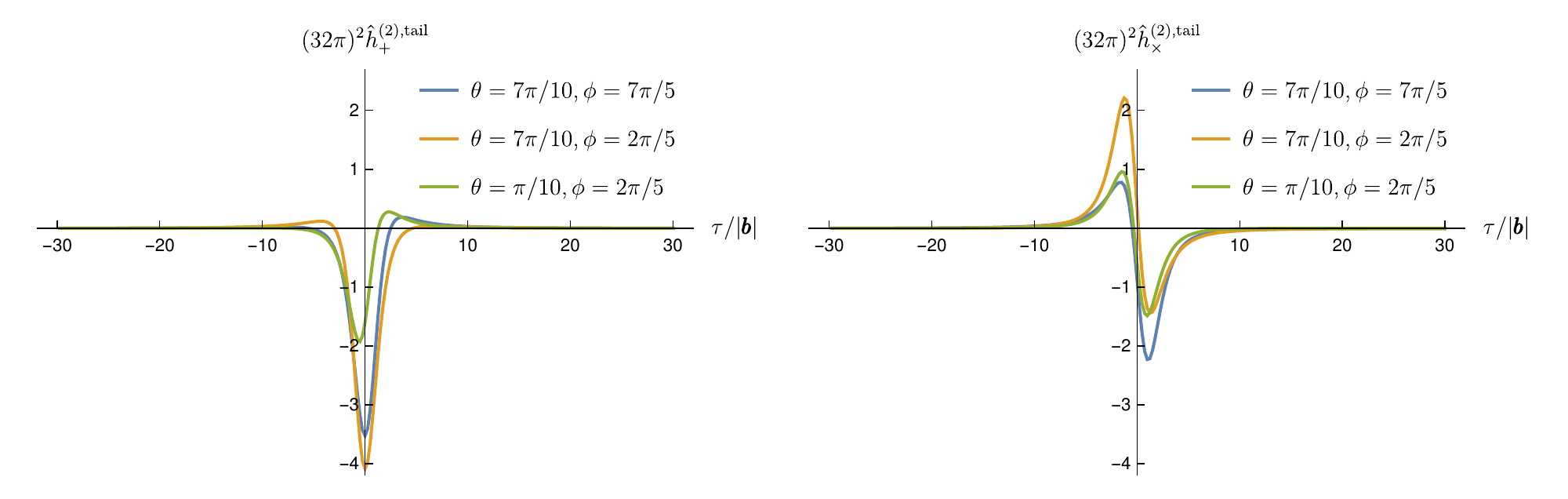}
    \caption{The amplitude-tail contribution to the NLO waveform in GR with the COM velocity $v=1/5$ and IR cutoff $\Lambda|\pmb{b}|=5\times10^{-10}$.}
    \label{fig:NLO_GR_tail}
\end{figure}

\begin{figure}
    \centering
    \includegraphics[width=\textwidth]{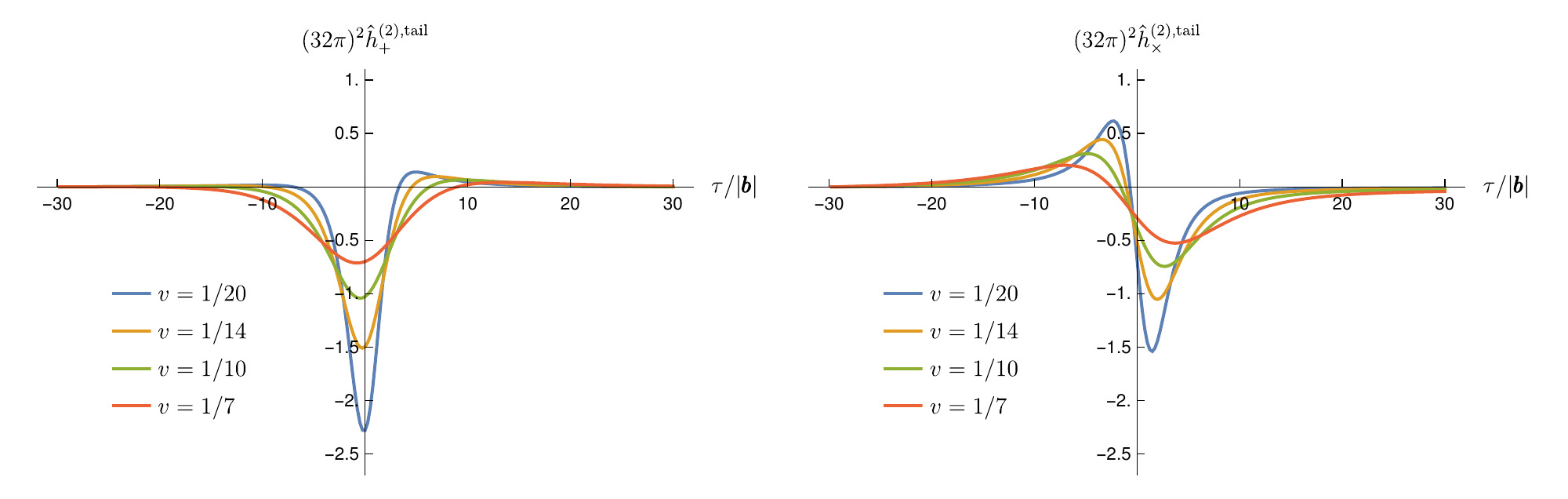}
    \caption{The amplitude-tail contribution to NLO waveform in GR on the COM velocities at the fixed angle $\theta=7\pi/10$ and $\phi=7\pi/5$, and IR cut off $\Lambda|\pmb{b}|=5\times10^{-10}$.}
    \label{fig:NLO_GR_tail_2}
\end{figure}

\begin{figure}
    \centering
    \includegraphics[width=\textwidth]{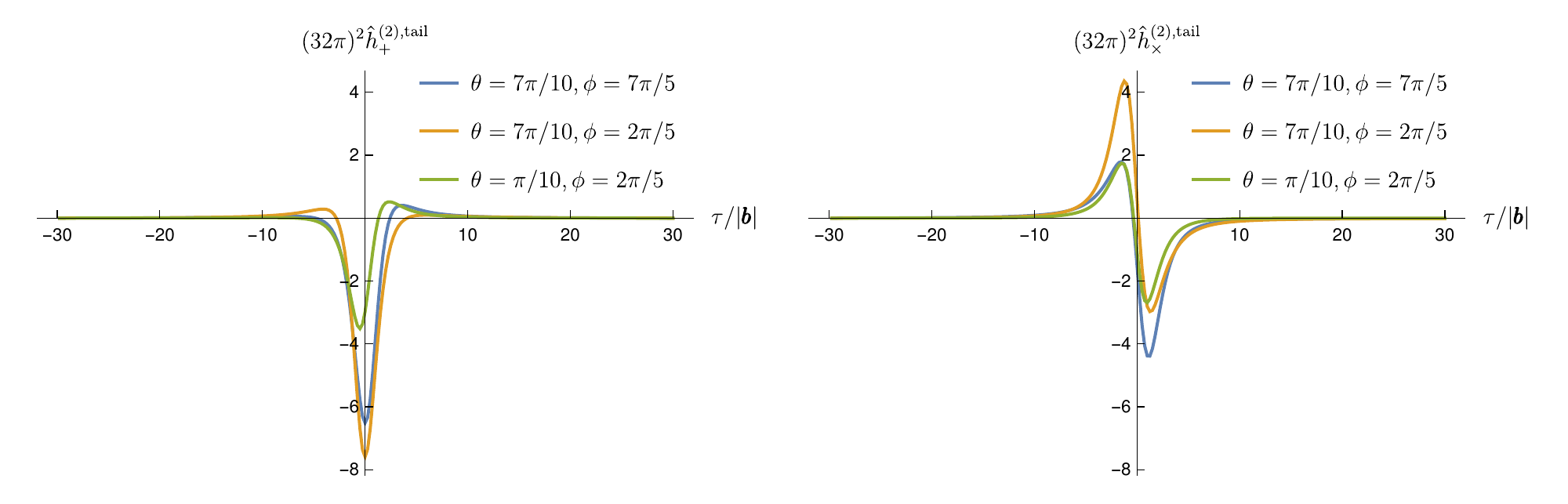}
    \caption{The amplitude-tail contribution to the NLO waveform in $\mathcal{N}=8$ supergravity with the COM velocity $v=1/5$ and IR cutoff $\Lambda|\pmb{b}|=5\times 10^{-10}$.}
    \label{fig:NLO_Neq8_tail}
\end{figure}

\begin{figure}
    \centering
    \includegraphics[width=\textwidth]{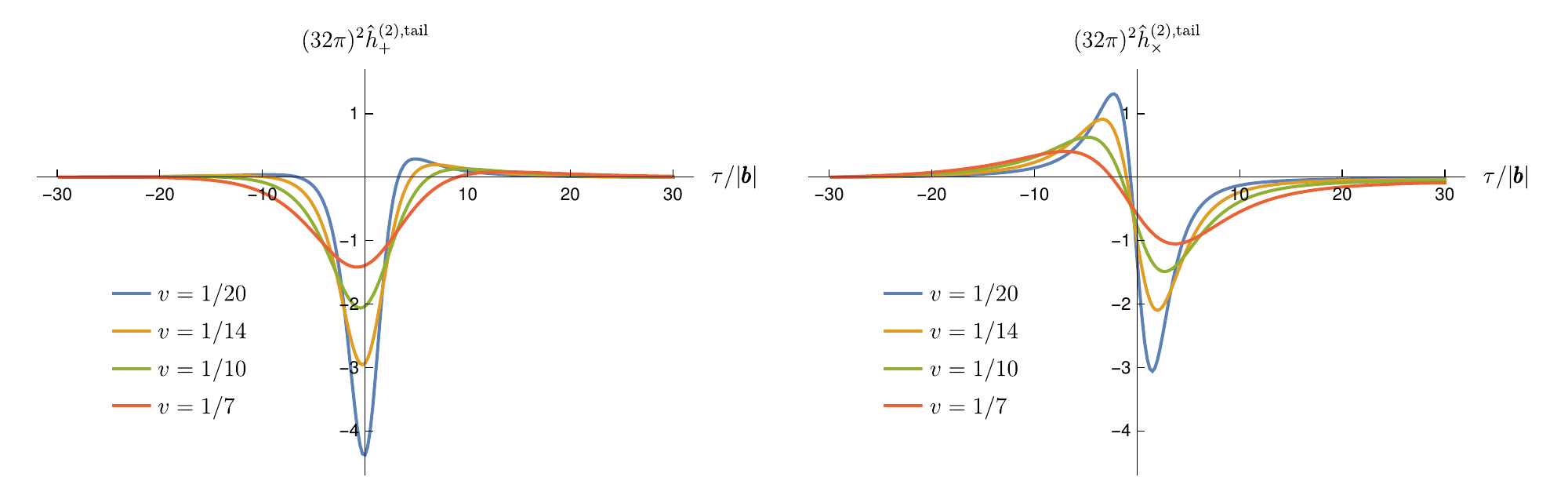}
    \caption{The amplitude-tail contribution to NLO waveform in $\mathcal{N}=8$ supergravity on the COM velocities at the fixed angle $\theta=7\pi/10$ and $\phi=7\pi/5$, and IR cut off $\Lambda|\pmb{b}|=5\times  10^{-10}$.}
    \label{fig:NLO_Neq8_tail_2}
\end{figure}

We also need to choose a value for the cutoff $\Lambda$ defining the infrared soft virtual gravitons.\footnote{The choice of $\Lambda$ is arbitrary and the waveform is invariant under $\Lambda\rightarrow\Lambda'$ after a concurrent redefinition of the observation time, $\tau \rightarrow \tau  - G (E_1+E_2-(\bm p_1+\bm p_2)\cdot \bm n)\log(\Lambda'{}^2/\Lambda^2)$ in \cref{tredef}. Nevertheless, one needs to make a choice, which is related to the typical frequency scale at the origin of the observation time~\cite{Porto:2012as}.} It is required to be well-inside the soft region defining the classical limit, $\Lambda\ll |\bm q|$. We may therefore set a dimensionless bound on the product of $\Lambda$ and the 
impact parameter $\bm b$, which is ${\cal O}(|\bm q|^{-1})$ being the Fourier-conjugate of the momentum transfer. 
We will choose $\Lambda|\bm{b}|=5\times 10^{-10}$.
Since in the classical limit $|\bm q|$ and the frequency $\omega$ of the outgoing graviton are of the same order, we may also relate $\Lambda$ 
and the lowest frequency accessible to a detector. 
The results for GR and $\mathcal{N}=8$ supergravity are plotted in \cref{fig:NLO_GR_tail,fig:NLO_GR_tail_2,fig:NLO_Neq8_tail,fig:NLO_Neq8_tail_2}, and we have chosen the same observation angles and velocities as before.

We note the absence of a memory contribution from the gravitational-wave tail, in agreement with the vanishing soft limit of the imaginary part of the one-loop five-point amplitude, as well as the close similarities of the velocity dependence of the tail and ${\overline M}_{5, \text{1 loop}}^{0,\text{2MPI, cl}}$ to the two gravitational-wave polarizations at fixed angle. In both contributions a higher amplitude of the wave corresponds to lower-velocity scattering, in agreement with the intuition that for fixed impact parameter lower-velocity particles experience a larger acceleration.
Moreover, the angular dependence of the tail and of ${\overline M}_{5, \text{1 loop}}^{0,\text{2MPI, cl}}$ contributions to ${\hat h}^{(2)}_+$ are also similar in shape and velocity dependence.
In contrast, their contributions ${\hat h}^{(2)}_\times$ are quite different, cf. e.g. the right panels of \cref{fig:NLO_GR_2,fig:NLO_GR_tail_2}. 
While the gravitational-wave tail contributions may be changed somewhat by varying the cutoff $\Lambda$, a difference between $(\theta, \phi)<(\pi/2, \pi)$ and $(\theta, \phi)>(\pi/2, \pi)$ persists. It is tempting to speculate that such a difference might offer an observable signature distinguishing the gravitational-wave tail from the local-in-time effects.

\section{Cut contribution}
\label{cutcontrib}
 
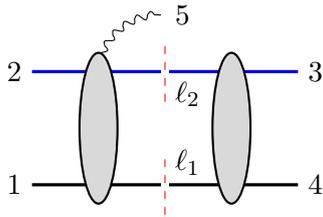
\begin{figure}
    \centering
    \begin{tikzpicture}
        \pgfmathsetmacro{\w}{1.75}
        \node [left=0pt] (p1) at (-\w,0) {$1$};
        \node [right=0pt] (p4) at (\w,0) {$4$};
        \node [left=0pt] (p2) at (-\w,\h) {$2$};
        \node [right=0pt] (p3) at (\w,\h) {$3$};
        \node [right=0pt] (p5) at (0,{3*\h/2}) {$5$};
        \draw [very thick] (p1) -- (p4);
        \draw [very thick,blue] (p2) -- (p3);
        \draw [graviton] (-\w/2,\h) to[bend left=45] (p5.west);
        \filldraw [fill=gray!30!white,thick] (-\w/2,\h/2) ellipse (0.25cm and 1cm);
        \filldraw [fill=gray!30!white,thick] (\w/2,\h/2) ellipse (0.25cm and 1cm);
        \draw [line width=3pt,draw=white] (0,{3*\h/2 + 0.2}) -- (0,-\h/2 - 0.2);
        \draw [dashed,draw=red] (0,-0.4) -- ++(0,0.8);
        \draw [dashed,draw=red] (0,\h-0.4) -- ++(0,0.8);
        \node at (0,0) [anchor=south west] {$\lm_1$};
        \node at (0,\h) [anchor=north west] {$\lm_2$};
    \end{tikzpicture}
    \caption{The cut that contributes to the $\opT$-bilinear matrix element of the NLO waveform.}
    \label{fig:2MPR}
\end{figure}

As discussed in \cref{KMOCloopSummary}, the NLO waveform received two contributions: (1) from the virtual one-loop five-point amplitude and (2) from a bilinear-in-$\opT$ cut term shown in \cref{fig:2MPR} 
whose importance was emphasized in Ref.~\cite{Caron-Huot:2023vxl}. 
In previous sections we computed the former and discussed in detail its contribution to the waveform. 
In this section we summarize the calculation of the latter and its contribution to the gravitational-wave memory in GR and in ${\cal N}=8$ supergravity. We also include plots of the complete waveform.

As mentioned earlier, in the full quantum theory the cut contribution may be extracted by simply taking the $s_{12}$ cut of the virtual amplitude. In the classical limit, the matter propagators of the amplitude combine to become principal-valued (PV) propagators. Since such propagators do not exhibit a singularity, one may naively conclude that this cut vanishes.
However, writing out the PV propagator as a sum two Feynman propagators reveals that the required $s_{12}$ cut is nonvanishing and the absence of the singularity of the PV propagator is due to the difference of the momenta of the two propagators being ${\cal O}(q)$. Thus, a less naive conclusion is that the $s_{12}$-channel cut picks out one of the two terms of the PV propagator and cancels it when subtracted from the amplitude.
This cancellation was first pointed out in Ref.~\cite{Caron-Huot:2023vxl}.

\subsection{The full waveform integrand}\label{fullexpress}

We computed the cut directly using the same strategy as the previous sections. Its general expression is 
\begin{align}\label{eq:cut1L}
\langle p_3 p_4 |\opT^\dagger \hat a_{hh}(k)\opT|p_1 p_2\rangle \Big|_{1\text{ loop}} &=\int 
    \frac{d^d\ell_1}{(2\pi)^d} \frac{d^2\ell_2}{(2\pi)^d} \hdelta(\lm_1^2-m_1^2)\hdelta(\lm_2^2-m_2^2) \nonumber
\\
&\quad  \times 
 \Mred_{5,\text{tree}}(p_1, \lm_1, p_2, \lm_2, k^{\varepsilon})
\Mred_{4,\text{tree}}(-\lm_1,p_4, -\lm_2,p_3)\ ,
\end{align}
where the tree amplitudes include the momentum-conservng delta functions. 
Taking them to belong to the theory of interest (i.e. either GR or ${\cal N}=8$ SG), the calculation proceeds in the standard way, by expanding in the soft limit and paying attention to the cut conditions imposed by the delta functions. As discussed in Ref.~\cite{Caron-Huot:2023vxl}, the classical part of the cut receives contribitions from the classical tree-level amplitudes and the second-order term in the expansion of the on-shell conditions. 

To write out the result, it is convenient to introduce a new class of integrals
\begin{align}
    I^{\text{cut,cut}}_{a_1,a_2,a_3}=\int\frac{\dd^d\lm}{(2\pi)^d}\frac{i\,\hdelta(2\ub_1\cdot\lm)\hdelta(2\ub_2\cdot\lm)}{[\lm^{2}]^{a_1}[(\lm+q_2)^{2}]^{a_2}[(\lm-q_1)^{2}]^{a_3}} \ .
\end{align}
We refer the super-classical and classical part of $\langle p_3 p_4 |\opT^\dagger \hat a_{hh}(k)\opT|p_1 p_2\rangle$ as $M_{5,\text{cut}}^{\text{sc. and cl.}}$. After reducing the cut to a master-integral basis, the general structure of its super-classical and classical terms is
\begin{align}
%    i\Mred_{5,\text{1 loop cut}} 
    M_{5,\text{cut}}^{\text{sc. and cl.}}
    &= (a_{4}+b_4+c_4)I^{\text{cut,cut}}_{0,1,0} + (a_{5}+b_5+c_5)I^{\text{cut,cut}}_{0,0,1} + (a_{7}+b_7+c_7) I^{\text{cut,cut}}_{1,1,0} \\
    & 
    \quad + (a_{8}+b_8+c_8) I^{\text{cut,cut}}_{1,0,1} + (a_{9}+b_9+c_9) I^{\text{cut,cut}}_{0,1,1} + (a_{10}+b_{10}+c_{10}) I^{\text{cut,cut}}_{1,1,1}\,, \nonumber
\end{align}
where $a_i\sim\frac{m_1^3m_2^3}{\hbar^2}$ are superclassical contributions and $b_i\sim\frac{m_1^3m_2^2}{\hbar}$ and $c_i\sim\frac{m_1^2m_2^3}{\hbar}$ are classical. The coefficients $b_i$ and $c_i$ are related by 
\begin{align}
    \ub_1\leftrightarrow \ub_2\,,\quad \mb_1\leftrightarrow \mb_2\,,\quad q_1\leftrightarrow q_2 \ .
\end{align}
We have verified that the coefficients $a_i$ are such that the superclassical term cancel against those of the amplitude, in agreement with the general arguments in \cref{kmoconeloop}. The remaining integrals cancel one of the two terms in the PV propagator, as shown in Ref.~\cite{Caron-Huot:2023vxl} and outlined earlier in this section.

Rather than use this cancellation, it is convenient for us to evaluate the integrals directly, for incoming $p_1$ and $p_2$. The result for the classical part $M_{5,\text{cut}}^\text{cl.}$ of the matrix element $\langle p_3 p_4 |\opT^\dagger \hat a_{hh}(k)\opT|p_1 p_2\rangle$ at one-loop order is
\begin{align}
    i M_{5,\text{cut}}^{\text{cl.}} &= \frac{i\,\kappa^2}{32\pi}\frac{\sigma(\sigma^2-Y)}{(\sigma^2-1)^{3/2}}(k\cdot p_1+k\cdot p_2)\left(\frac{1}{\epsilon}-\frac{1}{2}\log\frac{16\pi^2(k\cdot u_1)^2(k\cdot u_2)^2}{(\sigma^2-1)^2\mu^4}\right)\Mred_{5, \text{tree}}^{\text{cl.}}\nonumber\\
    & \quad - \frac{i \, \kappa^5}{256\pi} \frac{(2\sigma^2-X)^2}{(\sigma^2-1)^{3/2}}(k\cdot p_1+k\cdot p_2)A^{\text{cut}}_{\text{UV}}\left(\frac{1}{\epsilon}-\frac{1}{2}\log\frac{16\pi^2(k\cdot u_1)^2(k\cdot u_2)^2}{(\sigma^2-1)^2\mu^4}\right)\nonumber\\
    & \quad + i \, \kappa^5 \left\{A^{\text{cut}}_{\text{rat}} + A_1^{\text{cut}} \log\frac{k\cdot u_1}{k\cdot u_2}+A_2^{\text{cut}}\log\frac{(k\cdot u_1)(k\cdot u_2)}{-q_1^2(\sigma^2-1)}\right.\nonumber\\
    &\qquad\qquad\quad\left.+A_3^{\text{cut}}\log\frac{(k\cdot u_1)(k\cdot u_2)}{-q_2^2(\sigma^2-1)} +A_4^{\text{cut}}\arccosh \sigma \right\}
    \label{cutpart}
\end{align}
where $(X,Y)=(1,3/2)$ for GR and $(X,Y)=(0,2)$ for $\mathcal{N}=8$. The expressions of the relevant integrals are included in \cref{sec:addmasterint}. 
The $1/\epsilon$ on the first line is the IR divergence, while the one on the second line has a UV origin. 
In both GR and $\mathcal{N}=8$, it is
\begin{align}
    A^{\text{cut}}_{\text{UV}} = \frac{m_1^2m_2^2(u_1\cdot f\cdot u_2)^2[(k\cdot u_1)^2+\sigma(k\cdot u_1)(k\cdot u_2)+(k\cdot u_2)^2]}{(k\cdot u_1)^3(k\cdot u_2)^3} \ .
\end{align}
The UV contribution is independent of the momentum transfer and corresponds to a contact interaction in impact parameter space. We will therefore discard it. 
The explicit expressions of the coefficients $A^{\text{cut}}_i$ are included in two ancillary files, {\tt GR\_Cut\_Coeffs.m} and {\tt Neq8\_Cut\_Coeffs.m}, for GR and ${\cal N}=8$ supergravity, respectively.

The matrix element \eqref{one_and_two} determining the waveform, which combines the virtual amplitude and the cut (bilinear-in-$T$) contribution, is
\begin{align}
\label{fullmatrixelement}
  (-i)
\langle p_1- q_1, p_2- q_2 |\opS^\dagger \hat a_{hh}( k)\opS|p_1 p_2\rangle \Big|_{1\text{ loop}}=  \left(\Mred_{5, \text{1 loop}}^{\text{2MPI}, \text{cl.}} - i   \Mred_{5,\text{cut}}^{\text{cl.}}\right)\delta( k-q_1-q_2) \,,
\end{align}
where we have put back the momentum conservation delta function. Comparing with \cref{one_and_two}, one can read off the expression for $\mathcal{W}_{\text{amp}}^{\text{1 loop}}$ and $\mathcal{W}_{\text{cut}}^{\text{1 loop}}$. In particular, we have $\mathcal{W}_{\text{cut}}^{\text{1 loop}}(k)=-i\Mred_{5,\text{cut}}^{\text{cl.}}\delta(k-q_1-q_2)$.
Combining \cref{integratedAmplitudeX} and the first line of \cref{cutpart} leads to the full IR divergence of the waveform,
\begin{align}
(\Mred_{5, \text{1 loop}}^{\text{2MPI}, \text{cl.}} - i   M_{5,\text{cut}}^{\text{cl.}} )\Big|_{\text{IR}}=
    -\frac{i\,\kappa^2}{32\pi}\frac{1}{\epsilon}(k\cdot p_1+k\cdot p_2)\left(1+\frac{\sigma(\sigma^2-Y)}{(\sigma^2-1)^{3/2}}\right) \Mred_{5, \text{tree}}^{\text{cl.}} \ ,
\end{align}
which, for the case of GR, reproduces the calculation of Ref.~\cite{Caron-Huot:2023vxl}.\footnote{A possible interpretation of the IR divergence is based on the departure of the incoming trajectories from straight lines due to gravitational interactions~\cite{Caron-Huot:2023vxl}.} However, we note that, unlike the universal Weinberg IR divergence for the amplitudes, the cut contribution is sensitive to the details of the theory through the numerical coefficient $Y=(3/2, 0)$ for GR and ${\cal N}=8$ SG, respectively. In ${\cal N}=8$ SG the IR divergence receives contributions from the other light fields of the theory. We note that this IR divergence is compatible with \cref{eq:waveformIR} and thus can be removed by a time shift as described in \cref{IRdivANDwaveform}.

\subsection{The complete waveform}

Having computed the bilinear-in-$\opT$ contribution, we now proceed to complete the discussion in \cref{sec:Waveforms} and assemble the NLO waveform. We first discuss the gravitational-wave memory and then the time-domain waveform.

\subsubsection{Memory}
\label{sec:memory_full}

As we discussed previously, the amplitude contribution to the gravitational-wave memory is given in eq.~\eqref{Deltah} and is completely determined by the real part of the virtual amplitude. 
The cut contributes only to the imaginary part of the matrix element
$\langle p_1- q_1, p_2- q_2 |\opS^\dagger \hat a_{hh}(\tilde k)\opS|p_1 p_2\rangle^{0} \big|_{1\text{ loop}}$ and its contribution to the memory is
\begin{align}
\Delta(h_{+}^{\infty}+ih^{\infty}_{\times}) \big|_{\text{cut}} &= \lim_{\omega\rightarrow 0^+} (-i\omega) \Theta(\omega) \mathcal{W}_{\text{cut}}^\text{1 loop}\Big|_{k=(\omega,\omega\pmb{n})} \nonumber\\
&\quad + \lim_{\omega\rightarrow 0^-} (-i\omega)\Theta(-\omega)\mathcal{W}_{\text{cut}}^{\text{1 loop}\,*}\Big|_{k=(|\omega|,|\omega|\pmb{n})} \ .
\end{align}
We evaluate the Fourier transforms by direct integration using the $z_{i}$ coordinate system in \cref{eq:qbasis} as in~\cite{Cristofoli:2021vyo}.
The final result is
\begin{align}
\Delta(\hat{h}_{+}^{(2)}+i\hat{h}_{\times}^{(2)})\big|_{\text{cut}} &=
- \frac{(2\sigma^2-X)^2m_1m_2}{512\pi^2(\sigma^2-1)\Mtot^3}\left[\frac{m_2(u_1\cdot \tilde f \cdot \tilde{b})^2}{(u_1\cdot \tilde k)^3}+\frac{m_1(u_2\cdot \tilde f\cdot\tilde{b})^2}{(u_2\cdot \tilde k)^3}\right. \nonumber\\
     %& \qquad +\frac{1}{2}(S'+S)\frac{(u_1\cdot f\cdot u_2)^2}{\sigma^2-1}\left(\frac{m_1}{(u_2\cdot k_5)^3}+\frac{m_2}{(u_1\cdot k_5)^3}\right) \nonumber\\
    & \qquad - \left.\frac{(u_1\cdot \tilde f\cdot u_2)^2}{2(\sigma^2-1)(u_1\cdot \tilde k)(u_2\cdot \tilde k)}\left(\frac{\sigma m_1+m_2}{u_2\cdot \tilde k}+\frac{\sigma m_2+m_1}{u_1\cdot \tilde k}\right)\right] 
    \label{memory_cut_contrib2}
\end{align}
where $X=1$ and $X=0$ for GR and $\mathcal{N}=8$ respectively. The asymptotic metric at NLO follows then from eqs.~\eqref{eq:norm_h}, \eqref{fullmetric} and \eqref{memory_cut_contrib2}.

We can now compare the memory contribution to the asymptotic metric with existing predictions for the GR memory \cite{Sahoo:2021ctw}, which writes the universal terms in the small frequency expansion of the asymptotic metric in terms of the classical incoming and outgoing momenta. The leading soft (memory) term is
\begin{align}
\varepsilon^{\mu}\varepsilon^{\nu}h_{\mu\nu}  
%&= \frac{\varepsilon^{\mu}\varepsilon^{\nu}\mathcal{A}_{\mu\nu} }{\omega} + {\cal O}(\omega^0) 
&= \frac{i\,\mathcal{A}}{\omega} + \mathcal{O}(\omega^0)\,,\nonumber
\\
%\varepsilon^{\mu}\varepsilon^{\nu}\mathcal{A}(k,p,\Delta p)_{\mu\nu} 
\mathcal{A}(p,\Delta p)
&= - \frac{(\varepsilon\cdot p_3)^2}{\tilde k\cdot p_3}
- \frac{(\varepsilon\cdot p_4)^2}{\tilde k\cdot p_4} + \frac{(\varepsilon\cdot p_1)^2}{\tilde k\cdot p_1}+ \frac{(\varepsilon\cdot p_2)^2}{\tilde k\cdot p_2} \ .
\end{align}
In this expression, the outgoing matter momenta are given in terms of the incoming momenta and the (classical) impulse; through ${\cal O}(G^2)$ they are
\begin{align}
p_4 = p_1 + \Delta p 
\qquad
p_3 = p_2 - \Delta p
\qquad
\Delta p = G \Delta p^{(0)} +  G^2 \Delta p^{(1)}+\dots \ ,
\label{P3P4}
\end{align}
so $\mathcal{A}$ becomes
%\begin{align}
%\varepsilon^{\mu}\varepsilon^{\nu} \mathcal{A}(k,p,\Delta p)_{\mu\nu}&= G \varepsilon^{\mu}\varepsilon^{\nu}\mathcal{A}^{(0)}(k, p, \Delta p)_{\mu\nu} + G^2 \varepsilon^{\mu}\varepsilon^{\nu}\mathcal{A}^{(1)}(k, p, \Delta p)_{\mu\nu} + {\cal O}(G^3)
%\end{align}
%
\begin{align}
    \mathcal{A}(p,\Delta p) = G \mathcal{A}^{(0)}(p,\Delta p) + G^2 \mathcal{A}^{(1)}(p,\Delta p) + \mathcal{O}(G^3)
\end{align}
where
%
%\begin{align}
% \varepsilon^{\mu}\varepsilon^{\nu}\mathcal{A}^{(0)}(k, p, \Delta p)_{\mu\nu}   &= \,
% \tilde k\cdot \Delta p^{(0)}\left[
% \frac{(\varepsilon\cdot p_1)^2 }{(p_1\cdot \tilde k)^2}
% -\frac{(\varepsilon\cdot p_2)^2 }{(p_2\cdot \tilde k)^2}\right]
 %\nonumber\\
 %&+\,
%+ \varepsilon\cdot \Delta p^{(0)}\left[
% \frac{2\varepsilon\cdot p_2 }{p_2\cdot \tilde k}
% -\frac{2\varepsilon\cdot p_1 }{p_1\cdot \tilde k}\right]
%\end{align}
\begin{align}\label{GoneS}
 \mathcal{A}^{(0)}(p,\Delta p)   &= \,
 \tilde k\cdot \Delta p^{(0)}\left[
 \frac{(\varepsilon\cdot p_1)^2 }{(p_1\cdot \tilde k)^2}
 -\frac{(\varepsilon\cdot p_2)^2 }{(p_2\cdot \tilde k)^2}\right]
 %\nonumber\\
 %&+\,
+ \varepsilon\cdot \Delta p^{(0)}\left[
 \frac{2\varepsilon\cdot p_2 }{p_2\cdot \tilde k}
 -\frac{2\varepsilon\cdot p_1 }{p_1\cdot \tilde k}\right] \\
\label{GtwoS2}
\mathcal{A}^{(1)}(p, \Delta p) &= \tilde k\cdot \Delta p^{(1)}\left[
 \frac{(\varepsilon\cdot p_1)^2 }{(p_1\cdot \tilde k)^2}
 -\frac{(\varepsilon\cdot p_2)^2 }{(p_2\cdot \tilde k)^2}\right]
 +\,
 \varepsilon\cdot \Delta p^{(1)}\left[
 \frac{2\varepsilon\cdot p_2 }{p_2\cdot \tilde k}
 -\frac{2\varepsilon\cdot p_1 }{p_1\cdot \tilde k}\right]    
 \nonumber\\
&\quad -(k\cdot \Delta p^{(0)})^2\left[\frac{(\varepsilon\cdot p_1)^2}{(\tilde k \cdot p_1)^3} + \frac{(\varepsilon\cdot p_2)^2}{(\tilde k \cdot p_2)^3}\right]
-(\varepsilon\cdot \Delta p^{(0)})^2\left[\frac{1}{\tilde k \cdot p_1} + \frac{1}{\tilde k \cdot p_2}\right]
\nonumber\\
&\quad +2(k\cdot \Delta p^{(0)})(\varepsilon\cdot \Delta p^{(0)})\left[\frac{(\varepsilon\cdot p_1)}{(\tilde k \cdot p_1)^2} + \frac{(\varepsilon\cdot p_2)}{(\tilde k \cdot p_2)^2}\right] \ .
\end{align}
The tree-level and one-loop GR impulses entering \cref{GtwoS2} were computed in Ref.~\cite{Herrmann:2021tct} and, in our conventions,\footnote{The momentum transfer $q$ in \cite{Herrmann:2021tct} is defined to be the opposite of ours, 
$q_\text{EPRZ} = -q_\text{here}$. This in turn implies that the impact parameter vector has the opposite orientation from ours, $b_\text{EPRZ} = -b_\text{here}$.} are given by
\begin{align}\label{classicalmom}
\Big[\Delta p^{(0)}\Big]_\mu &= -\frac{m_1 m_2}{\sqrt{-b^2}} \frac{2(2\sigma^2-X)}{\sqrt{\sigma^2-1}} {\tilde b}_{\mu}\,,\nonumber
\\
\Big[\Delta p^{(1)}\Big]_\mu &= \Big[\Delta p^{(1,\perp)}\Big]_{\mu} + \Big[\Delta p^{(1,\parallel)}\Big]_{\mu}\,,
\end{align}
%where for the one-loop case, the 
where the ${\cal O}(G^2)$ transverse and longitudinal components of the impulse are
\begin{align}
\Big[\Delta p^{(1,\perp)}\Big]_{\mu} &=-\frac{m_1 m_2(m_1+m_2)}{-b^2}\frac{3\pi}{4}\frac{(5\sigma^2-1)X}{\sqrt{\sigma^2-1}}{\tilde b}_\mu
\nonumber\\
\Big[\Delta p^{(1,\parallel)}\Big]_\mu &=\frac{m_1^2 m_2^2}{-b^2}\frac{2(2\sigma^2-X)^2}{\sigma^2-1}\left[\frac{1}{m_1} \frac{(\sigma u_2 - u_1)_{\mu}}{\sigma^2-1} -\frac{1}{m_2}\frac{(\sigma u_1 - u_2)_{\mu}}{\sigma^2-1}\right] \,.
\end{align}
Upon substituting \cref{classicalmom} into \cref{GtwoS2}, one finds,
\begin{align}
\mathcal{A}^{(1)}(p, \Delta p) &= -\varepsilon^{\mu}\varepsilon^{\nu}\mathcal{S}(\tilde{k}, \Delta p^{(1,\perp)})_{\mu\nu}
-\frac{4 (2 \sigma^2-1)^2 m_1 m_2}{(\sigma^2-1)(-b^2)}
\left[
\frac{m_2(u_1\cdot \tilde{f}\cdot {\tilde b})^2}{(u_1\cdot \tilde{k})^3}+\frac{m_1(u_2\cdot \tilde{f}\cdot {\tilde b})^2}{(u_2\cdot \tilde{k})^3}
\right.
\nonumber
\\
&
\quad\left.-\, \frac{(u_1\cdot \tilde{f}\cdot u_2)^2}{(u_1\cdot \tilde{k})(u_2\cdot \tilde{k})}
\left(\frac{1}{(u_1\cdot \tilde{k})} \frac{m_1+\sigma m_2}{2(\sigma^2-1)}  
    + \frac{1}{(u_2\cdot \tilde{k})} \frac{m_2+\sigma m_1}{2(\sigma^2-1)}  \right)\right]
\end{align}
which matches the combined soft limit of the amplitude contribution~\eqref{NLOmemoryA} and the cut contribution~\eqref{memory_cut_contrib2} for both GR and $\mathcal{N}=8$. Therefore, we have verified that the memory given by the waveform reproduces the prediction from Ref.~\cite{Sahoo:2021ctw}.

\subsubsection{Time-domain waveform}

The evaluation of $\mathcal{W}_{\text{cut}}^{\text{1 loop}}$ follows the same steps as those leading to the evaluation of $\mathcal{W}_{\text{amp}}^{\text{1 loop}}$.
While the cut contribution is purely imaginary (for real polarization tensors), it contributes to both $h_{+}$ and $h_\times$ polarizations of the gravitational wave.

Instead of plotting the cut contribution separately, we directly combine it with the amplitude contribution discussed in \cref{amplitudeNLO} and plot the complete waveform. As in \cref{amplitudeNLO} we will choose the scale $\Lambda$ so that $ \Lambda|\pmb{b}| = 5 \times 10^{-10}$. Interestingly, and as may perhaps be anticipated by comparing the scales of \cref{fig:NLO_GR_1,fig:NLO_GR_tail}, for such values of $\Lambda$ the tail contribution dominates the waveform.  

\begin{figure}[tb]
\centering
\includegraphics[width=\textwidth]{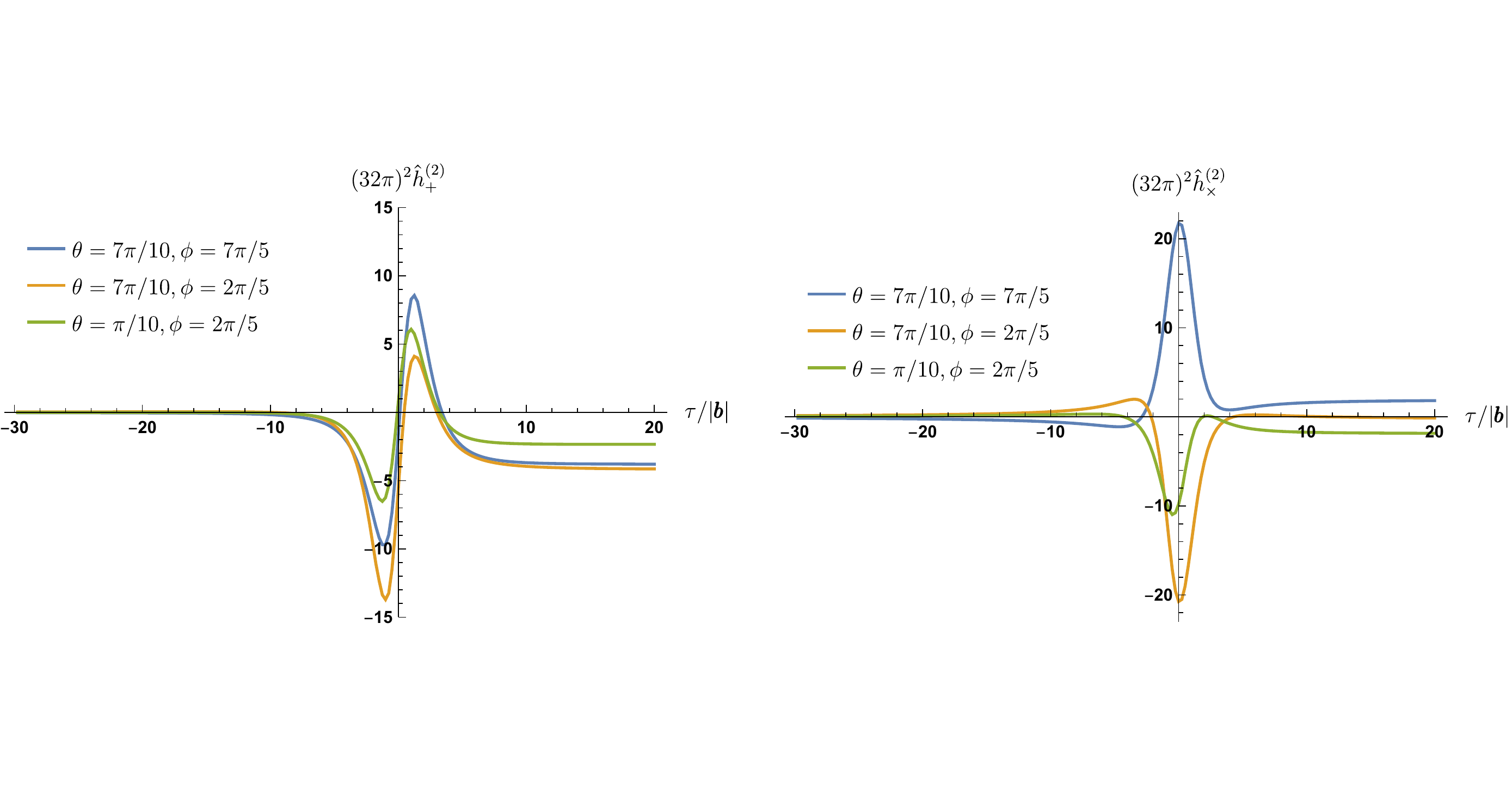}
\caption{The full GR NLO waveform at various angles. We work in the COM frame, with incoming particles moving along the $z$ axis with velocities $v=1/5$ and assume that $m_1=m_2$.} 
\label{NLO_GR_plot_fixv_full}
\end{figure}

\begin{figure}[tb]
\centering
\includegraphics[width=\textwidth]{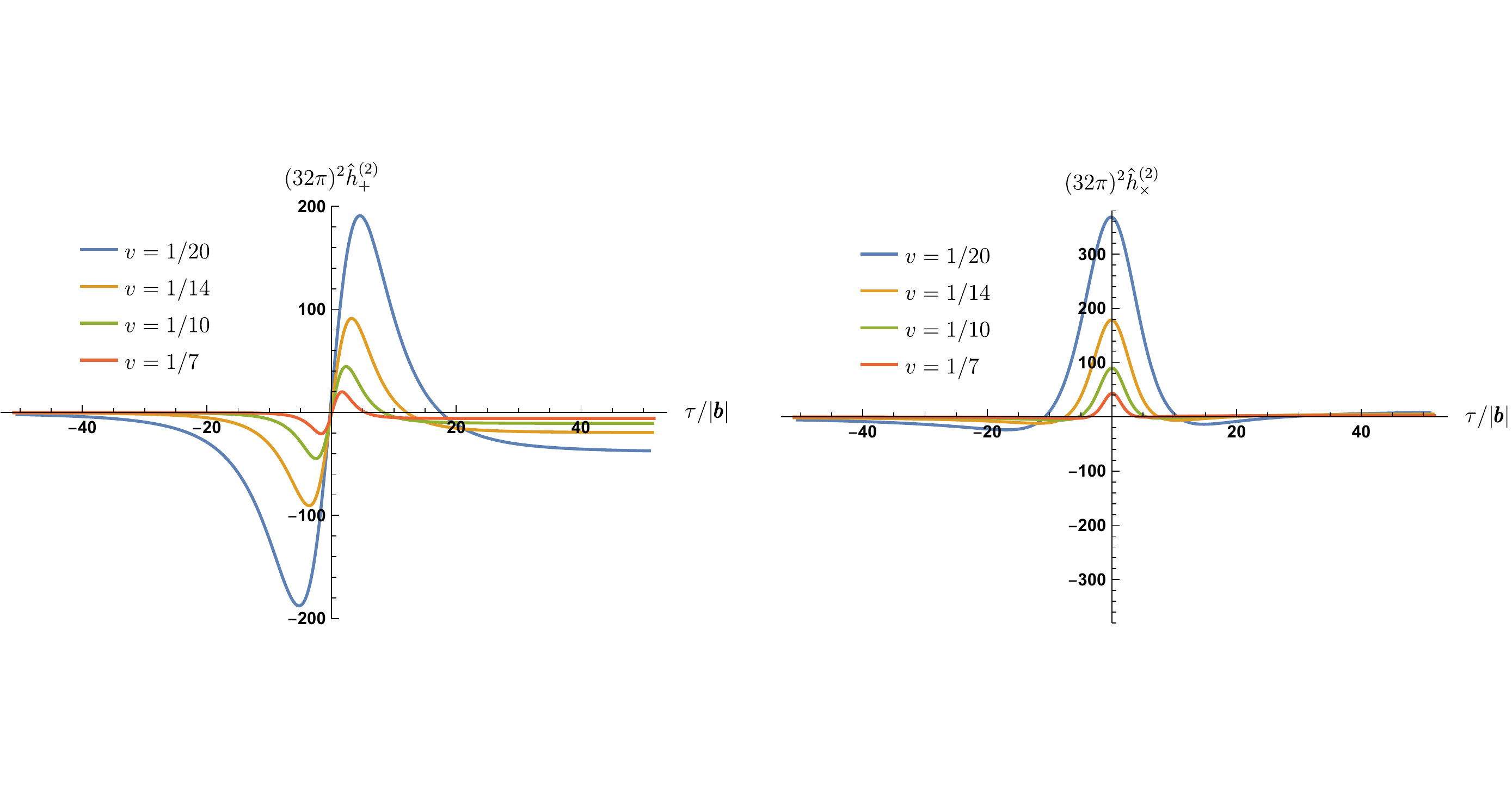}
\caption{The full GR NLO waveform in observation direction given by the angles $\theta = 7\pi/10$ and $\phi = 7\pi/5$ for various various velocities. We work in the COM frame, with incoming particles moving along the $z$ axis and assume that $m_1=m_2$.} 
\label{NLO_GR_plot_fixangle_full}
\end{figure}

We assume a center-of-mass frame and equal-mass particles moving initially along the $z$-axis. 
In \cref{NLO_GR_plot_fixv_full} we show the NLO $h_+$ and $h_\times$ polarizations for a selection of directions of observation specified by the polar angles $\theta$ and $\phi$ while \cref{NLO_GR_plot_fixangle_full} includes the two polarizations at fixed obervation direction but for various initial particle velocity (in units of the speed of light).
%

%%%%%%%%%%%%%% FIGURE %%%%%%%%%
%%%%%%%%%%%%%%%%%%%%%%%%%%%%%%%

%%%%%%%%%%%%%% FIGURE %%%%%%%%%
%%%%%%%%%%%%%%%%%%%%%%%%%%%%%%%

%%%%%%%%%%%%%% FIGURE %%%%%%%%%
\begin{figure}[tb]
\centering
\includegraphics[width=\textwidth]{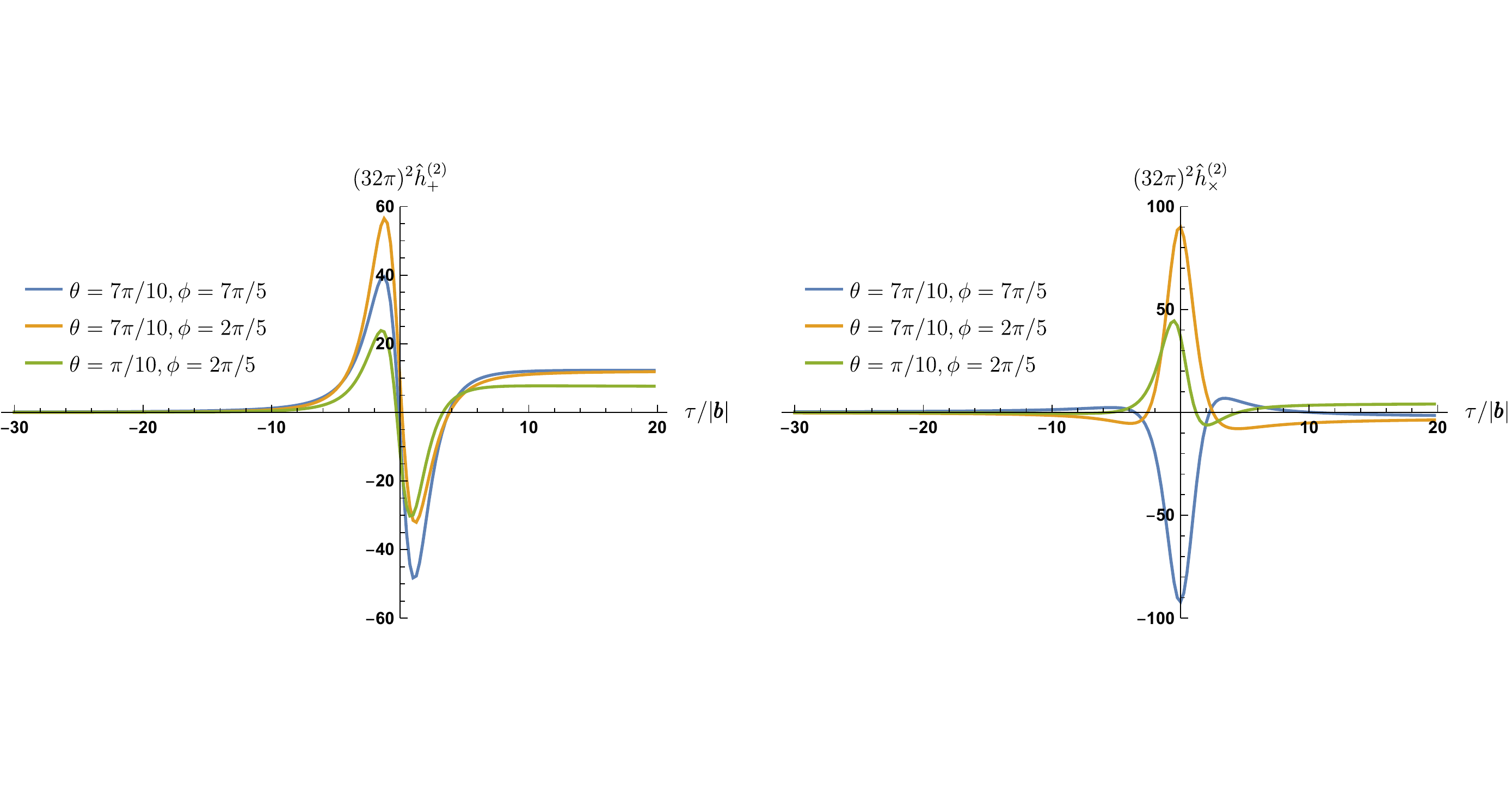}
\caption{The full ${\cal N}=8$ SG NLO waveform at various angles. We work in the COM frame, with incoming particles moving along the $z$ axis with velocities $v=1/5$ and assume that $m_1=m_2$.} 
\label{NLO_Neq8_plot_fixv_full}
\end{figure}
%%%%%%%%%%%%%%%%%%%%%%%%%%%%%%%

%%%%%%%%%%%%%% FIGURE %%%%%%%%%
\begin{figure}[tb]
\centering
\includegraphics[width=\textwidth]{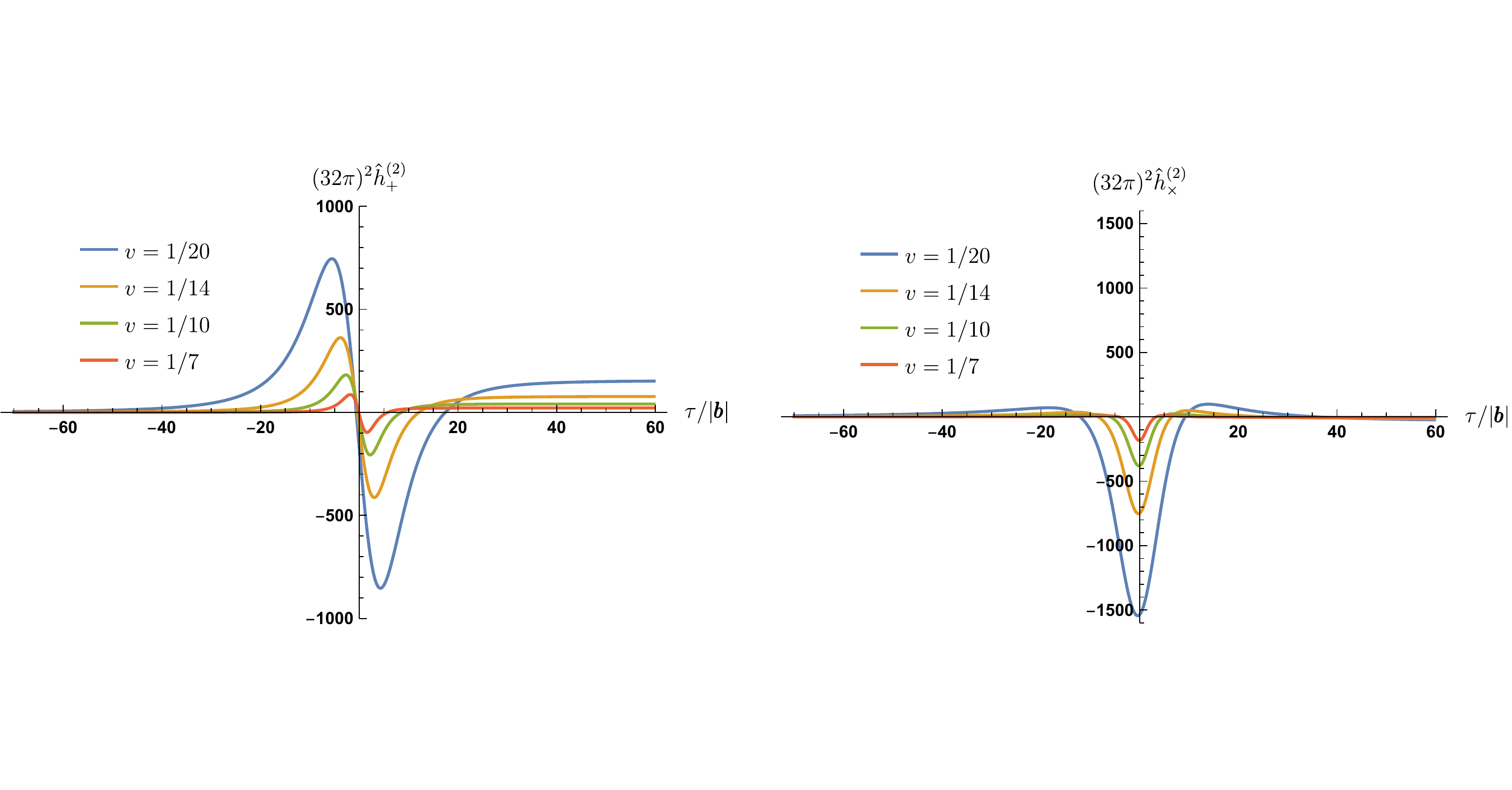}
\caption{The full ${\cal N}=8$  NLO waveform in observation direction given by the angles $\theta = 7\pi/10$ and $\phi = 7\pi/5$ for various various velocities. We work in the COM frame, with incoming particles moving along the $z$ axis and assume that $m_1=m_2$.} 
\label{NLO_Neq8_plot_fixangle_full}
\end{figure}
%%%%%%%%%%%%%%%%%%%%%%%%%%%%%%%

\Cref{NLO_Neq8_plot_fixv_full,NLO_Neq8_plot_fixangle_full} show the analogous plots for ${\cal N}=8$ supergravity. We note the presence of gravitational wave memory; as discussed in \cref{sec:memory_full}, it arises entirely from the cut contribution. Its analytic expression follows from \cref{GtwoS2} upon using the ${\cal N}=8$ SG impulses, i.e. 
$\Delta p^{(1,\perp)}= 0$ and $\Delta p^{(1,\parallel)}$ fixed by the on-shell condition in terms of the tree-level impulse  $\Delta p^{(0,\perp)}$.

\section{Conclusions}
\label{sec:Conclusions}

The observable-based formalism~\cite{Kosower:2018adc,Cristofoli:2021vyo} directly links scattering amplitudes and their cuts to observables of unbound binary systems in the classical regime. This framework naturally includes both conservative and dissipative effects, providing a means to find local observables such as the asymptotic gravitational waveform for a scattering event in addition to inclusive ones such as the impulse or the energy and angular momentum loss.
With the possible exception of singular momentum configurations, they are governed by four and five-point amplitudes~\cite{Cristofoli:2021jas}.
%
%We examined in detail this connection for the scattering waveform and found that, beyond leading order, the exponential form of amplitudes proposed in Refs.~\cite{Bern:2021dqo,Damgaard:2021ipf} implies that super-classical terms do not contribute to this observable.
%
%We examined in detail this connection for the scattering waveform and found that
%, beyond leading order, 
%the exponential form of amplitudes proposed in Refs.~\cite{Bern:2021dqo,Damgaard:2021ipf} implies that only (the suitably-defined) two-matter-particle-irreducible components of the amplitude can  contribute to this observable at least through one-loop order.
%
We examined in detail this connection for the scattering waveform and found that
%, beyond leading order, 
the exponential form of amplitudes proposed in Refs.~\cite{Bern:2021dqo,Damgaard:2021ipf} implies that only (the suitably-defined) two-matter-particle-irreducible components of the amplitude can contribute to the virtual part of this observable at least through one-loop order.

We computed these parts of the four-scalar-one-graviton amplitude at one-loop order in GR with minimally-coupled scalars and in ${\cal N}=8$ supergravity. We obtained the former using generalized unitarity, both by evaluating the classical part of the complete amplitude and by using a HEFT approach which effectively corresponds to truncating the tree amplitudes to classical order.
Comparing the two results and requiring that they agree 
%identifies a prescription for treating 
reveals that the HEFT uncut matter propagators
%, that they be treated 
should be interpreted as principal-valued.  
We obtained the integrand of the corresponding five-point one-loop amplitude in $\mathcal{N}=8$ supergravity
%, which describes the emission of gravitational radiation from the 
%scattering of two half-BPS black holes, 
by using the double copy of the corresponding integrand in ${\cal N}=4$ super-Yang-Mills theory and dimensional reduction.
For both theories we computed directly the relevant cut contribution as well.

Upon reduction to master integrals, the five-point classical amplitudes exhibit novel features compared to their four-point counterparts. 
For example, the outgoing graviton momentum injects a scale in some of the scaleless (and thus vanishing) integrals in the four-point amplitude, implying that they can contribute to the five-point classical amplitude.

The one-loop master integrals are sufficiently simple so we evaluated them by direct integration for physical kinematics.\footnote{They may also be evaluated as classical limits of the known one-loop master integrals that appear in the quantum theory.}
A distinguishing feature of the (2MPI part of the) classical five-point amplitude and of the cut contribution to the waveform is that they exhibit infrared-divergent phases. The structure of the amplitude's phase divergence is governed by Weinberg's classic result \cite{Weinberg:1965nx} and the classical nature of the amplitude; we argue that at this order -- and perhaps to all orders in perturbation theory -- it can be absorbed as a suitable shift in the definition of the retarded time. Consequently, the observation of the gravitational waveform is sensitive only to differences of retarded times rather than the absolute time of the process. This is consistent with the basic assumption of scattering theory that observations are carried out at infinite times.
We also gave a precise relation between the gravitational-wave memory and the soft limit of the corresponding scattering amplitude, along the lines of Refs.~\cite{Strominger:2014pwa, Sahoo:2021ctw}, demonstrating a close relation between the amplitude's contribution to the memory and the scattering angle to at least ${\cal O}(G^3)$. 
Interestingly, this implies that the amplitude's contribution to the gravitational-wave memory vanishes at ${\cal O}(G^2)$ for half-BPS black holes in $\mathcal{N}=8$ supergravity, which is confirmed by the explicit calculation. 
The cut however yields a nonzero memory both in general relativity and in $\mathcal{N}=8$ supergravity.

Even though our results are analytic in frequency and momentum space as well as in time-momentum space, we resorted to numerical integration to evaluate the one-dimensional integral completing the transformation to time-impact parameter space.
A complete analytic evaluation of the waveform, which we leave for future work, would involve obtaining an analytic expression for this last one-dimensional integral. The result would resolve certain numerical features that we encountered at special observation angles and would open the door to the evaluation of interesting inclusive and local observables, such as the energy and angular momentum loss at ${\cal O}(G^4)$ and the radiated energy spectrum $E(\omega)$ through traditional general relativity methods~\cite{Bonga:2018gzr, Thorne:1980ru, DeWitt:2011nnj, Peters:1964zz}.
The classical one-loop five-point amplitude constructed here allows the computation of inclusive quantities at 4PM order through the KMOC formalism, along the lines of the 3PM calculations of Refs.~\cite{Herrmann:2021lqe, Herrmann:2021tct, Manohar:2022dea}. 

While the next-to-leading order is the current state of the art for relativistic scattering waveforms, it is natural to think about higher orders. The experience with ${\cal N}$-extended supergravity loop calculations~\cite{Bern:2009kd, Carrasco:2011mn, Bern:2013uka, Bern:2014sna, Bern:2017ucb, Bern:2018jmv} suggests that the construction of the relevant classical integrand through double copy and generalized unitarity should scale well to higher orders and the results will provide tests of the HEFT approach to classical amplitudes beyond those already available. We expect, however, that the evaluation of the resulting master integrals will benefit from advanced techniques such as differential equations and method of regions, as was done for the full quantum two-loop five-point amplitude of ${\cal N}=8$ supergravity in Ref.~\cite{Abreu:2019rpt}. 
%Going to higher order could also help us better understand radiation-reaction, 
%which already appears at one-loop~\cite{Elkhidir:2023dco}, and other physical 
%phenomena that could appear. 

The current approach to analytic and semi-analytic binary inspiral waveforms makes use of the effective one-body formalism~\cite{Buonanno:1998gg,Buonanno:2000ef}, whose input is an off-shell 
binary Hamiltonian and radiation reaction forces. It would be interesting to identify features of gravitational waveforms that can be analytically 
continued between bound and unbound systems, in analogy with inclusive observables discussed in Refs.~\cite{Kalin:2019rwq, Kalin:2019inp, Cho:2021arx}.
We may expect that a detailed understanding of the analytic structure of the waveforms, both in the bound and unbound case, will be important.

\acknowledgments

We thank A.~Elkhidir, D.~O'Connell, M.~Sergola, I.~Vazquez-Holm, and A.~Brandhuber, G.~Brown, G.~Chen, S.~De Angelis,  J.~Gowdy, G.~Travaglini and L.~Bohnenblust, H.~Ita, M.~Kraus and J.~Schlenk for coordination on ongoing work. 
We also thank N. Arkani-Hamed, R. Akhoury, J. Berman, R. Britto, T.~Damour, S.~DeAngelis, L. Dixon, H. Elvang, S. Mizera, A.~Sen,  C.-H. Shen and G.~Sterman for stimulating discussions.
RR and FT also thank Z.~Bern, E. Herrmann, J. Parra-Martinez, M.~Ruf, C.-H. Shen, and M.~Zeng for collaboration on related topics, and especially M.~Ruf for providing valuable comments on our draft.
RR and FT would like to thank the Kavli Institute for Theoretical Physics at the University of California, Santa Barbara, for hospitality during the program ``High-Precision Gravitational Waves''. Their research there was supported in part by the National Science Foundation under Grant No. NSF PHY-1748958.
AH is supported by a Rackham Predoctoral Fellowship from the University of Michigan.
RR and FT are supported by the U.S. Department of Energy (DOE) under award number DE-SC00019066.

\appendix

\section{Self-consistency of HEFT}
\label{sec:selfconheft}

In this appendix, we focus on terms containing two-matter-particle cuts on which the HEFT integrand factorizes into lower point HEFT integrands. We argue such terms require quantum information about the HEFT tree amplitudes used in the unitarity construction.

We will use the $\hbar$ counting as introduced in \cref{rescaling}. Consider a cut of a four-scalar amplitude containing $k$ HEFT tree amplitudes, each with $\tilde{n}_{g, i}$ internal gravitons $n_{g,i}$ external gravitons, and $n_{\phi,i}$ scalars; call $\tilde{N}_g=\frac{1}{2}\sum_{i=1}^k \tilde{n}_{g, i}$ and $n_g=\sum_{i=1}^{k}n_{g,i}$ respectively the total number of internal and external gravitons, and $N_{\phi}=\frac{1}{2}\sum_{i=1}^k n_{\phi,i}-2$ the total number of internal scalars. Restoring the integration measure and the matter and graviton propagators, this cut scales as
\begin{align}\label{eq:classcial_scaling_2}
    \underbrace{\Bigg[\prod_{i=1}^{k}\hbar^{-3(\frac{1}{2}n_{\phi,i}-1)-\frac{1}{2}\tilde{n}_{g,i}-\frac{1}{2}n_{g,i}}\Bigg]}_{\text{HEFT tree amplitudes}}\underbrace{\vphantom{\Bigg[}\hbar^{-2\tilde{N}_g-N_\phi}\hbar^{4L}}_{\text{propagators and measures}}=\hbar^{-3-\frac{1}{2}n_{g}}\hbar^{\tilde{N}_g+1-k}\,,
\end{align}
where we have used \cref{classical_scaling} and the loop number $L=\tilde{N}_g+N_\phi-k+1$. This cut scales classically if
\begin{equation}
\tilde{N}_g+1-k = 0 \ .
\label{classcial_scaling_cut}
\end{equation}
Consistency requires that if we cut any line the result should still scale classically. Let us consider cutting a pair of matter lines. The resulting eight-scalar amplitude scales as
\begin{equation}
\hbar^{-9-\frac{1}{2}n_g}\hbar^{\tilde{N}_g+1-k} \hbar^{8-a} \,.
\end{equation}
Here we have isolated the classical scaling $\hbar^{-9}$ of an eight-scalar amplitude. 
If the graph remains connected, then we remove two matter propagators and two integral measures from \cref{eq:classcial_scaling_2}, which leads to $a=8$. On the other hand, if the graph is now disconnected, then we only need to remove two matter propagators and one integral measure instead, which leads to $a=4$.
Thus, if $a=8$ -- i.e. if the cut of the eight-scalar amplitude is connected -- then it has classical scaling upon use of \cref{classcial_scaling_cut} and the truncation to classical order is iteratively consistent.
In contrast, if $a=4$ -- i.e. if the cut of the eight-scalar amplitude is disconnected -- the cut is suppressed by $\hbar^4$, which means that the disconnected components contain quantum contributions. 

Repeating the calculation above for any pair of matter lines, it follows that any 2MPR contribution constructible by generalized unitarity with a classical scaling requires quantum information about some amplitude contributing to it. Since the HEFT approach prescribes that tree amplitudes be truncated to classical orders, the $\hbar$ scaling of 2MPR diagrams and cuts with classical vertices is super-classical, with one factor of $\hbar^{-1}$ for each present two-matter-particle cut. On the other hand, if we explicitly include some quantum operators in the Lagrangian and keep higher-order terms in the HEFT tree amplitudes, we will get classical 2MPR contributions. Again, the above argument only applies to two-matter-particle cuts on which the HEFT integrand factorizes into lower point integrands.

%However, these will be consistently subtracted in the KMOC formalism. Thus, only 2MPI cuts -- and consequently only 2MPI diagrams -- should contribute to classical scattering observables to any loop order.
%

\section{Evaluating box master integrals}\label{sec:integrals}

In this appendix, we derive all the box master integrals listed in \cref{sec:box}. We will repeatedly use the bubble integral with a cut matter propagator and a generic propagator in $d=4-2\epsilon$,
\begin{align}\label{eq:bub_general}
    \int\dd^{4-2\epsilon}\lm\frac{\delta(\ub_2\cdot\lm)}{(\lm-Q_1)^2-m^2+\ie}=-\frac{\pi^{(3-2\epsilon)/2}\Gamma(\frac{2\epsilon-1}{2})}{[-(\ub_2\cdot Q_1)^2+m^2-\ie]^{(2\epsilon-1)/2}}\,,
\end{align}
where $Q_1$ is a generic vector.

\subsection{\texorpdfstring{$I_{1,1,1,0}$}{I(1,1,1,0)}}

We first compute the box integral with three graviton propagators,
\begin{align}\label{eq:I1110step1}
    I_{1,1,1,0}=\int\frac{\dd^{d}\lm}{(2\pi)^{d}}\frac{\hdelta(2\ub_2\cdot\lm)}{\lm^2(\lm+q_2)^2[(\lm-q_1)^2+\ie]}
    &=\int_{0}^{1}\dd x\int\frac{\dd^d{\lm}}{(2\pi)^d}\frac{\hdelta(2\ub_2\cdot\lm)}{\lm^2[(\lm-q_1+xk_5)^2+\ie]^2}\,,
\end{align}
where we have introduced a Feynman parameter to combine the two massless propagators $(\lm+q_2)^2$ and $(\lm-q_1)^2$ into a single propagator. We next use an IBP identity to reduce the double propagator $[(\lm-q_1+xk_5)^2+\ie]^2$,
\begin{align}\label{eq:IBPbox1}
    & \int_{0}^{1}\dd x\int\frac{\dd^d\lm}{(2\pi)^d}\frac{\hdelta(2\ub_2\cdot\lm)}{\lm^2[(\lm-q_1+xk_5)^2+\ie]^2}\nonumber\\
    &=-\frac{d-3}{2(\ub_2\cdot q_1)^2}\int_{0}^{1}\dd x\frac{1}{(1-x)^2[(1-x)q_1^2+xq_2^2]}\int\frac{\dd^d\lm}{(2\pi)^d}\frac{\hdelta(2\ub_2\cdot\lm)}{(\lm-q_1+xk_5)^2+\ie}\nonumber\\
    &\quad-(d-4)\int_{0}^{1}\dd x\frac{1}{(1-x)q_1^2+xq_2^2}\int\frac{\dd^d\lm}{(2\pi)^d}\frac{\hdelta(2\ub_2\cdot\lm)}{\lm^2[(\lm-q_1+xk_5)^2+\ie]}\,.
\end{align}
The integral in the last line is finite in $d=4$,\footnote{One can show this by using
\begin{align*}
    \int \dd^4\lm \frac{\delta(\ub_2\cdot\lm)}{\lm^2[(\lm-Q_1)^2-m^2]}&=\frac{2\pi^2}{\sqrt{(\ub_2\cdot Q_1)^2-m^2-Q_1^2}}\arcsin\sqrt{1+\frac{(\ub_2\cdot Q_1)^2-m^2}{-Q_1^2}}
\end{align*}
for the $\lm$ integral. The remaining $x$ integral is convergent.
} such that the $(d-4)$ prefactor makes this term only contribute to $\mathcal{O}(\epsilon)$ order. Thus both the IR divergence and the finite part are given by the first term of \cref{eq:IBPbox1},
\begin{align}\label{eq:I1110ie}
    I_{1,1,1,0}& =-\frac{d-3}{2(\ub_2\cdot q_1)^2}\int_{0}^{1}\dd x\frac{1}{(1-x)^2[(1-x)q_1^2+xq_2^2]}\int\frac{\dd^d\lm}{(2\pi)^d}\frac{\hdelta(2\ub_2\cdot\lm)}{(\lm-q_1+xk_5)^2}+\mathcal{O}(\epsilon)\nonumber\\
    &=\frac{\pi^{5/2-\epsilon}\Gamma(\epsilon+1/2)}{(2\pi)^{4-2\epsilon}[-(\ub_2\cdot q_1)^2-\ie]^{1/2+\epsilon}}\int_{0}^{1}\frac{\dd x}{(1-x)^{1+2\epsilon}[(1-x)q_1^2+xq_2^2]}+\mathcal{O}(\epsilon)\\
    &=\frac{1}{32 \pi q_2^2\sqrt{-(\ub_2\cdot q_1)^2-\ie}}\left[-\frac{1}{\eIR}+\log(-4\pi(\ub_2\cdot q_1)^2-\ie)+2\log\frac{q_2^2}{q_1^2}\right]+\mathcal{O}(\epsilon)\,.\nonumber
\end{align}
The $\ie$ prescription suggests the analytic continuation $\log(-x-\ie)=\log(x)-i\pi$ for $x>0$ and $\sqrt{-(\ub_2\cdot q_1)^2-\ie}=-i(\ub_2\cdot q_1)$, which gives 
\begin{align}
    I_{1,1,1,0}
    &=\frac{1}{32q_2^2(\ub_2\cdot q_1)}+\frac{i}{32 \pi q_2^2(\ub_2\cdot q_1)}\left[-\frac{1}{\eIR}+\log(4\pi(\ub_2\cdot q_1)^2)+2\log\frac{q_2^2}{q_1^2}\right]+\mathcal{O}(\epsilon)\,.
\end{align}
The result agrees with \cref{eq:I1110re} in the main text.

\subsection{\texorpdfstring{$I^{\pm}_{1,1,0,1}$}{I(1,1,0,1)}}\label{I1101}

We next consider the box integrals with one uncut matter propagator. We start with
\begin{align}
    I^{+}_{1,1,0,1}&=\int\frac{\dd^{d}\lm}{(2\pi)^d}\frac{\hdelta(2\ub_2\cdot\lm)}{\lm^2(\lm+q_2)^2(2\ub_1\cdot\lm+\ie)}\nonumber\\
    &=\int_{0}^{\infty}\dd x\int\frac{\dd^d\lm}{(2\pi)^d}\frac{\hdelta(2\ub_2\cdot\lm)}{(\lm+q_2)^2[(\lm+x\ub_1)^2-x^2+\ie]^2}\,,
\end{align}
where we have used the \cref{eq:x_parameter} to combine $\lm^2$ and $2\ub_1\cdot\lm+\ie$.
The IBP identity can reduce the double propagator
\begin{align}\label{eq:I1101ibp}
    &\int_{0}^{\infty}\dd x\int\frac{\dd^d\lm}{(2\pi)^d}\frac{\hdelta(2\ub_2\cdot\lm)}{(\lm+q_2)^2[(\lm+x\ub_1)^2-x^2+\ie]^2}\nonumber\\
    &=\frac{d-3}{2(y^2-1)}\int_{0}^{\infty}\dd x\frac{1}{x^2(2x \ub_1\cdot q_2-q_2^2-\ie)}\int\frac{\dd^d\lm}{(2\pi)^d}\frac{\hdelta(2\ub_2\cdot\lm)}{(\lm+x \ub_1)^2-x^2+\ie}\nonumber\\
    &\quad +(d-4)\int_{0}^{\infty}\dd x\frac{1}{2x \ub_1\cdot q_2-q_2^2}\int\frac{\dd^d\lm}{(2\pi)^d}\frac{\hdelta(2\ub_2\cdot\lm)}{\lm^2[(\lm-q_2+x \ub_1)^2-x^2+\ie]}\,.
\end{align}
The integral proportional to $d-4$ contributes to $\mathcal{O}(\epsilon)$ order, for a reason similar to the discussion below \cref{eq:IBPbox1}. Now using \cref{eq:bub_general} for the bubble integral in the second line of \cref{eq:I1101ibp}, we get the IR divergence and the finite part of this integral,
\begin{align}\label{eq:I1101p}
    I^{+}_{1,1,0,1}
    %&=\int\frac{\dd^d\lm}{(2\pi)^d} \frac{\hdelta(\ub_2\cdot\lm)}{\lm^2(\lm+q_2)^2(2\ub_1\cdot\lm+\ie)}\nonumber\\
    &=\frac{d-3}{2(y^2-1)}\int_{0}^{\infty}\dd x\frac{1}{x^2(2x \ub_1\cdot q_2-q_2^2-\ie)}\int\frac{\dd^d\lm}{(2\pi)^d}\frac{\hdelta(2\ub_2\cdot\lm)}{(\lm+x \ub_1)^2-x^2+\ie}+\mathcal{O}(\epsilon)\nonumber\\
    &=-\frac{\pi^{5/2-\epsilon}\Gamma(\epsilon+1/2)}{(2\pi)^{4-2\epsilon}[-(y^2-1)-\ie]^{1/2+\epsilon}}\int_{0}^{\infty}\frac{\dd x}{x^{1+2\epsilon}(2x \ub_1\cdot q_2-q_2^2-\ie)}+\mathcal{O}(\epsilon)\nonumber\\
    &=\frac{i}{32 \pi q_2^2\sqrt{y^2-1}}\left[-\frac{1}{\eIR}+\log(-\pi (y^2-1)-\ie)-2\log\frac{\ub_1\cdot q_2-\ie}{-q_2^2}\right]+\mathcal{O}(\epsilon)\nonumber\\
    &=\frac{i}{32 \pi q_2^2\sqrt{y^2-1}}\left[-\frac{1}{\eIR}+\log(-\pi (y^2-1)-\ie)-2\log\frac{\ub_1\cdot q_2}{-q_2^2}\right]+\mathcal{O}(\epsilon)\,.
\end{align}
Here we can omit the $\ie$ prescription in $\log\frac{\ub_1\cdot q_2}{-q_2^2}$ because the argument is positive. To compute $I^{-}_{1,1,0,1}$, we first redefine the loop momentum to flip the sign of $\ie$,
\begin{align}\label{eq:Im1101}
    I^{-}_{1,1,0,1}=\int\frac{\dd^d\lm}{(2\pi)^{d}}\frac{\hdelta(2\ub_2\cdot\lm)}{\lm^2(\lm+q_2)^2(2\ub_1\cdot\lm-\ie)}=-\int\frac{\dd^d\lm}{(2\pi)^d}\frac{\hdelta(2\ub_2\cdot\lm)}{\lm^2(\lm-q_2)^2(2\ub_1\cdot\lm+\ie)}\,.
\end{align}
We can directly obtain the result of this integral by sending $q_2\rightarrow -q_2$ in \cref{eq:I1101p} while preserving the sign of $\ie$, 
\begin{align}
    I^{-}_{1,1,0,1}&=-\frac{i}{32 \pi q_2^2\sqrt{y^2-1}}\left[-\frac{1}{\eIR}+\log(-\pi (y^2-1)-\ie)-2\log\frac{-\ub_1\cdot q_2-\ie}{-q_2^2}\right]+\mathcal{O}(\epsilon)\nonumber\\
    &=\frac{1}{16q_2^2\sqrt{y^2-1}}-I^{+}_{1,1,0,1} +\mathcal{O}(\epsilon)\,.
\end{align}
Therefore, the combination $I^{+}_{1,1,0,1}+I^{-}_{1,1,0,1}$ is free of IR divergences,
\begin{align}
    I^{+}_{1,1,0,1}+I^{-}_{1,1,0,1}=\frac{1}{16q_2^2\sqrt{y^2-1}}\,,
\end{align}
which reproduces \cref{eq:I1101pv}.

\subsection{\texorpdfstring{$I^{\pm}_{1,0,1,1}$}{I(1,0,1,1)}}

We apply the same strategy to compute the last set of box integrals,
\begin{align}
    & I^{\pm}_{1,0,1,1}=\int\frac{\dd^{d}\lm}{(2\pi)^d}\frac{\hdelta(2\ub_2\cdot\lm)}{\lm^2(\lm-q_1)^2(2\ub_1\cdot\lm\pm\ie)}\,.
\end{align}
The only difference here is that the combination $I^{+}_{1,0,1,1}+I^{-}_{1,0,1,1}$ is now complex, with the imaginary part given by further cutting $(\lm-q_1)^2$. Starting with $I^{+}_{1,0,1,1}$, we first apply \cref{eq:x_parameter} to combine $(\lm-q_1)^2$ and $2\ub_1\cdot\lm+\ie$, followed by using an IBP identity to get,
\begin{align}
    I^+_{1,0,1,1}&=\int_{0}^{\infty}\dd x\int\frac{\dd^d\lm}{(2\pi)^d}\frac{\hdelta(2\ub_2\cdot\lm)}{\lm^2[(\lm-q_1+x \ub_1)^2-x^2+\ie]^2}\equiv I_a^+ + I_b^+ \,,
\end{align}
where $I_a^{+}$ and $I_b^{-}$ are defined as
\begin{align}
    I_a^+&=-\frac{d-3}{2q_1^2}\int_{0}^{\infty}\dd x\frac{1}{(y^2-1)x^2-2(\ub_2\cdot q_1)y x+(\ub_2\cdot q_1)^2}\nonumber\\
    &\qquad\qquad \times\int\frac{\dd^d\lm}{(2\pi)^d}\frac{\hdelta(2\ub_2\cdot\lm)}{(\lm-q_1+x \ub_1)^2-x^2+\ie}\,,\nonumber\\
    I_b^+&=-\frac{d-4}{q_1^2}\int_{0}^{\infty}\dd x\int\frac{\dd^d\lm}{(2\pi)^d}\frac{\hdelta(2\ub_2\cdot\lm)}{\lm^2[(\lm-q_1+x\ub_1)^2-x^2+\ie]}\,.
\end{align}
Similar to \cref{eq:Im1101}, we can rewrite $I^{-}_{1,0,1,1}$ as
\begin{align}
    I^{-}_{1,0,1,1}=\int\frac{\dd^{d}\lm}{(2\pi)^d}\frac{\hdelta(2\ub_2\cdot\lm)}{\lm^2(\lm-q_1)^2(2\ub_1\cdot\lm-\ie)}&=-\int\frac{\dd^{d}\lm}{(2\pi)^d}\frac{\hdelta(2\ub_2\cdot\lm)}{\lm^2(\lm+q_1)^2(2\ub_1\cdot\lm+\ie)}\nonumber\\
    &\equiv-(I^-_a+I^-_b)\,,
\end{align}
where $I^{-}_{a,b}=I^{+}_{a,b}\big|_{q_1\rightarrow -q_1}$. Therefore, we have
\begin{align}
    I^{+}_{1,0,1,1}+I^{-}_{1,0,1,1}=(I^+_a-I^-_a)+(I^+_b-I^-_b)\,,
\end{align}
such that we only need to compute the combination $I^{+}_a-I^{-}_a$ and $I^{+}_b-I^{-}_b$.

If we use \cref{eq:bub_general} with $d=4$ to integrate out the bubble in $I_a^\pm$, we will find that the resulting $x$ integral is divergent at $x\rightarrow\infty$. However, by taking the difference $I^+_a-I^-_a$, we get a finite result since the range of $x$ gets truncated. Interestingly, the result is proportional to the triangle integral \eqref{eq:I0011pv},
\begin{align}
    I_a^+-I_a^-=\frac{1}{q_1^2}(I^{+}_{0,0,1,1}+I^{-}_{0,0,1,1})=\frac{1}{16 q_1^2 \sqrt{y^2-1}}+\frac{i}{8\pi q_1^2\sqrt{y^2-1}}\arccosh(y)\,.
\end{align}
To compute $I^+_b-I^-_b$, we introduce another Feynman parameter $w$ and perform the $\lm$ integral in $d=4$. The result is finite due to the same truncation on the range of $x$,
\begin{align}
    I^+_b-I^-_b=-\frac{d-4}{16\pi^2 q_1^2}\int_{-\alpha}^{\alpha}\dd x\int_{0}^{1}\frac{\dd w}{[-w^2(x^2+\beta)(y^2-1)-w(1-w)q_1^2-\ie]^{1/2}}\,,
\end{align}
such that $I^+_b-I^-_b=0$ in $d=4$. Therefore, we have reproduced \cref{eq:I1011re},
\begin{align}\label{eq:I1011pv}
    I^{+}_{1,0,1,1}+I^{-}_{1,0,1,1}=I^+_a-I^-_a=\frac{1}{16 q_1^2 \sqrt{y^2-1}}+\frac{i}{8\pi q_1^2\sqrt{y^2-1}}\arccosh(y)\,.
\end{align}

\subsection{\texorpdfstring{$I^{\pm}_{0,1,1,1}$}{I(0,1,1,1)}}

We first use the same Feynman parameterization as in \cref{eq:I1110step1} to bring together the two massless propagators,
\begin{align}
    I^{+}_{0,1,1,1}&=\int\frac{\dd^d\lm}{(2\pi)^d}\frac{\hdelta(2\ub_2\cdot\lm)}{(\lm+q_2)^2(\lm-q_1)^2(2\ub_1\cdot\lm+\ie)}\nonumber\\
    &=\int_{0}^{1}\dd x\int\frac{\dd^d\lm}{(2\pi)^d}\frac{\hdelta(2\ub_2\cdot\lm)}{(\lm-q_1+xk_5)^4(2\ub_1\cdot\lm+\ie)}\,.
\end{align}
The resultant double propagator $(\lm-q_1+xk_5)^4$ is then reduced by the IBP relation,
\begin{align}\label{eq:I0111_reduce}
    &\int_{0}^{1}\dd x\int\frac{\dd^d\lm}{(2\pi)^d}\frac{\hdelta(2\ub_2\cdot\lm)}{(\lm-q_1+xk_5)^4(2\ub_1\cdot\lm+\ie)}\nonumber\\
    &=\frac{d-3}{4(\ub_2\cdot q_1)^2}\int_{0}^{1}\dd x\frac{x(\ub_1\cdot q_2)+(1-x)y(\ub_2\cdot q_1)}{(1-x)^2\mathcal{F}(x)}\int\frac{\dd^d\lm}{(2\pi)^d}\frac{\hdelta(2\ub_2\cdot\lm)}{(\lm-q_1+xk_5)^2+\ie}\nonumber\\
    &\quad+\frac{d-4}{2}\int_{0}^{1}\dd x\frac{y^2-1}{\mathcal{G}(x)}\int\frac{\dd^d\lm}{(2\pi)^d}\frac{\hdelta(2\ub_2\cdot\lm)}{(\lm-q_1+xk_5)^2(2\ub_1\cdot\lm+\ie)}\,,
\end{align}
where $\mathcal{G}(x)$ is a quadratic polynomial in $x$,
\begin{align}
    \mathcal{G}(x)=x^2(\ub_1\cdot q_2)^2+2x(1-x)y(\ub_1\cdot q_2)(\ub_2\cdot q_1)+(1-x)^2(\ub_2\cdot q_1)^2\,.
\end{align}
Under the principal value combination $I^{+}_{0,1,1,1}+I^{-}_{0,1,1,1}$, the first integral in \cref{eq:I0111_reduce} is doubled while the second integral is finite in $d=4$, which only contributes to $\mathcal{O}(\epsilon)$ due to the prefactor. We can now reproduce \cref{eq:I0111pv} through a direct integration,
\begin{align}
    I^{+}_{0,1,1,1}+I^{-}_{0,1,1,1}&=-\frac{\pi^{5/2-\epsilon}\Gamma(\epsilon+1/2)}{(2\pi)^{4-2\epsilon}[-(\ub_2\cdot q_1)^2-\ie]^{1/2+\epsilon}}\int_{0}^{1}\dd x\frac{x(\ub_1\cdot q_2)+(1-x)y(\ub_2\cdot q_1)}{(1+x)^{1+2\epsilon}\mathcal{F}(x)}\nonumber\\
    &=-\frac{1}{32(\ub_1\cdot q_2)(\ub_2\cdot q_1)}\\
    & \quad -\frac{i}{32\pi (\ub_1\cdot q_2)(\ub_2\cdot q_1)}\left[-\frac{1}{\eIR}+\log(4\pi(\ub_2\cdot q_1)^2)+2\log\frac{\ub_1\cdot q_2}{\ub_2\cdot q_1}\right]\,.\nonumber
\end{align}

\section{Master integrals determining the cut contribution}\label{sec:addmasterint}

In this appendix, we list all the double-cut master integrals,
\begin{align}
    I^{\text{cut,cut}}_{a_1,a_2,a_3}=\int\frac{\dd^d\lm}{(2\pi)^d}\frac{i\,\hdelta(2\ub_1\cdot\lm)\hdelta(2\ub_2\cdot\lm)}{[\lm^{2}]^{a_1}[(\lm+q_2)^{2}]^{a_2}[(\lm-q_1)^{2}]^{a_3}} \ ,
\end{align}
necessary for Sec.~\ref{fullexpress}. Due to the two delta functions, the dimension of the integral is reduced by two and the integrals are easy to evaluate directly using the same computation strategies as appendix \ref{sec:integrals}. Therefore, we simply summarize the results below:
\begin{align}
    \int\frac{\dd^d\lm}{(2\pi)^d} \frac{\hdelta(2\ub_1\cdot\lm)\hdelta(2\ub_2\cdot\lm)}{(\lm-q_1)^2} &= - \frac{1}{4\sqrt{y^2-1}}\frac{\Gamma(\epsilon)}{(4\pi)^{1-\epsilon}}\left(\frac{y^2-1}{(\ub_2\cdot q_1)^2}\right)^{\epsilon} \nonumber\\
    &= -\frac{1}{16\pi\sqrt{y^2-1}}\left(\frac{1}{\epsilon}-\gamma+\log(4\pi)-\log\frac{(\ub_2\cdot q_1)^2}{y^2-1}\right)
\end{align}

\begin{align}
    \int\frac{\dd^d\lm}{(2\pi)^d} \frac{\hdelta(2\ub_1\cdot\lm)\hdelta(2\ub_2\cdot\lm)}{(\lm+q_2)^2} &= - \frac{1}{4\sqrt{y^2-1}}\frac{\Gamma(\epsilon)}{(4\pi)^{1-\epsilon}}\left(\frac{y^2-1}{(\ub_1\cdot q_2)^2}\right)^{\epsilon} \nonumber\\
    &= -\frac{1}{16\pi\sqrt{y^2-1}}\left(\frac{1}{\epsilon}-\gamma+\log(4\pi)-\log\frac{(\ub_1\cdot q_2)^2}{y^2-1}\right)
\end{align}

\begin{align}
    \int\frac{\dd^d\lm}{(2\pi)^d}\frac{\hdelta(2\ub_1\cdot\lm)\hdelta(2\ub_2\cdot\lm)}{(\lm-q_1)^2(\lm+q_2)^2} = \frac{\arccosh(y)}{16\pi (\ub_1\cdot q_2)(\ub_2\cdot q_1)}
\end{align}

\begin{align}
    \int\frac{\dd^d\lm}{(2\pi)^d}\frac{\hdelta(2\ub_1\cdot\lm)\hdelta(2\ub_2\cdot\lm)}{\lm^2(\lm-q_1)^2} &= - \frac{1}{4\sqrt{y^2-1}}\frac{\Gamma(\epsilon)}{(4\pi)^{1-\epsilon}(-q_1^2)^{1+\epsilon}}{}_2F_{1}\Big[\begin{array}{cc} -\epsilon,1+\epsilon \\ 1-\epsilon
    \end{array},1+\frac{(\ub_2\cdot q_1)^2}{q_1^2(y^2-1)}\Big] \nonumber\\
    &= \frac{1}{16\pi q_1^2 \sqrt{y^2-1}}\left[\frac{1}{\epsilon}-\gamma+\log(4\pi) - \log\frac{q_1^4(y^2-1)}{(\ub_2\cdot q_1)^2}\right]
\end{align}

\begin{align}
    \int\frac{\dd^d\lm}{(2\pi)^d}\frac{\hdelta(2\ub_1\cdot\lm)\hdelta(2\ub_2\cdot\lm)}{\lm^2(\lm+q_2)^2} &= - \frac{1}{4\sqrt{y^2-1}}\frac{\Gamma(\epsilon)}{(4\pi)^{1-\epsilon}(-q_2^2)^{1+\epsilon}}{}_2F_{1}\Big[\begin{array}{cc} -\epsilon,1+\epsilon \\ 1-\epsilon
    \end{array},1+\frac{(\ub_1\cdot q_2)^2}{q_2^2(y^2-1)}\Big] \nonumber\\
    &= \frac{1}{16\pi q_2^2 \sqrt{y^2-1}}\left[\frac{1}{\epsilon}-\gamma+\log(4\pi) - \log\frac{q_2^4(y^2-1)}{(\ub_1\cdot q_2)^2}\right]
\end{align}

\noindent
Some of the integrals listed here have also been evaluated in Ref.~\cite{Caron-Huot:2023vxl}.

%%%%%%%%%%%%%%%%%%%%%%%%%%%%
\bibliographystyle{JHEP}
\bibliography{Draft.bib}

\providecommand{\href}[2]{#2}\begingroup\raggedright\begin{thebibliography}{100}

\bibitem{LIGOScientific:2021djp}
{\scshape LIGO Scientific, Virgo, KAGRA} collaboration, \emph{{GWTC-3: Compact
  Binary Coalescences Observed by LIGO and Virgo During the Second Part of the
  Third Observing Run}},  \href{https://arxiv.org/abs/2111.03606}{{\ttfamily
  2111.03606}}.

\bibitem{Buonanno:1998gg}
A.~Buonanno and T.~Damour, \emph{{Effective one-body approach to general
  relativistic two-body dynamics}},
  \href{https://doi.org/10.1103/PhysRevD.59.084006}{\emph{Phys. Rev. D}
  {\bfseries 59} (1999) 084006}
  [\href{https://arxiv.org/abs/gr-qc/9811091}{{\ttfamily gr-qc/9811091}}].

\bibitem{Damour:2000we}
T.~Damour, P.~Jaranowski and G.~Schaefer, \emph{{On the determination of the
  last stable orbit for circular general relativistic binaries at the third
  postNewtonian approximation}},
  \href{https://doi.org/10.1103/PhysRevD.62.084011}{\emph{Phys. Rev. D}
  {\bfseries 62} (2000) 084011}
  [\href{https://arxiv.org/abs/gr-qc/0005034}{{\ttfamily gr-qc/0005034}}].

\bibitem{Buonanno:2000ef}
A.~Buonanno and T.~Damour, \emph{{Transition from inspiral to plunge in binary
  black hole coalescences}},
  \href{https://doi.org/10.1103/PhysRevD.62.064015}{\emph{Phys. Rev. D}
  {\bfseries 62} (2000) 064015}
  [\href{https://arxiv.org/abs/gr-qc/0001013}{{\ttfamily gr-qc/0001013}}].

\bibitem{Damour:2001tu}
T.~Damour, \emph{{Coalescence of two spinning black holes: an effective
  one-body approach}},
  \href{https://doi.org/10.1103/PhysRevD.64.124013}{\emph{Phys. Rev. D}
  {\bfseries 64} (2001) 124013}
  [\href{https://arxiv.org/abs/gr-qc/0103018}{{\ttfamily gr-qc/0103018}}].

\bibitem{Damour:2014sva}
T.~Damour and A.~Nagar, \emph{{New effective-one-body description of coalescing
  nonprecessing spinning black-hole binaries}},
  \href{https://doi.org/10.1103/PhysRevD.90.044018}{\emph{Phys. Rev. D}
  {\bfseries 90} (2014) 044018}
  [\href{https://arxiv.org/abs/1406.6913}{{\ttfamily 1406.6913}}].

\bibitem{Pretorius:2005gq}
F.~Pretorius, \emph{{Evolution of binary black hole spacetimes}},
  \href{https://doi.org/10.1103/PhysRevLett.95.121101}{\emph{Phys. Rev. Lett.}
  {\bfseries 95} (2005) 121101}
  [\href{https://arxiv.org/abs/gr-qc/0507014}{{\ttfamily gr-qc/0507014}}].

\bibitem{Pretorius:2007nq}
F.~Pretorius, \emph{{Binary Black Hole Coalescence}},
  \href{https://arxiv.org/abs/0710.1338}{{\ttfamily 0710.1338}}.

\bibitem{Blanchet:2013haa}
L.~Blanchet, \emph{{Gravitational Radiation from Post-Newtonian Sources and
  Inspiralling Compact Binaries}},
  \href{https://doi.org/10.12942/lrr-2014-2}{\emph{Living Rev. Rel.} {\bfseries
  17} (2014) 2} [\href{https://arxiv.org/abs/1310.1528}{{\ttfamily
  1310.1528}}].

\bibitem{Schafer:2018kuf}
G.~Sch\"afer and P.~Jaranowski, \emph{{Hamiltonian formulation of general
  relativity and post-Newtonian dynamics of compact binaries}},
  \href{https://doi.org/10.1007/s41114-018-0016-5}{\emph{Living Rev. Rel.}
  {\bfseries 21} (2018) 7} [\href{https://arxiv.org/abs/1805.07240}{{\ttfamily
  1805.07240}}].

\bibitem{Goldberger:2004jt}
W.D.~Goldberger and I.Z.~Rothstein, \emph{{An Effective field theory of gravity
  for extended objects}},
  \href{https://doi.org/10.1103/PhysRevD.73.104029}{\emph{Phys. Rev. D}
  {\bfseries 73} (2006) 104029}
  [\href{https://arxiv.org/abs/hep-th/0409156}{{\ttfamily hep-th/0409156}}].

\bibitem{Goldberger:2006bd}
W.D.~Goldberger and I.Z.~Rothstein, \emph{{Towers of Gravitational Theories}},
  \href{https://doi.org/10.1142/S0218271806009698}{\emph{Gen. Rel. Grav.}
  {\bfseries 38} (2006) 1537}
  [\href{https://arxiv.org/abs/hep-th/0605238}{{\ttfamily hep-th/0605238}}].

\bibitem{Goldberger:2009qd}
W.D.~Goldberger and A.~Ross, \emph{{Gravitational radiative corrections from
  effective field theory}},
  \href{https://doi.org/10.1103/PhysRevD.81.124015}{\emph{Phys. Rev. D}
  {\bfseries 81} (2010) 124015}
  [\href{https://arxiv.org/abs/0912.4254}{{\ttfamily 0912.4254}}].

\bibitem{Porto:2012as}
R.A.~Porto, A.~Ross and I.Z.~Rothstein, \emph{{Spin induced multipole moments
  for the gravitational wave amplitude from binary inspirals to 2.5
  Post-Newtonian order}},
  \href{https://doi.org/10.1088/1475-7516/2012/09/028}{\emph{JCAP} {\bfseries
  09} (2012) 028} [\href{https://arxiv.org/abs/1203.2962}{{\ttfamily
  1203.2962}}].

\bibitem{Blanchet:1994ez}
L.~Blanchet and B.S.~Sathyaprakash, \emph{{Detecting the tail effect in
  gravitational wave experiments}},
  \href{https://doi.org/10.1103/PhysRevLett.74.1067}{\emph{Phys. Rev. Lett.}
  {\bfseries 74} (1995) 1067}.

\bibitem{Porto:2016pyg}
R.A.~Porto, \emph{{The effective field theorist\textquoteright{}s approach to
  gravitational dynamics}},
  \href{https://doi.org/10.1016/j.physrep.2016.04.003}{\emph{Phys. Rept.}
  {\bfseries 633} (2016) 1} [\href{https://arxiv.org/abs/1601.04914}{{\ttfamily
  1601.04914}}].

\bibitem{Levi:2018nxp}
M.~Levi, \emph{{Effective Field Theories of Post-Newtonian Gravity: A
  comprehensive review}},
  \href{https://doi.org/10.1088/1361-6633/ab12bc}{\emph{Rept. Prog. Phys.}
  {\bfseries 83} (2020) 075901}
  [\href{https://arxiv.org/abs/1807.01699}{{\ttfamily 1807.01699}}].

\bibitem{Khalil:2022ylj}
M.~Khalil, A.~Buonanno, J.~Steinhoff and J.~Vines, \emph{{Energetics and
  scattering of gravitational two-body systems at fourth post-Minkowskian
  order}}, \href{https://doi.org/10.1103/PhysRevD.106.024042}{\emph{Phys. Rev.
  D} {\bfseries 106} (2022) 024042}
  [\href{https://arxiv.org/abs/2204.05047}{{\ttfamily 2204.05047}}].

\bibitem{Kovacs:1stInSeries}
S.J.~Kovacs and K.S.~Thorne, \emph{{The Generation of Gravitational Waves. 1.
  Weak-field sources}}, \href{https://doi.org/10.1086/153783}{\emph{Astrophys.
  J.} {\bfseries 200} (1975) 245}.

\bibitem{Crowley:1977us}
R.J.~Crowley and K.S.~Thorne, \emph{{The Generation of Gravitational Waves. 2.
  The Postlinear Formalism Revisited}},
  \href{https://doi.org/10.1086/155397}{\emph{Astrophys. J.} {\bfseries 215}
  (1977) 624}.

\bibitem{Kovacs:1977uw}
S.J.~Kovacs and K.S.~Thorne, \emph{{The Generation of Gravitational Waves. 3.
  Derivation of Bremsstrahlung Formulas}},
  \href{https://doi.org/10.1086/155576}{\emph{Astrophys. J.} {\bfseries 217}
  (1977) 252}.

\bibitem{Kovacs:1978eu}
S.J.~Kovacs and K.S.~Thorne, \emph{{The Generation of Gravitational Waves. 4.
  Bremsstrahlung}}, \href{https://doi.org/10.1086/156350}{\emph{Astrophys. J.}
  {\bfseries 224} (1978) 62}.

\bibitem{Amati:1990xe}
D.~Amati, M.~Ciafaloni and G.~Veneziano, \emph{{Higher Order Gravitational
  Deflection and Soft Bremsstrahlung in Planckian Energy Superstring
  Collisions}}, \href{https://doi.org/10.1016/0550-3213(90)90375-N}{\emph{Nucl.
  Phys. B} {\bfseries 347} (1990) 550}.

\bibitem{Ciafaloni:2018uwe}
M.~Ciafaloni, D.~Colferai and G.~Veneziano, \emph{{Infrared features of
  gravitational scattering and radiation in the eikonal approach}},
  \href{https://doi.org/10.1103/PhysRevD.99.066008}{\emph{Phys. Rev. D}
  {\bfseries 99} (2019) 066008}
  [\href{https://arxiv.org/abs/1812.08137}{{\ttfamily 1812.08137}}].

\bibitem{Damour:2020tta}
T.~Damour, \emph{{Radiative contribution to classical gravitational scattering
  at the third order in $G$}},
  \href{https://doi.org/10.1103/PhysRevD.102.124008}{\emph{Phys. Rev. D}
  {\bfseries 102} (2020) 124008}
  [\href{https://arxiv.org/abs/2010.01641}{{\ttfamily 2010.01641}}].

\bibitem{Kalin:2020mvi}
G.~K\"alin and R.A.~Porto, \emph{{Post-Minkowskian Effective Field Theory for
  Conservative Binary Dynamics}},
  \href{https://doi.org/10.1007/JHEP11(2020)106}{\emph{JHEP} {\bfseries 11}
  (2020) 106} [\href{https://arxiv.org/abs/2006.01184}{{\ttfamily
  2006.01184}}].

\bibitem{Kalin:2020fhe}
G.~K\"alin, Z.~Liu and R.A.~Porto, \emph{{Conservative Dynamics of Binary
  Systems to Third Post-Minkowskian Order from the Effective Field Theory
  Approach}}, \href{https://doi.org/10.1103/PhysRevLett.125.261103}{\emph{Phys.
  Rev. Lett.} {\bfseries 125} (2020) 261103}
  [\href{https://arxiv.org/abs/2007.04977}{{\ttfamily 2007.04977}}].

\bibitem{Dlapa:2021npj}
C.~Dlapa, G.~K\"alin, Z.~Liu and R.A.~Porto, \emph{{Dynamics of binary systems
  to fourth Post-Minkowskian order from the effective field theory approach}},
  \href{https://doi.org/10.1016/j.physletb.2022.137203}{\emph{Phys. Lett. B}
  {\bfseries 831} (2022) 137203}
  [\href{https://arxiv.org/abs/2106.08276}{{\ttfamily 2106.08276}}].

\bibitem{Herrmann:2021tct}
E.~Herrmann, J.~Parra-Martinez, M.S.~Ruf and M.~Zeng, \emph{{Radiative
  classical gravitational observables at $ \mathcal{O} $(G$^{3}$) from
  scattering amplitudes}},
  \href{https://doi.org/10.1007/JHEP10(2021)148}{\emph{JHEP} {\bfseries 10}
  (2021) 148} [\href{https://arxiv.org/abs/2104.03957}{{\ttfamily
  2104.03957}}].

\bibitem{Riva:2021vnj}
M.M.~Riva and F.~Vernizzi, \emph{{Radiated momentum in the post-Minkowskian
  worldline approach via reverse unitarity}},
  \href{https://doi.org/10.1007/JHEP11(2021)228}{\emph{JHEP} {\bfseries 11}
  (2021) 228} [\href{https://arxiv.org/abs/2110.10140}{{\ttfamily
  2110.10140}}].

\bibitem{Mougiakakos:2022sic}
S.~Mougiakakos, M.M.~Riva and F.~Vernizzi, \emph{{Gravitational Bremsstrahlung
  with Tidal Effects in the Post-Minkowskian Expansion}},
  \href{https://doi.org/10.1103/PhysRevLett.129.121101}{\emph{Phys. Rev. Lett.}
  {\bfseries 129} (2022) 121101}
  [\href{https://arxiv.org/abs/2204.06556}{{\ttfamily 2204.06556}}].

\bibitem{Riva:2022fru}
M.M.~Riva, F.~Vernizzi and L.K.~Wong, \emph{{Gravitational bremsstrahlung from
  spinning binaries in the post-Minkowskian expansion}},
  \href{https://doi.org/10.1103/PhysRevD.106.044013}{\emph{Phys. Rev. D}
  {\bfseries 106} (2022) 044013}
  [\href{https://arxiv.org/abs/2205.15295}{{\ttfamily 2205.15295}}].

\bibitem{Manohar:2022dea}
A.V.~Manohar, A.K.~Ridgway and C.-H.~Shen, \emph{{Radiated Angular Momentum and
  Dissipative Effects in Classical Scattering}},
  \href{https://doi.org/10.1103/PhysRevLett.129.121601}{\emph{Phys. Rev. Lett.}
  {\bfseries 129} (2022) 121601}
  [\href{https://arxiv.org/abs/2203.04283}{{\ttfamily 2203.04283}}].

\bibitem{Jakobsen:2022psy}
G.U.~Jakobsen, G.~Mogull, J.~Plefka and B.~Sauer, \emph{{All things retarded:
  radiation-reaction in worldline quantum field theory}},
  \href{https://doi.org/10.1007/JHEP10(2022)128}{\emph{JHEP} {\bfseries 10}
  (2022) 128} [\href{https://arxiv.org/abs/2207.00569}{{\ttfamily
  2207.00569}}].

\bibitem{Jakobsen:2021smu}
G.U.~Jakobsen, G.~Mogull, J.~Plefka and J.~Steinhoff, \emph{{Classical
  Gravitational Bremsstrahlung from a Worldline Quantum Field Theory}},
  \href{https://doi.org/10.1103/PhysRevLett.126.201103}{\emph{Phys. Rev. Lett.}
  {\bfseries 126} (2021) 201103}
  [\href{https://arxiv.org/abs/2101.12688}{{\ttfamily 2101.12688}}].

\bibitem{Jakobsen:2021lvp}
G.U.~Jakobsen, G.~Mogull, J.~Plefka and J.~Steinhoff, \emph{{Gravitational
  Bremsstrahlung and Hidden Supersymmetry of Spinning Bodies}},
  \href{https://doi.org/10.1103/PhysRevLett.128.011101}{\emph{Phys. Rev. Lett.}
  {\bfseries 128} (2022) 011101}
  [\href{https://arxiv.org/abs/2106.10256}{{\ttfamily 2106.10256}}].

\bibitem{Kalin:2022hph}
G.~K\"alin, J.~Neef and R.A.~Porto, \emph{{Radiation-reaction in the Effective
  Field Theory approach to Post-Minkowskian dynamics}},
  \href{https://doi.org/10.1007/JHEP01(2023)140}{\emph{JHEP} {\bfseries 01}
  (2023) 140} [\href{https://arxiv.org/abs/2207.00580}{{\ttfamily
  2207.00580}}].

\bibitem{Kalin:2019rwq}
G.~K\"alin and R.A.~Porto, \emph{{From Boundary Data to Bound States}},
  \href{https://doi.org/10.1007/JHEP01(2020)072}{\emph{JHEP} {\bfseries 01}
  (2020) 072} [\href{https://arxiv.org/abs/1910.03008}{{\ttfamily
  1910.03008}}].

\bibitem{Kalin:2019inp}
G.~K\"alin and R.A.~Porto, \emph{{From boundary data to bound states. Part II.
  Scattering angle to dynamical invariants (with twist)}},
  \href{https://doi.org/10.1007/JHEP02(2020)120}{\emph{JHEP} {\bfseries 02}
  (2020) 120} [\href{https://arxiv.org/abs/1911.09130}{{\ttfamily
  1911.09130}}].

\bibitem{Bini:2020hmy}
D.~Bini, T.~Damour and A.~Geralico, \emph{{Sixth post-Newtonian
  nonlocal-in-time dynamics of binary systems}},
  \href{https://doi.org/10.1103/PhysRevD.102.084047}{\emph{Phys. Rev. D}
  {\bfseries 102} (2020) 084047}
  [\href{https://arxiv.org/abs/2007.11239}{{\ttfamily 2007.11239}}].

\bibitem{Cho:2021arx}
G.~Cho, G.~K\"alin and R.A.~Porto, \emph{{From boundary data to bound states.
  Part III. Radiative effects}},
  \href{https://doi.org/10.1007/JHEP04(2022)154}{\emph{JHEP} {\bfseries 04}
  (2022) 154} [\href{https://arxiv.org/abs/2112.03976}{{\ttfamily
  2112.03976}}].

\bibitem{Kosower:2018adc}
D.A.~Kosower, B.~Maybee and D.~O'Connell, \emph{{Amplitudes, Observables, and
  Classical Scattering}},
  \href{https://doi.org/10.1007/JHEP02(2019)137}{\emph{JHEP} {\bfseries 02}
  (2019) 137} [\href{https://arxiv.org/abs/1811.10950}{{\ttfamily
  1811.10950}}].

\bibitem{Cristofoli:2021vyo}
A.~Cristofoli, R.~Gonzo, D.A.~Kosower and D.~O'Connell, \emph{{Waveforms from
  amplitudes}}, \href{https://doi.org/10.1103/PhysRevD.106.056007}{\emph{Phys.
  Rev. D} {\bfseries 106} (2022) 056007}
  [\href{https://arxiv.org/abs/2107.10193}{{\ttfamily 2107.10193}}].

\bibitem{UnitarityMethod}
Z.~Bern, L.J.~Dixon, D.C.~Dunbar and D.A.~Kosower, \emph{{One loop $n$-point
  gauge theory amplitudes, unitarity and collinear limits}},
  \href{https://doi.org/10.1016/0550-3213(94)90179-1}{\emph{Nucl. Phys.}
  {\bfseries B425} (1994) 217}
  [\href{https://arxiv.org/abs/hep-ph/9403226}{{\ttfamily hep-ph/9403226}}].

\bibitem{BernMorgan}
Z.~Bern and A.G.~Morgan, \emph{{Massive loop amplitudes from unitarity}},
  \href{https://doi.org/10.1016/0550-3213(96)00078-8}{\emph{Nucl. Phys.}
  {\bfseries B467} (1996) 479}
  [\href{https://arxiv.org/abs/hep-ph/9511336}{{\ttfamily hep-ph/9511336}}].

\bibitem{Fusing}
Z.~Bern, L.J.~Dixon, D.C.~Dunbar and D.A.~Kosower, \emph{{Fusing gauge theory
  tree amplitudes into loop amplitudes}},
  \href{https://doi.org/10.1016/0550-3213(94)00488-Z}{\emph{Nucl. Phys.}
  {\bfseries B435} (1995) 59}
  [\href{https://arxiv.org/abs/hep-ph/9409265}{{\ttfamily hep-ph/9409265}}].

\bibitem{Bern:1997sc}
Z.~Bern, L.J.~Dixon and D.A.~Kosower, \emph{{One loop amplitudes for e+ e- to
  four partons}},
  \href{https://doi.org/10.1016/S0550-3213(97)00703-7}{\emph{Nucl. Phys. B}
  {\bfseries 513} (1998) 3}
  [\href{https://arxiv.org/abs/hep-ph/9708239}{{\ttfamily hep-ph/9708239}}].

\bibitem{Britto:2004nc}
R.~Britto, F.~Cachazo and B.~Feng, \emph{{Generalized unitarity and one-loop
  amplitudes in N=4 super-Yang-Mills}},
  \href{https://doi.org/10.1016/j.nuclphysb.2005.07.014}{\emph{Nucl. Phys. B}
  {\bfseries 725} (2005) 275}
  [\href{https://arxiv.org/abs/hep-th/0412103}{{\ttfamily hep-th/0412103}}].

\bibitem{Bern:2007ct}
Z.~Bern, J.J.M.~Carrasco, H.~Johansson and D.A.~Kosower, \emph{{Maximally
  supersymmetric planar Yang-Mills amplitudes at five loops}},
  \href{https://doi.org/10.1103/PhysRevD.76.125020}{\emph{Phys. Rev. D}
  {\bfseries 76} (2007) 125020}
  [\href{https://arxiv.org/abs/0705.1864}{{\ttfamily 0705.1864}}].

\bibitem{BCJ}
Z.~Bern, J.J.M.~Carrasco and H.~Johansson, \emph{{New relations for
  gauge-theory amplitudes}},
  \href{https://doi.org/10.1103/PhysRevD.78.085011}{\emph{Phys. Rev.}
  {\bfseries D78} (2008) 085011}
  [\href{https://arxiv.org/abs/0805.3993}{{\ttfamily 0805.3993}}].

\bibitem{BCJLoop}
Z.~Bern, J.J.M.~Carrasco and H.~Johansson, \emph{{Perturbative quantum gravity
  as a double copy of gauge theory}},
  \href{https://doi.org/10.1103/PhysRevLett.105.061602}{\emph{Phys. Rev. Lett.}
  {\bfseries 105} (2010) 061602}
  [\href{https://arxiv.org/abs/1004.0476}{{\ttfamily 1004.0476}}].

\bibitem{KLT}
H.~Kawai, D.C.~Lewellen and S.H.H.~Tye, \emph{{A relation between tree
  amplitudes of closed and open strings}},
  \href{https://doi.org/10.1016/0550-3213(86)90362-7}{\emph{Nucl. Phys.}
  {\bfseries B269} (1986) 1}.

\bibitem{Kotikov:1990kg}
A.V.~Kotikov, \emph{{Differential equations method: New technique for massive
  Feynman diagrams calculation}},
  \href{https://doi.org/10.1016/0370-2693(91)90413-K}{\emph{Phys. Lett. B}
  {\bfseries 254} (1991) 158}.

\bibitem{Bern:1993kr}
Z.~Bern, L.J.~Dixon and D.A.~Kosower, \emph{{Dimensionally regulated pentagon
  integrals}}, \href{https://doi.org/10.1016/0550-3213(94)90398-0}{\emph{Nucl.
  Phys. B} {\bfseries 412} (1994) 751}
  [\href{https://arxiv.org/abs/hep-ph/9306240}{{\ttfamily hep-ph/9306240}}].

\bibitem{Remiddi:1997ny}
E.~Remiddi, \emph{{Differential equations for Feynman graph amplitudes}},
  \href{https://doi.org/10.1007/BF03185566}{\emph{Nuovo Cim. A} {\bfseries 110}
  (1997) 1435} [\href{https://arxiv.org/abs/hep-th/9711188}{{\ttfamily
  hep-th/9711188}}].

\bibitem{Gehrmann:1999as}
T.~Gehrmann and E.~Remiddi, \emph{{Differential equations for two loop four
  point functions}},
  \href{https://doi.org/10.1016/S0550-3213(00)00223-6}{\emph{Nucl. Phys. B}
  {\bfseries 580} (2000) 485}
  [\href{https://arxiv.org/abs/hep-ph/9912329}{{\ttfamily hep-ph/9912329}}].

\bibitem{Chetyrkin:1981qh}
K.G.~Chetyrkin and F.V.~Tkachov, \emph{{Integration by Parts: The Algorithm to
  Calculate beta Functions in 4 Loops}},
  \href{https://doi.org/10.1016/0550-3213(81)90199-1}{\emph{Nucl. Phys. B}
  {\bfseries 192} (1981) 159}.

\bibitem{Smirnov:2008iw}
A.V.~Smirnov, \emph{{Algorithm FIRE -- Feynman Integral REduction}},
  \href{https://doi.org/10.1088/1126-6708/2008/10/107}{\emph{JHEP} {\bfseries
  10} (2008) 107} [\href{https://arxiv.org/abs/0807.3243}{{\ttfamily
  0807.3243}}].

\bibitem{Smirnov:2019qkx}
A.V.~Smirnov and F.S.~Chuharev, \emph{{FIRE6: Feynman Integral REduction with
  Modular Arithmetic}},
  \href{https://doi.org/10.1016/j.cpc.2019.106877}{\emph{Comput. Phys. Commun.}
  {\bfseries 247} (2020) 106877}
  [\href{https://arxiv.org/abs/1901.07808}{{\ttfamily 1901.07808}}].

\bibitem{Caron-Huot:2023vxl}
S.~Caron-Huot, M.~Giroux, H.S.~Hannesdottir and S.~Mizera, \emph{{What can be
  measured asymptotically?}},
  \href{https://arxiv.org/abs/2308.02125}{{\ttfamily 2308.02125}}.

\bibitem{Bern:2021dqo}
Z.~Bern, J.~Parra-Martinez, R.~Roiban, M.S.~Ruf, C.-H.~Shen, M.P.~Solon et~al.,
  \emph{{Scattering Amplitudes and Conservative Binary Dynamics at ${\cal
  O}(G^4)$}}, \href{https://doi.org/10.1103/PhysRevLett.126.171601}{\emph{Phys.
  Rev. Lett.} {\bfseries 126} (2021) 171601}
  [\href{https://arxiv.org/abs/2101.07254}{{\ttfamily 2101.07254}}].

\bibitem{Damgaard:2021ipf}
P.H.~Damgaard, L.~Plante and P.~Vanhove, \emph{{On an exponential
  representation of the gravitational S-matrix}},
  \href{https://doi.org/10.1007/JHEP11(2021)213}{\emph{JHEP} {\bfseries 11}
  (2021) 213} [\href{https://arxiv.org/abs/2107.12891}{{\ttfamily
  2107.12891}}].

\bibitem{Brandhuber:2021kpo}
A.~Brandhuber, G.~Chen, G.~Travaglini and C.~Wen, \emph{{A new gauge-invariant
  double copy for heavy-mass effective theory}},
  \href{https://doi.org/10.1007/JHEP07(2021)047}{\emph{JHEP} {\bfseries 07}
  (2021) 047} [\href{https://arxiv.org/abs/2104.11206}{{\ttfamily
  2104.11206}}].

\bibitem{Brandhuber:2021eyq}
A.~Brandhuber, G.~Chen, G.~Travaglini and C.~Wen, \emph{{Classical
  gravitational scattering from a gauge-invariant double copy}},
  \href{https://doi.org/10.1007/JHEP10(2021)118}{\emph{JHEP} {\bfseries 10}
  (2021) 118} [\href{https://arxiv.org/abs/2108.04216}{{\ttfamily
  2108.04216}}].

\bibitem{Arkani-Hamed:2008owk}
N.~Arkani-Hamed, F.~Cachazo and J.~Kaplan, \emph{{What is the Simplest Quantum
  Field Theory?}}, \href{https://doi.org/10.1007/JHEP09(2010)016}{\emph{JHEP}
  {\bfseries 09} (2010) 016} [\href{https://arxiv.org/abs/0808.1446}{{\ttfamily
  0808.1446}}].

\bibitem{Caron-Huot:2018ape}
S.~Caron-Huot and Z.~Zahraee, \emph{{Integrability of Black Hole Orbits in
  Maximal Supergravity}},
  \href{https://doi.org/10.1007/JHEP07(2019)179}{\emph{JHEP} {\bfseries 07}
  (2019) 179} [\href{https://arxiv.org/abs/1810.04694}{{\ttfamily
  1810.04694}}].

\bibitem{Mafra:2014gja}
C.R.~Mafra and O.~Schlotterer, \emph{{Towards one-loop SYM amplitudes from the
  pure spinor BRST cohomology}},
  \href{https://doi.org/10.1002/prop.201400076}{\emph{Fortsch. Phys.}
  {\bfseries 63} (2015) 105} [\href{https://arxiv.org/abs/1410.0668}{{\ttfamily
  1410.0668}}].

\bibitem{He:2017spx}
S.~He, O.~Schlotterer and Y.~Zhang, \emph{{New BCJ representations for one-loop
  amplitudes in gauge theories and gravity}},
  \href{https://doi.org/10.1016/j.nuclphysb.2018.03.003}{\emph{Nucl. Phys. B}
  {\bfseries 930} (2018) 328}
  [\href{https://arxiv.org/abs/1706.00640}{{\ttfamily 1706.00640}}].

\bibitem{Weinberg:1965nx}
S.~Weinberg, \emph{{Infrared photons and gravitons}},
  \href{https://doi.org/10.1103/PhysRev.140.B516}{\emph{Phys. Rev.} {\bfseries
  140} (1965) B516}.

\bibitem{Brandhuber:2023hhy}
A.~Brandhuber, G.R.~Brown, G.~Chen, S.~De~Angelis, J.~Gowdy and G.~Travaglini,
  \emph{{One-loop Gravitational Bremsstrahlung and Waveforms from a Heavy-Mass
  Effective Field Theory}},  \href{https://arxiv.org/abs/2303.06111}{{\ttfamily
  2303.06111}}.

\bibitem{Elkhidir:2023dco}
A.~Elkhidir, D.~O'Connell, M.~Sergola and I.A.~Vazquez-Holm, \emph{{Radiation
  and Reaction at One Loop}},
  \href{https://arxiv.org/abs/2303.06211}{{\ttfamily 2303.06211}}.

\bibitem{Cheung:2018wkq}
C.~Cheung, I.Z.~Rothstein and M.P.~Solon, \emph{{From Scattering Amplitudes to
  Classical Potentials in the Post-Minkowskian Expansion}},
  \href{https://doi.org/10.1103/PhysRevLett.121.251101}{\emph{Phys. Rev. Lett.}
  {\bfseries 121} (2018) 251101}
  [\href{https://arxiv.org/abs/1808.02489}{{\ttfamily 1808.02489}}].

\bibitem{Bern:2019nnu}
Z.~Bern, C.~Cheung, R.~Roiban, C.-H.~Shen, M.P.~Solon and M.~Zeng,
  \emph{{Scattering Amplitudes and the Conservative Hamiltonian for Binary
  Systems at Third Post-Minkowskian Order}},
  \href{https://doi.org/10.1103/PhysRevLett.122.201603}{\emph{Phys. Rev. Lett.}
  {\bfseries 122} (2019) 201603}
  [\href{https://arxiv.org/abs/1901.04424}{{\ttfamily 1901.04424}}].

\bibitem{Bern:2019crd}
Z.~Bern, C.~Cheung, R.~Roiban, C.-H.~Shen, M.P.~Solon and M.~Zeng, \emph{{Black
  Hole Binary Dynamics from the Double Copy and Effective Theory}},
  \href{https://doi.org/10.1007/JHEP10(2019)206}{\emph{JHEP} {\bfseries 10}
  (2019) 206} [\href{https://arxiv.org/abs/1908.01493}{{\ttfamily
  1908.01493}}].

\bibitem{Luna:2017dtq}
A.~Luna, I.~Nicholson, D.~O'Connell and C.D.~White, \emph{{Inelastic Black Hole
  Scattering from Charged Scalar Amplitudes}},
  \href{https://doi.org/10.1007/JHEP03(2018)044}{\emph{JHEP} {\bfseries 03}
  (2018) 044} [\href{https://arxiv.org/abs/1711.03901}{{\ttfamily
  1711.03901}}].

\bibitem{Beneke:1997zp}
M.~Beneke and V.A.~Smirnov, \emph{{Asymptotic expansion of Feynman integrals
  near threshold}},
  \href{https://doi.org/10.1016/S0550-3213(98)00138-2}{\emph{Nucl. Phys. B}
  {\bfseries 522} (1998) 321}
  [\href{https://arxiv.org/abs/hep-ph/9711391}{{\ttfamily hep-ph/9711391}}].

\bibitem{Georgi:1990um}
H.~Georgi, \emph{{An Effective Field Theory for Heavy Quarks at Low-energies}},
  \href{https://doi.org/10.1016/0370-2693(90)91128-X}{\emph{Phys. Lett. B}
  {\bfseries 240} (1990) 447}.

\bibitem{Luke:1992cs}
M.E.~Luke and A.V.~Manohar, \emph{{Reparametrization invariance constraints on
  heavy particle effective field theories}},
  \href{https://doi.org/10.1016/0370-2693(92)91786-9}{\emph{Phys. Lett. B}
  {\bfseries 286} (1992) 348}
  [\href{https://arxiv.org/abs/hep-ph/9205228}{{\ttfamily hep-ph/9205228}}].

\bibitem{Neubert:1993mb}
M.~Neubert, \emph{{Heavy quark symmetry}},
  \href{https://doi.org/10.1016/0370-1573(94)90091-4}{\emph{Phys. Rept.}
  {\bfseries 245} (1994) 259}
  [\href{https://arxiv.org/abs/hep-ph/9306320}{{\ttfamily hep-ph/9306320}}].

\bibitem{Manohar:2000dt}
A.V.~Manohar and M.B.~Wise, \emph{{Heavy quark physics}}, vol.~10 (2000).

\bibitem{Damgaard:2019lfh}
P.H.~Damgaard, K.~Haddad and A.~Helset, \emph{{Heavy Black Hole Effective
  Theory}}, \href{https://doi.org/10.1007/JHEP11(2019)070}{\emph{JHEP}
  {\bfseries 11} (2019) 070}
  [\href{https://arxiv.org/abs/1908.10308}{{\ttfamily 1908.10308}}].

\bibitem{Aoude:2020onz}
R.~Aoude, K.~Haddad and A.~Helset, \emph{{On-shell heavy particle effective
  theories}}, \href{https://doi.org/10.1007/JHEP05(2020)051}{\emph{JHEP}
  {\bfseries 05} (2020) 051}
  [\href{https://arxiv.org/abs/2001.09164}{{\ttfamily 2001.09164}}].

\bibitem{Haddad:2020tvs}
K.~Haddad and A.~Helset, \emph{{The double copy for heavy particles}},
  \href{https://doi.org/10.1103/PhysRevLett.125.181603}{\emph{Phys. Rev. Lett.}
  {\bfseries 125} (2020) 181603}
  [\href{https://arxiv.org/abs/2005.13897}{{\ttfamily 2005.13897}}].

\bibitem{Akhoury:2013yua}
R.~Akhoury, R.~Saotome and G.~Sterman, \emph{{High Energy Scattering in
  Perturbative Quantum Gravity at Next to Leading Power}},
  \href{https://doi.org/10.1103/PhysRevD.103.064036}{\emph{Phys. Rev. D}
  {\bfseries 103} (2021) 064036}
  [\href{https://arxiv.org/abs/1308.5204}{{\ttfamily 1308.5204}}].

\bibitem{Kosower:2022yvp}
D.A.~Kosower, R.~Monteiro and D.~O'Connell, \emph{{The SAGEX Review on
  Scattering Amplitudes, Chapter 14: Classical Gravity from Scattering
  Amplitudes}},  \href{https://arxiv.org/abs/2203.13025}{{\ttfamily
  2203.13025}}.

\bibitem{Newman:1961qr}
E.~Newman and R.~Penrose, \emph{{An Approach to gravitational radiation by a
  method of spin coefficients}},
  \href{https://doi.org/10.1063/1.1724257}{\emph{J. Math. Phys.} {\bfseries 3}
  (1962) 566}.

\bibitem{Herrmann:2021lqe}
E.~Herrmann, J.~Parra-Martinez, M.S.~Ruf and M.~Zeng, \emph{{Gravitational
  Bremsstrahlung from Reverse Unitarity}},
  \href{https://doi.org/10.1103/PhysRevLett.126.201602}{\emph{Phys. Rev. Lett.}
  {\bfseries 126} (2021) 201602}
  [\href{https://arxiv.org/abs/2101.07255}{{\ttfamily 2101.07255}}].

\bibitem{Anastasiou:2002yz}
C.~Anastasiou and K.~Melnikov, \emph{{Higgs boson production at hadron
  colliders in NNLO QCD}},
  \href{https://doi.org/10.1016/S0550-3213(02)00837-4}{\emph{Nucl. Phys. B}
  {\bfseries 646} (2002) 220}
  [\href{https://arxiv.org/abs/hep-ph/0207004}{{\ttfamily hep-ph/0207004}}].

\bibitem{Anastasiou:2002qz}
C.~Anastasiou, L.J.~Dixon and K.~Melnikov, \emph{{NLO Higgs boson rapidity
  distributions at hadron colliders}},
  \href{https://doi.org/10.1016/S0920-5632(03)80168-8}{\emph{Nucl. Phys. B
  Proc. Suppl.} {\bfseries 116} (2003) 193}
  [\href{https://arxiv.org/abs/hep-ph/0211141}{{\ttfamily hep-ph/0211141}}].

\bibitem{Anastasiou:2003yy}
C.~Anastasiou, L.J.~Dixon, K.~Melnikov and F.~Petriello, \emph{{Dilepton
  rapidity distribution in the Drell-Yan process at NNLO in QCD}},
  \href{https://doi.org/10.1103/PhysRevLett.91.182002}{\emph{Phys. Rev. Lett.}
  {\bfseries 91} (2003) 182002}
  [\href{https://arxiv.org/abs/hep-ph/0306192}{{\ttfamily hep-ph/0306192}}].

\bibitem{Kinoshita:1962ur}
T.~Kinoshita, \emph{{Mass singularities of Feynman amplitudes}},
  \href{https://doi.org/10.1063/1.1724268}{\emph{J. Math. Phys.} {\bfseries 3}
  (1962) 650}.

\bibitem{Lee:1964is}
T.D.~Lee and M.~Nauenberg, \emph{{Degenerate Systems and Mass Singularities}},
  \href{https://doi.org/10.1103/PhysRev.133.B1549}{\emph{Phys. Rev.} {\bfseries
  133} (1964) B1549}.

\bibitem{Dirac:1955uv}
P.A.M.~Dirac, \emph{{Gauge invariant formulation of quantum electrodynamics}},
  \href{https://doi.org/10.1139/p55-081}{\emph{Can. J. Phys.} {\bfseries 33}
  (1955) 650}.

\bibitem{Kulish:1970ut}
P.P.~Kulish and L.D.~Faddeev, \emph{{Asymptotic conditions and infrared
  divergences in quantum electrodynamics}},
  \href{https://doi.org/10.1007/BF01066485}{\emph{Theor. Math. Phys.}
  {\bfseries 4} (1970) 745}.

\bibitem{Grammer:1973db}
G.~Grammer, Jr. and D.R.~Yennie, \emph{{Improved treatment for the infrared
  divergence problem in quantum electrodynamics}},
  \href{https://doi.org/10.1103/PhysRevD.8.4332}{\emph{Phys. Rev. D} {\bfseries
  8} (1973) 4332}.

\bibitem{Ware:2013zja}
J.~Ware, R.~Saotome and R.~Akhoury, \emph{{Construction of an asymptotic S
  matrix for perturbative quantum gravity}},
  \href{https://doi.org/10.1007/JHEP10(2013)159}{\emph{JHEP} {\bfseries 10}
  (2013) 159} [\href{https://arxiv.org/abs/1308.6285}{{\ttfamily 1308.6285}}].

\bibitem{Chung:1965zza}
V.~Chung, \emph{{Infrared Divergence in Quantum Electrodynamics}},
  \href{https://doi.org/10.1103/PhysRev.140.B1110}{\emph{Phys. Rev.} {\bfseries
  140} (1965) B1110}.

\bibitem{Cristofoli:2021jas}
A.~Cristofoli, R.~Gonzo, N.~Moynihan, D.~O'Connell, A.~Ross, M.~Sergola et~al.,
  \emph{{The Uncertainty Principle and Classical Amplitudes}},
  \href{https://arxiv.org/abs/2112.07556}{{\ttfamily 2112.07556}}.

\bibitem{Britto:2021pud}
R.~Britto, R.~Gonzo and G.R.~Jehu, \emph{{Graviton particle statistics and
  coherent states from classical scattering amplitudes}},
  \href{https://doi.org/10.1007/JHEP03(2022)214}{\emph{JHEP} {\bfseries 03}
  (2022) 214} [\href{https://arxiv.org/abs/2112.07036}{{\ttfamily
  2112.07036}}].

\bibitem{Bern:2021yeh}
Z.~Bern, J.~Parra-Martinez, R.~Roiban, M.S.~Ruf, C.-H.~Shen, M.P.~Solon et~al.,
  \emph{{Scattering Amplitudes, the Tail Effect, and Conservative Binary
  Dynamics at O(G4)}},
  \href{https://doi.org/10.1103/PhysRevLett.128.161103}{\emph{Phys. Rev. Lett.}
  {\bfseries 128} (2022) 161103}
  [\href{https://arxiv.org/abs/2112.10750}{{\ttfamily 2112.10750}}].

\bibitem{Blanchet:1992br}
L.~Blanchet and T.~Damour, \emph{{Hereditary effects in gravitational
  radiation}}, \href{https://doi.org/10.1103/PhysRevD.46.4304}{\emph{Phys. Rev.
  D} {\bfseries 46} (1992) 4304}.

\bibitem{Blanchet:1993ec}
L.~Blanchet and G.~Schaefer, \emph{{Gravitational wave tails and binary star
  systems}}, \href{https://doi.org/10.1088/0264-9381/10/12/026}{\emph{Class.
  Quant. Grav.} {\bfseries 10} (1993) 2699}.

\bibitem{Carrasco:2021bmu}
J.J.M.~Carrasco and I.A.~Vazquez-Holm, \emph{{Extracting Einstein from the
  loop-level double-copy}},
  \href{https://doi.org/10.1007/JHEP11(2021)088}{\emph{JHEP} {\bfseries 11}
  (2021) 088} [\href{https://arxiv.org/abs/2108.06798}{{\ttfamily
  2108.06798}}].

\bibitem{Kosmopoulos:2020pcd}
D.~Kosmopoulos, \emph{{Simplifying D-dimensional physical-state sums in gauge
  theory and gravity}},
  \href{https://doi.org/10.1103/PhysRevD.105.056025}{\emph{Phys. Rev. D}
  {\bfseries 105} (2022) 056025}
  [\href{https://arxiv.org/abs/2009.00141}{{\ttfamily 2009.00141}}].

\bibitem{KoemansCollado:2019ggb}
A.~Koemans~Collado, P.~Di~Vecchia and R.~Russo, \emph{{Revisiting the second
  post-Minkowskian eikonal and the dynamics of binary black holes}},
  \href{https://doi.org/10.1103/PhysRevD.100.066028}{\emph{Phys. Rev. D}
  {\bfseries 100} (2019) 066028}
  [\href{https://arxiv.org/abs/1904.02667}{{\ttfamily 1904.02667}}].

\bibitem{Bern:2004cz}
Z.~Bern, L.J.~Dixon and D.A.~Kosower, \emph{{Two-loop g ---\ensuremath{>} gg
  splitting amplitudes in QCD}},
  \href{https://doi.org/10.1088/1126-6708/2004/08/012}{\emph{JHEP} {\bfseries
  08} (2004) 012} [\href{https://arxiv.org/abs/hep-ph/0404293}{{\ttfamily
  hep-ph/0404293}}].

\bibitem{Parra-Martinez:2020dzs}
J.~Parra-Martinez, M.S.~Ruf and M.~Zeng, \emph{{Extremal black hole scattering
  at $\mathcal{O}(G^3)$: graviton dominance, eikonal exponentiation, and
  differential equations}},
  \href{https://doi.org/10.1007/JHEP11(2020)023}{\emph{JHEP} {\bfseries 11}
  (2020) 023} [\href{https://arxiv.org/abs/2005.04236}{{\ttfamily
  2005.04236}}].

\bibitem{Passarino:1978jh}
G.~Passarino and M.J.G.~Veltman, \emph{{One Loop Corrections for e+ e-
  Annihilation Into mu+ mu- in the Weinberg Model}},
  \href{https://doi.org/10.1016/0550-3213(79)90234-7}{\emph{Nucl. Phys. B}
  {\bfseries 160} (1979) 151}.

\bibitem{Bjerrum-Bohr:2006xbk}
N.E.J.~Bjerrum-Bohr, D.C.~Dunbar, H.~Ita, W.B.~Perkins and K.~Risager,
  \emph{{The No-Triangle Hypothesis for N=8 Supergravity}},
  \href{https://doi.org/10.1088/1126-6708/2006/12/072}{\emph{JHEP} {\bfseries
  12} (2006) 072} [\href{https://arxiv.org/abs/hep-th/0610043}{{\ttfamily
  hep-th/0610043}}].

\bibitem{vanNeerven:1983vr}
W.L.~van Neerven and J.A.M.~Vermaseren, \emph{{LARGE LOOP INTEGRALS}},
  \href{https://doi.org/10.1016/0370-2693(84)90237-5}{\emph{Phys. Lett. B}
  {\bfseries 137} (1984) 241}.

\bibitem{Bern:1992em}
Z.~Bern, L.J.~Dixon and D.A.~Kosower, \emph{{Dimensionally regulated one loop
  integrals}}, \href{https://doi.org/10.1016/0370-2693(93)90400-C}{\emph{Phys.
  Lett. B} {\bfseries 302} (1993) 299}
  [\href{https://arxiv.org/abs/hep-ph/9212308}{{\ttfamily hep-ph/9212308}}].

\bibitem{Damour:2016gwp}
T.~Damour, \emph{{Gravitational scattering, post-Minkowskian approximation and
  Effective One-Body theory}},
  \href{https://doi.org/10.1103/PhysRevD.94.104015}{\emph{Phys. Rev. D}
  {\bfseries 94} (2016) 104015}
  [\href{https://arxiv.org/abs/1609.00354}{{\ttfamily 1609.00354}}].

\bibitem{Damour:2017zjx}
T.~Damour, \emph{{High-energy gravitational scattering and the general
  relativistic two-body problem}},
  \href{https://doi.org/10.1103/PhysRevD.97.044038}{\emph{Phys. Rev. D}
  {\bfseries 97} (2018) 044038}
  [\href{https://arxiv.org/abs/1710.10599}{{\ttfamily 1710.10599}}].

\bibitem{Sahoo:2021ctw}
B.~Sahoo and A.~Sen, \emph{{Classical soft graviton theorem rewritten}},
  \href{https://doi.org/10.1007/JHEP01(2022)077}{\emph{JHEP} {\bfseries 01}
  (2022) 077} [\href{https://arxiv.org/abs/2105.08739}{{\ttfamily
  2105.08739}}].

\bibitem{Strominger:2014pwa}
A.~Strominger and A.~Zhiboedov, \emph{{Gravitational Memory, BMS
  Supertranslations and Soft Theorems}},
  \href{https://doi.org/10.1007/JHEP01(2016)086}{\emph{JHEP} {\bfseries 01}
  (2016) 086} [\href{https://arxiv.org/abs/1411.5745}{{\ttfamily 1411.5745}}].

\bibitem{Bonga:2018gzr}
B.~Bonga and E.~Poisson, \emph{{Coulombic contribution to angular momentum flux
  in general relativity}},
  \href{https://doi.org/10.1103/PhysRevD.99.064024}{\emph{Phys. Rev. D}
  {\bfseries 99} (2019) 064024}
  [\href{https://arxiv.org/abs/1808.01288}{{\ttfamily 1808.01288}}].

\bibitem{Thorne:1980ru}
K.S.~Thorne, \emph{{Multipole Expansions of Gravitational Radiation}},
  \href{https://doi.org/10.1103/RevModPhys.52.299}{\emph{Rev. Mod. Phys.}
  {\bfseries 52} (1980) 299}.

\bibitem{DeWitt:2011nnj}
B.~DeWitt, \emph{{Bryce DeWitt's Lectures on Gravitation}}, vol.~826, Springer
  (2011),
  \href{https://doi.org/10.1007/978-3-540-36911-0}{10.1007/978-3-540-36911-0}.

\bibitem{Peters:1964zz}
P.C.~Peters, \emph{{Gravitational Radiation and the Motion of Two Point
  Masses}}, \href{https://doi.org/10.1103/PhysRev.136.B1224}{\emph{Phys. Rev.}
  {\bfseries 136} (1964) B1224}.

\bibitem{Bern:2009kd}
Z.~Bern, J.J.~Carrasco, L.J.~Dixon, H.~Johansson and R.~Roiban, \emph{{The
  Ultraviolet Behavior of N=8 Supergravity at Four Loops}},
  \href{https://doi.org/10.1103/PhysRevLett.103.081301}{\emph{Phys. Rev. Lett.}
  {\bfseries 103} (2009) 081301}
  [\href{https://arxiv.org/abs/0905.2326}{{\ttfamily 0905.2326}}].

\bibitem{Carrasco:2011mn}
J.J.M.~Carrasco and H.~Johansson, \emph{{Five-Point Amplitudes in N=4
  Super-Yang-Mills Theory and N=8 Supergravity}},
  \href{https://doi.org/10.1103/PhysRevD.85.025006}{\emph{Phys. Rev. D}
  {\bfseries 85} (2012) 025006}
  [\href{https://arxiv.org/abs/1106.4711}{{\ttfamily 1106.4711}}].

\bibitem{Bern:2013uka}
Z.~Bern, S.~Davies, T.~Dennen, A.V.~Smirnov and V.A.~Smirnov,
  \emph{{Ultraviolet Properties of N=4 Supergravity at Four Loops}},
  \href{https://doi.org/10.1103/PhysRevLett.111.231302}{\emph{Phys. Rev. Lett.}
  {\bfseries 111} (2013) 231302}
  [\href{https://arxiv.org/abs/1309.2498}{{\ttfamily 1309.2498}}].

\bibitem{Bern:2014sna}
Z.~Bern, S.~Davies and T.~Dennen, \emph{{Enhanced ultraviolet cancellations in
  $\mathcal N=5$ supergravity at four loops}},
  \href{https://doi.org/10.1103/PhysRevD.90.105011}{\emph{Phys. Rev. D}
  {\bfseries 90} (2014) 105011}
  [\href{https://arxiv.org/abs/1409.3089}{{\ttfamily 1409.3089}}].

\bibitem{Bern:2017ucb}
Z.~Bern, J.J.M.~Carrasco, W.-M.~Chen, H.~Johansson, R.~Roiban and M.~Zeng,
  \emph{{Five-loop four-point integrand of $N=8$ supergravity as a generalized
  double copy}}, \href{https://doi.org/10.1103/PhysRevD.96.126012}{\emph{Phys.
  Rev. D} {\bfseries 96} (2017) 126012}
  [\href{https://arxiv.org/abs/1708.06807}{{\ttfamily 1708.06807}}].

\bibitem{Bern:2018jmv}
Z.~Bern, J.J.~Carrasco, W.-M.~Chen, A.~Edison, H.~Johansson, J.~Parra-Martinez
  et~al., \emph{{Ultraviolet Properties of $\mathcal N = 8$ Supergravity at
  Five Loops}}, \href{https://doi.org/10.1103/PhysRevD.98.086021}{\emph{Phys.
  Rev. D} {\bfseries 98} (2018) 086021}
  [\href{https://arxiv.org/abs/1804.09311}{{\ttfamily 1804.09311}}].

\bibitem{Abreu:2019rpt}
S.~Abreu, L.J.~Dixon, E.~Herrmann, B.~Page and M.~Zeng, \emph{{The two-loop
  five-point amplitude in $ \mathcal{N} $ = 8 supergravity}},
  \href{https://doi.org/10.1007/JHEP03(2019)123}{\emph{JHEP} {\bfseries 03}
  (2019) 123} [\href{https://arxiv.org/abs/1901.08563}{{\ttfamily
  1901.08563}}].

\end{thebibliography}\endgroup
%%%%%%%%%%%%%%%%%%%%%%%%%%%%
%%%%%%%%%%%%%%%%%%%%%%%%%%%%
%%%%%%%%%%%%%%%%%%%%%%%%%%%%
%%%%%%%%%%%%%%%%%%%%%%%%%%%%
%%%%%%%%%%%%%%%%%%%%%%%%%%%%

\end{document}